\documentclass[letterpaper,11pt]{article}
\pdfoutput=1

\usepackage{jheppub}
\usepackage{mathtools,amssymb,braket,mathrsfs,tikz,subfig,xspace,xcolor,float,color,siunitx,comment,afterpage,multirow,array,hhline,amsmath,amsthm,framed,colortbl, bm}
\usepackage[countmax]{subfloat}

\newcommand{\be}{\begin{equation}}
\newcommand{\ee}{\end{equation}}

\newcommand{\E}{\mathcal{E}}

\newcommand{\M}{\mathcal{M}}
\newcommand{\X}{\mathcal{X}}
\newcommand{\EMD}{\text{EMD}}

\renewcommand{\O}{\mathcal{O}}

\usepackage{array}

\usepackage{comment}

\usepackage{url}
\usepackage{xspace}
\usepackage{dsfont}
\usepackage{amssymb}
\usepackage{amsmath}
\usepackage{graphicx}
\usepackage{hhline}
\usepackage{physics}

\usepackage{color}
\definecolor{mit-red}{rgb}{0.64,.12,0.2}
\definecolor{darkred}{rgb}{1.0,0.1,0.1}
\definecolor{darkgreen}{rgb}{0.1,0.7,0.1}
\definecolor{darkblue}{rgb}{0.1,0.1,1.0}

\usepackage{amsmath}

\DeclareMathOperator*{\argmin}{argmin}
\DeclareMathOperator*{\paramin}{(arg)min}

\DeclareRobustCommand{\Sec}[1]{Sec.~\ref{sec:#1}}
\DeclareRobustCommand{\Secs}[2]{Secs.~\ref{sec:#1} and \ref{sec:#2}}
\DeclareRobustCommand{\App}[1]{App.~\ref{app:#1}}
\DeclareRobustCommand{\Apps}[2]{Apps.~\ref{app:#1} and \ref{app:#2}}
\DeclareRobustCommand{\Tab}[1]{Table~\ref{tab:#1}}
\DeclareRobustCommand{\Tabs}[2]{Tables~\ref{tab:#1} and \ref{tab:#2}}
\DeclareRobustCommand{\Fig}[1]{Fig.~\ref{fig:#1}}
\DeclareRobustCommand{\Figs}[2]{Figs.~\ref{fig:#1} and \ref{fig:#2}}
\DeclareRobustCommand{\Eq}[1]{Eq.~(\ref{eq:#1})}
\DeclareRobustCommand{\Eqs}[2]{Eqs.~(\ref{eq:#1}) and (\ref{eq:#2})}
\DeclareRobustCommand{\Refer}[1]{Ref.~\cite{#1}}
\DeclareRobustCommand{\Refers}[1]{Refs.~\cite{#1}}
\DeclareRobustCommand{\Def}[1]{Def.~\ref{def:#1}}

\newtheorem{definition}{Definition}

\def\GeV{\text{GeV}}

\newcommand{\FastJet}{{\sc FastJet}\xspace}

\newcommand{\Adam}{{\sc Adam}\xspace}

\newcommand{\Shaper}{\textsc{Shaper}\xspace}

\begin{document}

% Magic to fix footnote behavior.
\count\footins = 1000
\interfootnotelinepenalty=10000
\setlength{\footnotesep}{0.6\baselineskip}

%Change tables
\renewcommand{\arraystretch}{1.3}

\title{SHAPER:  Can You Hear the Shape of a Jet?}

\preprint{MIT-CTP 5535}

\author[a,b]{Demba Ba,}
\author[a,b,c,d]{Akshunna S. Dogra,}
\author[b,e]{Rikab Gambhir,}
\author[f]{Abiy Tasissa,}
\author[b,e]{Jesse Thaler}

\affiliation[a]{John A. Paulson School of Engineering and Applied Sciences, Harvard University, Western Ave, Cambridge, MA, U.S.A.}
\affiliation[b]{The NSF AI Institute for Artificial Intelligence and Fundamental Interactions, U.S.A.}
\affiliation[c]{Department of Mathematics, Imperial College London, Queen's Gate, London, U.K.}
\affiliation[d]{EPSRC CDT in Mathematics of Random Systems: Analysis, Modelling and Simulation, Oxford, U.K.}
\affiliation[e]{Center for Theoretical Physics, Massachusetts Institute of Technology, Massachusetts Ave, Cambridge, MA 02139, U.S.A.}
\affiliation[f]{Department of Mathematics, Tufts University, College Ave, Medford, MA 02155, U.S.A.}

\emailAdd{demba@seas.harvard.edu}
\emailAdd{adogra@nyu.edu}
\emailAdd{rikab@mit.edu}
\emailAdd{Abiy.Tasissa@tufts.edu}
\emailAdd{jthaler@mit.edu}

\abstract{
The identification of interesting substructures within jets is an important tool for searching for new physics and probing the Standard Model at colliders.
Many of these substructure tools have previously been shown to take the form of optimal transport problems, in particular the Energy Mover's Distance (EMD).
In this work, we show that the EMD is in fact \emph{the} natural structure for comparing collider events, which accounts for its recent success in understanding event and jet substructure. 
We then present a Shape Hunting Algorithm using Parameterized Energy Reconstruction (\Shaper), which is a general framework for defining and computing shape-based observables. 
\Shaper generalizes $N$-jettiness from point clusters to any extended, parametrizable shape.
This is accomplished by efficiently minimizing the EMD between events and parameterized manifolds of energy flows representing idealized shapes, implemented using the dual-potential Sinkhorn approximation of the Wasserstein metric.
We show how the geometric language of observables as manifolds can be used to define novel observables with built-in infrared-and-collinear safety.
We demonstrate the efficacy of the \Shaper framework by performing empirical jet substructure studies using several examples of new shape-based observables.
}

\maketitle

\section{Introduction}

Collisions at the Large Hadron Collider (LHC) produce events with hundreds of particles in the final state, which must be carefully analyzed to extract information about the underlying physics.
In order to make sense of these high-dimensional data, increasingly sophisticated observables are required that are well understood at both  the theoretical and experimental levels.
Event shape~\cite{PhysRevLett.39.1587, Barber:1979bj, Dasgupta:2003iq, Dissertori:2008cn} and jet shape~\cite{Almeida:2008yp, Gur-Ari:2011cjr} observables have played an important role in refining our understanding of the structure of high energy collisions, by relating hadronic final states to perturbatively accessible partonic degrees of freedom.
Many shape observables, such as event thrust~\cite{BRANDT196457, PhysRevLett.39.1587, DERUJULA1978387} and jet angularities~\cite{ Berger:2003iw, Berger:2004xf}, have been computed to next-to-next-to-next-to leading log (N$^3$LL) accuracy~\cite{Becher:2008cf} and next-to-next-to leading log accuracy N$^2$LL~\cite{Banfi:2014sua} in $e^+e^-$ collisions, respectively.
Shape observables have been extensively measured and used to search for new physics signatures~\cite{Althoff:1983ew,Abrams:1989ez,Li:1989sn,Buskulic:1995aw,Adriani:1992gs, Braunschweig:1990yd,Abe:1994mf,Heister:2003aj,Abdallah:2003xz,Achard:2004sv, Abbiendi:2004qz, Abdesselam:2010pt}. 

It was shown in \Refer{2020} that many of these event shapes and jet shapes can be cast as optimal transport problems, using the Energy Mover's Distance (EMD).
The EMD was introduced in \Refer{Komiske_2019} in order to provide a quantitative measure of the ``distance'' between two collider events, $\mathcal{E}$ and $\mathcal{E}'$.
The EMD is based off the ``earth mover's distance'' from computer vision~\cite{192468, 10.5555/938978.939133, Rubner2004TheEM, Pele2008ALT, tangentEMD}, which itself is a special case of the Wasserstein metric~\cite{wasserstein1969markov, dobrushin1970prescribing}.
The EMD has since seen many uses in collider physics applications, such as in building a metrized latent space of events~\cite{Komiske_2019, Komiske:2019jim, Collins:2021pld, Park:2022zov} and in event/jet tagging and classification~\cite{CrispimRomao:2020ejk, Cai:2020vzx,Cai:2021hnn}.
The EMD has also been used to define a novel shape observable, the event isotropy~\cite{Cesarotti:2020hwb, Cesarotti_2021, ATLAS:2022jwu}, which probes how ``uniform'' an event $\E$ looks by comparing it to the idealized isotropic event $\mathcal{U}$.  

In this paper, we seek to explain the effectiveness of the Wasserstein metric, by showing that it is the unique metric on collider events that both is continuous and respects the detector geometry \emph{faithfully}.
As shown in \Refer{Komiske_2019}, continuity encodes the collider physics concept of infrared-and-collinear (IRC) safety.
Geometric faithfulness is, to our knowledge, a new concept for the collider community, which allows statements to be made about spatial distributions of energy within events. 
After advocating for Wasserstein geometry, we then generalize the notion of event shapes and jet shapes, motivated by the EMD.
Our framework -- called \Shaper\ -- not only allows observables to be defined that probe \emph{any} geometric substructure of events and jets in an IRC-safe way, analogous to the event isotropy probing the uniform structure of events, but also allows those observables to be numerically estimated.

In particular, we:
\begin{enumerate}

    \item \textbf{Motivate the Wasserstein Metric:}
    In \Sec{EMD}, we show that the Wasserstein metric is \emph{the} natural structure for building shape-based observables for collider physics, justifying its success in Refs.~\cite{2020, Komiske_2019, Komiske:2019jim, Collins:2021pld, Park:2022zov, CrispimRomao:2020ejk, Cai:2020vzx,Cai:2021hnn, Cesarotti:2020hwb, Cesarotti_2021, ATLAS:2022jwu} and beyond.
    By adopting a measure-theoretic language for energy flows, we show that the EMD is \emph{not} an ad-hoc structure to impose on the space of collider events, but rather the only structure that faithfully respects the detector geometry and continuity on the space of events.
    Further details of this argument are provided in \Apps{measure_theory}{construction}.

    \item \textbf{Use Optimal Transport to Define Shapes:}
    In \Sec{hearing}, we build off the work in \Refer{2020}, where it was shown that several well-known shape observables can be described in the form:
    \begin{align}
    \mathcal{O}_\M(\E) &\equiv \min_{\E_\theta \in \M} {\rm EMD}^{(\beta, R)}(\E, \E_\theta), \label{eq:shape_observable_intro} \\
    \theta_\M(\E) &\equiv \argmin_{\E_\theta \in \M} {\rm EMD}^{(\beta, R)}(\E, \E_\theta), \label{eq:shape_parameter_intro} 
    \end{align}
    where $\M$ is a parameterized manifold of energy flows that define the shape, $R$ sets a length scale for the shape, and $\beta$ is a distance weighting exponent. Importantly, both the observable $\O_\M$ and the optimal shape parameters $\theta_\M$ can be separately extracted from the EMD. 
    We extend this construction to define many new shape observables, by greatly expanding the class of manifolds $\M$ considered, which can be constructed as explicit geometric shapes. 
    We develop a prescription for defining new custom shape observables by parameterizing probability distributions, and even prescriptions for composing old shape observables together to form new ones.

    \item \textbf{Introduce \textsc{SHAPER}:} In \Sec{SHAPER}, we introduce \Shaper, or \textbf{S}hape \textbf{H}unting \textbf{A}lgorithm using \textbf{P}arameterized \textbf{E}nergy \textbf{R}econstruction.  This is a computational framework for defining shapes and evaluating \Eqs{shape_observable_intro}{shape_parameter_intro} on data.
    \Shaper leverages the Sinkhorn approximation of the Wasserstein metric~\cite{sinkhorn_1966, cuturi2013sinkhorn, CLASON2021124432, feydy2019interpolating}, which enables fast numerical calculation and even gradient estimation with respect to entire events, enabling easy and efficient optimization. See \textsc{NEEMo}~\cite{Kitouni:2022qyr} for an alternative gradient-based Wasserstein estimator.
    
    \item \textbf{Evaluate Empirical Examples:}
    To demonstrate the potential of \Shaper, in \Sec{empirical} we define and evaluate several observables on a top and QCD jet benchmark dataset~\cite{Butter:2017cot,Kasieczka:2019dbj}.
    These shape observables can be used to extract dynamic jet radii and jet energies, and even non-radially-symmetric structures, such as jet eccentricity.
    In particular, we show empirical examples of our new observables for jet substructure analysis and automated pileup removal.
\end{enumerate}

Generalized shape observables defined using \Shaper can be used to probe interesting collider signatures.
For example, \Shaper can be used to build specialized jet algorithms with dynamic radii\footnote{See e.g.~\cite{Krohn:2009zg, Mackey:2015hwa, Mukhopadhyaya:2023rsb, Larkoski:2023nye} for other examples of dynamic jet radii.} and even dynamic pileup mitigation.
This can be viewed as a generalization of $k$-means type clustering algorithms, such as \textsc{XCone}~\cite{Stewart:2015waa}: rather than finding $k$ points that best approximate an event, shape observables can be used to find the $k$ geometric structures that best describe the event.
This means that it is possible to design specialized jet algorithms that select for e.g.\ elliptical or non-isotropic jets, or that even probe the soft and collinear structure of jets separately.
This can prove especially useful, for example, in boosted top or heavy vector boson decays that produce ``fat jets'' with multi-pronged substructure, which may not be described well by circular or isotropic patterns. 
We comment on further phenomenological studies in \Sec{conclusions}.

\section{The Unreasonable Effectiveness of Wasserstein in Collider Physics}\label{sec:EMD}

In this section, we aim to answer the question, ``If I had never heard of event or jet shapes before, how could I have come up with them myself?'' 
Our discussion builds off the work of \Refers{Komiske_2019, 2020}, wherein the EMD was introduced as a new language for event and jet shape observables.
Here, we show that the EMD is \emph{the} natural language for event and jet shapes. 
We do this by showing that the EMD is the unique metric between a geometric shape $\E'$ and an event $\E$ that encodes IRC safety (through its topological features) and faithfully respects the geometry of the detector.
This section is largely self-contained, and readers primarily interested in the construction of new shape observables can skip to \Sec{hearing}.

We begin with a review of event shapes and jet shapes, noting that they all share a general form -- they can all be written as a minimization of a universal loss function between the event and a parameterized set of ``idealized'' events, which can be interpreted as geometric shapes.
We show that if the universal loss function is both IRC-safe and reduces to the ground metric on the detector geometry (that is, it \emph{faithfully} lifts the ground metric, without distorting extended objects), then the universal loss function must indeed be the Wasserstein metric. More details of this construction can be found in \Apps{measure_theory}{construction}.

%\Rikab{Note for Jesse/Feedback-givers: My goal with this section is to (1) add some mathematical structure to the observation in Hidden Geometry that common event shapes are geometric, (2) use this structure to generalize the concept of an event shape (with shape parameters), and (3) justify \textit{why} Wasserstein/OT works for this, which is something that left me unsatisfied about the Metric Space and Hidden Geometry papers. Would appreciate feedback on how well I accomplished this!}

\subsection{Event Shapes, Jet Shapes, and Geometric Shapes}\label{sec:shapes}

We begin with a review of event shapes and jet shapes. \emph{Event shapes} are observables that probe the geometric distribution of energy in events. Many different event shapes, such as thrust~\cite{BRANDT196457, PhysRevLett.39.1587, DERUJULA1978387}, spherocity~\cite{Georgi:1977sf}, broadening~\cite{Larkoski:2014uqa}, and $N$-jettiness~\cite{Stewart:2010tn, Stewart:2015waa}, have been defined and extensively studied over the years in the context of $e^+e^-$ collisions~\cite{Dasgupta:2003iq}, with analogues studied in the context of $pp$ collisions~\cite{Banfi_2004, Banfi:2010xy}.
We may also include jet algorithms, such as \textsc{XCone}~\cite{Stewart:2015waa} and sequential recombination algorithms ($k_T$~\cite{Catani:1993hr, Ellis:1993tq}, Cambridge-Aachen~\cite{Dokshitzer:1997in, Wobisch:1998wt}, and anti-$k_T$~\cite{Cacciari:2008gp}), in this list.
Similarly, \emph{jet shapes} probe the geometric distribution of energy within individual jets rather than the global event.
Examples of commonly studied jet shapes include the \textit{integrated} jet shape\footnote{A note about nomenclature: The original ``jet shape'' refers to the observable $\Psi(r/R)$, the radial jet energy fraction~\cite{Ellis:1992qq, PhysRevLett.70.713}. However, the word has been hijacked by \Refer{Ellis_2010} to refer to observables consisting of weighted sums of particle momenta. We later justify the name ``shapes'' by showing how this corresponds to fitting actual geometric shapes to event data.}~\cite{Ellis:1992qq, PhysRevLett.70.713}, angularities~\cite{ Berger:2003iw, Berger:2004xf}, and $N$-subjettiness~\cite{Thaler:2010tr}. While there are significant theoretical complications when considering the difference between event and jet shapes, such as the introduction of non-global logarithms~\cite{Dasgupta:2001sh, Banfi:2010pa}, for our purposes, we will treat event shapes and jet shapes interchangeably.

Following the definition in \Refer{Ellis_2010}, an event shape (jet shape) is an IRC-safe weighted sum over the four-momenta of the particles in an event (jet).
These observables probe the geometric distribution of energy in an event (jet), and typically depend on the \emph{detector ground metric}, $d(x,y)$, which defines distances between points on the detector.
Expressions for common event and jet shapes can be found in \Tabs{event_shapes}{jet_shapes}, respectively.
We note that all the above listed event shapes can be written in the generic form:
\begin{align}
    \mathcal{O}(p_1,...,p_M) = \min_{\theta \in \M}F\left(\sum_{i=1}^M E_i \, \phi_\theta(x_i) \right), \label{eq:eventshape}
\end{align}
where $F$ and $\phi_\theta$ are generic functions, and the observable may involve a minimization (or maximization) over auxiliary parameters $\theta$ living in some constrained manifold $\M$. The choice of $F$, $\phi_\theta$, and $\M$ define the event shape.
Not every shape requires an optimization -- for instance, while recoil-free jet angularities require an optimization over possible jet axes, it is also possible to define angularities with respect to a fixed axis~\cite{2020}.
In this case, the minimization may be written over the trivial manifold isomorphic to $\M = \{0\}$.
It is also common to divide by the total energy scale $E_{\rm tot}$, or some other hard scale, as this reduces the sensitivity of the event (jet) shape on experimental jet energy scale uncertainties~\cite{Banfi:2010xy}. %
Unless otherwise stated, we will normalize our events such that $E_{\rm tot} = 1$ without loss of generality.

% ###### Event Shapes Table ######
\begin{table}[t]
\centering
\small\begin{tabular}{rcl}
\hline
\hline
 Event Shape & Description  & Expression  \\ 
 \hline\hline
Thrust~\cite{BRANDT196457, PhysRevLett.39.1587, DERUJULA1978387} & How Pencil-Like? & $t(\mathcal E)=2\min_{\hat{n}}\left(\sum_i E_i (1-|\hat{n}_i \cdot \hat{n}|)\right)$\\
Spherocity~\cite{Georgi:1977sf} & How Tranverse-Planar?& $s(\mathcal E)= \min_{\hat{n}}\left(\sum_i E_i  |\hat{n}_i \times \hat{n}|\right)^2$  \\ 
Broadening~\cite{Larkoski:2014uqa} & How 2-Pronged?& $b(\mathcal E)= \min_{\hat{n}_1,\hat{n}_2 }\left(\sum_i E_i \min(d_{i1}, d_{i2})\right)$  \\ 
$N$-jettiness~\cite{Stewart:2010tn, Stewart:2015waa} & How $N$-particle like? & $\mathcal T_N^{(\beta)} (\mathcal E)= \min_{\hat{n}_1,...,\hat{n}_N }\left(\sum_i E_i \min(R^\beta, d_{i1}^\beta, ..., d_{iN}^\beta)\right)$\\
Isotropy~\cite{Cesarotti:2020hwb} & How Uniform? & $\mathcal I^{(\beta)}(\mathcal E)= \min_{\mathcal{U} \in \M}\left(\EMD^{(\beta, R)}(\E, \mathcal{U})\right)$\\
\hline
\textsc{XCone}~\cite{Stewart:2015waa} & Which $N$-particles? & $\hat{n}_i(\mathcal E)= \argmin_{\hat{n}_1,...,\hat{n}_N }\left(\sum_i E_i \min(R^\beta, d_{i1}^\beta, ..., d_{iN}^\beta)\right)$\\
S. Recomb.~\cite{Catani:1993hr, Ellis:1993tq, Dokshitzer:1997in, Wobisch:1998wt,Cacciari:2008gp} & Clustering History? & $d_{ij}^N (\mathcal E)= \min(E_i^{2p}, E_j^{2p})\frac{d_{ij}^2}{R^2};\,\, d_{iR}^N (\mathcal E)= E_i^{2p}$\\
\hline
\hline
\end{tabular}
\caption{
Common event shapes and jet algorithms studied in collider physics. Note that most of these observables take the general form of \Eq{eventshape}. Here, we do not necessarily normalize energies.}
\label{tab:event_shapes}
\end{table}

% ###### Jet Shapes Table ######
\begin{table}[t]
\centering
\small\begin{tabular}{rcl}
\hline
\hline
 Jet Shape & Description  & Expression  \\ 
 \hline\hline
 Angularities~\cite{ Berger:2003iw, Berger:2004xf}& Angular Moments? & $\lambda_\beta(\mathcal J) = \sum_i E_i d_{iJ}^\beta$\\
 & ... Recoil Free? & $\lambda_\beta(\mathcal J) = \min_{\hat{n}}\left(\sum_i E_i d_{in}^\beta\right)$\\
$N$-subjettiness~\cite{Thaler:2010tr} & How $N$-Particle Like? & $\mathcal T_N^{(\beta)} (\mathcal J)= \min_{\hat{n}_1,...,\hat{n}_N }\left(\sum_i E_i \min(d_{i1}^\beta, ..., d_{iN}^\beta)\right)$\\
Int. Shape~\cite{Ellis:1992qq, PhysRevLett.70.713} & Radial Energy CDF? & $\psi_{\mathcal J}(r/R) = \left(\sum_i E_i \Theta(r - d_{iJ})\right)/\left(\sum_i E_i \Theta(R - d_{iJ})\right) $\\
\hline
\hline
\end{tabular}
\caption{Common jet shapes studied in collider physics. Note that most of these observables take the general form of \Eq{eventshape}. The notation $d_{iJ}$ refers to the distance from particle $i$ to the jet axis. Here, we do not necessarily normalize energies. }
\label{tab:jet_shapes}
\end{table}

We propose to write \Eq{eventshape} in a universal form, such that the event shape is instead specified solely by the choice of $\M$:
\begin{align}
    \O_\M(\E) = \min_{\theta \in \M}\left[\mathcal{L}(\E;\theta)\right], \label{eq:universal_loss}
\end{align}
where $\mathcal{L}$ is a universal loss function.
All geometrical information about the event shape is then contained in the construction of $\M$.
To emphasize this, we will adopt the notation $\O_\M$ for these observables, to remind us that the observable is defined through the choice of $\M$.

The task is now to determine what universal $\mathcal{L}$ reproduces all event and jet shape observables -- we will argue in \Sec{wasserstein} that $\mathcal{L}$ must be the Wasserstein metric.
To begin, we may rewrite \Eq{universal_loss} in a more suggestive form. We note that for all of the event and jet shapes in \Tabs{event_shapes}{jet_shapes}, there is always some optimal $\E^*$, not necessarily unique, such that $\mathcal{O}_\M(\E^*) = 0$. 
For example, the $\E^*$ for thrust is a perfectly back-to-back event, the $\E^*$ for $N$-subjettiness is an  event with exactly $N$ particles, and so on. 
Thus, it is convenient to rewrite \Eq{universal_loss}, such that the minimization is over a space of events $\E_\theta$, and that $\mathcal{L} = 0$ is achieved when $\E = \E_\theta$, where  $\theta$ parameterizes the space of all $\E^*$'s:
\begin{align}
    \mathcal{O}_\M(\E) = \min_{\E_\theta \in \M}\left[\mathcal{L}(\E;\E_\theta)\right]. \label{eq:geometric_shapes}
\end{align}
\Eq{geometric_shapes} provides a nice geometric intuition for event and jet shapes.
We can interpret $\mathcal{O}_\M(\E)$ as the answer to ``How close, in event space, is my event to looking like an optimal $\E^*$?''.
Importantly, the $\E^*$'s do not have to be physically realized events -- they can be \emph{any} radiation pattern measured on the detector wall, even continuous ones. 
For example, we can take $\E^*$ to be events with a radiation pattern that look like the interior of a hexagon -- then the event shape $\mathcal{O}_\M(\E)$ is a measure of how far $\E$ is, in ``event space'', from an idealized hexagonal event.
By taking our idealized events $\E^*$ to have radiation patterns resembling literal geometric shapes, living in the parameterized manifold $\M$, the observable $\O_\M(\E)$ can be used as a measure of how much $\E$ ``looks like'' the shape of interest.

\subsection{Measure-ing the Energy Flow}\label{sec:energyflow}

In order to make progress in determining the universal loss function in \Eq{geometric_shapes}, we must first understand the IRC-safe information available for us to use within the events $\E$ and $\E_\theta$. 
This information is represented by the $\emph{energy flow}$ of the event. 
We first briefly review energy flows, before proposing a new definition of the energy flow as a measure theoretic quantity, which enables a useful language for discussing ``idealized'' events such as those discussed in \Sec{shapes}.

The \emph{energy flow} $\E$ of an event is the distribution of energy within the event.
At a very high level, in a collider experiment, one has a detector with geometry $\mathcal{X}$ infinitely far away from the collision site -- for instance, in \textit{pp} collisions such as those at the LHC, one uses a cylindrical detector $\mathcal{X} = [y_{\rm min}, y_{\rm max}] \times S^1$, where $y\in [y_{\rm min}, y_{\rm max}]$ is the rapidity and $\phi \in S^1$ is the azimuthal angle.
After a collision, particles hit the detector at a site with coordinate $x_i \in \mathcal{X}$, where the energy $E_i$ is recorded by a calorimeter.
The energy flow $\E$ for an event with $M$ particles of energies $E_i$ measured at locations $x_i$ is given by:
\begin{align}\label{eq:energyflow}
    \E(x) = \sum_{i=1}^M E_i \, \delta(x-x_i).
\end{align}

The energy flow quantifies the total amount of energy measured at position  $x$, which can be thought of as an idealized calorimeter cell. 
Assuming that the particles are massless, this is the complete \emph{accessible}\footnote{Here, we take \emph{accessible} to mean the calorimeter information after infinite time has passed, preserving no timing information. We implicitly assume that the calorimeter's detector response is linear, so that two photons entering the same calorimeter cell cannot be distinguished from a single photon with their summed energy, though even if the response is nonlinear, one cannot distinguish how many photons entered the calorimeter cell from the total energy alone.} kinematic information about the event, which therefore allows us to consider an event and its energy flow interchangeably.
In the context of hadron colliders, the transverse momentum $p_T$ is often used in place of the energy $E$.
In this paper, however, we focus on energies to save on notational complexity, as the story is relatively unchanged when switching to $p_T$.

The energy flow operator is well-understood theoretically and in some cases, can even be computed analytically~\cite{Tkachov:1995kk,Sveshnikov:1995vi,Korchemsky:1997sy,Basham:1978zq,Cherzor:1997ak,Tkachov:1999py,Korchemsky:1999kt,Belitsky:2001ij,Berger:2002jt,Bauer:2008dt,Hofman:2008ar,Mateu:2012nk,Belitsky:2013xxa}. In terms of field-theoretic quantities, the energy flow is given as:
\begin{align}
    \E(x) = \lim_{r\to \infty} \int_{-\infty}^\infty dt\, n_i \,T^{0i}(t, rn_i), \label{eq:stress_energy}
\end{align}
where $n_i$ is the unit 3-vector corresponding to the detector coordinate $x$.
We assume for this work that the spectrum of energies is non-negative -- that is, for all $x$, $\E(x) \geq 0$.

\begin{figure}[t]
    \centering
    \includegraphics[width=0.85\textwidth,trim={7cm 7cm 7cm 7cm}, clip]{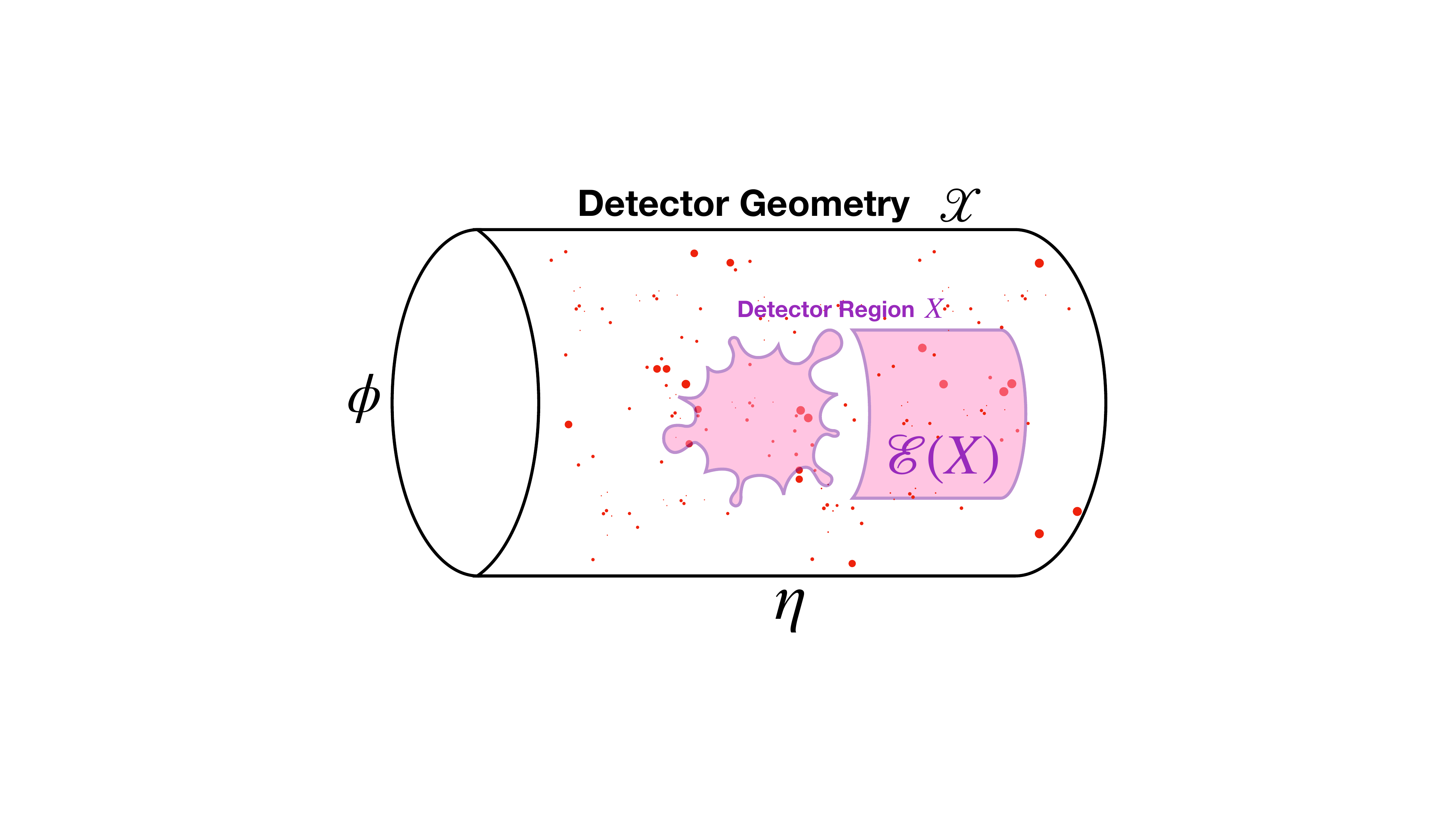}
    \caption{An illustration of the energy flow $\mathcal{E}(X)$, which is the total amount of radiation captured inside a subset $X$ (in purple) of the total detector geometry $\mathcal{X}$. The red dots represent particles that hit the detector wall, with their size proportional to their energy. }
    \label{fig:energyflowmeasure}
\end{figure}

We now propose a natural generalization of the energy flow that captures its salient properties and is key to enabling our geometric analysis:
\begin{definition}\label{def:energyflow}
The energy flow $\E$ of an event in a detector geometry $\mathcal{X}$ is a (positive) \textbf{measure} over subsets  $X \subseteq \mathcal{X}$, such that $\E(\mathcal{X}) = E_{\rm tot}$, the total energy of the event.
\end{definition}
\noindent In this new language, the energy flow $\E(X)$ is the total energy measured in \emph{any} region $X \subseteq \mathcal{X}$ of the detector, rather than just probing a localized point $x\in \mathcal{X}$.
The region $X$ can be an extended set and does not need to be connected.
\Fig{energyflowmeasure} illustrates an example of this on a cylindrical collider.
This generalized notion of energy flow $\E(X)$ reduces to the usual energy flow $\E(x)$, which we now refer to as the \emph{energy flow density}, and can be written as:
    \begin{align}
        \E(X) = \int_X dx\, \E(x). \label{eq:energyflowdensity}
    \end{align}
A particle measured at $x_i$ will contribute energy to $\E(X)$ only if $x_i \in X$, which can be seen by carrying out the integration over the $\delta$-functions in \Eq{energyflow}.\footnote{We will assume that energy flows can always be written as the integral of an associated energy flow density. Note that the energy flow density depends on the choice of coordinates $x$ used on $\mathcal{X}$.}
Unlike \Eq{energyflow}, however, we do not restrict energy flows to just a finite sum of localized $\delta$-functions -- they can be continuous, extended deposits of energy!
In \Sec{hearing}, we will see energy flows with continuous energy distributions are key to defining generalized shape observables.
We will refer to energy flows whose densities can be represented by a finite sum of weighted $\delta$-functions (as in, for example, \Eq{energyflow}) as \emph{atomic measures} or \emph{atomic flows}. We will often write atomic measures as $\E \sim \sum_i E_i\, \delta_{x_i}$ for notational simplicity.

Under \Def{energyflow}, energy flows inherit a very rich and natural mathematical structure. The most important operation for our purposes is the integral of a function $\phi:\mathcal{X}\to\mathbb{R}$ against an energy flow $\E$, which we denote $\expval{\E, \phi}$, defined as:
\begin{align}
    \expval{\E, \phi} &\equiv \int_\mathcal{X}dx\, \E(x) \, \phi(x) \\
    &= \sum_i E_i\, \phi(x_i) \text{ for atomic flows.}
\end{align}
This operation can be thought of as the energy-weighted expectation value of the random variable $\phi$ under the distribution $\E$.
A brief review of this, and other salient measure-theoretic concepts and definitions we call upon in this paper, is presented in \App{measure_theory}.

\subsection{Geometrizing IRC Safety}\label{sec:irc}

Infrared and colinear (IRC) safety is an incredibly powerful constraint on the form of observables -- it ensures not only that an observable is well-defined in perturbation theory, but also that the observable is robust to detector effects.
Using the language developed in \Sec{energyflow}, IRC safety becomes a topological statement on the space of energy flows, which we may use to place constraints on the potential form of the universal loss function $\mathcal{L}$ of \Eq{geometric_shapes}.

An observable $\mathcal{O}$ is IRC safe if it satisfies:\footnote{There are several different statements of IRC-safety with different limit structures, each with different pathologies. A brief discussion of this can be found in Sec. 2.1 of \Refer{2020}.}
\begin{itemize}
    \item \textbf{Infrared safety}: For any event atomic $\E$, adding or removing an $\epsilon$-soft emission to $\E$ leaves $\O$ unchanged as $\epsilon\to0$.
    \item \textbf{Collinear safety}: For any atomic event $\E$, splitting any particle into two particles at the same location with the same total energy leaves $\O$ unchanged. Moreover, translating either particle by an $\epsilon$-small displacement leaves $\O$ unchanged as $\epsilon\to 0$.
\end{itemize}
Essentially, IRC safety means that observables should not change significantly if we change $\E$ by slightly adjusting particle energies and positions.
As with energy flows, we propose a generalization of IRC safety that captures all its salient features:
\begin{definition}
An observable $\O$ is IRC safe if it is continuous with respect to the weak* topology on energy flows.
\end{definition}
A function $f$ on energy flows is continuous to the weak* topology if, for any sequence of energy flows $\E_n$ that converges to $\E$, the function $F(\E_n)$ converges to $F(\E)$ (see \App{measure_theory} for more details). Note that this is actually a slightly \emph{weaker} constraint than the one considered in \Refer{2020}, which defines IRC safety through the metric topology induced by the EMD -- the definition here does not require a metric on the space of events, or even a metric on the detector space, only a notion of continuity. In fact, there is a large class of metrics one can place on the space of events to metrize the weak* topology, not just the EMD. 

An interesting consequence of this definition is that if an observable $\O$ is IRC safe, then $\O(\E)$ for \emph{any} energy flow $\E$ can be arbitrarily well-approximated by atomic energy flows. This implies, for example, that a continuous circle can be arbitrarily well approximated by a finite number of points arranged in a ring -- this is makes possible to not only encode continuous distributions numerically, but also to make broad statements about the behavior of IRC safe observables by considering their action only on simple atomic energy flows.

In order to be IRC safe, our universal loss function $\mathcal{L}(\E,\E')$ must be continuous in both of its arguments. 
This is very restrictive, and immediately implies that $\mathcal{L}$ cannot have any terms that are discontinuous in either energy or distance, e.g.\ terms like $E^{-1}$ or $d(x,y)^{-1}$, or any term of the form $\expval{\E, \phi}$ for noncontinuous $\phi$.
Recalling the discussion in \Sec{shapes} that $\mathcal{L}$ quantifies how close in the space of energy flows $\E$ and $\E'$ are, it is convenient (though not strictly necessary) to use $\mathcal{L}$ to \emph{metrize} the weak* topology -- that is, if an observable $\O$ is continuous with respect to the metric topology induced by $\mathcal{L}$, then it is also continuous with respect to the weak* topology, and therefore is IRC safe. 
This allows the same universal loss function $\mathcal{L}$ to be used both to define shape observables and to define IRC safety.\footnote{This is not required however -- there are many different choices of $\mathcal{L}$ that may be used to give the same definition of IRC safety, e.g.\ maximum mean discrepancies, which does not necessarily have to be the same function $\mathcal{L}$ whose minimum defines shapes as in \Eq{geometric_shapes}}
This is convenient, since it captures the very intuitive notion that if two events geometrically look similar (that is, $\mathcal{L}$ is small), then IRC-safe observables evaluated on them should also be the same.

\subsection{The Importance of Being Faithful}\label{sec:wasserstein}

In order to encode geometric information about energy distributions, the universal loss function $\mathcal{L}$ of \Eq{geometric_shapes} must explicitly depend on the detector ground metric, $d(x,y)$. While there are many metrics on the space of measures that encode geometric information while also being IRC-safe (as defined in \Sec{irc}), a natural choice is the family of Wasserstein metrics, which we denote $\mathcal{L}(\E, \E') = \EMD^{(\beta,R)}(\E, \E')$ (for \emph{Earth-Mover's} or \emph{Energy-Movers Distance}, which we will use synonymously). We show in this section that unlike other potential candidates for $\mathcal{L}$, the Wasserstein metric will never warp distances between shapes -- that is, the Wasserstein metric lifts the ground metric of the detector \emph{faithfully}. A constructive proof of this can be found in \App{construction}.

The EMD between two measures $\E$ and $\E'$ is given by:
\begin{align} \label{eq:EMD}
    \EMD^{(\beta,R)}(\E, \E') = \min_{\pi \in \M(\mathcal{X}\times\mathcal{X})}\left[\frac{1}{\beta R^\beta}\expval{\pi, d(x,y)^{\beta}}\right] + |\Delta E_{\rm tot}|, \nonumber \\
    \pi(\mathcal{X}, Y) \leq \E'(Y) , \quad \pi(X, \mathcal{X}) \leq \E(X), \quad \pi(\mathcal{X}, \mathcal{X}) = \min(E_{\rm tot}, E'_{\rm tot}),
\end{align}
where $d(x,y)$ is the ground metric between points $x$ and $y$ on $\mathcal{X}$, and $\mathcal{M}(\mathcal{X} \times \mathcal{X})$ is the space of all positive measures on $\mathcal{X} \times \mathcal{X}$. The parameter $R > 0$ sets a distance scale for the EMD, and sets the relative scale for the two terms in \Eq{EMD}. The parameter $\beta \geq 1$ sets the distance norm.\footnote{In order to satisfy the triangle inequality and be a true metric, the first term in the EMD must be raised to the $1/\beta$ power, and either  $2R$ should exceed the largest value of $d(x,y)$~\cite{Komiske_2019,2020} or $\Delta  E_{\rm tot}$ must be guaranteed to be zero. In this paper, we will not need the triangle inequality, so we will not do this.} Note that our definition of the EMD differs from \Refers{Komiske_2019,2020} by a factor of $\beta$, which we do to match the conventions of the \textsc{geomloss}~\cite{feydy2019interpolating} package. The additional energy difference term, $|\Delta E_{\rm tot}| $, contributes whenever the two energy flows do not have the same total energy.

The Wasserstein metric is special in that it \emph{faithfully} lifts the ground metric, $d$. To lift the ground metric means that the Wasserstein metric reduces to $d(x,y)^\beta$ when evaluated on two point measures $\E \sim \delta_x$ and $\E' \sim \delta_y$ -- that is, the Wasserstein metric preserves distances between points. Moreover, to do so \emph{faithfully} means that the Wasserstein metric preserves distances for entire extended shapes: if $\E$ is \emph{any} measure, and $\E'$ is the same as $\E$ whose density is translated by a vector $t$ (that is, $\E'$ has corresponding density $\E(x-t)$), then the metric between them is simply $d(0,t)^\beta$.

To see this explicitly, we can compare to two other potential candidates for our universal loss function $\mathcal{L}$: the class of Maximum Mean Discrepancies (MMDs)~\cite{ramdas2017wasserstein} and Chamfer distances~\cite{10.5555/1622943.1622971}, which can respectively be written as:\footnote{We choose the MMD and Chamfer distance for comparison because, as shown in \App{construction}, the most general loss that is symmetric, IRC safe, and lifts the ground metric (though not necessarily faithfully) has terms individually resembling the Wasserstein, MMD, and Chamfer distances.}
\begin{align}
    \text{MMD}^{(\beta)}(\E, \E') &= -\frac{1}{2\beta}\int_{\X\times\X}dx\, dy\, d(x,y)^\beta\, \left(\E(x)-\E'(x)\right)\left(\E(y)-\E'(y)\right),\label{eq:MMD} \\
    \text{CD}^{(\beta)}(\E, \E') &= \frac{1}{2\beta}\int_{\X}dx\,\min_{y\in{\rm Supp}(\E')}[d(x,y)^\beta]\,\E(x) + \frac{1}{2\beta}\int_{\X}dy\,\min_{x\in{\rm Supp}(\E)}[d(x,y)^\beta]\,\E'(y).\label{eq:chamfer}
\end{align}
These candidate loss functions are IRC safe, translationally invariant, and even lift the ground metric, but importantly, they do not do so faithfully. To see this, we choose the following example energy flows:
\begin{align}
    \E(x) &\sim \frac{1}{2}\delta_0 + \frac{1}{2}\delta_a, \quad \E'(x) \sim \frac{1}{2}\delta_t + \frac{1}{2}\delta_{a+t},
\end{align}
where $a$ and $t$ are arbitrary vectors (where $\E$ consists of 2 points separated by a vector $a$, and $\E'$ is $\E$ translated by a vector $t$).
A direct computation using Eqs. \eqref{eq:EMD}, \eqref{eq:MMD}, and  \eqref{eq:chamfer} (with $d(x,y) = |x-y|$) yields:
\begin{align}
    \EMD^{(\beta,R=1)}(\E,\E') &= \frac{1}{\beta}|t|^\beta, \\
    \text{MMD}^{(\beta)}(\E, \E') &= \frac{1}{2\beta}\left(|t|^\beta + \frac{1}{2}|t-a|^\beta + \frac{1}{2}|t+a|^\beta -|a|^\beta \right),\label{eq:MMD_computation}\\
    \text{CD}^{(\beta)}(\E, \E') &= \frac{1}{2\beta}\left(\min\left[|t|^\beta, |t+a|^\beta\right] + \min\left[|t|^\beta, |t-a|^\beta\right] \right).\label{eq:chamfer_computation}
\end{align}
While \Eqs{MMD_computation}{chamfer_computation} do indeed reduce to $\sim|t|^\beta$ when $a\to0$ (that is, when $\E$ reduces to a single point), in general the MMD and Chamfer distance effectively ``distort'' the shape.\footnote{Note that this is avoided for the MMD in the $\beta = 2$ case. This is in contrast to the Wasserstein metric, which is faithful for \emph{any} $\beta \geq 1$. While some observables are defined with $\beta = 2$, this is not sufficient for all observables, and in particular we will focus on observables with $\beta = 1$, which is the only ``true'' metric satisfying the triangle inequality.}
When $\beta = 1$, for instance, $\E'$ appears slightly closer to $\E$, as measured using either MMD or the Chamfer distance, than its total displacement $|t|$ -- energy in the interior of an extended distribution gets effectively ``screened'' by the energy in the rest of the distribution!
Not only does this distortion ruin our ability to think of our observables as measuring the geometric distribution of energy in the detector, the screening effect also induces a practical issue, as it causes vanishing gradients when trying to optimize over $\mathcal{L}$~\cite{feydy2019interpolating} (in this case, by minimizing $t$, for example). 
The Wasserstein metric does not suffer these problems, making it the natural choice for our universal loss function.

\section{Hearing Shapes} \label{sec:hearing}

Having constructed the Wasserstein metric and EMD in \Sec{EMD} for event and jet shapes, we next generalize \Eqs{shape_observable_intro}{shape_parameter_intro}, which were originally introduced in \Refer{2020} as a common form for many well-known observables. 
We treat \Eqs{shape_observable_intro}{shape_parameter_intro} as definitions for \emph{shape observables} $\O_\M$ and \emph{shape parameters} $\theta_\M$, which together are the natural generalization of event and jet shapes.
Moreover, we show how the manifold of energy flows $\M$ can be chosen to construct new observables that probe specific geometric structures.
We provide a prescription for building $\M$, which defines the shape observable, as well as prescriptions for \emph{composing} shape observables together, allowing new shape observables to be defined from simpler ones in a geometrically intuitive way.\footnote{Like music, shapes are composed using the \Shaper paradigm by splicing and overlaying together smaller shapes, after which it may be ``heard'' by evaluating on event data -- a posthoc rationalization for the title of this paper.} 

This section proceeds as follows.
First, we define generalized shape observables and shape parameters, and discuss their properties. 
Next, we discuss our prescription for shape composition, which can be used to define shape observables and parameters that probe complex geometric structure.
Finally, we use our prescription to construct a large (but importantly, inexhaustive) suite of novel shape observables for jet substructure analysis to serve as an example of what can be done with this framework.
Several examples of emperical studies using these new shape observables, evalulated using the \Shaper framework defined in \Sec{SHAPER}, can be found in \Sec{empirical}.

\subsection{Shape Observables and Shape Parameters}\label{sec:observables}

We define a \emph{shape observable} $\mathcal{O}_\M$ as follows:

\begin{definition}
A shape observable $\mathcal{O}_\M$, with associated shape parameters $\theta_\M$, on an energy flow $\E$ is any function of the form:
\begin{align}
    \mathcal{O}(\E) &\equiv \min_{\E_\theta \in \M} {\rm EMD}^{(\beta, R)}(\E, \E_\theta), \label{eq:shape_observable}\\
    \theta(\E) &\equiv \argmin_{\E_\theta \in \M} {\rm EMD}^{(\beta, R)}(\E, \E_\theta), \label{eq:shape_parameter}
\end{align}
where $\M$ is a manifold of positive measures on the detector space $\mathcal{X}$, and ${\rm EMD}^{(\beta, R)}$ is the $\beta$-Wasserstein distance with length scale $R$.
\end{definition}

Importantly, we return \emph{both} the minimum EMD value, $\mathcal{O}(\E)$, \emph{and} the parameters of the shape that produced the minimum EMD, $\theta(\E)$. For a manifold $\M$ of generic parameterized shapes, we refer the former as the ``\emph{shapiness}'' of $\E$, and the latter as the associated ``\emph{shape parameters}'' of $\E$. For example, if $\M$ is the manifold of $N$-(sub)jet events, then $\mathcal{O}(\E)$ is the $N$-(sub)jettiness of $\E$, and $\theta(\E)$ are the (sub)jet parameters of $\E$, highlighting that \Eq{shape_observable} is really a generalization of the $N$-jettiness. Intuitively, $\mathcal{O}(\E)$ answers the question, ``How much like a \textit{shape} does my event look like?'', while $\theta(\E)$ answers the question, ``Which \textit{shape} does my event look most like?''.

For a fixed choice of $\beta$ and scale $R$ defining the EMD, shape observables are completely specified by the choice of the manifold of energy flows $\M$. This choice specifies the class of shapes being considered. Practically speaking, this manifold can be defined by choosing a set of coordinates $\theta$ on the manifold, that parameterize a set of constrained energy flows $\E_\theta$. Since, as established in \Sec{energyflow}, energy flows are positive measures on the detector space, $\M$ can be built as a (weighted) parameterized probability distribution $p_\theta$, which is realized by a finite sampling procedure for weighted points on $\mathcal{X}$.\footnote{Because energy flow densities depend on the choice of coordinates, the choice of $p_\theta$ corresponding to the class of shapes prescribed by $\M$ is not unique. For example, to sample a Gaussian distribution, one can sample uniformly weight points with positions distributed as a Gaussian, or sample uniformly spaced points with Gaussian weights depeding on their position, or some mixture of both.}

In \Refer{2020}, the manifold corresponding to several event and jet shapes were listed. However, the power of our framework is that \emph{any} manifold of parameterized energy flows defines a valid shape observable, and moreover, this observable directly probes intuitive geometric information. By defining, for example, $\M$ to be the manifold of energy flows resembling uniform rings of energy, uniform disks of energy, or even uniform ellipses of energy, the corresponding observables directly quantify the diskiness, circliness, and ellipsiness of events. To our knowledge, the event isotropy~\cite{Cesarotti:2020hwb} is the first observable of this form with no known alternative formulation, as it quantifies the ``uniforminess'' of events. Examples of analyses using these custom observables can be found in \Sec{empirical}.

Below, we list some useful properties of all shape observables:

\begin{enumerate}
    \item \textbf{Monotonicity}: If $\mathcal{O}_1$ and $\mathcal{O}_2$ are defined by manifolds $\M_1$ and $\M_2$, for a fixed choice of $\beta$ and $R$, then $\M_1 \supseteq \M_2$ implies that for all energy flows $\E$, we have $\mathcal{O}_1(\E) \leq \mathcal{O}_2(\E)$. This captures the monotonic nature of observables such as $N$-jettiness, which satisfy $\tau_{N+1} \leq \tau_N$.
    \item \textbf{Closure}: $\mathcal{O}(\E) = 0$ if and only if $\E \in \M$. It then follows that the shape parameters $\theta(\E)$ are such that $\E_\theta = \E$. In particular, this implies that for $\M_\E$, the space of \emph{all} energy flows, $\O(\E) = 0$ for all events. 
    \item \textbf{Approximation Bounds}:  For any $L$-Lipschitz function $\phi$ on $\mathcal{X}$, the ``optimal shape'' $\E_\theta$ corresponding to $\E$ satisfies $\frac{1}{RL}|\expval{\E,\phi}-\expval{\E_\theta, \phi}| \leq \mathcal{O}(\E)$~\cite{Komiske_2019}. That is, the optimal shape $\E_\theta$ can be used to approximate additive Lipschitz observables on $\E$, up to a known bounded error. 
    \item  \textbf{Upper Bounds}: If $\mathcal{X}$ is bounded by a maximum distance scale $R_{\rm max}$, and both $\E$ and all energy flows on $\M$ satisfy $E_{\rm tot} = 1$, then $\mathcal{O}(\E)$ is bounded above by $\left(\frac{R_{\rm max}}{R}\right)^\beta$.\footnote{This is not necessarily the least upper bound, which depends specifically on the choice of $\M$.} This can be seen by considering the extreme case where $\E$ and $\E_\theta$ are singleton points located $R_{\rm max}$ away from each other. Any configuration other than this will yield a lower EMD. This makes it always possible to normalize $\mathcal{O}(\E) \in [0,1]$. 
\end{enumerate}

We call a manifold of energy flows $\M$ \emph{balanced} if all of the energy flows are all normalized (i.e. $E_{\rm tot} = 1)$, and unbalanced otherwise. Similarly, we call shape observable $\O_\M$ balanced if $\M$ is balanced \emph{and} it is only evaluated on normalized energy flows. For a balanced observable, the choice of scale $R$ constitutes only a change of units for the distance metric, and is unimportant.\footnote{Though we will see in \Sec{SHAPER} that $R$ matters again for our numerical approximations.} Moreover, in the limit $R\xrightarrow{}\infty$, the quantity $R^\beta \mathcal{O}(\E)$ is only finite if $E_{\rm tot} = E_{\theta \rm tot}$~\cite{2020}. In this limit, the $\paramin$ can be written over the submanifold $\M_{E} \subseteq \M$, which is the submanifold events with the same total energy as $\E$. Thus, the $R\to\infty$ limit effectively forces unbalanced shape observables to be balanced. We will primarily consider balanced shape observables for the remainder of this paper, though there are many important unbalanced observables (e.g. $N$-jettiness) that one may consider.

\subsection{Composing Shapes}\label{sec:composing}

Given two shape observables $\mathcal{O}_1$ and $\mathcal{O}_2$, with the same choice of $\beta$ and $R$, we can define a new \emph{composite observable} $\mathcal{O} = \mathcal{O}_1\oplus\mathcal{O}_2$. As with all shape observables above, we define the composite shape observable $\O$ by specifying the corresponding manifold $\M$. We consider two possible scenarios to define $\M$:

\begin{itemize}
    \item \textbf{Case 1 (Balanced)}: If both $\mathcal{O}_1$ and $\mathcal{O}_2$ are balanced shape observables,  then we define $\M$ to be the manifold of all energy flows $\E$ of the form $\E = z_1\E_1 + z_2\E_2$, where $\E_1 \in \M_1$ and $\E_2 \in \M_2$, and $(z_1, z_2)$ is a point in $\Delta_1$, the 1-simplex.\footnote{The $N$-simplex, $\Delta_N$, is the set of all points $z_1,...,z_{N+1} \geq 0$ such that $\sum_{i=1}^{N+1}z_i = 1$.} 
    \item \textbf{Case 2 (Totally Unbalanced)}: If the energy flows on $\M_1$ and $\M_2$ have unconstrained $E_{\rm tot}$'s (that is, $E_{\rm tot}$ is an independent free parameter for each manifold), then we define $\M$ to be the manifold of all energy flows $\E$ of the form $\E = \E_1 + \E_2$, where $\E_1 \in \M_1$ and $\E_2 \in \M_2$.
\end{itemize}

There exist other possible cases, such as if the manifolds only admit a small range of $E_{\rm tot}$ values, but we will not consider them in this work.\footnote{There is no natural way to define $\M$ from $\M_1$ and $\M_2$; this is a choice that we make, and other choices could be considered. For example, one may consider defining $\M$ to instead be the manifold of energy flows $\E = \E_1 \E_2$, which is a measure on $\mathcal{X}\times\mathcal{X}$, and could potentially be useful for measuring factorization properties. We leave the study of this class of shapes for potential future work.} In Case 1, the 1-simplex $\Delta_1$ introduces 2 additional (constrained) parameters to the shape, $z_1$ and $z_2$, that ensure that all energy flows are still normalized to 1, so that $\mathcal{O}$ is balanced. Here, $\M$ can be realized by generating points according to the parameterized sampling procedures for $\M_1$ and $\M_2$, and multiplying the energy weights of the generated particles by $z_1$ and $z_2$, respectively. The values of $z_1$ and $z_2$ control the relative contribution of each base shape to the composition. In Case 2, the total energy remains unconstrained, and the relative importance of each base shape to the contribution is controlled by the ratio of the $E_{1 \rm tot}$ and $E_{2 \rm tot}$ parameters. Note that monotonicity implies that, for both Case 1 and Case 2, $\mathcal{O}(\E) \leq \mathcal{O}_1(\E)$ and $\mathcal{O}(\E) \leq \mathcal{O}_2(\E)$. 

It is easy to extend these definitions to a composition of $N$ observables, $\mathcal{O} = \bigoplus_{i=1}^N \mathcal{O}_i$. For the balanced case, $\M$ comprises of energy flows of the form $\E = \sum_{i = 1}^N x_i \E_i$ for $\E_i \in \M_i$ and $z_i \in \Delta_{N-1}$. In the totally unbalanced case, $\M$ comprises energy flows of the form $\E = \sum_{i = 1}^N \E_i$ for $\E_i \in \M_i$. When the $\mathcal{O}_i$ are all copies of the same shape observable $\mathcal{O}$, we define the notation $N\times\mathcal{O}$, and name the resulting composite shape observable the ``$N$-shapiness''.

Given this language, we can understand several existing shape observables as composites of basic ``shape'' building blocks. For instance, we can define an observable, the (sub)pointiness $\tau_1(\E)$, corresponding to the manifold of weighted (normalized) $\delta$-function measures on $\mathcal{X}$. Geometrically, this observable to fitting a single weighted (normalized) point to an event $\E$, the former with a floating energy weight. We can then define the $N$-(sub)pointiness of $\E$ to be $\tau_N(\E) = N \times \tau_1(\E)$ -- this is exactly the $N$-(sub)jettiness of $\E$, with $N$-subjettiness corresponding to balanced composition, and $N$-jettiness corresponding to totally unbalanced composition.

Shape composition provide a novel avenue for understanding event and jet substructure. For example, the observable $\tau_N \oplus \mathcal{I}$, where $\mathcal{I}$ is the event isotropy, can be thought of as a ``pileup-corrected'' $N$-(sub)jettiness, where a uniform background is subtracted off. One can use this observable to probe the percentage of energy deposited in hard jets versus a uniform background due to pileup~\cite{Soyez:2018opl} and underlying event effects~\cite{PhysRevD.65.092002, Agocs:2010ft}. In particular, the shape parameter $\theta(\E) = z_1(\E)$ is an estimate of the percentage of energy in $E$ due to hard jets. In \Sec{pileup}, we consider perform empirical studies of these types of shape observables.

\subsection{Examples of Novel Shapes}\label{sec:custom_shapes}

In this subsection, we list (some) potentially phenomenologically interesting shape observables, all of which are defined for the first time in this work, that can be constructed using the prescription outlined above. For all these observables, we consider the detector geometry to be a rectangular patch of a cylinder, $\mathcal{X} = [y_{\rm min}, y_{\rm max}] \times [\phi_{\rm min}, \phi_{\rm max}]$, though one could consider the full cylinder as well. These observables are summarized in \Tab{custom_shapes}, and we show empirical examples of these observables in action in \Sec{empirical}. We consider balanced observables with $\beta = 1$, leaving unbalanced observables for future work.

\begin{table}[t]
\centering
\begin{tabular}{|c|c|c|c|}
\hline\hline
\bf Sec. & \bf Shape  &  \bf Specification & \bf Illustration  \\
\hline\hline
    & &  &  \multirow{5}{*}{\raisebox{-7em}{\includegraphics[scale=0.09,page=5]{Figures/shape_illustrations.pdf}}} \\
 \ref{sec:n_ringiness} & {\bf Ringiness} & {\bf Manifold of Rings} & \\ 
 & $\mathcal{O}_R$  & $\E_{x_0,R_0}(x) = \frac{1}{2\pi R_0}$ for $|x-x_0| = R_0$&  \\
&  & $x_0$ = Center, $R_0$ = Radius & \\
& &  &  \\ \hline
& &  &  \multirow{5}{*}{\raisebox{-7em}{\includegraphics[scale=0.09,page=4]{Figures/shape_illustrations.pdf}}} \\
 \ref{sec:n_diskiness} & {\bf Diskiness} & {\bf Manifold of Disks} & \\ 
 & $\mathcal{O}_D$  & $\E_{x_0,R_0}(x) = \frac{1}{\pi R_0^2}$ for $|x-x_0| \leq R_0$&  \\
&  & $x_0$ = Center, $R_0$ = Radius & \\
& &  &  \\ \hline
& &  &  \multirow{5}{*}{\raisebox{-7em}{\includegraphics[scale=0.09,page=3]{Figures/shape_illustrations.pdf}}} \\
 \ref{sec:n_ellipsiness} & {\bf Ellipsiness} & {\bf Manifold of Ellipses} & \\ 
 & $\mathcal{O}_E$  & $\E_{x_0,a,b,\varphi}(x) = \frac{1}{\pi ab}$ for $x \in \text{Ellipse}_{x_0, a, b, \varphi}$&  \\
&  & $x_0$ = Center, $a,b$ = Semi-axes, $\varphi$ = Tilt & \\
& &  &  \\ \hline
& &  &  \multirow{5}{*}{\raisebox{-7em}{\includegraphics[scale=0.09,page=2]{Figures/shape_illustrations.pdf}}} \\
 \ref{sec:plus_pointiness} & {\bf (Ellipse} & {\bf Composite Shape}  & \\ 
 & {\bf $+$Point)iness}  & $\mathcal{O}_E \oplus \tau_1$&  \\
&  & Fixed to same center $x_0$ & \\
& &  &  \\ \hline
& &  &  \multirow{5}{*}{\raisebox{-7em}{\includegraphics[scale=0.09, page=1]{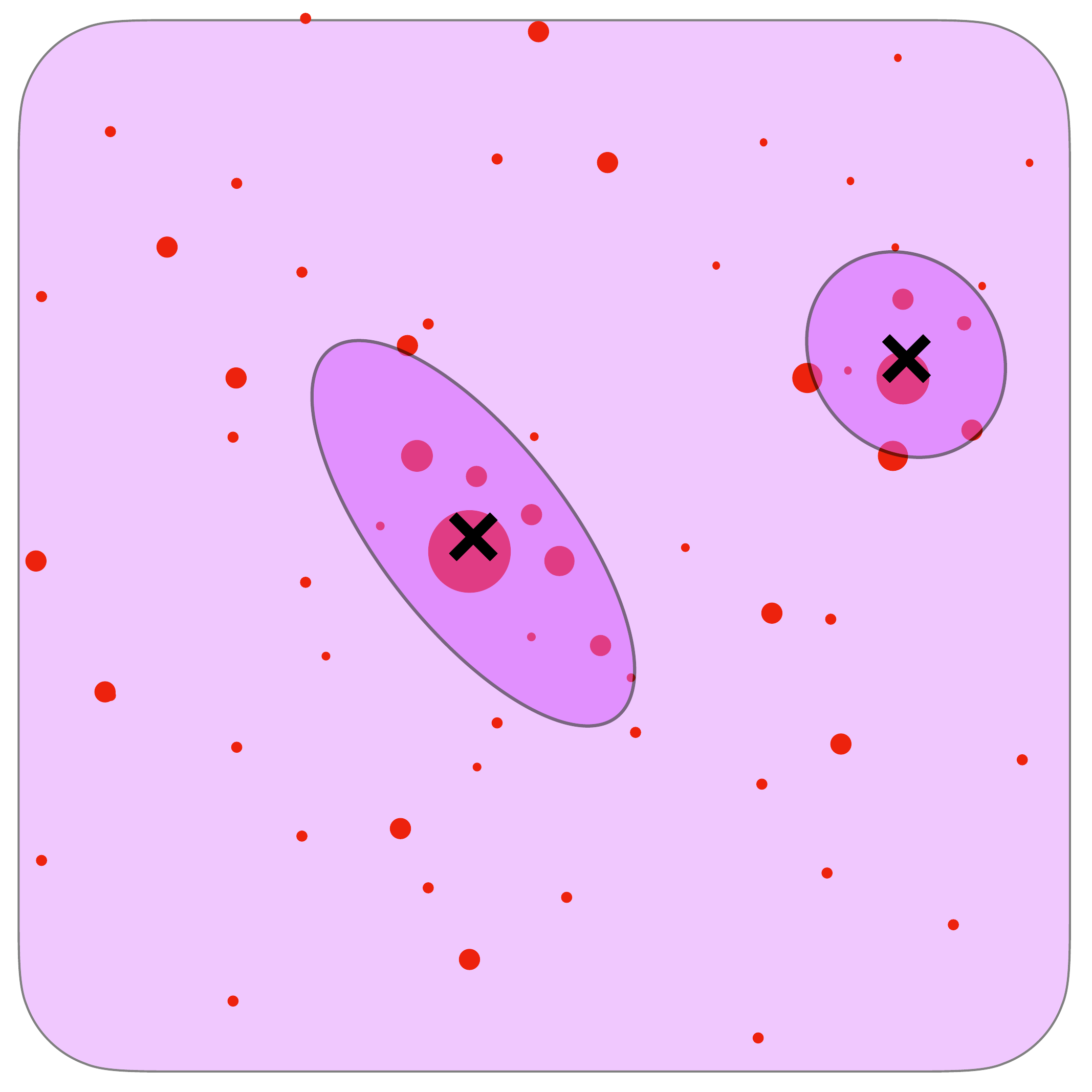}}} \\
 \ref{sec:plus_pileup} & {\bf N-(Ellipse} & {\bf Composite Shape}  & \\ 
 & {\bf $+$Point)iness}  & $N\times(\mathcal{O}_E \oplus \tau_1)\oplus\mathcal{I}$&  \\
&  {\bf $+$Pileup}&  & \\
& & & \\
\hline\hline
\end{tabular}
\caption{\label{tab:custom_shapes}
Custom observables, defined using the \Shaper prescription, designed to probe jet substructure at increasing levels of complexity. For each observable, the manifold parameterization is given, either explicitly, or as a composition of previously defined objects. Here, $\tau_1$ is the one-pointiness (1-subjettiness), and $\mathcal{I}$ is the event isotropy. More details on these types of observables, plus explicit construction of sampling functions, can be found in \Sec{custom_shapes}.
}
\end{table}
\afterpage{\clearpage}

\subsubsection{$N$-Ringiness}\label{sec:n_ringiness}

We first consider a simple shape observable: \emph{ringiness}, which probes how ring-like an event is. We begin by defining the manifold of all ring-like energy flows $\M_R$, which consist of energy flows corresponding to energy flow densities of the form:
\begin{align}
    \E_{x_0, R_0}(x) =\begin{cases} 
      \frac{1}{2\pi R_0} & |x-x_0| = R \\
      0 & \text{otherwise}   
      \end{cases},
\end{align}
where the parameters $x_0$ and $R_0$ correspond to the center and radius of a ring, respectively. To build a sampler, we use the so-called \emph{reparameterization trick}~\cite{https://doi.org/10.48550/arxiv.1506.02557}. Drawing $N$ samples from the unit uniform distribution $\phi \sim U(0,1)$, the distribution of points:
\begin{align}
    x = R_0\left(\cos(2\pi\phi), \sin(2\pi\phi)\right) + x_0, \label{eq:ring_sampler}
\end{align}
where each particle has weight $\frac{1}{N}$, is a realization of $\E_{x_0, R_0}(x)$. 

The corresponding observable, $\mathcal{O}_R(\E)$, is the ringiness of the event $\E$. While most QCD jets are not expected to be ring-like, this observable can identify clumps of radiation scattered around a central point, as may be the case in a 3-pronged top quark decay. Additionally, observables that probe the boundary of a jet with an empty interior may prove useful in studies of the dead-cone effect \cite{Dokshitzer:1991fc, Dokshitzer:1991fd, PhysRevLett.69.3025} where collinear radiation is relatively suppressed.

Having defined ringiness, we can next define $N$-ringiness, which probes how much an event looks like $N$ rings, each of arbitrary center, radius, and weight. This observable is defined as $N\times \mathcal{O}_R$. Following the prescription outlined in \Sec{composing}, we can build $N$ weighted rings, by separately sampling \Eq{ring_sampler} for each ring's center and radius, and multiplying the weights by $z_i \in \Delta_{N-1}$. 

For numerical methods, one must set initial values for these parameters in an IRC-safe way. While in principle, the choice of initialization should make no difference, in practice, the presence of numerical effects and local minima make the choice of initialization important. In our initialization scheme for $N$-rings, we perform $k_T$ clustering to find $N$ subjets. The location of the subjets is then taken to be the ring center, and the subjet energy is taken to be the ring energy. We choose to initialize the radius of each ring to zero, so that the $N$-ringiness is guaranteed to only deviate from $N$-subjettiness only if it will make the event more ringlike, though it is also possible to initialize the radius to e.g.\ the distance of the $(N+1)$-th jet in the clustering history.

\subsubsection{$N$-Diskiness}\label{sec:n_diskiness}

Next, we define \emph{diskiness} $O_D$, which measures how much like a disk an event is. Similar to ringiness, we parameterize the manifold of energy flow densities:
\begin{align}
    \E_{x_0, R_0}(x) =\begin{cases} 
      \frac{1}{\pi R_0^2} & |x-x_0| \leq R \\
      0 & \text{otherwise}   
      \end{cases},
\end{align}
where $x_0$ and $R_0$ are the center and radius of the disk.
 
To build a sampler, as with ringiness, we draw $N$ samples from the unit uniform distribution $\phi \sim U(0,1)$, and also $N$ points $r \sim U(0,1)$. Then, the distribution of points:
\begin{align}
    x = \sqrt{r}R_0\left(\cos(2\pi\phi), \sin(2\pi\phi)\right) + x_0, \label{eq:disk_sampler}
\end{align}
where each particle has weight $\frac{1}{N}$, is a realization of a uniform disk. The random variable $r$ controls the radius of the point being sampled, and the square root is from a Jacobian factor to make the disk uniform. 

Given the diskiness, we can easily compose the $N$-diskiness, $N\times\O_D$. The $N$-diskiness is analogous to the $N$-jettiness/\textsc{XCone} jet algorithm, in that it returns the locations of conical clusters of particles. However, unlike \textsc{XCone}, the radius $R_0$ is a learned, rather than fixed parameter, allowing for dynamic jet radii.\footnote{In place of fixing the radius $R$, as in the \textsc{XCone} algorithm, one instead assumes that the jet energies are uniform across the disk, so the number of assumptions is conserved.} To initialize the $N$-diskiness, the exact same procedure is used as described in \Sec{n_ringiness} for $N$-ringiness. Note that there are many ways to modify the $N$-diskiness to produce similar observables -- for example, one can replace the uniform disks with Gaussians to probe different radiation patterns.

\subsubsection{$N$-Ellipsiness}\label{sec:n_ellipsiness}

Jets need not necessarily be circular! Indeed, many jet algorithms, such as the widely-used $k_T$~\cite{Cacciari:2008gp} and Cambridge-Aachen~\cite{Dokshitzer:1997in, Wobisch:1998wt} algorithms, do not return circular jets. Motivated by this, we define a generalization of diskiness, the \emph{ellipsiness} $\O_E$ of a jet. The manifold of ellipses is given by energy flow densities of the form:
\begin{align}
    \E_{x_0, a, b, \varphi}(\Vec{x}) =\begin{cases} 
      \frac{1}{\pi ab} & \left(\frac{(x-x_0)\cos\varphi + (y-y_0)\sin\varphi}{a}\right)^2 + \left(\frac{(y-y_0)\cos\varphi - (x-x_0)\sin\varphi}{b}\right)^2\leq 1 \\
      0 & \text{otherwise}   
      \end{cases},
\end{align}
$x_0$ is the center of the ellipse, $a$ and $b$ are the semi-major and semi-minor axes,\footnote{non-respectively; $a$ corresponds to the $x$-axis and $b$ to the $y$-axis, and we make no distinction here which is the major versus minor axis.} and $\varphi$ is the tilt of the $x$-axis. Here, we have restored vector notation $\Vec{x}$ to indicate that the $x$-and $y$-axes are treated differently. There are many equivalent alternate parameterizations of the ellipse, including in terms of its focal length $c = \sqrt{\max(a,b)^2 - \min(a,b)^2}$ and eccentricity $e = \sqrt{1 - \frac{\min(a,b)}{\max(a,b)}}$. Note that for $a = b$, the ellipse reduces to a disk, and the $\varphi$ parameter becomes redundant.

The sampling procedure for disks can be recycled for ellipses, with some small modifications. Given $N$ sampled points $\phi, r \sim U(0,1)$, the distribution:
\begin{align}
    x =  U_\varphi \cdot \left(a\sqrt{r}\cos(2\pi\phi ), b\sqrt{r}\sin(2\pi\phi)\right)^T + x_0, \label{eq:ellipse_sampler}
\end{align}
where $U_{\varphi}$ is the $2\times2$ rotation matrix corresponding to                                                                                                                                                                                                                                                                                                                                                                                                                                                                                                                                                                                                                                                                                                                                                                                                                                                                                                                                                                                                                                                                                                                                                                                                                                                                                                                                                                                                                                                                                                                                                                                                                                                                                                                                                                                                                                                                                                                                                                                                                                                                                                                                                                                                                                                                                                                                                                                                                                                                                                                                                                                                                                                                                                                                                                                                                                                                                                                                                                                                                                                                                                                                                                                                                                                                                                                                                                                                                                                                                                                                                                                                                                                                                                                                                                                                                                                                                                                                                                                                                                                                                                                                                                                                                                                                                                                                                                                                                                                                                                                                                                                                                                                                                                                                                                                                                                                                                                                                                                                                                                                                                                                                                                                                                                                                                                                                                                                                                                                                                                                                                                                                                                                                                                                                                                                                                                                                                                                                                                                                                                                                                                                                                                                                                                                                                                                                                                                                                                                                                                                                                                                                                                                                                                                                                                                                                                                                                                                                                                                                                                                                                                                                                                                                                                                                                                                                                                                                                                                                                                                                                                                                                                                                                                                                                                                                                                                                                                                                                                                                                                                                                                                                                                                                                                                                                                                                                                                                                                                                                                                                                                                                                                                                                                                                                                                                                                                                                                                                                                                                                                                                                                                                                                                                                                                                                                                                                                                                                                                                                                                                                                                                                                                                                                                                                                                                                                                                                                                                                                                                                                                                                                                                                                                                                                                                                                                                                                                                                                                                                                                                                                                                                                                                                                                                                                                                                                                                                                                                                                                                                                                                                                                                                                                                                                                                                                                                                                                                                                                                                                                                                                                                                                                                                                                                                                                                                                                                                                                                                                                                                                                                                                                                                                                                                                                                                                                                                                                                                                                                                                                                                                                                                                                                                                                                                                                                                                                                                                                                                                                                                                                                                                                                                                                                                                                                                                                                                                                                                                                                                                                                                                                                                                                                                                                                                                                                                                                                                                                                                                                                                                                                                                                                                                                                                                                                                                                                                                                                                                                                                                                                                                                                                                                                                                                                                                                                                                                                                                                                                                                                                                                                                                                                                                                                                                                                                                                                                                                                                                                                                                                                                                                                                                                                                                                                                                                                                                                                                                                                                                                                                                                                                                                                                                                                                                                                                                                                                                                                                                                                                                                                                                                                                                                                                                                                                                                                                                                                                                                                                                                                                                                                                                                                                                                                                                                                                                                                                                                                                                                                                                                                                                                                                                                                                                                                                                                                                                                                                                                                                                                                                                                                                                                                                                                                                                                                                                                                                                                                                                                                                                                                                                                                                                                                                                                                                                                                                                                                                                                                                                                                                                                                                                                                                                                                                                                                                                                                                                                                                                                                                                                                                                                                                                                                                                                                                                                                                                                                                                                                                                                                                                                                                                                                                                                                                                                                                                                                                                                                                                                                                                                                                                                                                                                                                                                                                                                                                                                                                                                                                                                                                                                                                                                                                                                                                                                                                                                                                                                                                                                                                                                                                                                                                                                                                                                                                                                                                                                                                                                                                                                                                                                                                                                                                                                                                                                                                                                                                                                                                                                                                                                                                                                                                                                                                                                                                                                                                                                                                                                                                                                                                                                                                                                                                                                                                                                                                                                                                                                                                                                                                                                                                                                                                                                                                                                                                                                                                                                                                                                                                                                                                                                                                                                                                                                                                                                                                                                                                                                                                                                                                                                                                                                                                                                                                                                                                                                                                                                                                                                                                                                                                                                                                                                                                                                                                                                                                                                                                                                                                                                                                                                                                                                                                                                                                                                                                                                                                                                                                                                                                                                                                                                                                                                                                                                                                                                                                                                                                                                                                                                                                                                                                                                                                                                                                                                                                                                                                                                                                                                                                                                                                                                                                                                                                                                                                                                                                                                                                                                                                                                                                                                                                                                                                                                                                                                                                                                                                                                                                                                                                                                                                                                                                                                                                                                                                                                                                                                                                                                                                                                                                                                                                                                                                                                                                                                                                                                                                                                                                                                                                                                                                                                                                                                                                                                                                                                                                                                                                                                                                                                                                                                                                                                                                                                                                                                                                                                                                                                                                                                                                                                                                                                                                                                                                                                                                                                                                                                                                                                                                                                                                                                                                                                                                                                                                                                                                                                                                                                                                                                                                                                                                                                                                                                                                                                                                                                                                                                                                                                                                                                                                                                                                                                                                                                                                                                                                                                                                                                                                                                                                                                                                                                                                                                                                                                                                                                                                                                                                                                                                                                                                                                                                                                                                                                                                                                                                                                                                                                                                                                                                                                                                                                                                                                                                                                                                                                                                                                                                                                                                                                                                                                                                                                                                                                                                                                                                                                                                                                                                                                                                                                                                                                                                                                                                                                                                                                                                                                                                                                                                                                                                                                                                                                                                                                                                                                                                                                                                                                                                                                                                                                                                                                                                                                                                                                                                                                                                                                                                                                                                                                                                                                                                                                                                                                                                                                                                                                                                                                                                                                                                                                                                                                                                                                                                                                                                                                                                                                                                                                                                                                                                                                                                                                                                                                                                                                                                                                                                                                                                                                                                                                                                                                                                                                                                                                                                                                                                                                                                                                                                                                                                                                                                                                                                                                                                                                                                                                                                                                                                                                                                                                                                                                                                                                                                                                   the angle $\varphi$ and each particle has weight $\frac{1}{N}$, is a realization of a uniform ellipse. We can then easily compose the $N$-ellipsiness, which can serve as a jet algorithm that finds non-circular jets. In particular, this shape observable allows for the eccentriciy $e$ of the clustered jets to be extracted, allowing one to quantify how far from circular each jet is. As with the $N$-ringiness and $N$-diskiness, the centers of the ellipses are chosen using the $k_T$ clustering algorithm. Both $a$ and $b$ are initialized to be zero, so that deviations from either $N$-subjettiness or $N$-diskiness occur if it makes the event more elliptical.

\subsubsection{... Plus Pointiness}\label{sec:plus_pointiness}

Energy is not uniformly distributed within a jet! Indeed, to leading order in perturbative QCD, much of a jet's radiation will be soft and/or collinear with respect to the emitting parton. We can probe this by composing together shapes that explicitly target soft and collinear radiation separately. To this end, we construct a set of new observables, the (\textit{shape}$+$point)iness, for $\textit{shape} \in \{\O_R, \O_D, \O_E\}$. This is defined using the shape composition prescription described in \Sec{composing}, as:
\begin{align}
    \mathcal{O}_i^{\tau} = \O_i \oplus \tau_1,
\end{align}
where $\tau_1$ is the 1-pointiness (equivalently, the 1-subjettiness), and $\O_i$ is any of $\{\O_R, \O_D, \O_E\}$ previously defined. Importantly, we fix the location of the $\delta$-function in $\tau_1$ to be $x_0$, though one may consider letting the location of the $\delta$-function float to define a recoil-free variant.\footnote{In the elliptical case, one may consider attaching $\delta$-functions to one or both of the focii instead of the center. We leave the study of variants of these observables to future work.} We can then extend this definition to compose the $N$-(shape$+$point)iness. 

When used as a jet algorithm, the $N$-(shape$+$point)iness provides a more physical picture of perturbative QCD than do the previously defined shapes. The base shapes, particularly disks and ellipses, capture wide-angle soft radiation, while the $\delta$-functions capture both hard and soft collinear radiation at the center of the (sub)jet. Moreover, within each shape-point pair, the floating parameters $z_1$ and $z_2$ tell us the fraction of radiation in the wide-angle and collinear sectors, which in principle can be calculated in and compared to perturbative QCD.

When initializing observables of this type, the initialization occurs as described in previous sections using the $k_T$ algorithm with all radii set to zero. However, we choose to split the $k_T$ energy equally between the shape and the $\delta$-function. Note that this is an IRC-safe choice, since at zero radius, the shape is indistinguishable from the $\delta$-function.

\subsubsection{... Plus Pileup}\label {sec:plus_pileup}

In hadron-hadron collisions, there are many sources of contamination in jets, including underlying event contributions from proton remnants~\cite{PhysRevD.65.092002, Agocs:2010ft}, and pileup due to simultaneous hadron collisions~\cite{Soyez:2018opl}. We will collectively refer to these sources of contamination as \emph{pileup} for simplicity. Pileup contamination biases and smears the ``true'' value of observables reconstructed from final state particles, driving the need for mitigation techniques. 

Pileup is approximately uniformly distributed in the rapidity-azimuth plane. This is exactly the shape probed by the event isotropy, $\mathcal{I}$. Thus, in order to protect shape observables against pileup contamination, we can compose them with the event isotropy, which will soak up radiation uniform in the plane. This defines the \emph{shapines$+$pileup} observable:
\begin{align}
    \O^\mathcal{I}_i = \O_i \oplus \mathcal{I},
\end{align}
where $\O_i$ is any shape observable, including those previously defined. As a departure from \Refer{Cesarotti:2020hwb}, we realize the uniform event by randomly sampling in the plane, rather than defining a grid. We also primarily focus on the $\beta = 1$ event isotropy. Unlike mitigation techniques such as area subtraction~\cite{Cacciari_2008, Cacciari_2008_2, Soyez:2012hv} or jet grooming~\cite{Larkoski:2014wba, Dasgupta_2013}, where an implicit assumption is made about the pileup energy density (either explicitly as an input $\rho$, or implicitly through a soft scale $z_{\rm cut}$), the shape observable $\O^\mathcal{I}_i$ makes no explicit energy scale assumptions.%
\footnote{Of course, via the choice of $\mathcal{I}$, we are still making an explicit assumption about the shape of the pileup distribution, even if the overall energy scale is learned.} The uniform energy weight, $z_2$, is optimized over, and so the observable ``learns'' its own pileup scale, which can then be extracted. We choose to initialize the pileup scale $z_2$ to zero, though one could choose any value of $z_2$ if they had a prior on the amount of pileup in events.

\subsubsection{... And More!}

This has \emph{not} been an exhaustive list -- one can use any manifold $\M$ of energy flows one can think of, with the only two limits being imagination and the ability to write down a sampling procedure. Other examples of shapes include polygons, hardcoded jet topologies (for example, two-pronged jets restricted to between $\Delta R = R_1$ and $R_2$ apart), Gaussian clusters, graph-based shapes, and so on. These observables can also be combined into more complex ones using shape composition. All of these can be constructed within the \Shaper framework (more details in \Sec{SHAPER}), and we encourage the community to use this prescription to develop their own observables.

\section{The \Shaper Framework}\label{sec:SHAPER}

Calculating the Wasserstein metric in \Eq{EMD} is notoriously difficult; if both events have $n$ particles, then the runtime needed by a brute force, generic Wasserstein solver can be as high as $\mathcal{O}(n^3\log{n})$ \cite{alt_17}. Generic solvers also make it difficult to extract the gradients of the metric with respect to one of the events, $\nabla_{\E}\EMD(\E, \E')$, which are necessary for performing gradient descent over the space of events in $\M$.  Fortunately, by using the (de-biased) Sinkhorn divergence, which uses an $\epsilon$-regularization to approximate the Wasserstein metric, the total costs can be lowered all the way down to $\mathcal{O}({n^2}\log{n})$~\cite{sink_n2ln(n), Sink_ICML_19, sink_nlogn, sink_n_comp, wass_n3_input_dist, feydy2019interpolating}.

In this section, we introduce the \textbf{S}hape \textbf{H}unting \textbf{A}lgorithm for \textbf{P}arameterized \textbf{E}nergy \textbf{R}econstruction -- or \Shaper\ -- to define and calculate shape observables. \Shaper is a {\sc Pytorch}-enabled~\cite{NEURIPS2019_9015} and parallelized computational framework for defining and composing shape observables and their corresponding energy flow manifolds, built using the \textsc{geomloss}~\cite{feydy2019interpolating} package. We start by outlining the \Shaper algorithm. Then, we provide details on the Sinkhorn divergence, before ending this section with implementation details. For the rest of this paper, we restrict ourselves to balanced observables, i.e. $E_{\theta, {\rm tot}} = E_{\rm tot} = 1$, leaving the unbalanced case for future work.

\subsection{The \Shaper Algorithm}\label{sec:algorithm}

We now describe how to perform the minimization $\paramin_{\E'_\theta \in \M}\left[\EMD(\E, \E'_\theta)\right]$ using \Shaper.\footnote{\textsc{NEEMo}~\cite{Kitouni:2022qyr} is another differentiable EMD estimator that works by parameterizing the space of Lipschitz-Kantorivich potentials.} The \Shaper algorithm for estimating shape observables on an event is as follows:

\begin{enumerate}
    \item \textbf{Define}: Following the prescription of \Sec{observables}, define a manifold $\M$ and coordinates $\theta$ parametrizing the manifold. Define the ground metric $d(x,y)$, the exponent $\beta$, and the radius $R$. This fully defines the observable $\O$. Build a sampling function $p_\theta$ that uses the parameters $\theta$ to transform some base distribution into a realization of the energy flows $\E'_\theta \in \M$. Finally, choose an approximation parameter $\epsilon \ll 1$ and an annealing parameter $\Delta \in (0, 1)$.
    \item \textbf{Initialize}: For each event $\E$, choose initial parameters $\theta$. This initialization should be done in an IRC-safe way. 
    \item \textbf{Compute the EMD}: Compute the de-biased Sinkhorn divergence, $S_\epsilon(\E,\E')$, as defined in \Sec{sinkhorn} below, as an estimate of the EMD. Save the corresponding de-biased Kantorovich potentials, $F$ and $G$. 
    \item \textbf{Gradient Update}: Perform the gradient update:
    \begin{align}
        \theta \xleftarrow{} \theta - \alpha \left(\sum_{j=1}^M G(y_j)\pdv{E'_j}{\theta} + \sum_{j=1}^ME'_j\pdv{G(y_j)}{y_j}\pdv{y_j}{\theta} \right),
    \end{align}
    where $\alpha$ is a learning rate hyper-parameter. The first term is the dependence of the EMD on particle energies due to $\theta$, and the second is the dependence due on particle positions due to $\theta$, both of which are implicit through the sampling function $p_\theta$. This step can be replaced with any other gradient descent optimizer.
    \item \textbf{Constrain}: If the manifold $\M$ is nontrivial, impose any necessary constraints on the coordinates $\theta$, such as wrapping angles between $-\pi$ and $\pi$, enforcing positivity, or a simplex projection. 
    \item \textbf{Converge}: Repeat Steps 3--5 until convergence. Return the final value of the $\EMD$ and the final $\theta$ parameters. 
\end{enumerate}
The \Shaper framework contains modules to aid or automate each of these steps, which we describe further in \Sec{implementation}.

\subsection{The Dual Formulation of Wasserstein}

Observe that the EMD in \Eq{EMD} falls into a generic class of problems called \emph{linear programs}. A linear program involves minimizing a function $\mathcal{L}(x) = \expval{c, x}$ over vectors $x$, where $c$ is some cost function linear in $x$. Furthermore, $x$ satisfies some linear constraint of the form $b = Ax$, and we additionally require $x \geq 0$. In our case, $x$ is the (flattened) transfer matrix $\pi$, $c$ is the (flattened) distance matrix $d^\beta$, $b = (\E, \E')$ are the energy flows, and $A$ is a matrix enforcing the simplex constraints on $\pi$. 

The theory of linear programs is well-studied \cite{gartner_06}. In particular, for every \emph{primal} linear program, there exists a \emph{dual} linear program, where the constraints and variables to be optimized switch roles, similar to the method of Lagrange multipliers. In the dual problem, one instead maximizes the function $\mathcal{L}(y) = \expval{b, y}$, subject to $A^T y \leq c$.\footnote{For problems showcasing strong duality, the existence of an optimal solution for the primal problem implies the existence of an optimal solution for the dual problem. The problems we consider in this work admit strong duality. See \Refer{Vilani_03} for a mathematically rigorous discussion.} For the Wasserstein metric, the dual formulation looks like:
\begin{align}
    \EMD(\E, \E')^{(\beta, R)} &= \max_{f,g:\mathcal{X}\to\mathbb{R}}\left[\expval{\E, f} + \expval{\E', g}\right], \text{ such that }f(x) + g(y) \leq \frac{1}{\beta R^\beta}d(x,y)^\beta, \label{eq:dual}
\end{align}
where $f$ and $g$ are known as the \emph{dual potentials} or \emph{Kantorovich potentials}. This formulation of the EMD is known as the \emph{Kantorovich–Rubinstein} metric \cite{Vilani_03}. 

In this form, the EMD has several nice properties. First, the arguments $\E$ and $\E'$ are explicit, rather than implicit in the form of constraints. This makes taking the gradient of the EMD with respect to either energy flow much easier. This property is incredibly useful for performing optimizations over energy flows, since it enables easy differentiation. Second, the optimization over an $MN$-dimensional object, $\pi_{ij}$, is replaced by an optimization over the $(M+N)$-dimensional object, $f_i$ and $g_j$, making the simplex constraint structure more apparent. It can be shown that the optimal choice of $f$ and $g$ actually saturates the bound in \Eq{dual} \cite{feydy_20}. Recalling that the ground metric satisfies $d(x,x) = 0$ for all $x$, we can see that the optimal $f, g$ pair satisfies $f(x) = - g(x)$. This allows us to rewrite the constraint as:
\begin{align}
    |f(x) - f(y)| \leq \frac{1}{\beta R^\beta}d(x,y)^\beta.
\end{align}
That is, $f$ is $\beta$-Hölder continuous. Note that for $\beta=1$, this reduces to Lipschitz continuity on $f$.

\subsection{Reviewing the Sinkhorn Divergence}\label{sec:sinkhorn}

The source of the difficulty in evaluating \Eq{dual} is the highly nonconvex optimization. To alleviate this, we introduce a regulator~\cite{cuturi2013sinkhorn, CLASON2021124432} to the dual Wasserstein metric:
\begin{align}
    {\rm OT}^{(\beta,R)}_\epsilon(\E, \E') = \max_{f, g:\mathcal{X}\xrightarrow[]{}\mathbb{R}} \Biggl[&\expval{\E, f} +\expval{\E',g} \nonumber\\&- \epsilon^\beta\log\expval{\E(x) \otimes \E'(y), e^{\left(\frac{1}{\epsilon^\beta}(f(x) + g(y) - \frac{1}{\beta R^\beta}d(x,y)^\beta)\right)}}\Biggr] , \label{eq:sinkhorn_dual}
\end{align}
where $\epsilon$ is a regulation parameter. The quantity ${\rm OT}^{(\beta,R)}_{\epsilon}(\E, \E')$ is known as the \emph{Sinkhorn divergence}~\cite{sinkhorn_1966} between measures $\E$ and $\E'$. It reduces to the EMD as $\epsilon\to0$.\footnote{In the $\epsilon \to \infty$ limit, we recover instead the maximum mean discrepancy (MMD) \cite{ramdas2017wasserstein}, another potential metric on collider events, which we have shown in \Sec{wasserstein} is not faithful.}  Notably, for any $\epsilon > 0$, the optimization over $f$ and $g$ is fully convex~\cite{feydy_20}, making the minimum significantly easier to evaluate. 

%The regularization term can be thought of as the thermodynamic free energy of a system at a temperature $k_BT = \epsilon^\beta$ consisting of states labeled by $(x,y)$ with ``energy'' $f(x) + g(y) - \frac{d(x,y)^\beta}{\beta R^\beta}$ and degeneracy $\E(x)\E'(y)$. Here, the term $d(x,y)$ can be thought of as an ``interaction potential'' between states $x$ and $y$ of two component subsystems.

Note that there are \emph{no} constraints on the functions $f$ and $g$ anymore. Instead, the maximum will only be achieved when $f(x) + g(y)$ is within order $\epsilon^\beta$ of $\frac{1}{\beta R^\beta}d(x,y)^\beta$, a softer version of the original simplex constraint. We can view the parameter $\epsilon$ as ``blurring'' the distance metric $d(x,y)$, where $\epsilon$ is a distance scale measured in units of $R$.\footnote{Even though the $R$ parameter is unimportant for calculating the exact Wasserstein metric for balanced observables, beyond defining a unit scale, its importance re-emerges when defining the blurring scale $\epsilon$.} 

% Given a solution $f, g$ to the dual formulation of the Sinkhorn divergence, we can reconstruct the optimal transport matrix (density) $\pi$ of the primal formulation:

% \begin{align}
%     \pi(x,y) = \E(x) \E'(y) e^{\frac{1}{\epsilon}(f(x)+g(y) - \frac{1}{\beta R^\beta}d(x,y)^\beta)}.
% \end{align}

As an unconstrained, convex minimization problem, we can estimate the Sinkhorn divergence using simple gradient descent by taking derivatives of \Eq{sinkhorn_dual} with respect to $f$ and $g$. Given two atomic measures, $\E$ and $\E'$, with $M$ and $N$ particles respectively, and an approximation parameter $\epsilon$, we estimate the Kantorovich potentials $f(x_i)$ and $g(y_j)$ that give us the Sinkhorn divergence using the following algorithm:
\begin{enumerate}
    \item \textbf{Initialize}: Initialize $f(x_i) = 0$ and $g(y_j)$ = 0.
    \item \textbf{Gradient Update}: Update $f$ and $g$ simultaneously as follows:
    \begin{align}
    f(x_i) &\xleftarrow{} -\epsilon^\beta \log( \sum_{j=1}^N E'_j e^{\left(\frac{1}{\epsilon^\beta}(g(y_j) - \frac{1}{\beta R^\beta}d(x_i,y_j)^\beta)\right)} ),\\
    g(y_j) &\xleftarrow{} -\epsilon^\beta \log( \sum_{i=1}^M E_i e^{\left(\frac{1}{\epsilon^\beta}(f(x_i) - \frac{1}{\beta R^\beta}d(x_i,y_j)^\beta)\right)} ).
    \end{align}
    \item \textbf{Converge}: Repeat Step 2 until convergence. Return the Kantorovich potentials $f$ and $g$, and the Sinkhorn Divergence \Eq{sinkhorn_dual} evaluated on these potentials.
\end{enumerate}
This algorithm is known to converge in finite time~\cite{sinkhorn_1966, sinkhorn1967diagonal, sinkhorn1967concerning}. The runtime of each iteration scales as approximately $\mathcal{O}((M+N)^2)$, and the algorithm converges in approximately $\frac{1}{\epsilon^\beta}$ iterations. This can be further improved to only $\log(\frac{1}{\epsilon^\beta}) / \log(\frac{1}{\Delta})$ iterations through the use of simulated annealing~\cite{KOSOWSKY1994477, Bertsekas}, with a parameter $\Delta \in (0,1)$. Beginning with a larger effective blurring radius $\epsilon' = 2R$, after every iteration of the Sinkhorn algorithm, we decrease $\epsilon' \xleftarrow{} \Delta \epsilon'$, until finally reaching $\epsilon' = \epsilon$. Intuitively, we start with a large blurring scale $R$, and slowly ``zoom in'' to a distance scale of $\epsilon$ to refine the estimate of the Sinkhorn divergence.

However, the Sinkhorn divergence is biased, meaning that it is not generically the case that ${\rm OT}_\epsilon(\E', \E') = 0$, which is an important property of the Wasserstein metric. We use therefore the de-biased Sinkhorn divergence, defined in \Refer{feydy2019interpolating}:
\begin{align}
    S_{\epsilon}(\E, \E') = {\rm OT}_\epsilon(\E, \E') - \frac{1}{2}{\rm OT}_\epsilon(\E, \E) - \frac{1}{2}{\rm OT}_\epsilon(\E', \E').
\end{align}
The de-biased Sinkhorn divergence satisfies $S_{\epsilon}(\E,\E) = 0$ by construction. This can be easily realized algorithmically by simply substituting new de-biased Kantorovich potentials $f \to F$ and $g \to G$, where
\begin{align}
    F(x) &= f(x) - \Tilde{f}(x), \\
    G(y) &= g(y) - \Tilde{g}(y).
\end{align}
Here, the notation $\Tilde{f}(x)$ refers to the first Kantorovich potential corresponding to $\text{OT}_{\epsilon}(\E,\E)$, and $\Tilde{g}(y)$ refers to the second Kantorovich potential corresponding to $\text{OT}_{\epsilon}(\E',\E')$. For the rest of this paper, we refer to the de-biased Sinkhorn divergence as simply the Sinkhorn divergence wherever there is no chance for confusion.

In addition to returning the Sinkhorn divergence, this algorithm also returns the Kantorovich potentials, which allow us access to approximate gradients of the EMD, which can be used for shape parameter optimization. The gradients of the EMD with respect to the input measures can be read off of \Eq{sinkhorn_dual}:
\begin{align}
    \nabla_{\E} \EMD(\E, \E') &= F \quad\Rightarrow{}\quad \begin{cases}
        \quad\nabla_{E_i} \EMD(\E, \E') = F(x_i),\\
       \quad\nabla_{x_i} \EMD(\E, \E') = E_i \nabla F(x_i),
    \end{cases} \nonumber\\
    \nabla_{\E'} \EMD(\E, \E') &= G \quad\Rightarrow{}\quad  \begin{cases}
        \quad\nabla_{E_j'} \EMD(\E, \E') = G(y_j),\\
       \quad\nabla_{y_j} \EMD(\E, \E') = E_j' \nabla G(y_j).
    \end{cases}
    \label{eq:emd_grads_2}
\end{align}

\subsection{Implementation Details}\label{sec:implementation}

Before turning to our case study, we discuss the specific details of the \Shaper implementation.
\Shaper uses the \textsc{geomloss} package as a backend for computing Sinkhorn divergences, as described in \Sec{sinkhorn}. By default, we use a relatively conservative annealing value of $\Delta = 0.9$, and choose $\epsilon^\beta = 10^{-3}$ for our estimates.\footnote{See \Secs{benchmarking_sinkhorn}{benchmarking_optimization} for studies involving the choice of $\epsilon$.} Currently, the \Shaper algorithm is implemented only for balanced observables, with either $\beta = 1$ or $2$.

Once the EMD and Kantorovich potentials are estimated using the \textsc{geomloss} backend, the gradient updates in Step 4 are then handled using automatic differentiation and backpropagation in \textsc{pytorch}. One may select from a suite of common machine learning optimizers to perform the gradient update -- by default, we use the \Adam optimizer~\cite{https://doi.org/10.48550/arxiv.1412.6980}, with a learning rate of 0.01. Steps 3--5 can be easily parallelized over batches of hundreds of events at once. This is accomplished by treating the parameters $\theta$ for each $\E_n$ to be completely independent. The Sinkhorn divergence can be computed on many events at once, and since the parameters are independent, we take advantage of highly parallelized \textsc{pytorch} operations to perform independent derivatives $\nabla_{\theta_n}$ of the combined batch loss, $\sum_n \EMD(\E_n, \E'_{\theta_n})$, all at once.

When using \Shaper, one must specify a maximum number of epochs, so that the program eventually halts even if convergence is not achieved. We set this number by default to be 500 epochs, though we observe that convergence happens far earlier than this. We define convergence through an early stopping procedure: if an event's EMD has not decreased in at least $N_{\max}$ epochs, stop early, and return the minimum $\EMD$ ever achieved during the training, and the parameters that achieved that minimum. When training on a large batch of events, we stop early when a certain fixed percentage (we choose 95\% by default) have hit this condition -- this is because there tends to be a handful of outlier events that take exceptionally long to converge. 
We choose $N_{\max} = 25$ epochs by default. Both this, and the batch stopping percentage, are adjustable user parameters.

To facilitate Steps 1 and 2 of the \Shaper algorithm, where the user input occurs, many common manifolds, such as sets of $N$ points, hypercubes, simplices, and so on, are already pre-built. These pre-built ``building-block'' manifolds are listed in \Tab{manifolds}.\footnote{Note that these manifolds are the \emph{parameters} from which shapes $\E_{\theta}$ can be defined, not the shapes themselves. For example, a circular shape is defined by a point parameter (the center) and a positive real parameter (the radius), whereas the circle manifold $S^1$ whenever an angle parameter is needed to define a shape.} From these, more complex manifolds can be easily constructed, such as the rings, disks, and ellipses described in \Sec{custom_shapes}. The $\oplus$ and $N\times$ operators make it very easy to define new composite observables from old ones and quickly build sophisticated shapes. Parameters can also be easily frozen for more customization. Furthermore, it is straightforward to define more building-block manifold as needed within the framework. Each of the observables described in \Sec{custom_shapes} is also pre-built into \Shaper. When defining custom manifolds that use the $N$-points as a building block, the custom shape will automatically use the same IRC-safe $k_T$-clustering initialization scheme as the observables in \Sec{custom_shapes}, though it is possible  to modify the initialization scheme as needed.

\begin{table}[tp]
\centering
\begin{tabular}{|c|c|c|}
\hline\hline
\bf Manifold & \bf Description  &  \bf Constraints  \\
\hline\hline
  Trivial & The set $\{0\}$ & None \\
  $N$-Points, $\mathcal{X}^N$ & Set of $N$ points, $x_i$, in $\mathbb{R}^2$ & None \\
  Positive Reals, $\mathbb{R}^N_{\geq 0}$ & Set of $N$ positive numbers, $R_i$ & Clipped to $R_i \geq 0$ \\
  Hypercube, $\Lambda^N$ & Set of $N$ numbers, $\lambda_i$, between 0 and 1 & Clipped to $0 \leq \lambda_i \leq 1$ \\
  Circle / Torus, $S^N$ & Set of $N$ angles, $\phi_i$ & Wrapped to $-\pi \leq \phi_i < \pi$ \\
  $N$-Simplex, $\Delta_N$ & Set of $N+1$ numbers, $z_i \geq 0$, summed to 1 & Simplex projection\\
\hline\hline
\end{tabular}
\caption{\label{tab:manifolds}
The building-block manifolds implemented in \Shaper, from which parameters can be defined and more complex manifolds can be built. For each manifold, a description is given, and constraint instructions are provided to project points onto the manifold.
Note that we do not distinguish here between ``positive'' reals versus ``non-negative'' reals, since for continuous parameterizations there is no difference between including zero and getting arbitrarily close to zero.}
\end{table}

Each manifold contains instructions on how to enforce parameter constraints (as needed for Step 5). For most manifolds, this usually involves clipping the values to be within a desired range, though for the simplex $\Delta_N$, which occurs in almost every shape for which energy weights must be normalized, the enforcement is nontrivial. We use a simplex projection algorithm, inspired by the $K$-Deep-Simplex framework~\cite{tankala2020k} and its nonlinear extension \cite{mueller2022geometric}, which solves the linear program~\cite{https://doi.org/10.48550/arxiv.1309.1541}:
\begin{align}
    \min_{z_i}\left[\sum_i^{N+1}|z_i-y_i|^2\right] \text{ such that } \sum_i^{N+1} z_i = 1, z_i \geq 0,
\end{align}
which finds the simplex $z_i \in \Delta_N$ closest to a set of unnormalized points $y_i$.

\section{Empirical Studies with Jets}\label{sec:empirical}

We now use the \Shaper framework and custom shape observable in example collider physics analyses.
We begin by benchmarking the \Shaper algorithm, testing the performance of the Sinkhorn divergence and the optimization procedure by comparing the jet isotropy and $N$-subjettiness calculated using \Shaper to other methods.
% We also show visualizations of the learned energy and position gradients on a test event.
%
Next, we calculate the ring, disk, and ellipse-based observables defined in \Sec{custom_shapes} for a dataset of top and QCD jets, showing visualizations of each shape and analyzing the learned EMD's and parameters. Finally, we explore the potential to use shape observables for automatic pileup removal.

\subsection{Dataset}\label{sec:dataset}

For our empirical studies, we use the top tagging benchmark of \Refers{Butter:2017cot,Kasieczka:2019dbj}, which is a dataset consisting of a top quark jet signal and a mixed light-quark/gluon jet background. These samples are generated in {\sc Pythia} 8.2.15~\cite{Sjostrand:2014zea} at 14 TeV, and then passed through {\sc Delphes} 3.3.2~\cite{deFavereau:2013fsa} to simulate the ATLAS detector. Jets are defined using the anti-$k_T$ algorithm~\cite{Cacciari:2008gp} in \FastJet 3.1.3~\cite{Cacciari:2011ma} with $R = 0.8$. Only the leading jet in any event is considered, and we select jets satisfying $p_{T,J} \in [475,525]\,\GeV$ and $|\eta_J| < 2$. The signal and background samples are generated using $t\Bar{t}$ and QCD dijet events respectively. For signal top jets, a top parton, plus its decay products, are required to be within $\Delta R = 0.8$ of the jet axis. All events are translated such that the jet axis is at $(0,\,0)$ on the rapidity-azimuth plane.

In this dataset, multiple parton interactions and pileup have not been included. To mock up the effects of pileup contamination in data, we add in pileup ``by hand''. To each event, we add in $N$ particles randomly distributed in an $R\times R = 0.8\times0.8$ square centered at the origin on the rapidity-azimuth plane, where $N$ is Poisson-distributed with a mean of 75. Each particle is given an energy weight randomly sampled from a normal distribution with mean $\frac{E_{\rm PU}}{N}$ and standard deviation $\frac{E_{\rm fluct}}{N}$, where we take $E_{\rm PU}$, which represents the total amount of pileup radiation, to be uniformly distributed between $50$ and $250$ GeV, and $E_{\rm fluct}$, which represents per-particle
fluctuations, to be $25$ GeV.\footnote{A floor of 0 GeV is set to avoid negative energies.} Many refinements of this simplistic mockup could be considered, but this suffices to show qualitative features of \Shaper and the shape observables defined in \Sec{custom_shapes}. For the purposes of calculating shape observables, all jets are normalized such that $E_{\rm tot} = 1$, though we save the original total energy of each jet for the purpose of restoring units.

\begin{figure*}[t]
    \centering
    \subfloat[]{
         \includegraphics[width=0.45\textwidth]{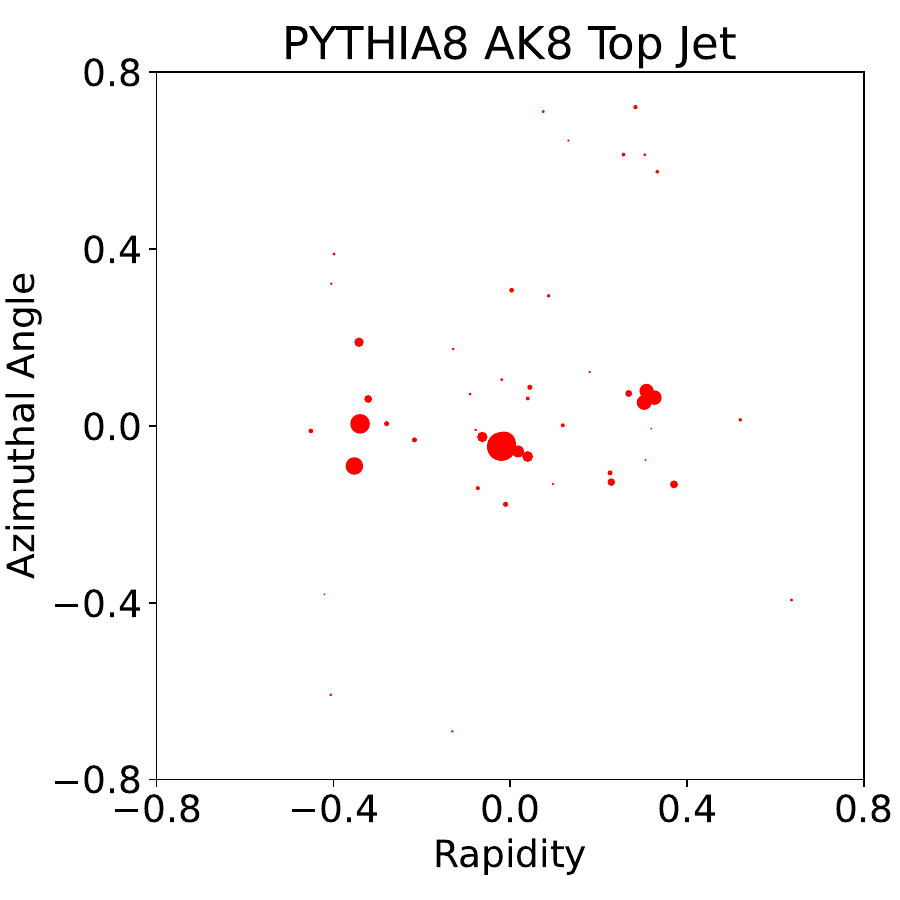}
        \label{fig:dataset_a}
    }
    $\quad$
    \subfloat[]{
        \includegraphics[width=0.45\textwidth]{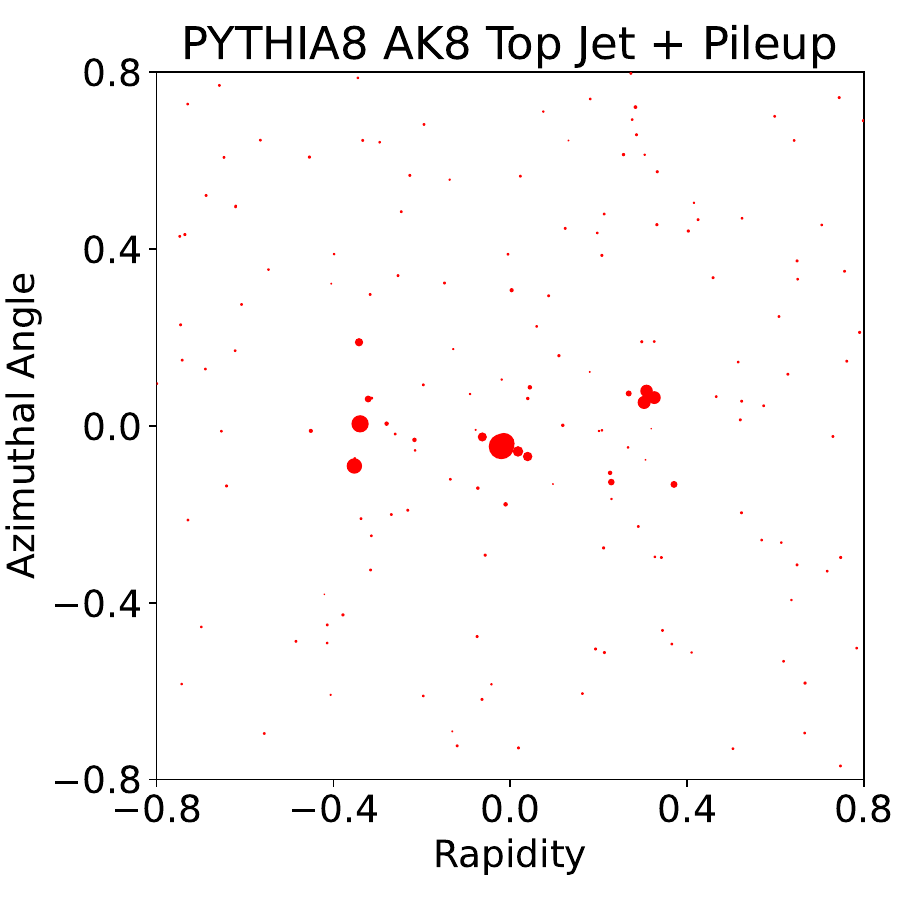}
        \label{fig:dataset_b}
    }
    \caption{
        Example of a top jet (signal) event in our dataset, (a) before and (b) after the pileup mock-up procedure described in \Sec{dataset}. Each red point is a single particle, and the size of the point is proportional to the relative energy of the particle.
        }
    \label{fig:dataset}
\end{figure*}

An example top jet is shown in \Fig{dataset}, before and after pileup is added, to illustrate this procedure.
This contamination procedure is performed for the benchmarking studies in \Secs{benchmarking_sinkhorn}{benchmarking_optimization} and the pileup studies in \Sec{pileup}. For the jet substructure studies in \Secs{jets}{jets_parameters}, we do not add any pileup, and instead we require that the jets have an invariant mass $m_J \in [145, 205]\,\GeV$ to more closely match the analysis conditions of \Refer{Thaler:2010tr}.

\subsection{Benchmarking Sinkhorn: Jet Isotropy}\label{sec:benchmarking_sinkhorn}

We first use \Shaper to compute the jet isotropy for the purposes of benchmarking the Sinkhorn divergence for runtime and accuracy. Jet isotropy is an ideal benchmark since the minimization is trivial; for balanced isotropy, the parameterized manifold consists only of a single event. Therefore, no gradient descent is necessary, and this is purely a test of the Sinkhorn approximation. This can be viewed as a proxy for the per-epoch runtime and accuracy of the \Shaper algorithm. 

We compute the $n\times n$ jet isotropy, which is the isotropy given by computing the EMD to the uniform event:
\begin{align}
    \mathcal{U}_{n\times n} \sim \text{ particles in an } R\times R  \text{ square, arranged in a uniform } n\times n \text{ grid,}
\end{align}
as defined in \Refer{Cesarotti:2020hwb}. We do this using \Shaper with many different values of $\epsilon$, and compare to the same calculation done using the Python Optimal Transport (POT)~\cite{flamary2021pot} implementation of the EMD, which was used in \Refers{Cesarotti:2020hwb, Cesarotti_2021}.

\begin{figure*}[tp]
    \centering
    \subfloat[]{
         \includegraphics[width=0.5\textwidth]{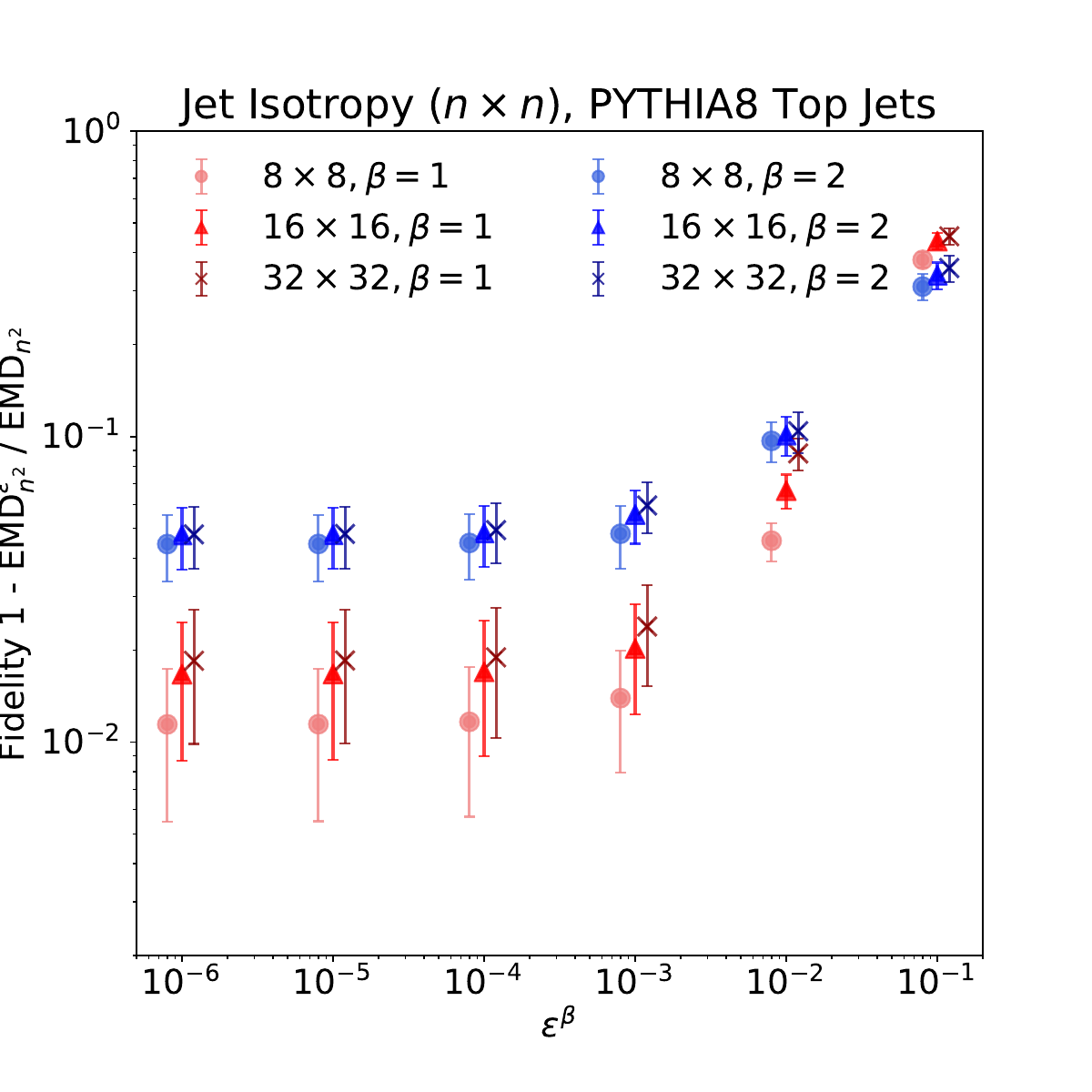}
        \label{fig:iso1}
    }
    \subfloat[]{
        \includegraphics[width=0.5\textwidth]{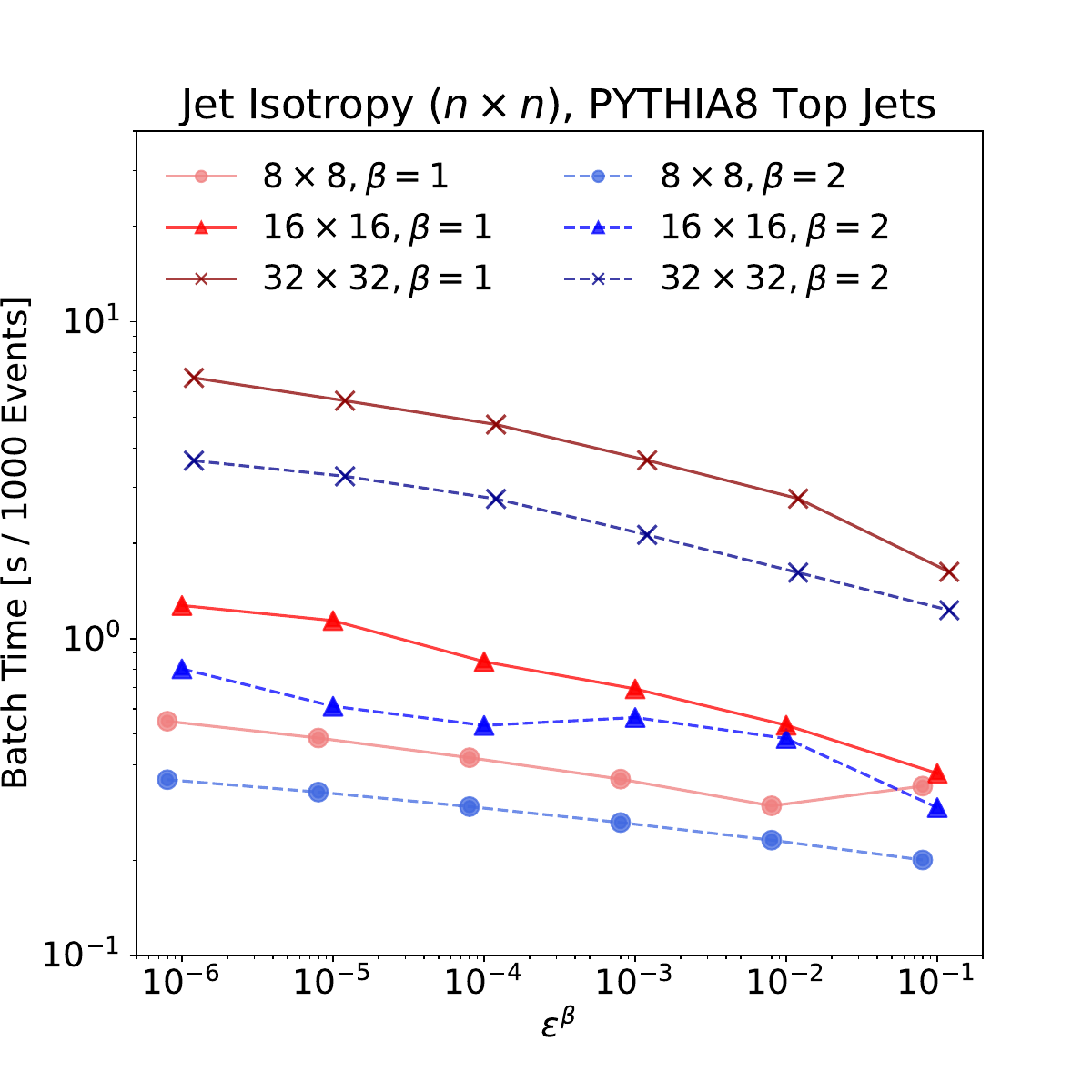}
        \label{fig:iso2}
    }
    \caption{
         The (a) fidelity and (b) runtime of \Shaper when computing the $n\times n$ jet isotropy,  for different values of $\epsilon$, $n$, and $\beta$. The fidelity is defined as the ratio of the Sinkhorn divergence to the ``true'' Wasserstein metric, as computed using the POT library, across a batch of 1000 events. The runtime is the total time to evaluate the Sinkhorn divergences for the entire 1000 event batch, as computed using a NVIDIA A100. An annealing parameter of $\Delta = 0.9$ is used globally. 
        }
    \label{fig:isotropy_benchmark}
\end{figure*}

\begin{figure*}[tp]
    \centering

        \includegraphics[width=0.5\textwidth]{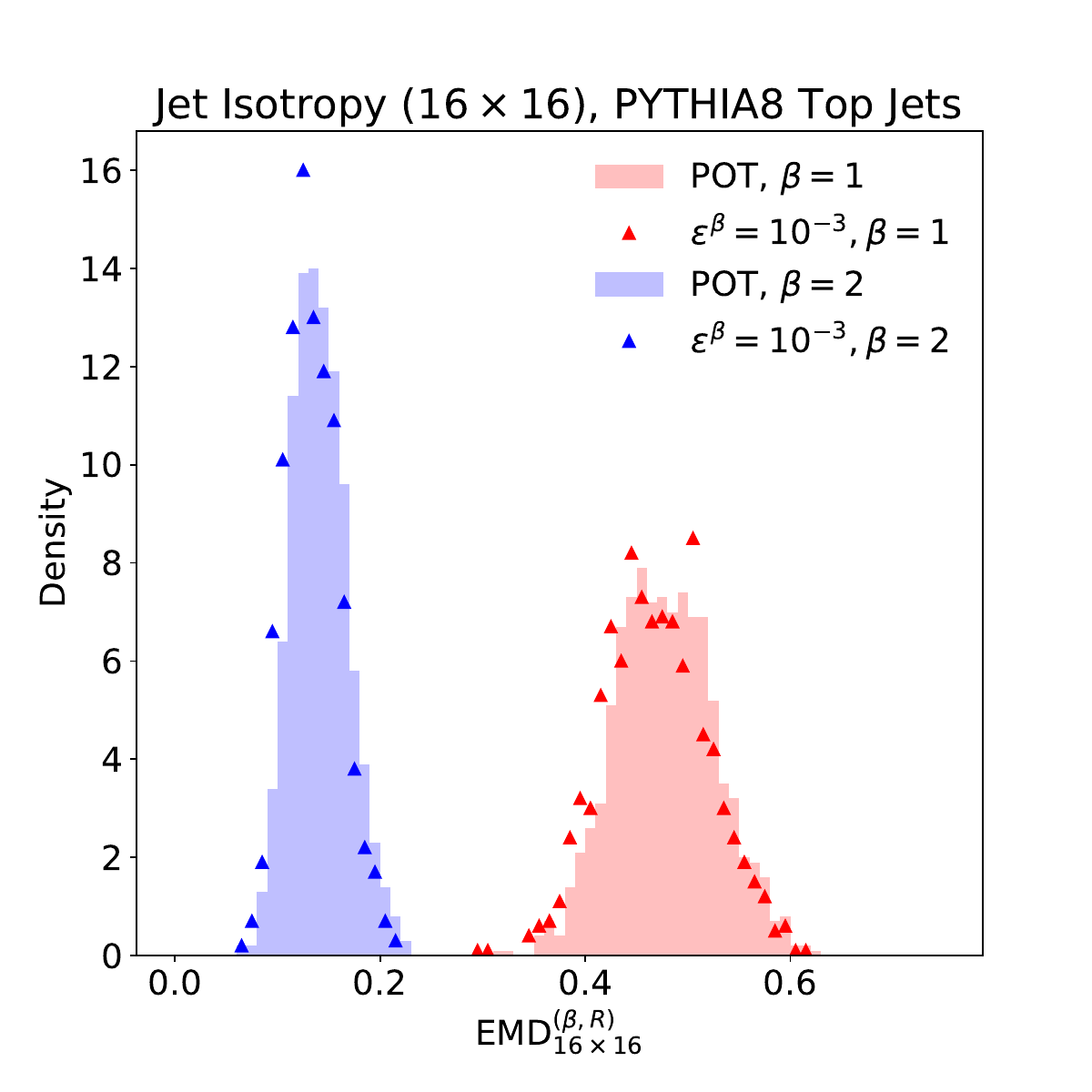}

    \caption{
       Distributions of the learned $16\times 16$ jet isotropy, for $\beta = 1$ (red) and $\beta = 2$ (blue), as calculated using the POT library (filled) and \Shaper (points) with $\epsilon^\beta = 10^{-3}$. 
        }
    \label{fig:isotropy_distributions}
\end{figure*}

In \Fig{isotropy_benchmark}, we show the results of a runtime vs.\ accuracy study. We compute the jet isotropy of 1000 top jets for several different values of $n$, $\epsilon$, and for $\beta = 1$ and $2$.
In \Fig{isotropy_distributions}, we show the learned $16\times16$ jet isotropies for $\epsilon^{\beta} = 10^{-3}$. For this experiment, we use a fixed (conservative) annealing parameter of $\Delta = 0.9$. \Shaper allows for events to be computed in parallelized batches; we run the entire computation in a single batch on a NVIDIA A100, and report the total runtime of the entire batch.\footnote{In principle, the only limiting factor to how many events \Shaper can process at once is the ability to fit everything on a single GPU. We find that we can run up to 10000 events in parallel on 32 GB of memory of a NVIDIA A100.}

We see from \Fig{isotropy_benchmark} that the accuracy of the Sinkhorn divergence, as an estimator for the Wasserstein metric, begins to saturate at $\epsilon^\beta = 10^{-3}$, and that there is no substantial gain from choosing a smaller $\epsilon$. Picking $\epsilon^\beta = 10^{-3}$ ensures percent level accuracy for $\beta  = 1$, and few-percent level accuracy for $\beta = 2$, which we can see visually in \Fig{isotropy_distributions}. Furthermore, for larger values of $n\times n$, the accuracy is mostly independent of $n$. We also observe from \Figs{isotropy_benchmark}{isotropy_distributions} that Sinkhorn tends to slightly underestimate the Wasserstein metric, which can be understood from the strictly negative $\epsilon^\beta$-regulator in \Eq{sinkhorn_dual}. Note that for $n = 16$, it takes under 1 second to process 1000 events -- this implies it is possible to process millions of events on the order of an hour, with further speedups possible by choosing a more aggressive value for the annealing parameter $\Delta$.

\subsection{Benchmarking Optimization: $N$-Subjettiness}\label{sec:benchmarking_optimization}

We perform a second benchmark investigation using $N$-subjettiness. Unlike the jet isotropy, $N$-subjettiness requires a nontrivial minimization. This allows us to use it to estimate the fidelity of the \Shaper algorithm's optimization step.

It is well known that the ratio $\tau_{32} \equiv \tau_3 / \tau_2$ is a good discriminant between top and QCD jets, as top jets tend to have 3 prongs more often than QCD jets, and thus have lower expected values of $\tau_{32}$~\cite{Thaler:2010tr}. We compute this ratio for several different values of $\epsilon$ to see if any discrimination power is lost (or gained) in the $\epsilon$-approximation. Within the \Shaper framework, the $N$-subjettiness is given by the following manifold of parameterized events:
\begin{align}
\label{eq:nsubjettiness}
    \E_{x_i, z_i}(x) = \sum_{i=1}^{N} z_i\, \delta(x-x_i), \qquad x_i \in \mathcal{X}, z_i \in \Delta_{N-1}.
\end{align}

\begin{figure*}[tp]
    \centering
    \subfloat[]{
         \includegraphics[width=0.5\textwidth]{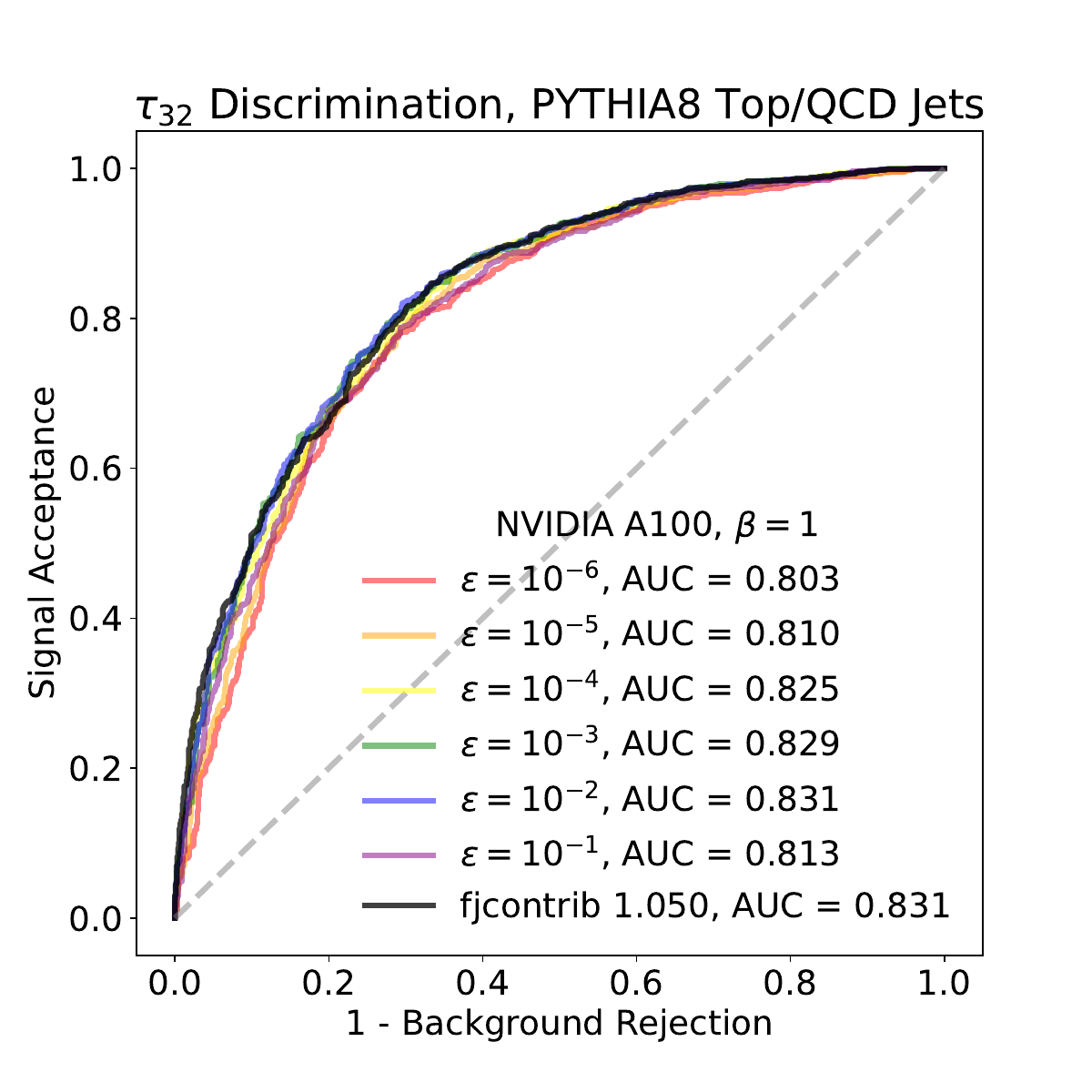}
        \label{fig:roc1}
    }
    \subfloat[]{
        \includegraphics[width=0.5\textwidth]{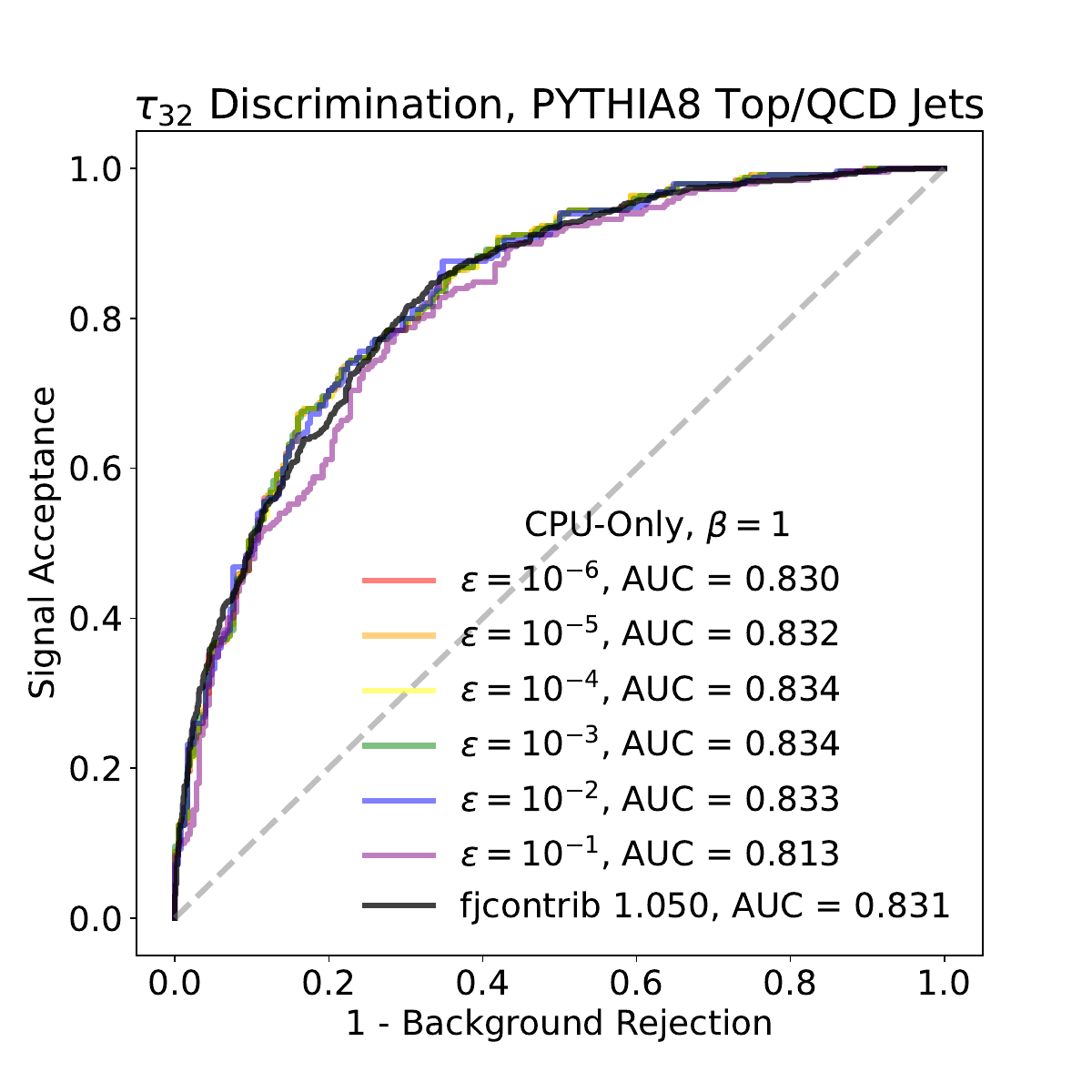}
        \label{fig:roc2}
    }
    \caption{
        A ROC curve showing the performance of $\tau_{32} = \tau_3 / \tau_2$ as a discriminator between top (signal) and QCD (background) jets, for several different values of $\epsilon$, as calculated using (a) an NVIDIA A100 and (b) only on CPU. A baseline curve, calculated using the $N$-subjettiness routines in \FastJet 3.4.0, is shown in black.}
    \label{fig:roc_benchmarks}
\end{figure*}

\begin{figure*}[htbp]
    \centering

        \includegraphics[width=0.5\textwidth]{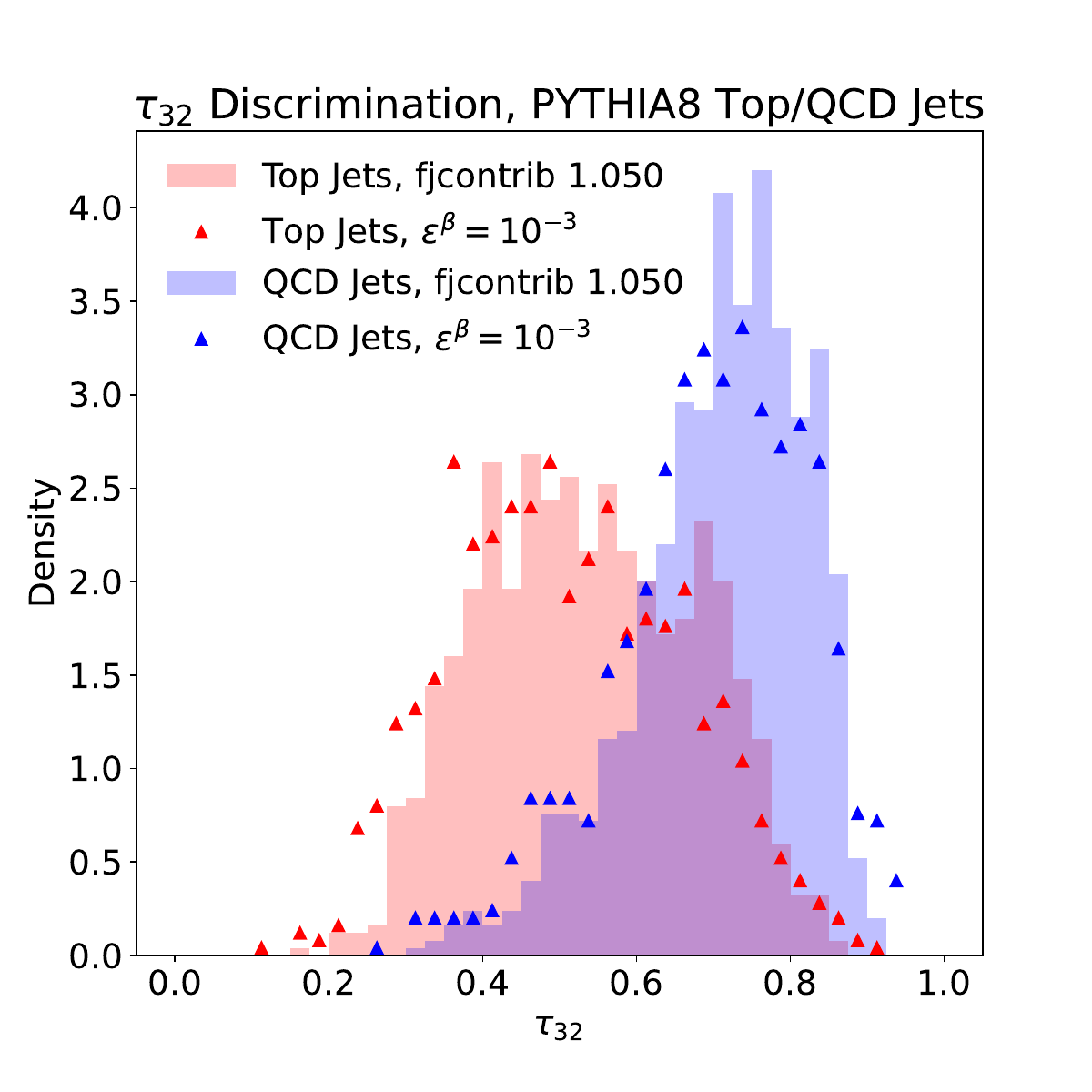}

    \caption{
       Distributions of the learned $\tau_{32}$, for top jets (red) and QCD jets (blue), as calculated using \FastJet 3.4.0 (filled) and \Shaper (points) with $\epsilon^\beta = 10^{-3}$. 
        }
    \label{fig:tau32_distributions}
\end{figure*}

As a baseline, we compute $N$-subjettiness using \FastJet 3.4.0 with \textsc{FJcontrib} 1.050.
The results of this study are shown in \Fig{roc1} as ROC curves, computed using an NVIDIA A100 GPU. We see that the Sinkhorn approximations have roughly the same discriminatory power as the baseline for $\epsilon \sim 10^{-3}$. In \Fig{tau32_distributions}, we show the distributions of $\tau_{32}$ for both datasets computed with \FastJet and \Shaper with $\epsilon = 10^{-3}$, and see good agreement between the two methods.
As with the isotropy study in \Sec{benchmarking_sinkhorn}, we observe that \Shaper tends to slightly underestimate the observable.
In order to gauge the impact of float precision on our estimates, we repeat ROC curve calculation using only a CPU, which is shown in \Fig{roc2}. For values of $\epsilon \ll 10^{-3}$, we see that on the GPU architecture, the performance actually begins to degrade due to the lower machine precision due to the accumulation of floating-point errors during the optimization, while it saturates on the CPU. Therefore, it is recommended to use $\epsilon^\beta \sim 10^{-3}$, as this is the most stable compromise between fidelity and machine precision.

\subsection{Hearing Gradients}

\Shaper can be used to not only estimate the shapiness of events, but also take derivatives of the shapiness with respect to the event. As discussed in \Sec{sinkhorn}, this is completely automatic, since the gradients with respect to the energy flow are given manifestly by the Kantorovich potentials, allowing us to see precisely how our EMD calculations depend on the energies and positions of particles in an event. 

Reading off of \Eq{emd_grads_2}, we obtain an expression for the gradient of the EMD with respect to the energy $E_i$ of particle $i$:
\begin{align}
    \nabla_{E_i}\EMD(\E, \E'_\theta) &= F(x_i) \label{eq:z_grad}.
\end{align}
If the gradient at particle $i$ is negative, then increasing the energy of that particle will decrease the EMD, making the event more $\M$-like. By adding ``ghost'' particles to $\E$, one can probe the energy dependence of the EMD from any point in $\E$.
Similarly, we can take the gradient the EMD with respect to the position $x_i$ of particle $i$:
\begin{align}
    \nabla_{x_i}\EMD(\E, \E'_\theta) &= E_i \nabla_{x_i}F(x_i) \label{eq:x_grad}.
\end{align}
This gradient (times $-1$) tells us where to move the particle $i$ to decrease the EMD.
Both the energy and position gradients answer the question, ``If I want to decrease the EMD (make my energy look \emph{more} like my shape), what should I do to a particle at site $x_i$?'' Moreover, because of the reparameterization invariance of the energy flow density, the energy and position gradients are both are valid ways to change the EMD: one can either change the energy of particles at the location $x_i$, move the particles at $x_i$ somewhere else, or some combination of both. 

\begin{figure*}[tp]
    \centering
    \subfloat[]{
         \includegraphics[width=0.56\textwidth]{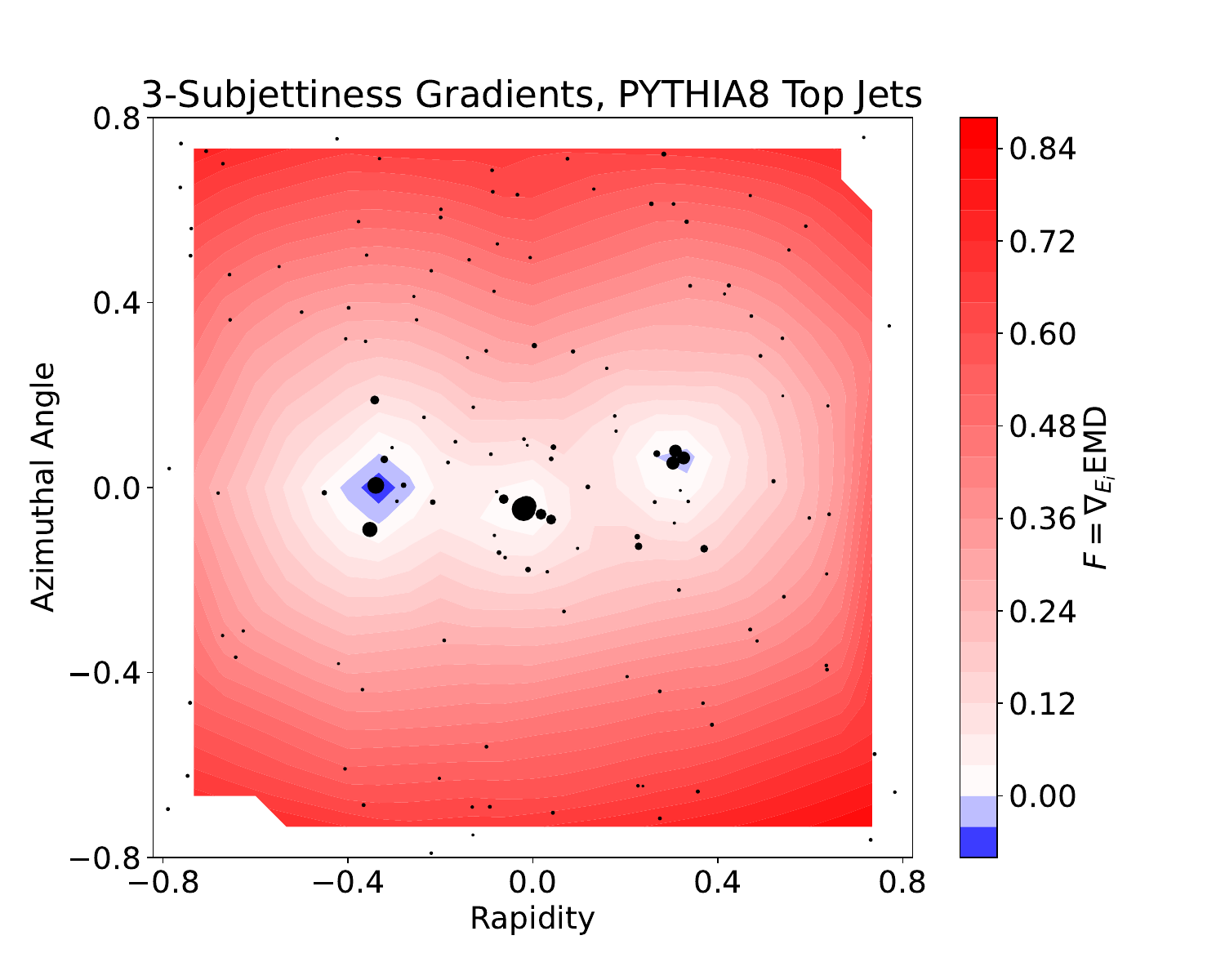}
        \label{fig:subjettiness_grads}
    }
    \subfloat[]{
        \includegraphics[width=0.44\textwidth]{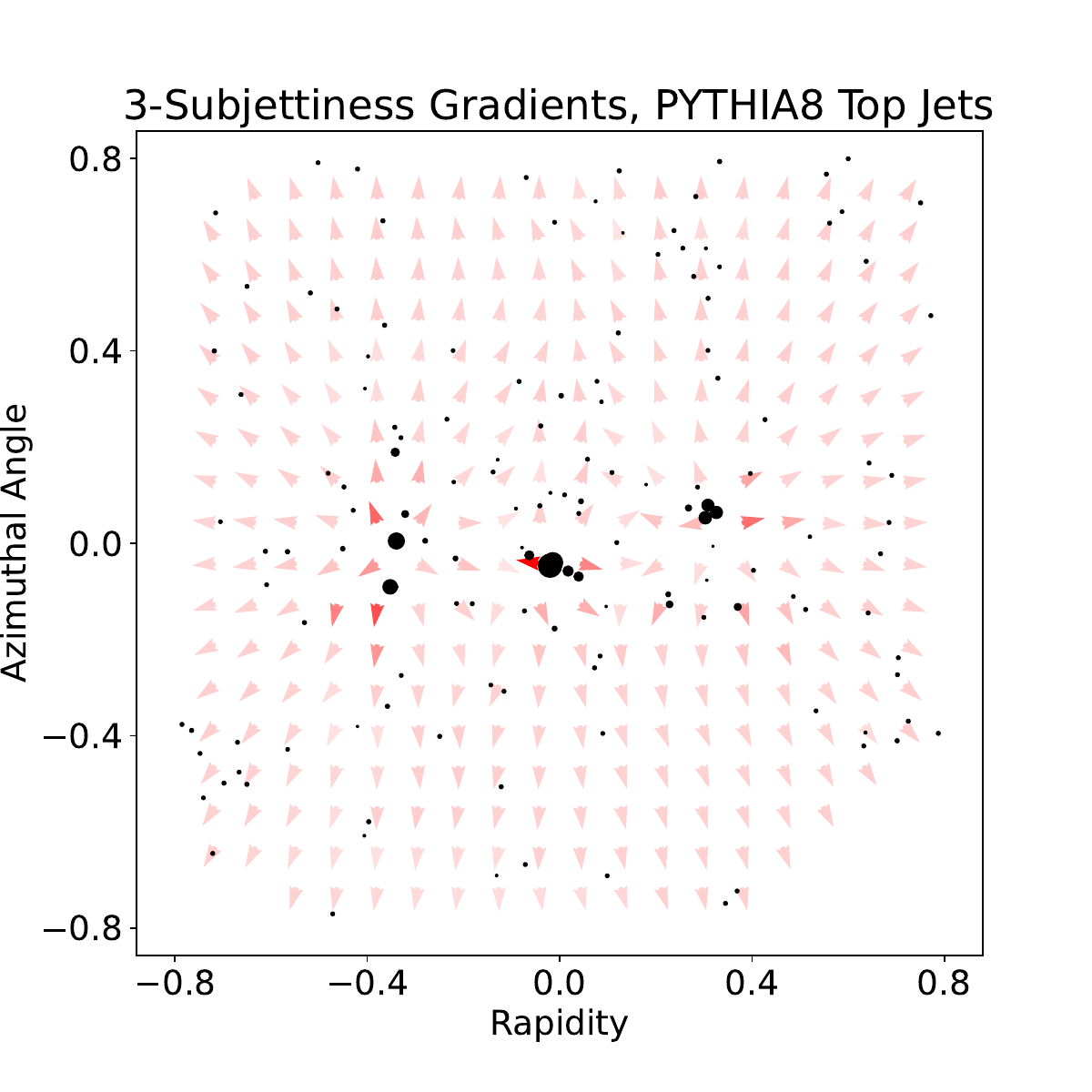}
        \label{fig:subjettiness_quiver}
    }
    \caption{
        For an example top jet, (a) gradients of the $3$-subjettiness with respect to the particle weights, $\nabla_{E_i} \EMD$, and (b) gradients off the $3$-subjettiness with respect to the particle positions, $\nabla_{x_i} \EMD$, are shown. In (a), increasing the particle energy in red regions will increase the $3$-subjettiness (look less like 3 subjets), and increasing particle energy in blue regions will decrease the $3$-subjettiness (look more like 3 subjets). In (b), moving particles along or against the arrows will increase or decrease the $3$-subjettiness, respectively, with the arrow's shading indicating the relative magnitude of the change. 
        }
    \label{fig:subjettiness_gradients}
\end{figure*}

\begin{figure*}[htp]
    \centering
    \subfloat[]{
         \includegraphics[width=0.56\textwidth]{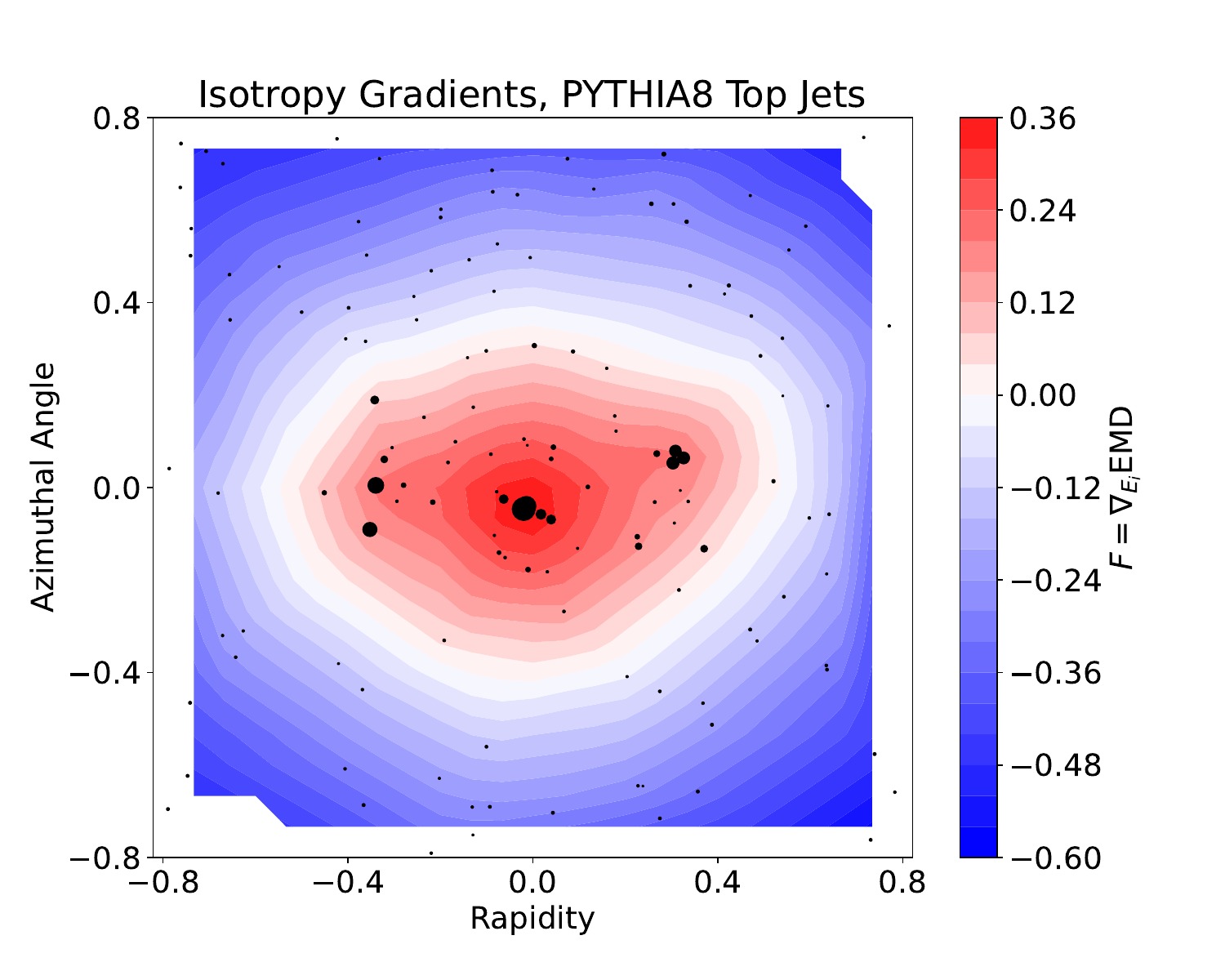}
        \label{fig:isotropy_grads}
    }
    \subfloat[]{
        \includegraphics[width=0.44\textwidth]{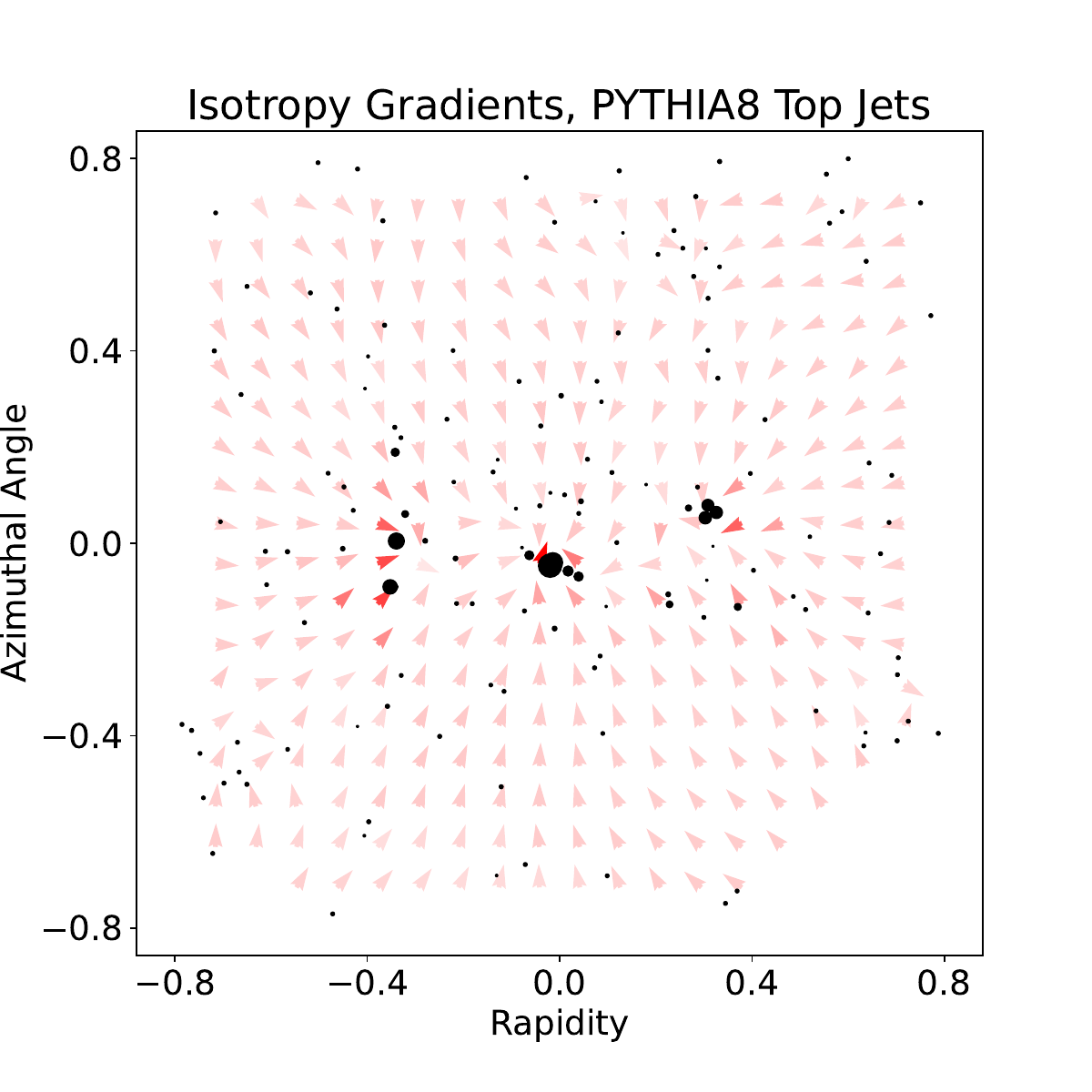}
        \label{fig:isotropy_quiver}
    }
    \caption{
        The same as \Fig{subjettiness_gradients}, but for the $16\times16$ jet isotropy with $\beta = 1$. 
        }
    \label{fig:isotropy_gradients}
\end{figure*}

In \Figs{subjettiness_gradients}{isotropy_gradients}, the gradients from \Eqs{z_grad}{x_grad} are plotted for an example top jet, for the $3$-subjettiness and $16\times16$,$ \beta = 1$ jet isotropy, respectively.
Using \Fig{subjettiness_gradients}, we can see what parts of the event contribute to the $3$-subjettiness -- the three large clusters contribute negatively to the EMD (make the event look more like 3 subjets), while the rest of the event contribute positively to the EMD (makes the event deviate from 3 subjets). Similarly, we see from \Fig{isotropy_gradients} that the overdensity of energy at the center of the event makes it less isotropic. In both figures, the vector quiver plot tells us which way particles should ``flow'' (against) to change the shape.

There are many potential applications of calculating gradients of a shape with respect to an energy flow. In experimental contexts, for example, one can use the fact that the gradients are practically instantaneous to compute to do easy Gaussian error propagation due to detector-induced uncertainties in particle energies and positions~\cite{CMS:2016lmd, ATLAS:2020cli}. This avoids having to do expensive re-sampling and recalculation of the event shape. Phenomenologically, these gradients can be used to probe the sensitivity of observables to certain radiation patterns. For example, the sensitivity of an observable to pileup can be measured by taking derivatives with respect to the pileup scale~\cite{Soyez:2012hv}, which can be numerically realized in \Shaper, a potential avenue for future work.

\subsection{Hearing Jets Ring (and Disk, and Ellipse)}\label{sec:jets}

We next use \Shaper to realize the custom shape observables defined in \Sec{custom_shapes}, starting with some visualizations of these shape observables on an example event.

The three base observables we consider are the ringiness, diskiness, and ellipsiness, as defined in \Sec{custom_shapes}. We calculate the shape observables $N\times\O$ (the $N$-ringiness, the $N$-diskiness, and $N$-ellipsiness) and $N\times\O^\tau$ (the $N$-(ring$+$point)iness, the $N$-(disk$+$point)iness, and $N$-(ellipse$+$point)iness), as defined in \Sec{custom_shapes} with $\beta = 1$.
We also consider, for comparison, the $N$-subjettiness, $\tau_N$, as defined in \Eq{nsubjettiness}. 
We use \Shaper to evaluate all of these shape observables on a single of top jet event from our dataset, restricted to $m_J \in [145, 205]\,\GeV$, though without any pileup contamination. Each extended shape is sampled with $100$ points, with $\epsilon = 10^{-3}$ and $\Delta = 0.9$. 

\begin{figure*}[tp]
    \centering
    \subfloat[]{
         \includegraphics[width=0.32\textwidth]{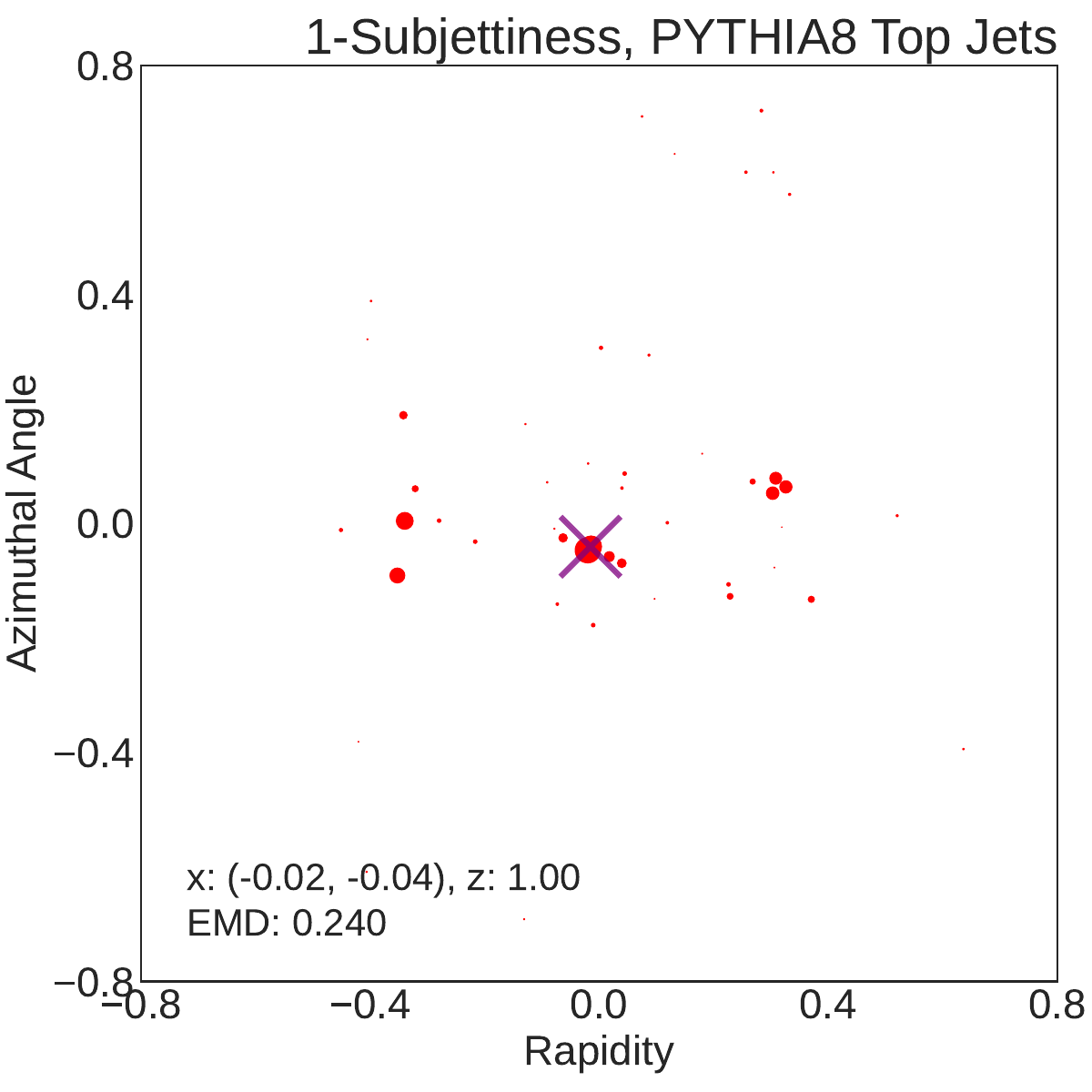}
        \label{fig:1_subjettiness}
    }
    \subfloat[]{
        \includegraphics[width=0.32\textwidth]{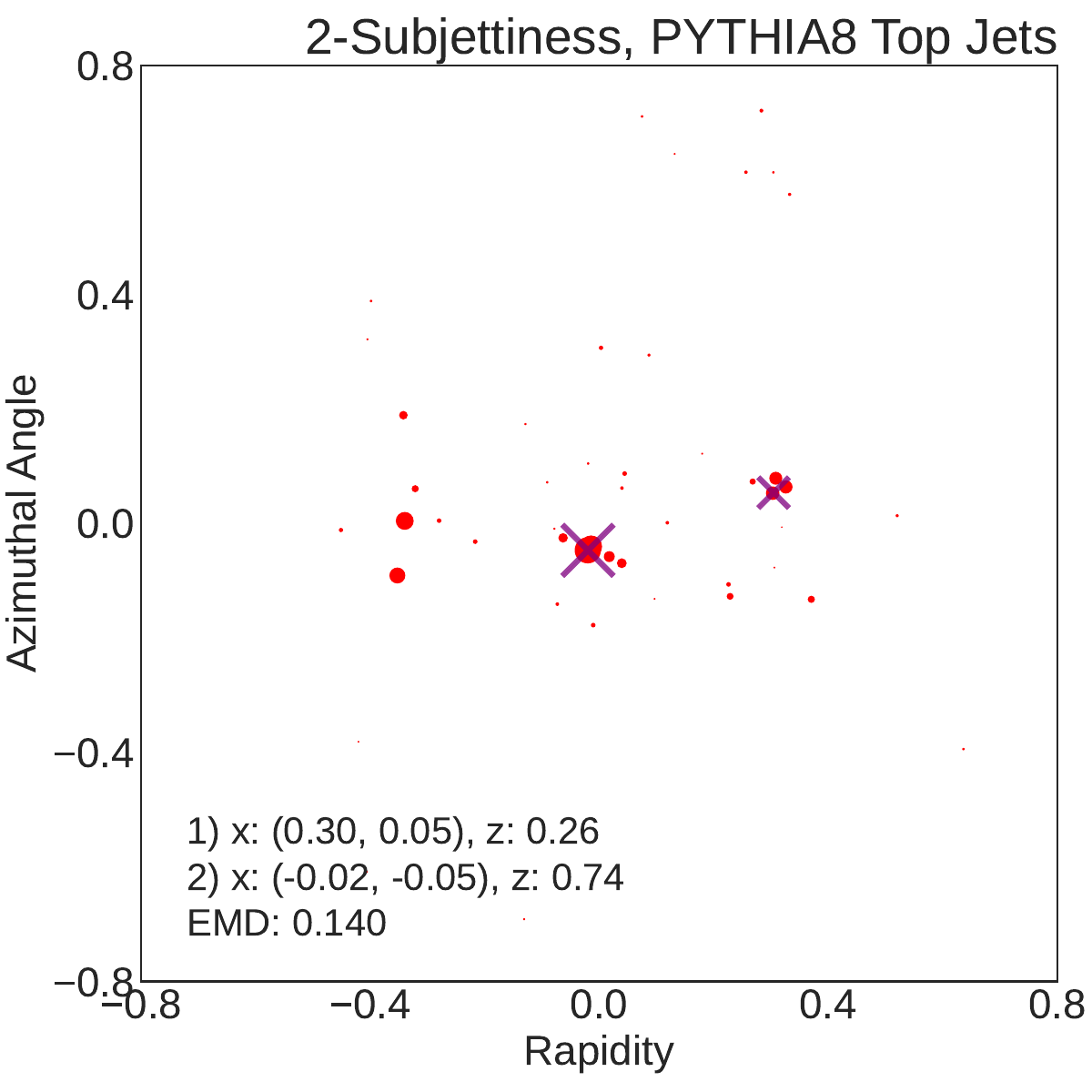}
        \label{fig:2_subjettiness}
    }
    \subfloat[]{
        \includegraphics[width=0.32\textwidth]{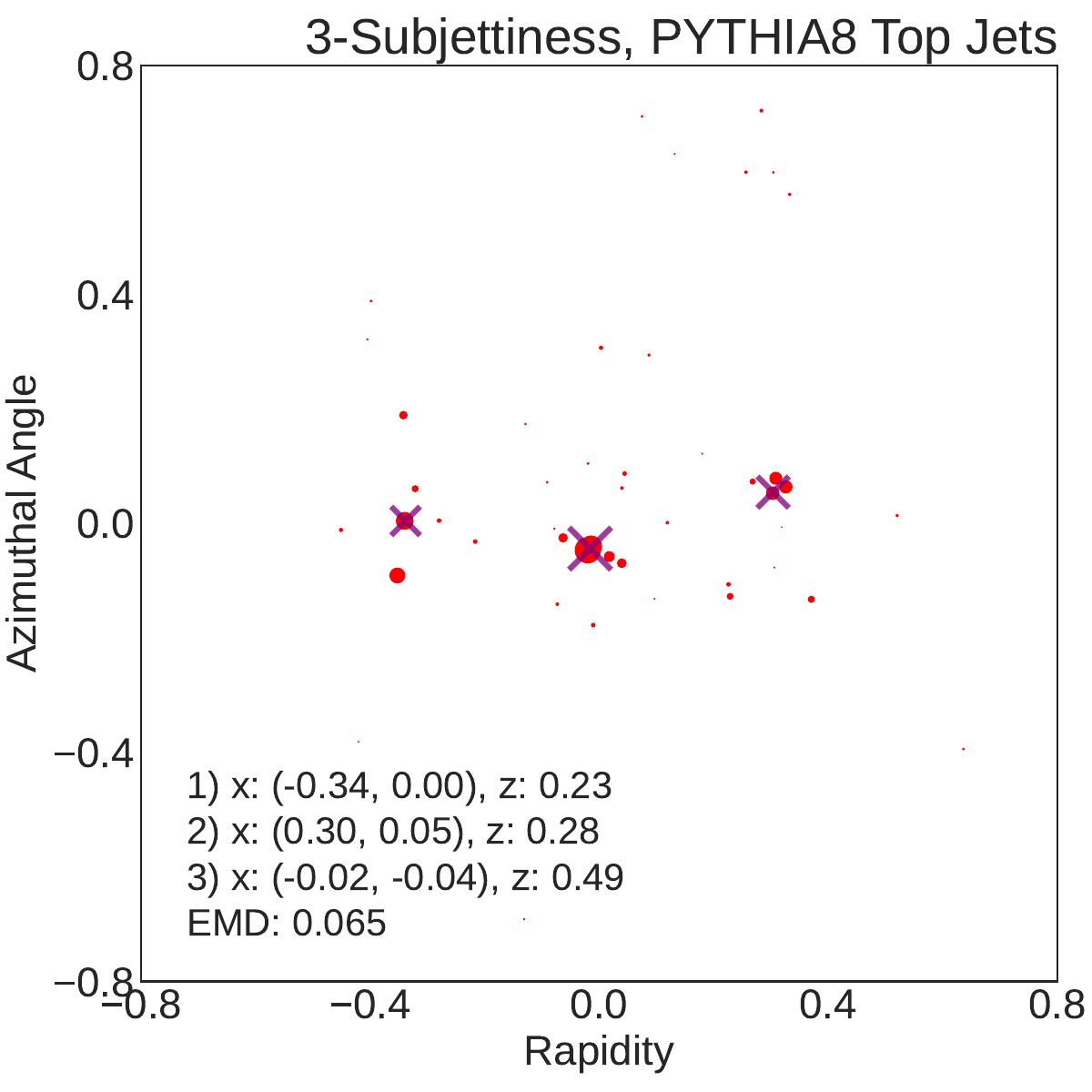}
        \label{fig:3_subjettiness}
    }
    \caption{
        The (a) 1-, (b) 2-, and (c) 3-subjettiness of an example top jet event. Subjets are represented by a purple ``$\times$'', with size proportional to the subjet's energy weight. 
        }
    \label{fig:n_subjettiness}
\end{figure*}

\begin{figure*}[tp]
    \centering
    \subfloat[]{
         \includegraphics[width=0.32\textwidth]{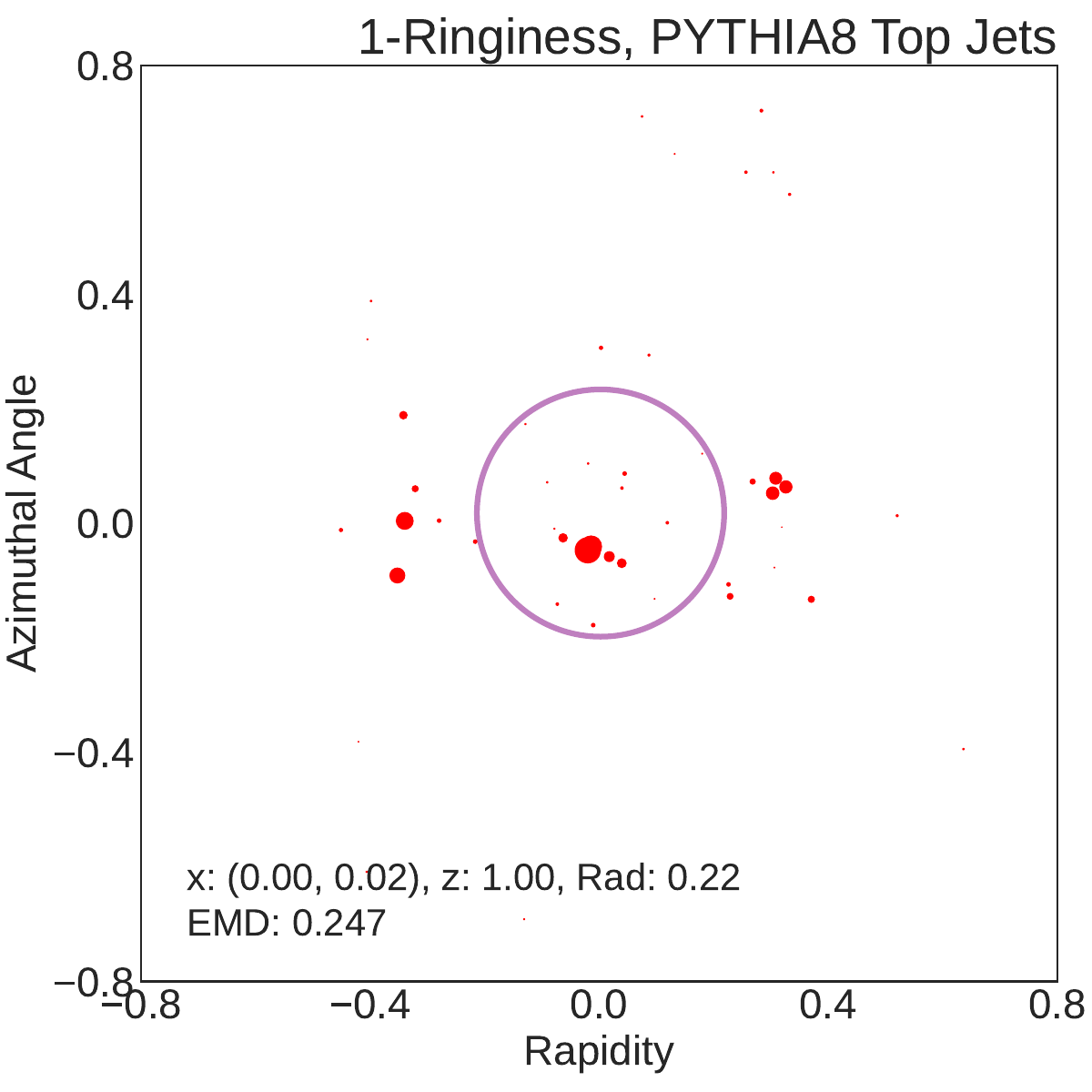}
        \label{fig:1_ringiness}
    }
    \subfloat[]{
        \includegraphics[width=0.32\textwidth]{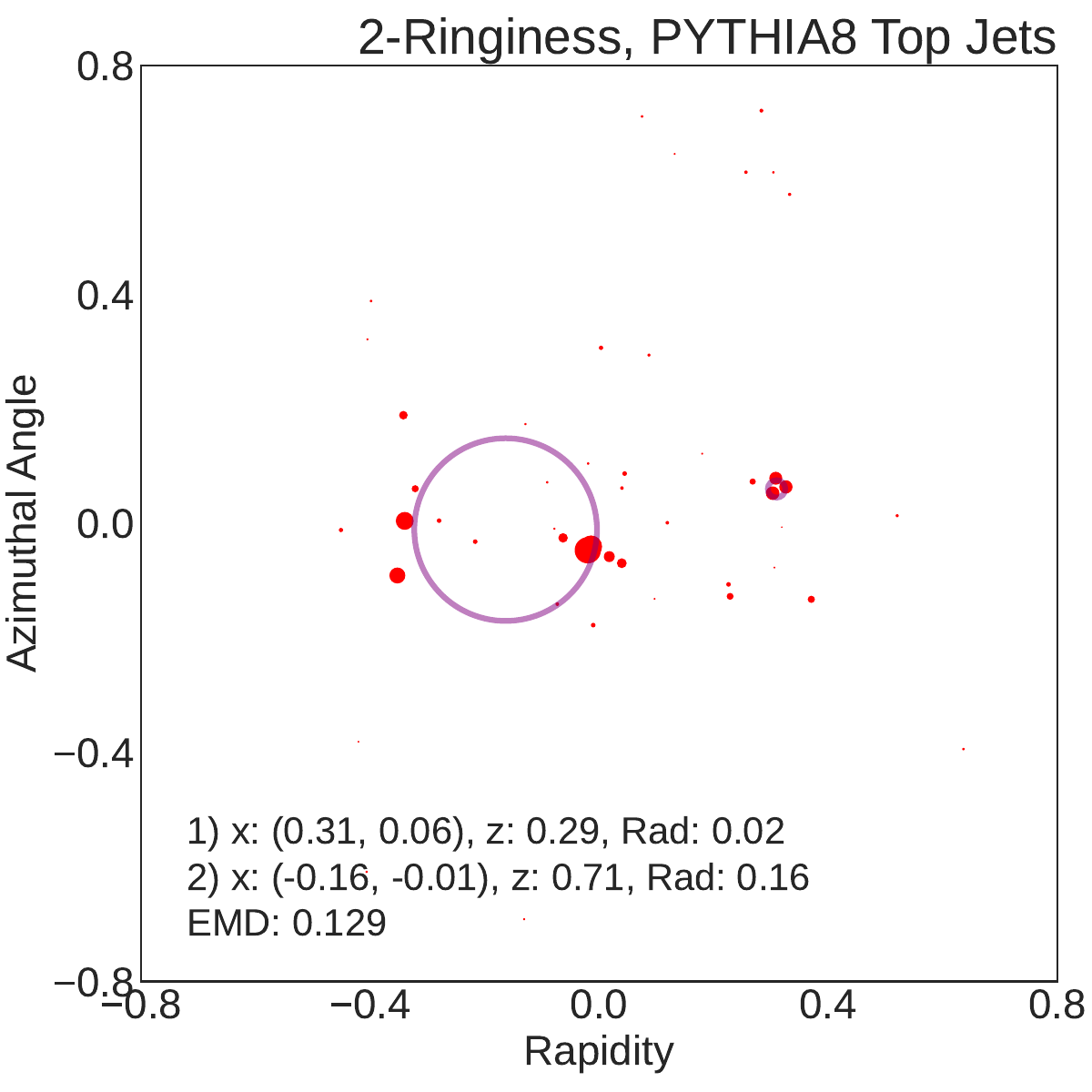}
        \label{fig:2_ringiness}
    }
    \subfloat[]{
        \includegraphics[width=0.32\textwidth]{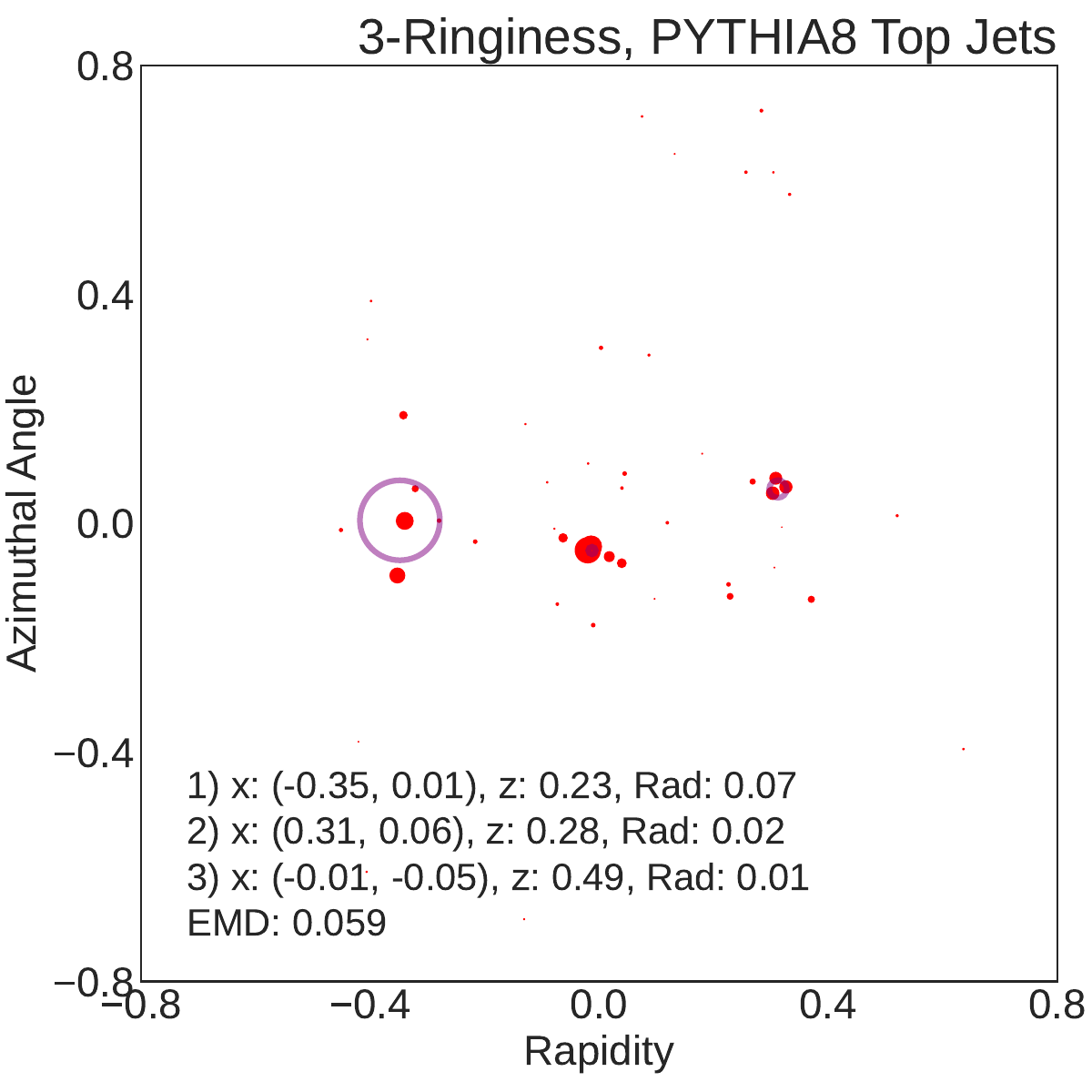}
        \label{fig:3_ringiness}
    }
    \vspace{1pt}
    \subfloat[]{
         \includegraphics[width=0.32\textwidth]{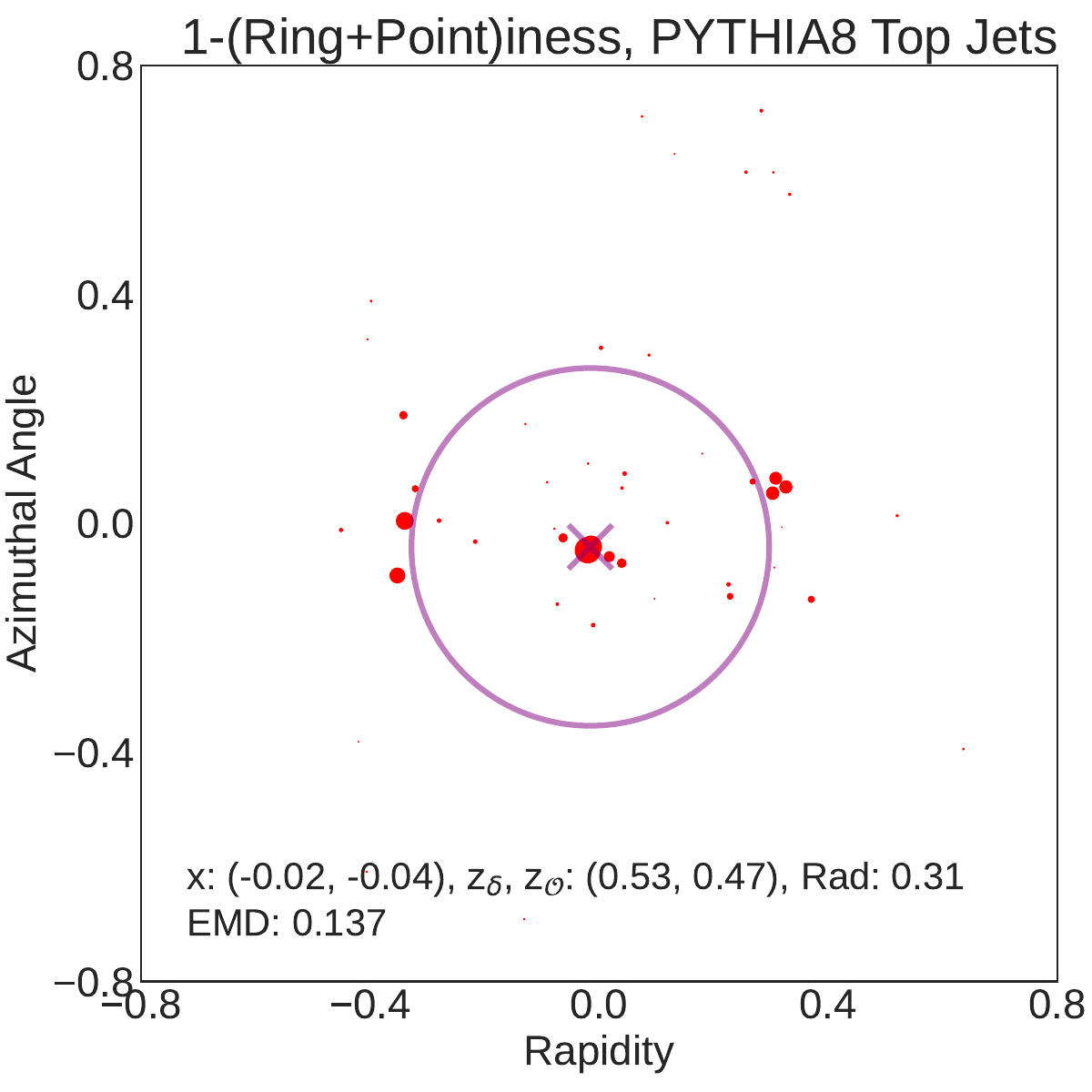}
        \label{fig:1_pringiness}
    }
    \subfloat[]{
        \includegraphics[width=0.32\textwidth]{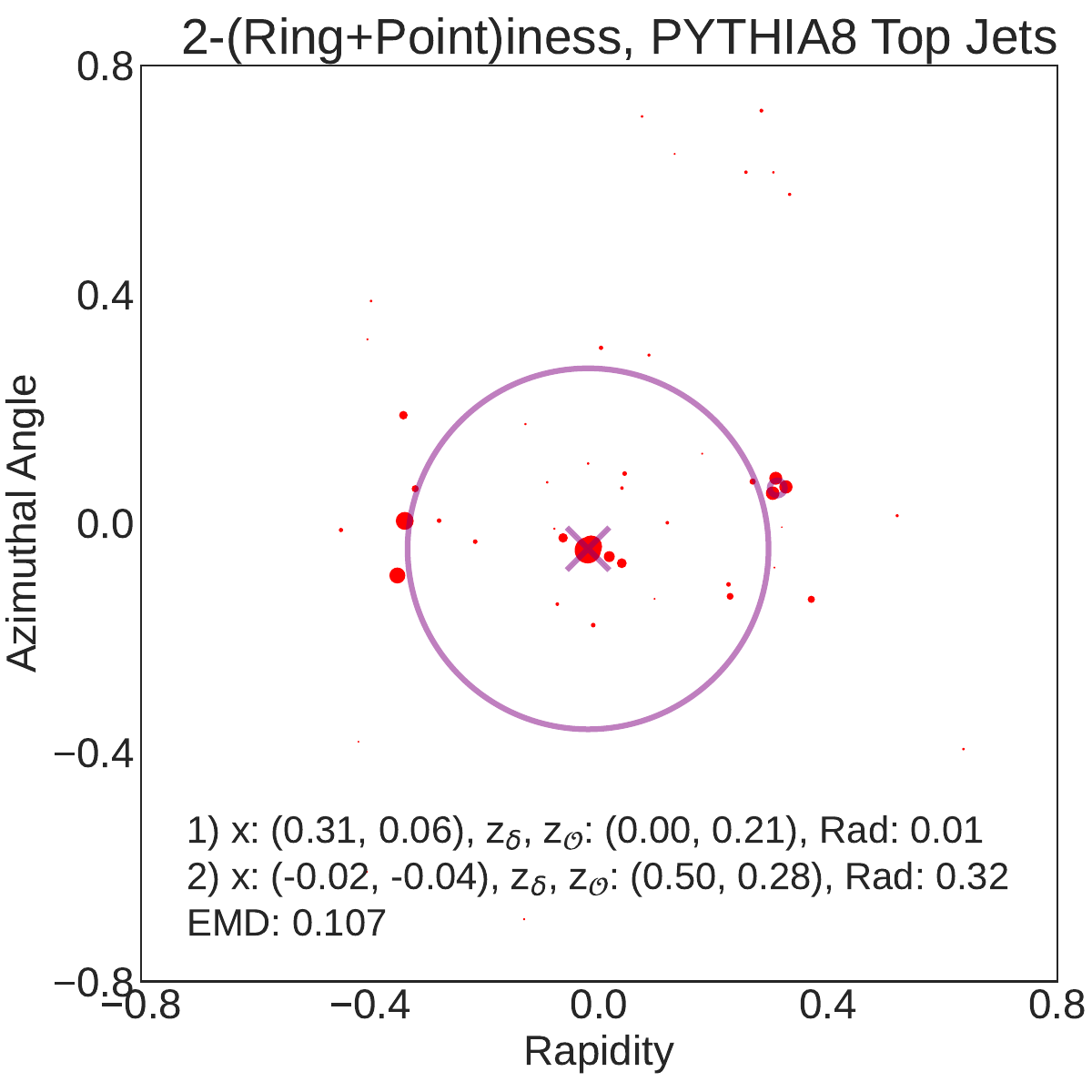}
        \label{fig:2_pringiness}
    }
    \subfloat[]{
        \includegraphics[width=0.32\textwidth]{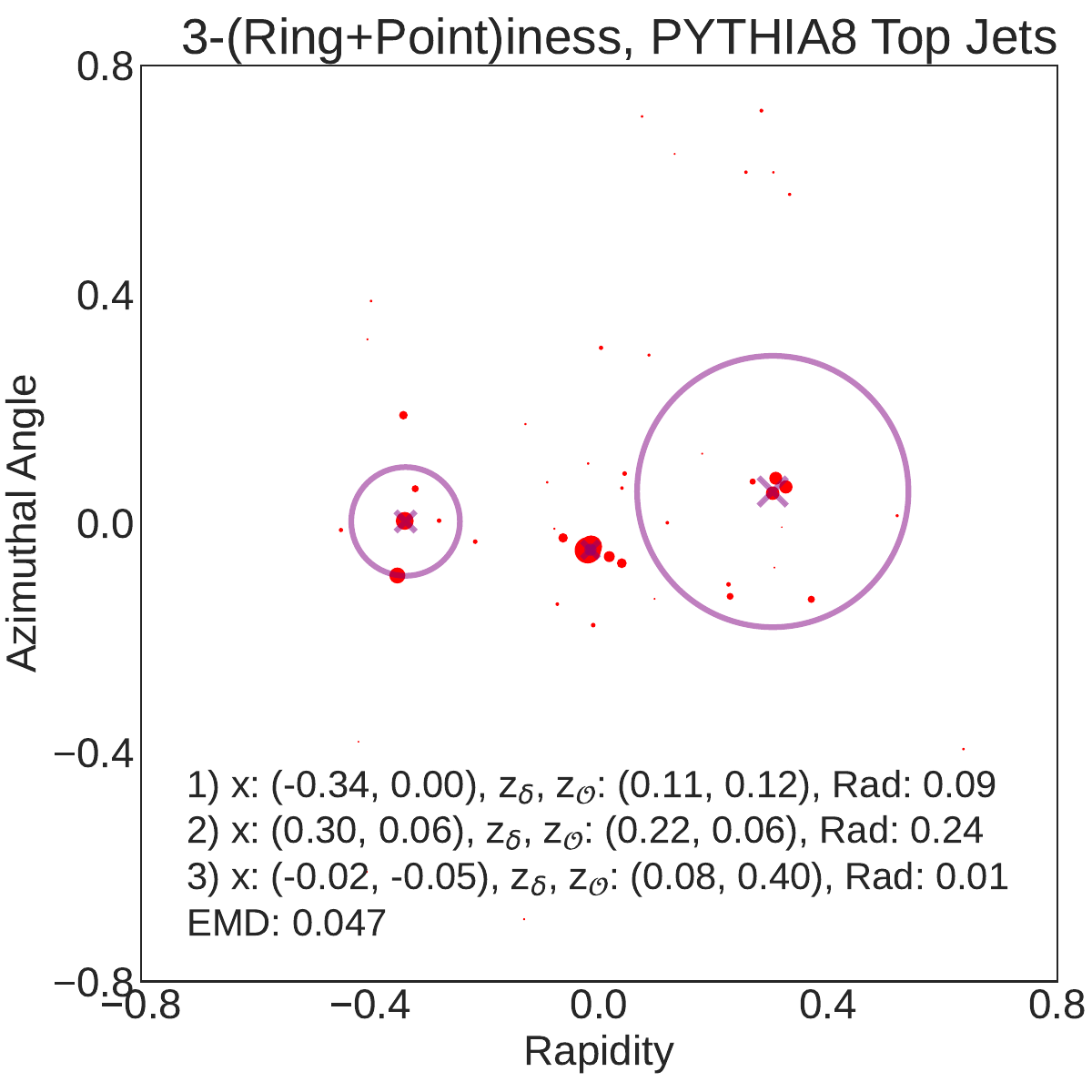}
        \label{fig:3_pringiness}
    }
    \caption{
        Top row: the (a) 1-, (b) 2-, and (c) 3-ringiness of an example top jet event. Bottom row: the (d) 1-, (e) 2-, and (f) 3-(ring$+$point)iness of the same top jet event. The point is represented by a ``$\times$'', with size proportional to its energy weight. 
        }
    \label{fig:n_ringiness}
\end{figure*}

\begin{figure*}[tp]
    \centering
    \subfloat[]{
         \includegraphics[width=0.32\textwidth]{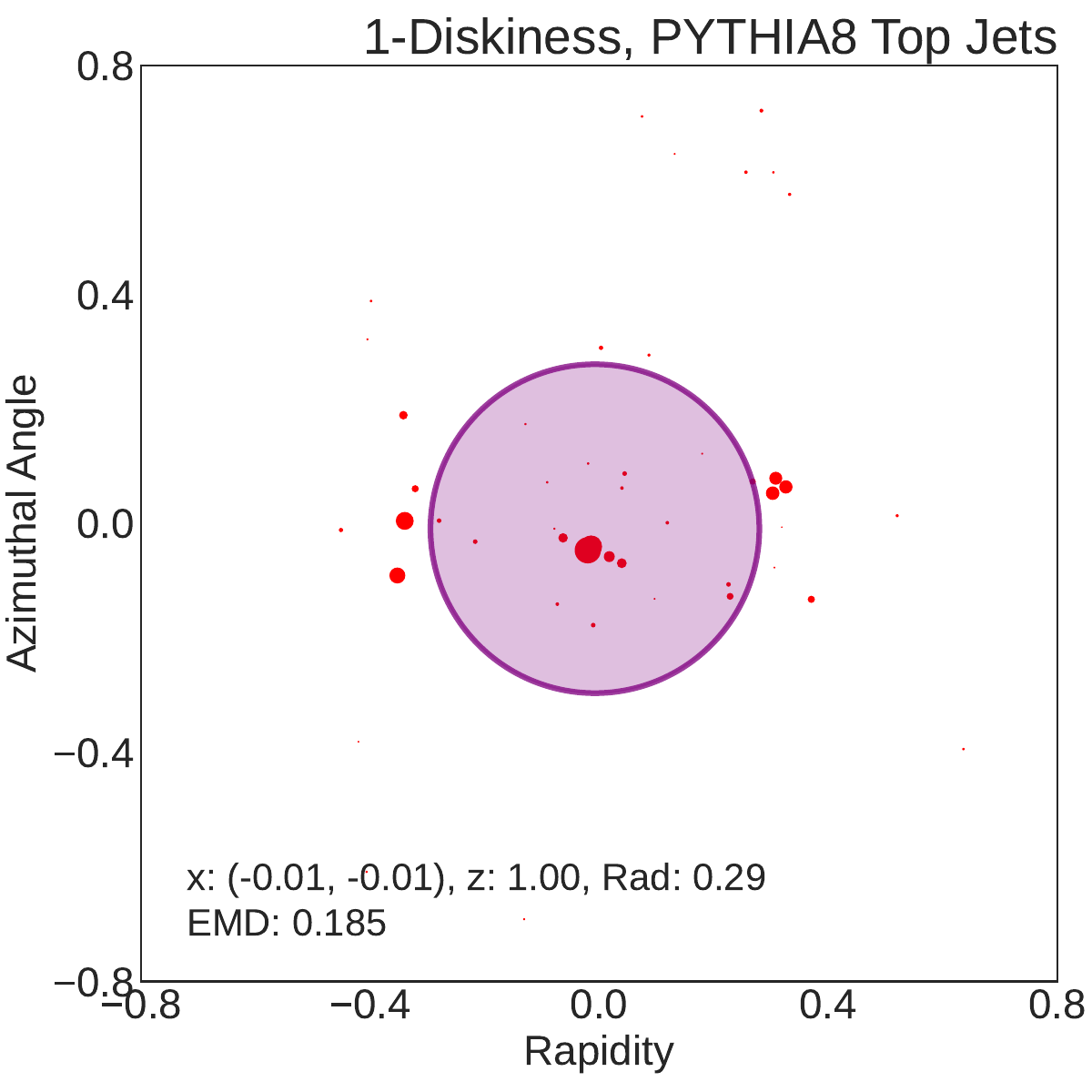}
        \label{fig:1_diskiness}
    }
    \subfloat[]{
        \includegraphics[width=0.32\textwidth]{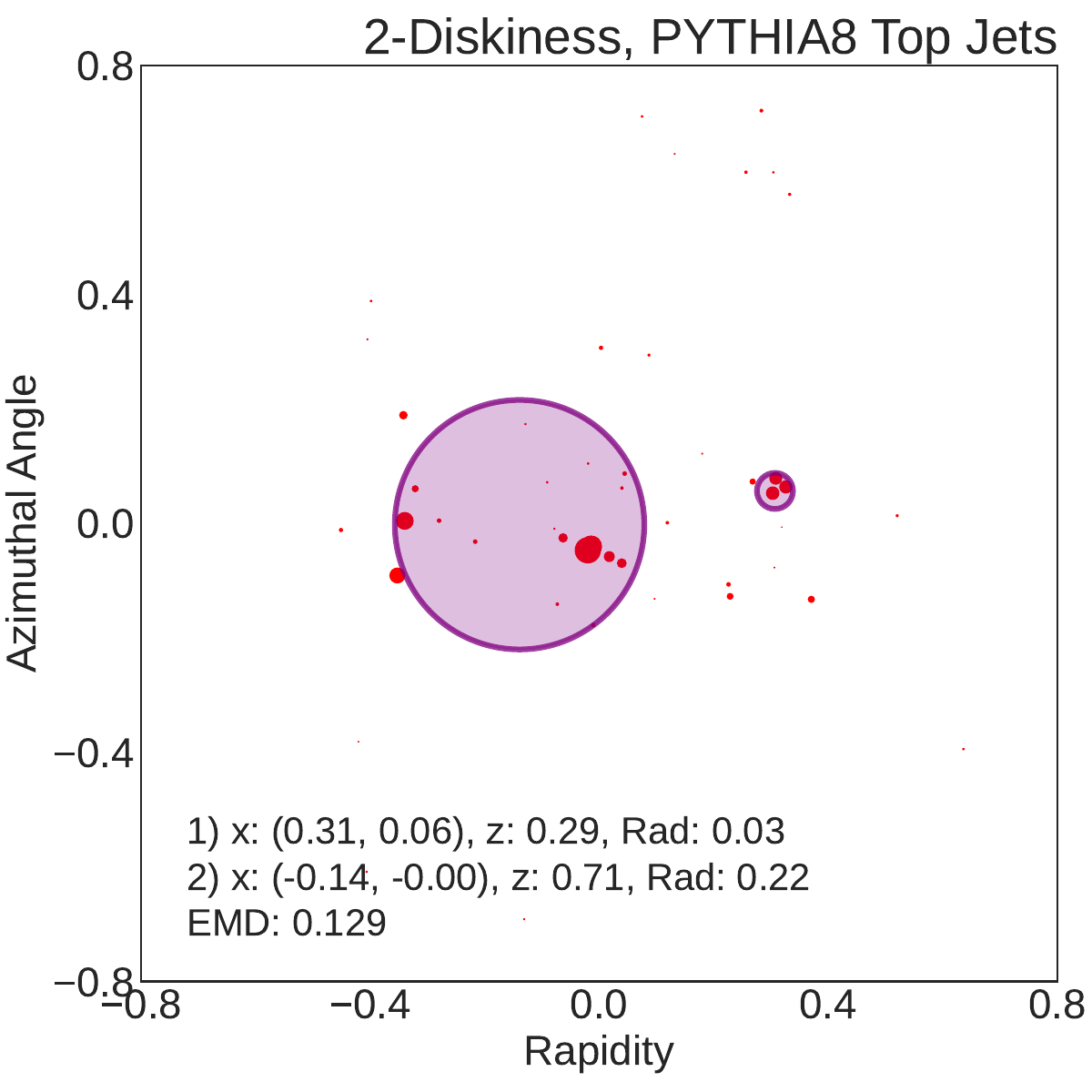}
        \label{fig:2_diskiness}
    }
    \subfloat[]{
        \includegraphics[width=0.32\textwidth]{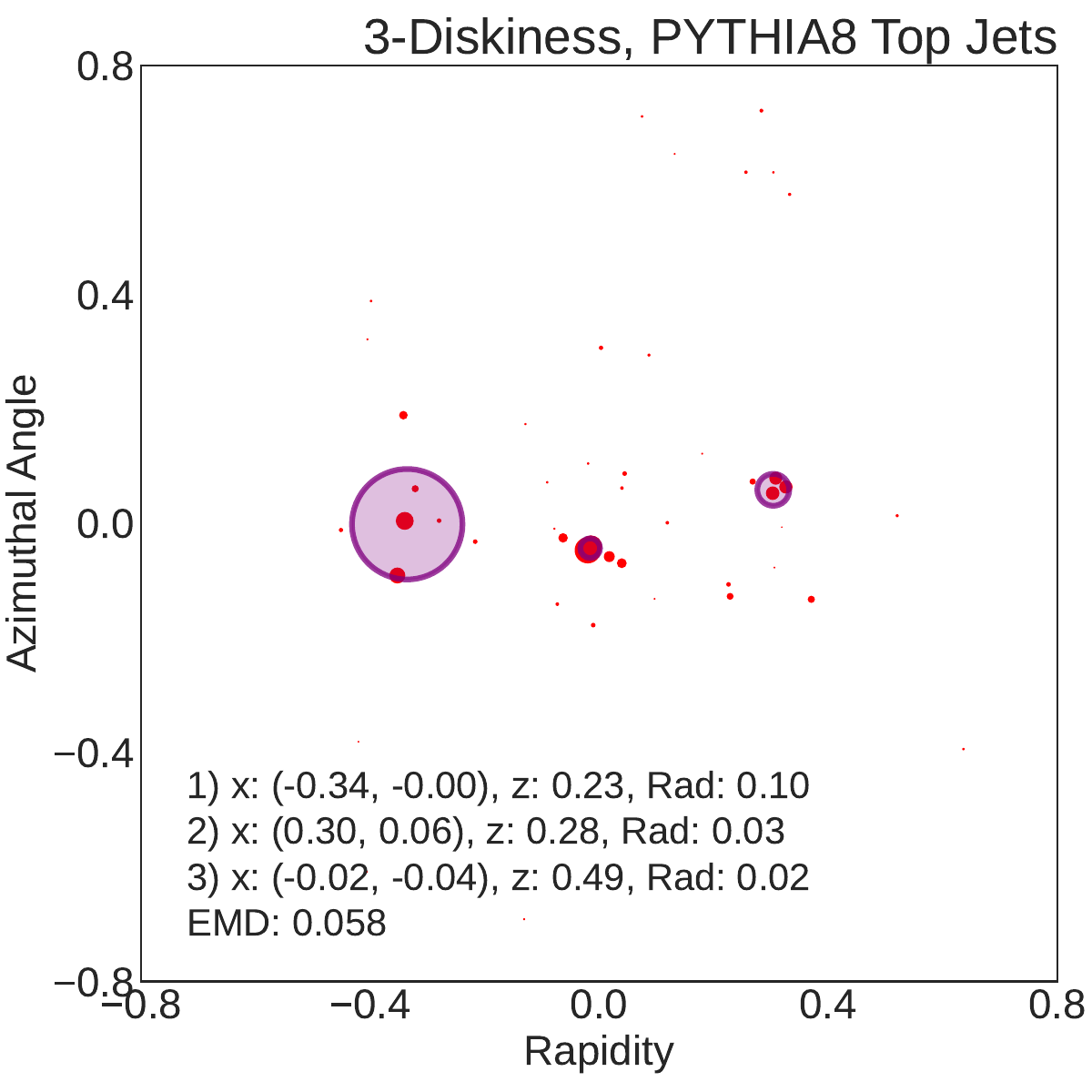}
        \label{fig:3_diskiness}
    }
    \vspace{1pt}
    \subfloat[]{
         \includegraphics[width=0.32\textwidth]{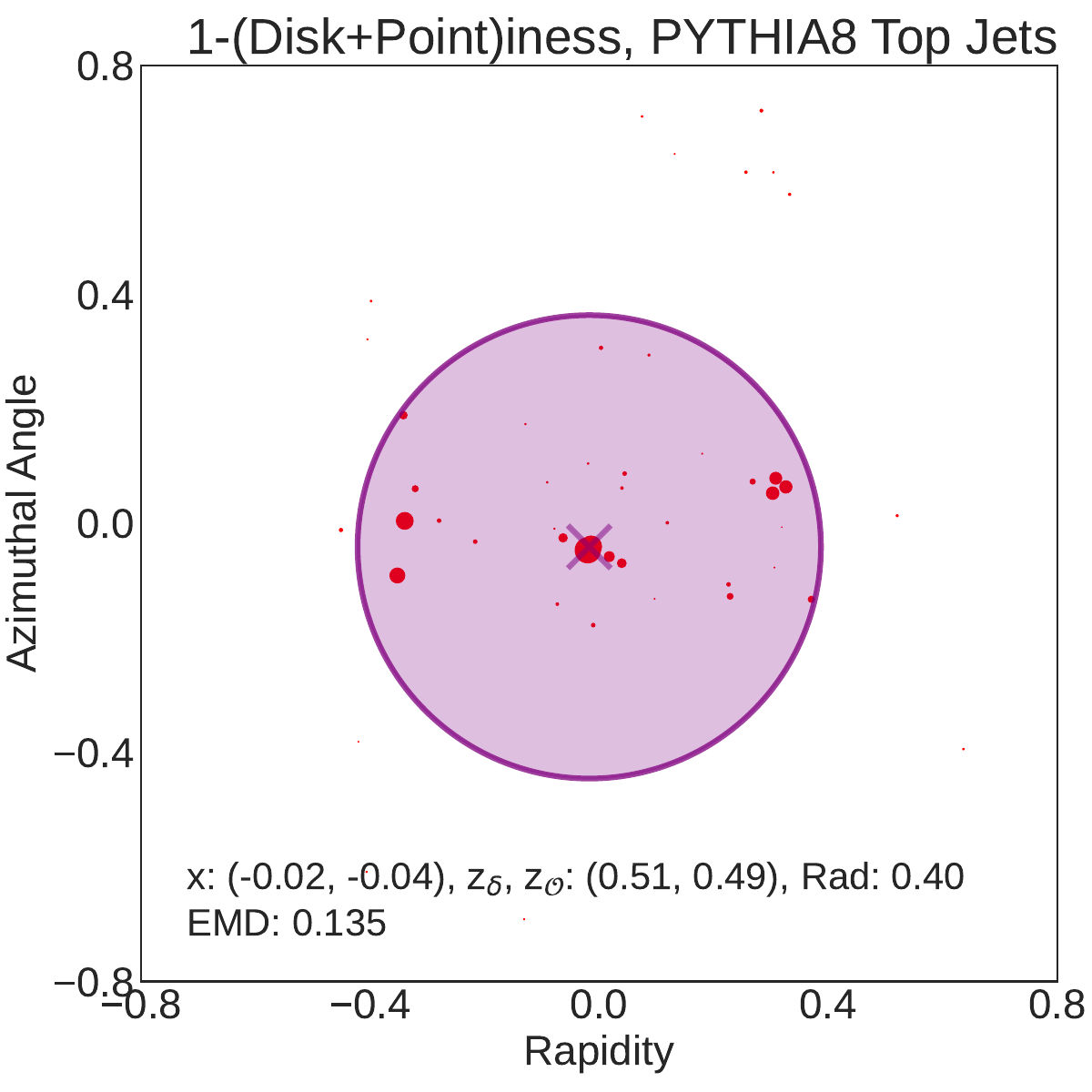}
        \label{fig:1_pdiskiness}
    }
    \subfloat[]{
        \includegraphics[width=0.32\textwidth]{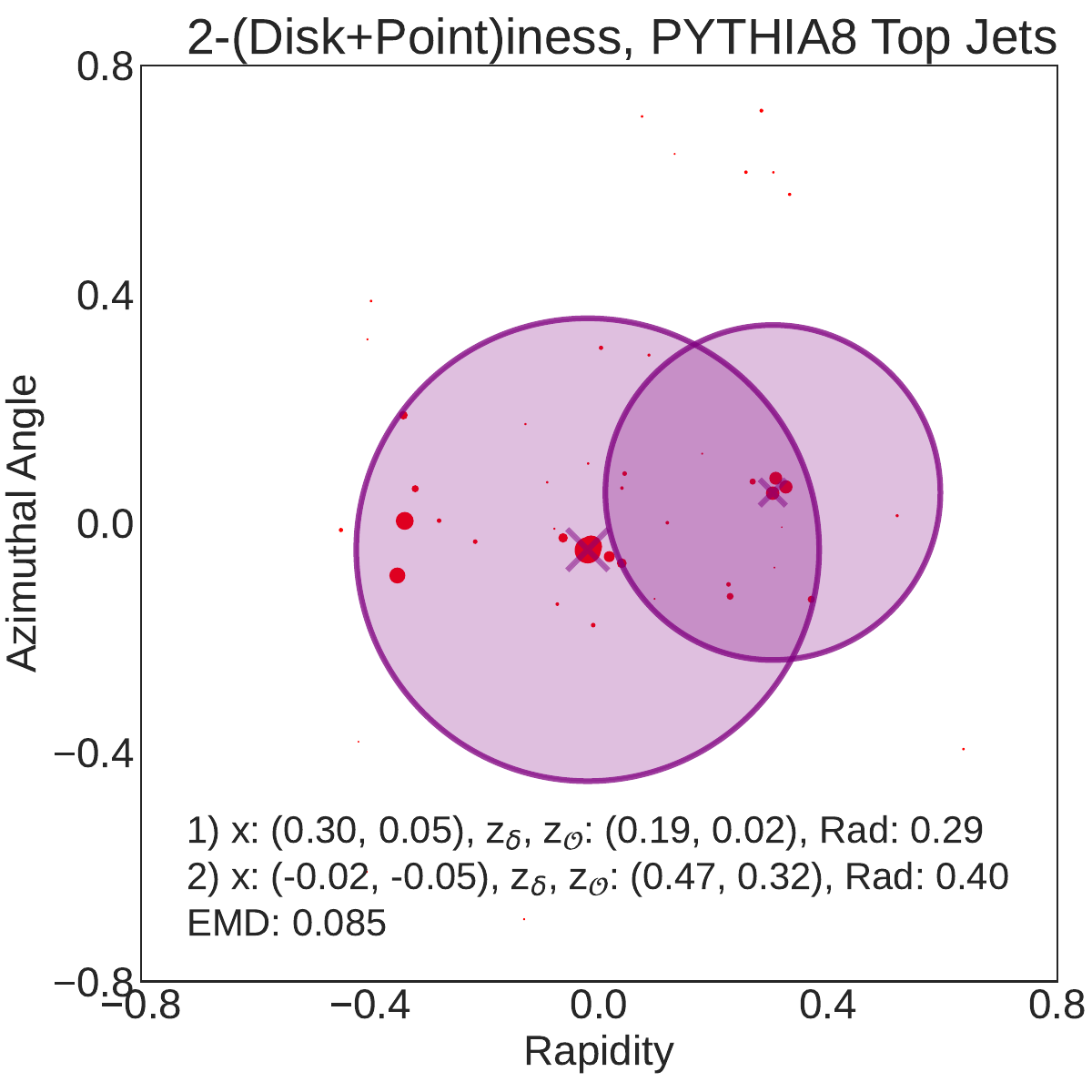}
        \label{fig:2_pdiskiness}
    }
    \subfloat[]{
        \includegraphics[width=0.32\textwidth]{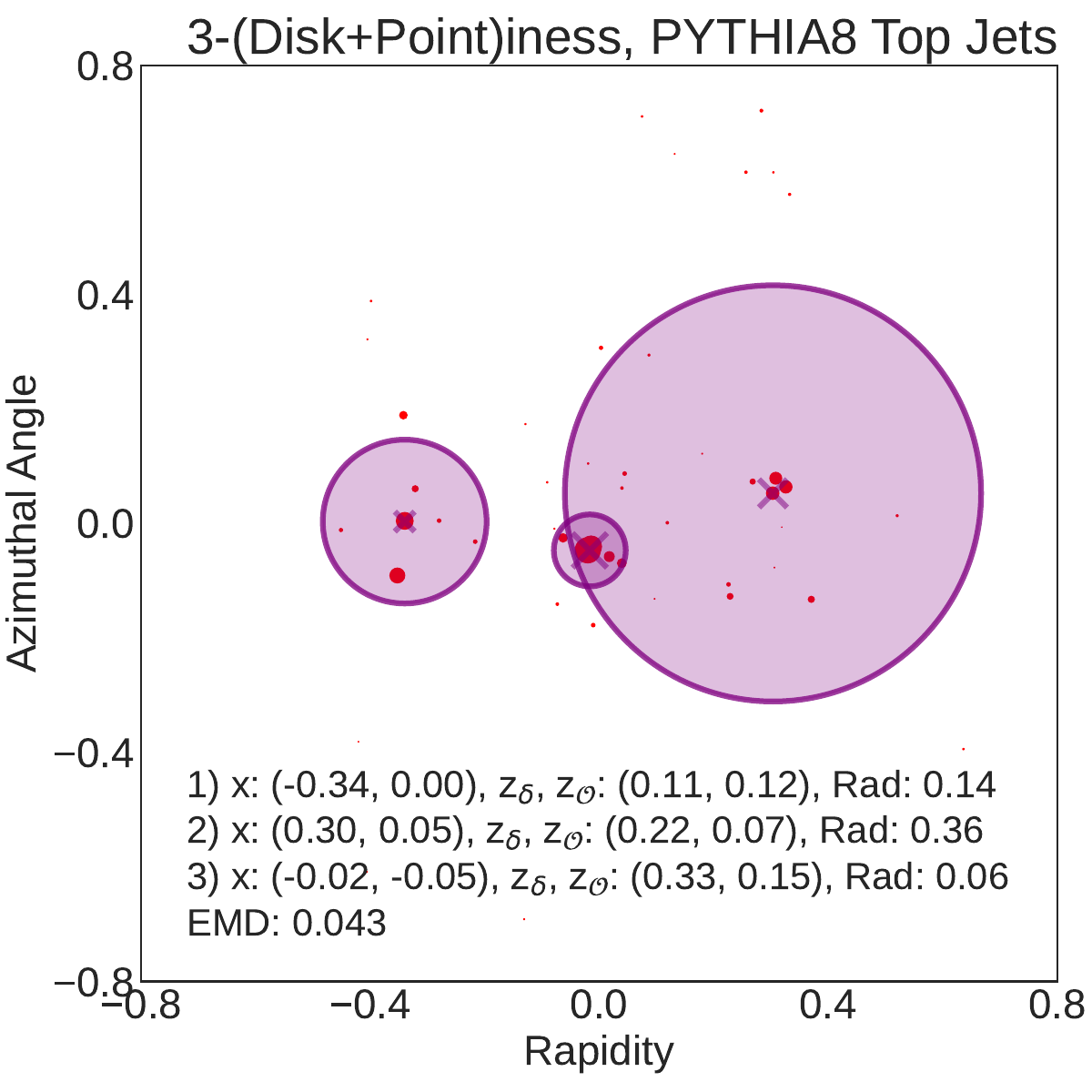}
        \label{fig:3_pdiskiness}
    }
    \caption{
        Top row: the (a) 1-, (b) 2-, and (c) 3-diskiness of an example top jet event. Bottom row: the (d) 1-, (e) 2-, and (f) 3-(disk$+$point)iness of the same top jet event. The point is represented by a ``$\times$'', with size proportional to its energy weight. 
        }
    \label{fig:n_diskiness}
\end{figure*}

\begin{figure*}[tp]
    \centering
    \subfloat[]{
         \includegraphics[width=0.32\textwidth]{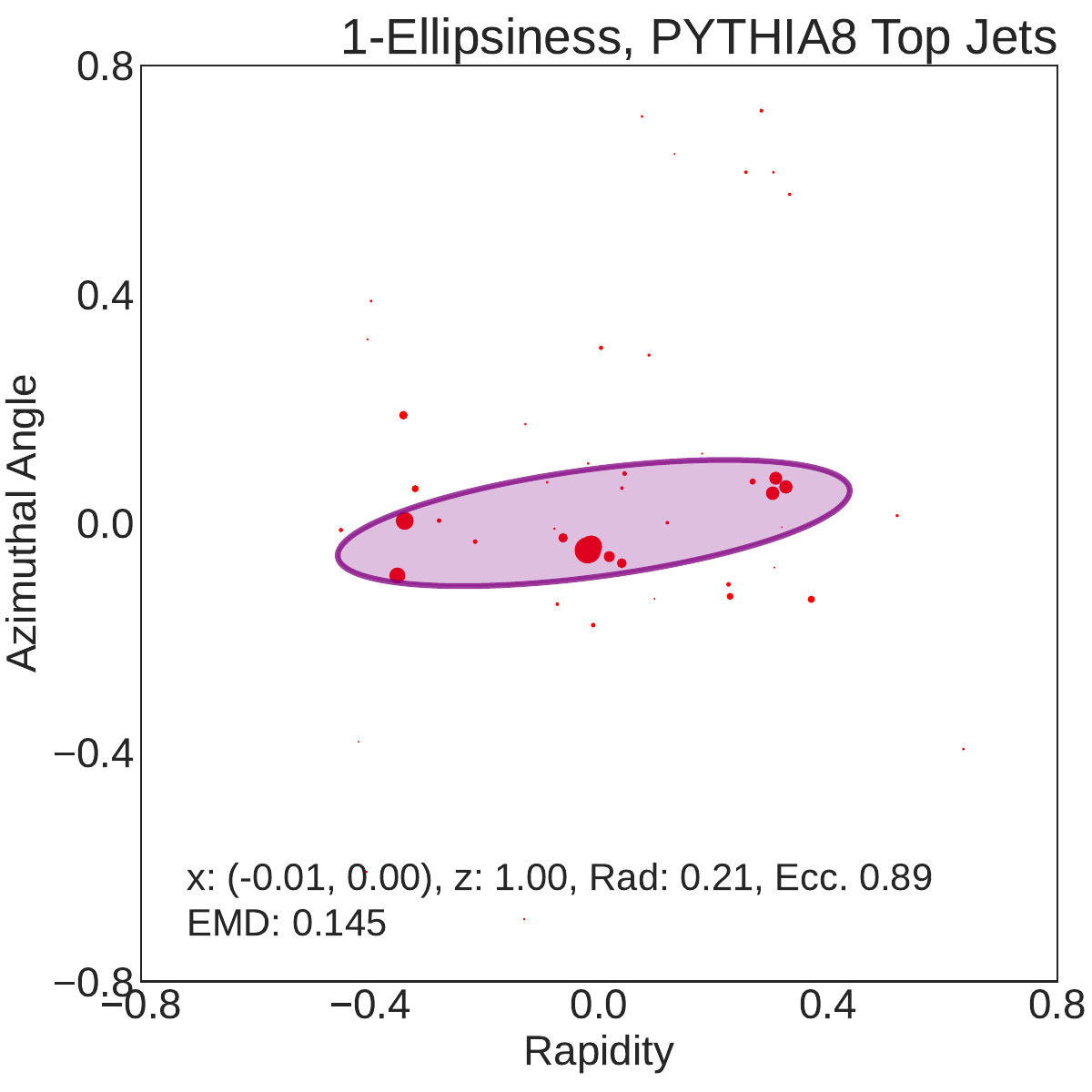}
        \label{fig:1_ellipseiness}
    }
    \subfloat[]{
        \includegraphics[width=0.32\textwidth]{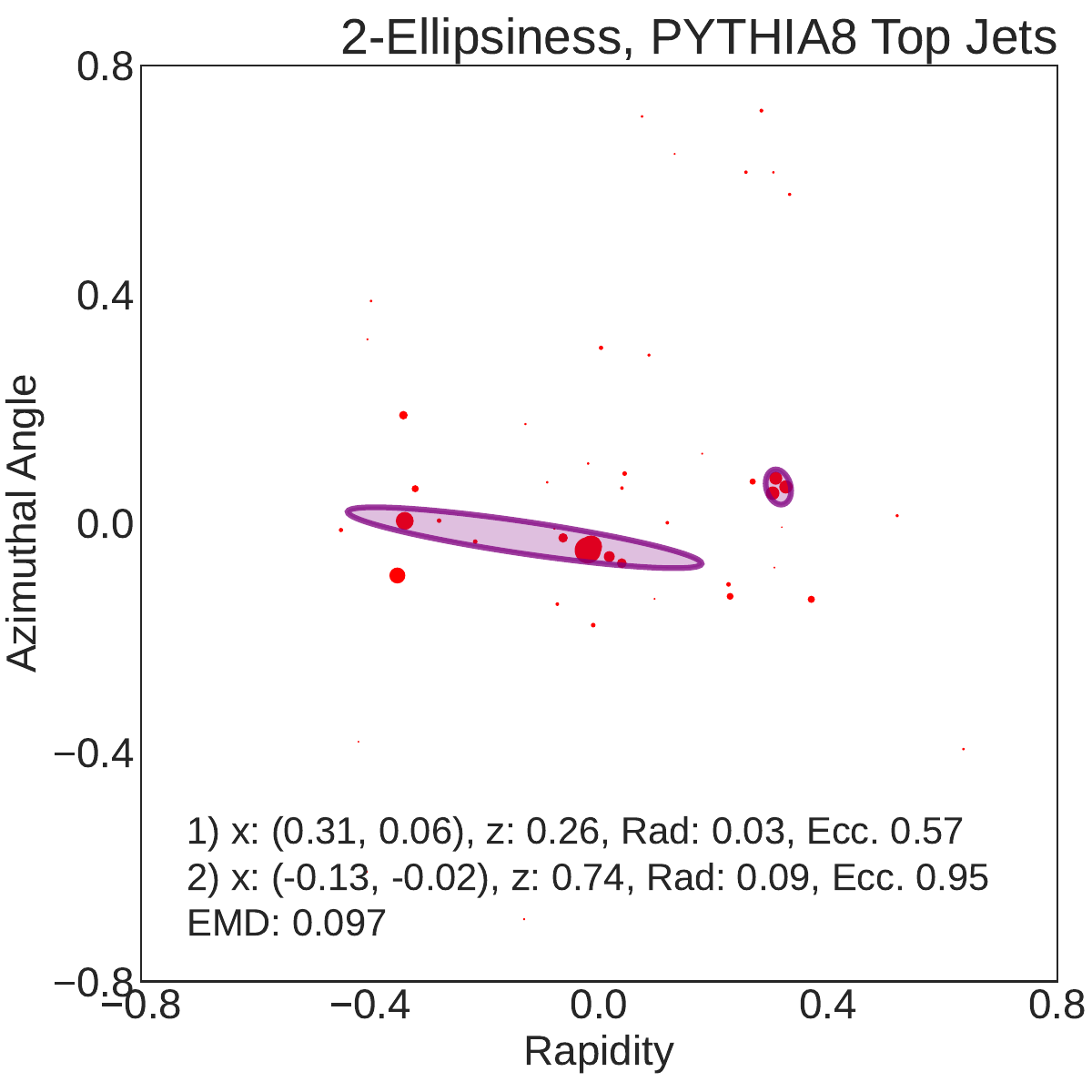}
        \label{fig:2_ellipseiness}
    }
    \subfloat[]{
        \includegraphics[width=0.32\textwidth]{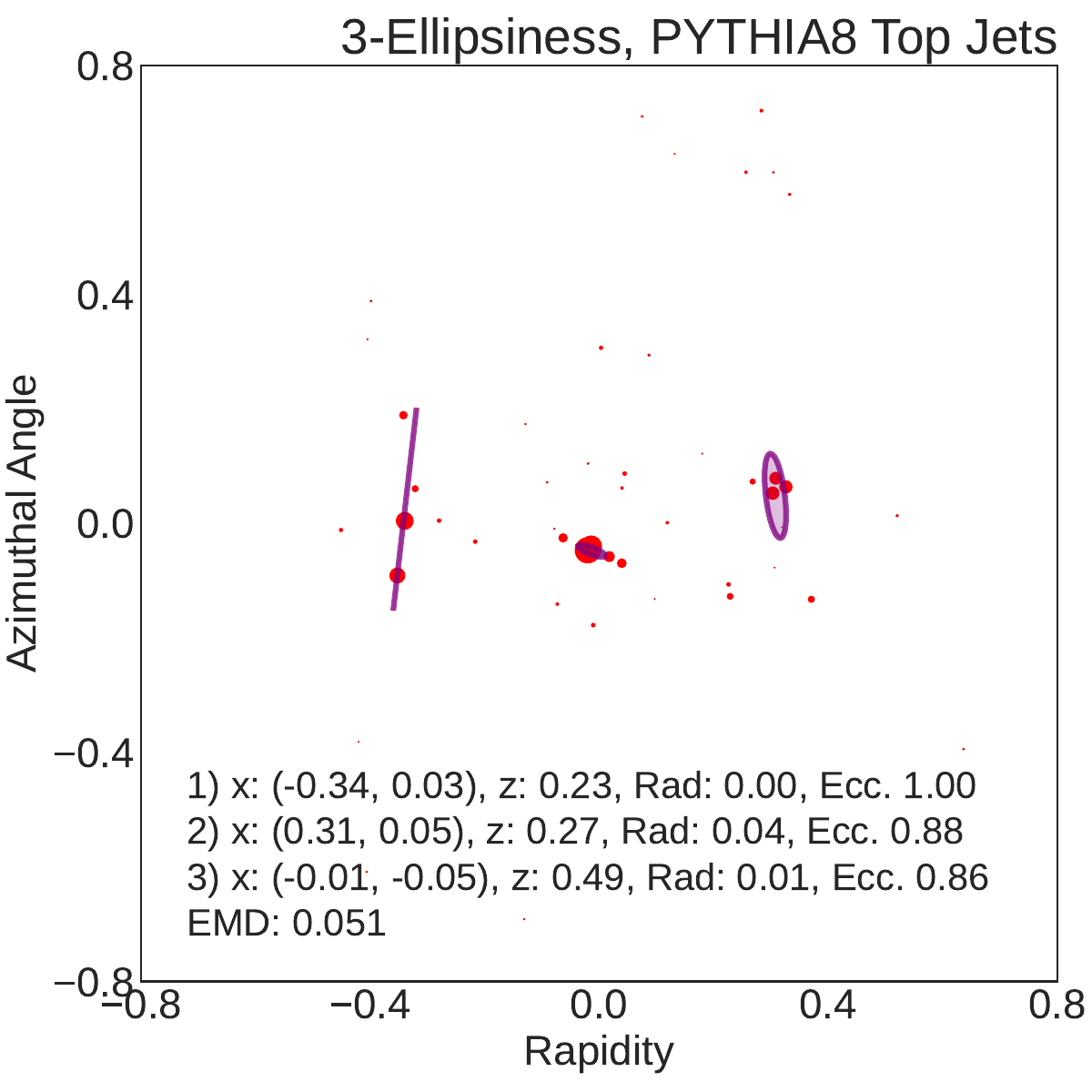}
        \label{fig:3_ellipseiness}
    }
    \vspace{1pt}
    \subfloat[]{
         \includegraphics[width=0.32\textwidth]{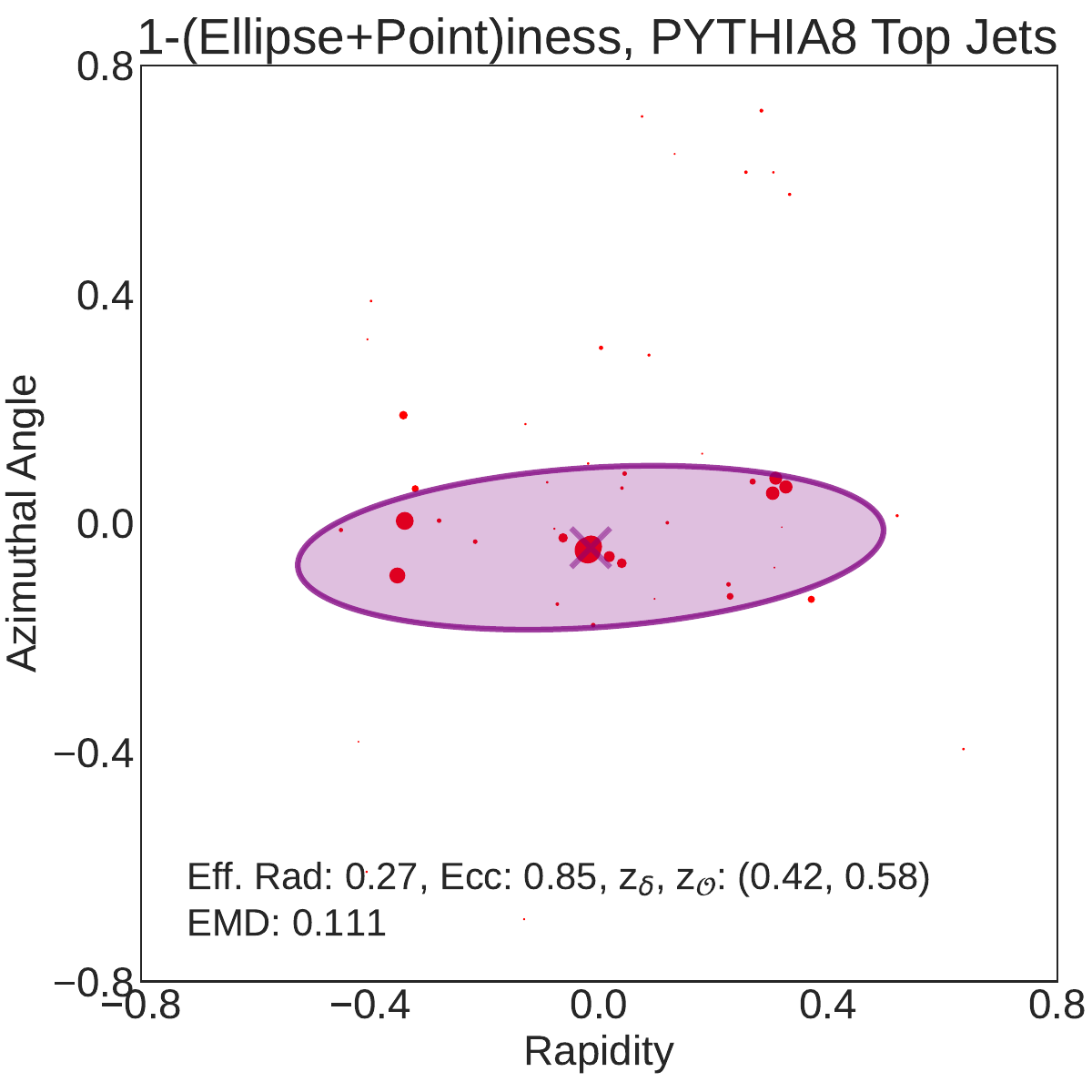}
        \label{fig:1_pellipseiness}
    }
    \subfloat[]{
        \includegraphics[width=0.32\textwidth]{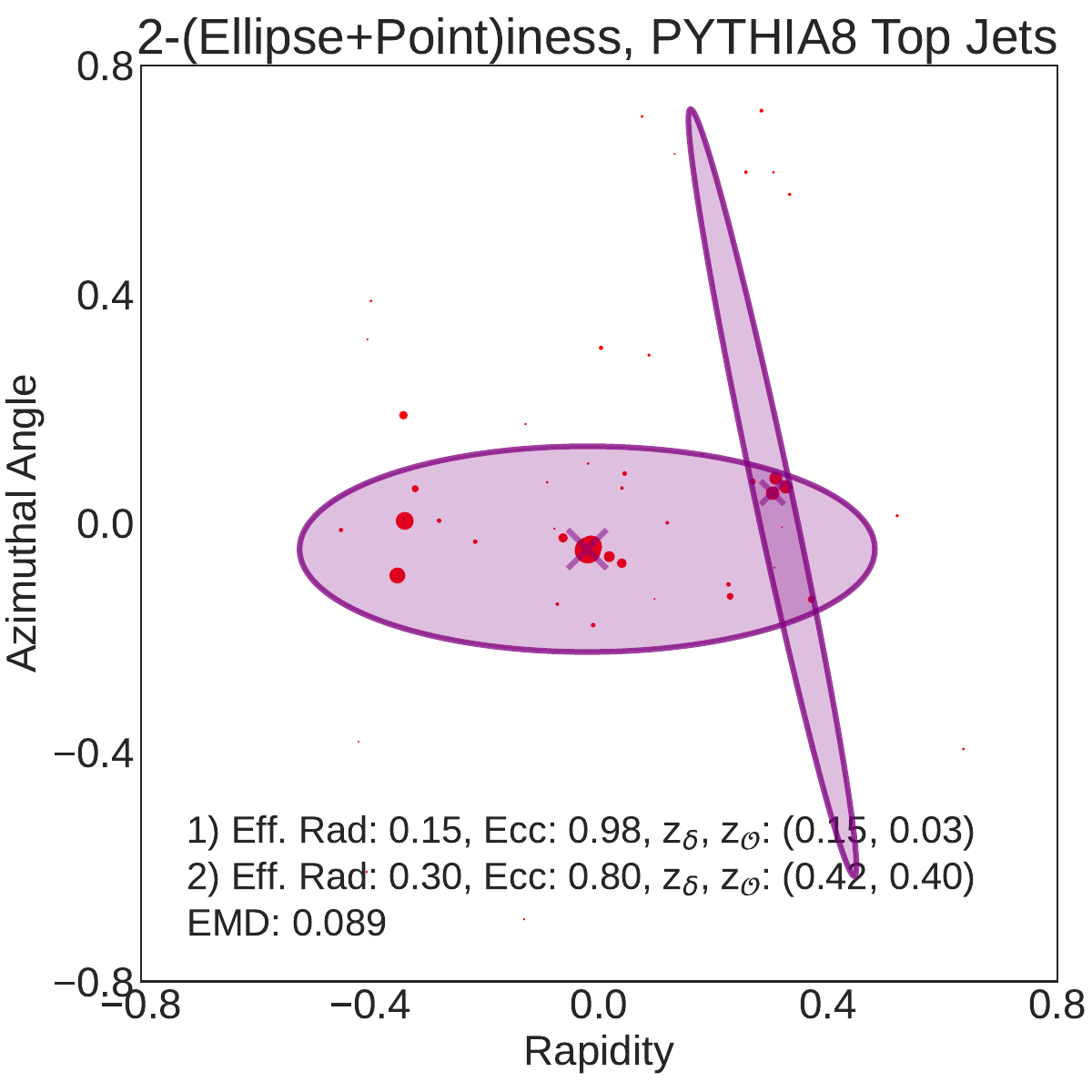}
        \label{fig:2_pellipseiness}
    }
    \subfloat[]{
        \includegraphics[width=0.32\textwidth]{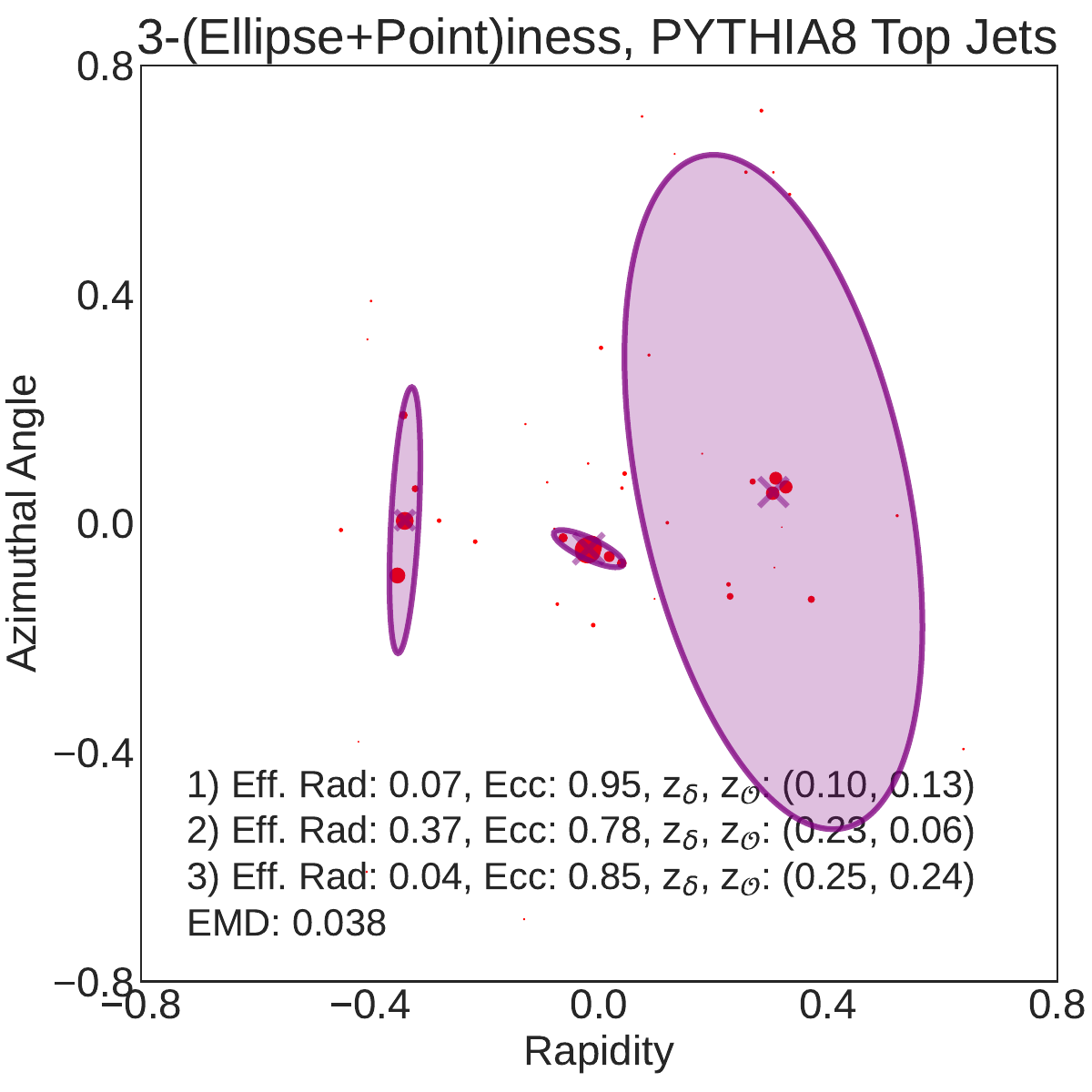}
        \label{fig:3_pellipseiness}
    }
    \caption{
        Top row: the (a) 1-, (b) 2-, and (c) 3-ellipsiness of an example top jet event. Bottom row: the (d) 1-, (e) 2-, and (f) 3-(ellipse$+$point)iness of the same top jet event. The point is represented by a ``$\times$'', with size proportional to its energy weight. The effective radius of the ellipses are given by the geometric mean of the two axes, $\sqrt{ab}$.
        }
    \label{fig:n_ellipsiness}
\end{figure*}

Geometric visualizations of each of the 21 event shapes, as evaluated on an example top jet, can be found in Figs. \ref{fig:n_subjettiness}, \ref{fig:n_ringiness}, \ref{fig:n_diskiness}, and \ref{fig:n_ellipsiness}. From these visualizations, we note some interesting qualitative features of these shape observables. First, the point variants of each shape correspond more closely to clusters of energy.
For example, while the $N=3$ uniform rings (\Fig{n_ringiness}), disks (\Fig{n_diskiness}), or ellipses (\Fig{n_ellipsiness}) do not necessarily capture the regions of highest energy, the point variants of each shape align very well with the $N$-subjettinesses of \Fig{n_subjettiness}.
Correspondingly, the EMD's of the shapes are significantly reduced for the point variants -- this suggests that this event in particular is not well modeled by uniform radiation profiles, but rather looks more like localized spikes with radiation clouds around them. Moreover, we can see that circular radiation clouds do not model the event as well as elliptical ones -- this is reflected in the $N$-ellipsiness in \Fig{n_ellipsiness}, which learns extremely eccentric line-like structures in an attempt to best model the event, which results in lower EMD's than the corresponding $N$-diskinesses in \Fig{n_diskiness}. We also note that the $N$-subjettiness and the point shape variants all qualitatively find the same jet centers, suggesting that these shapes can be treated as perturbations to $N$-subjettiness.

\subsection{Shapiness and Shape Parameters}\label{sec:jets_parameters}

Continuing the discussion in \Sec{jets}, we now use \Shaper to compute distributions of these shape observables on a large sample of top and QCD jets, restricted to $m_J \in [145, 205]\,\GeV$, though without any pileup contamination. In particular, we show the utility of both the \emph{shapiness} $\mathcal{O}(\E)$ and the \emph{shape parameters} $\theta(\E)$ in describing the geometry of jets.

For each histogram in this section, we calculate an AUC score, showing the efficacy of a cut on that observable as a top/QCD discriminant. Note that these AUCs are for cuts on a single feature -- the discrimination power can in principle be improved by transforming these features and combining many features per jet.

\begin{figure*}[tp]
    \centering
    \subfloat[]{
         \includegraphics[width=0.32\textwidth]{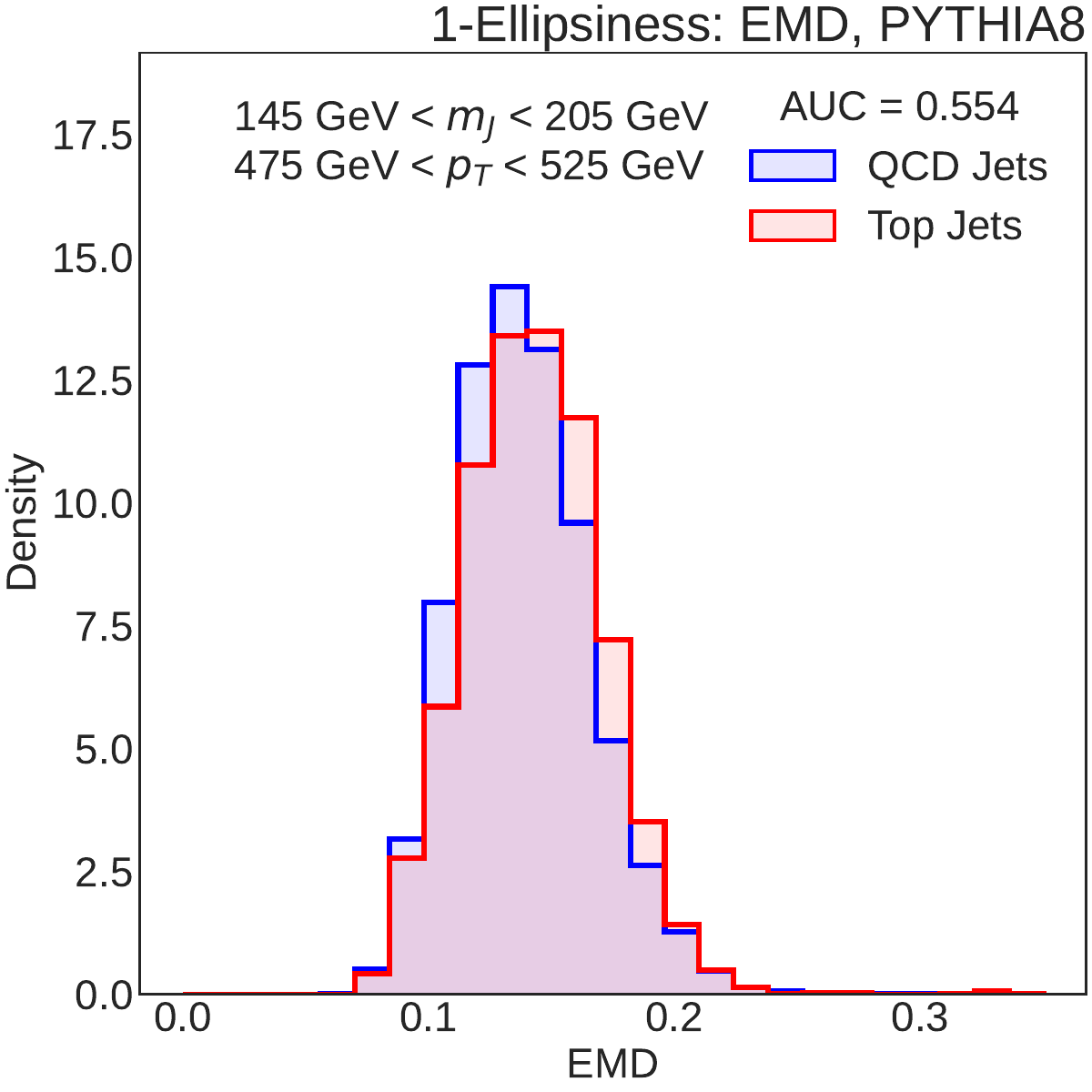}
        \label{fig:emd_ellipse1}
    }
    \subfloat[]{
         \includegraphics[width=0.32\textwidth]{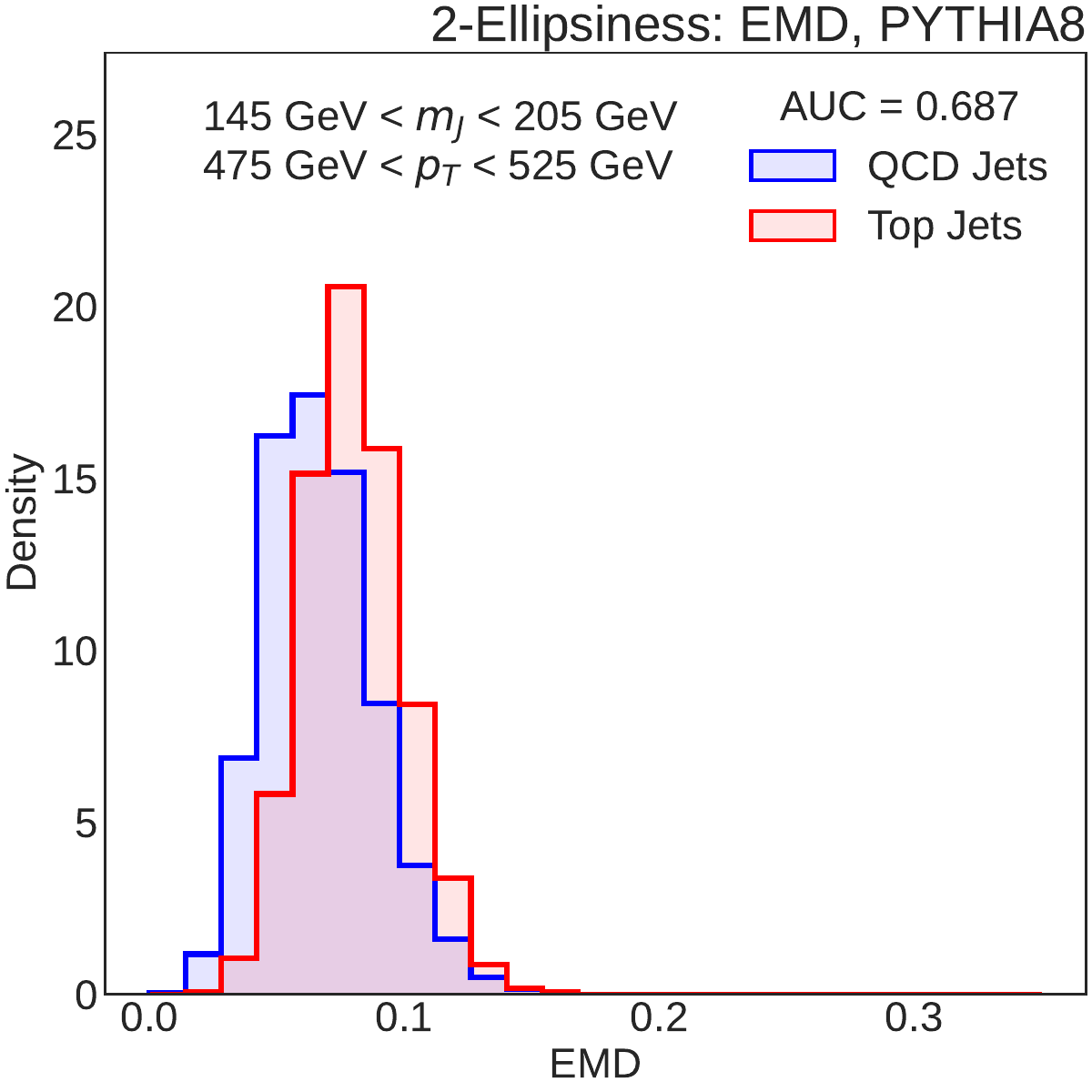}
        \label{fig:emd_ellipse2}
    }
    \subfloat[]{
         \includegraphics[width=0.32\textwidth]{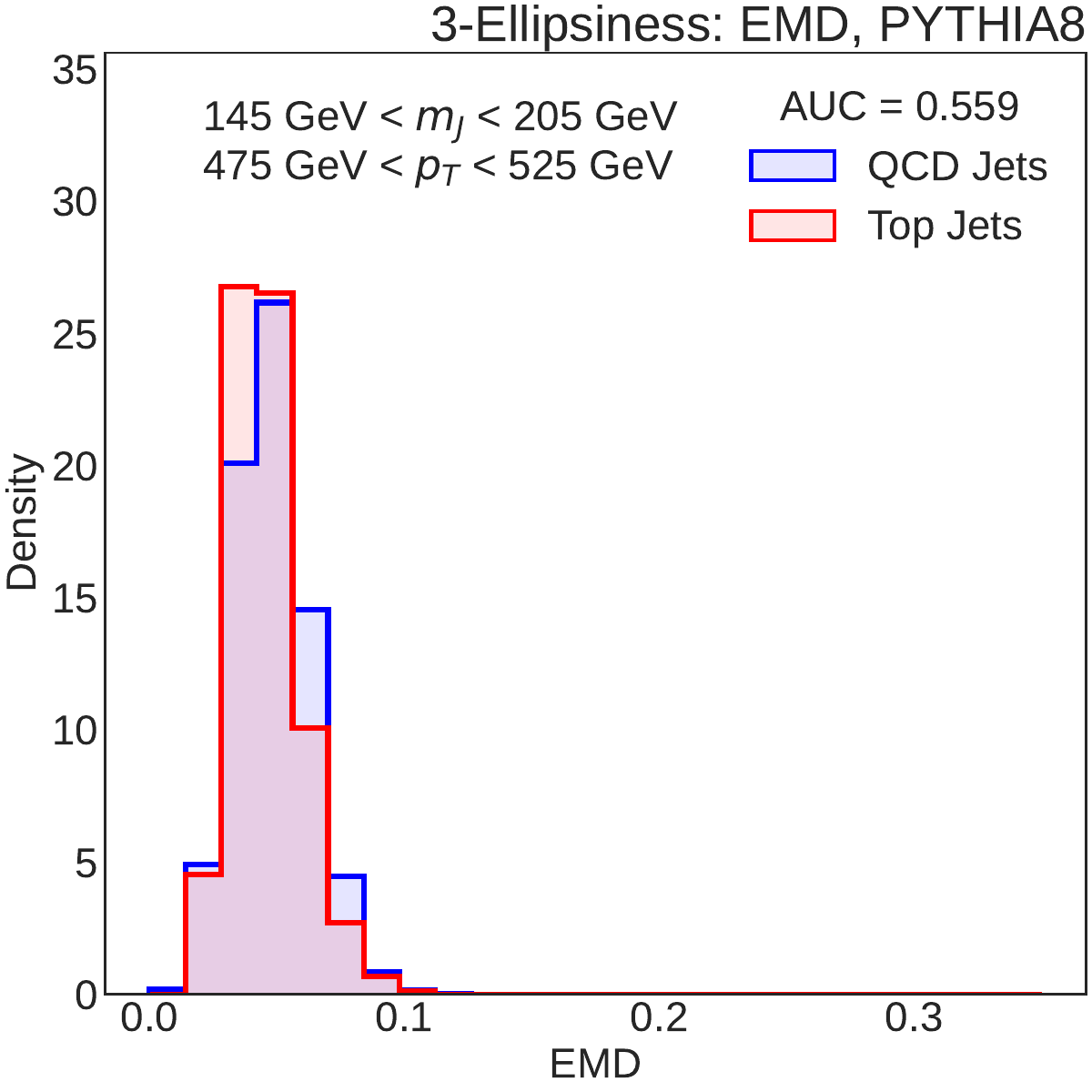}
        \label{fig:emd_ellipse3}
    }
    \vspace{1pt}
    \subfloat[]{
         \includegraphics[width=0.32\textwidth]{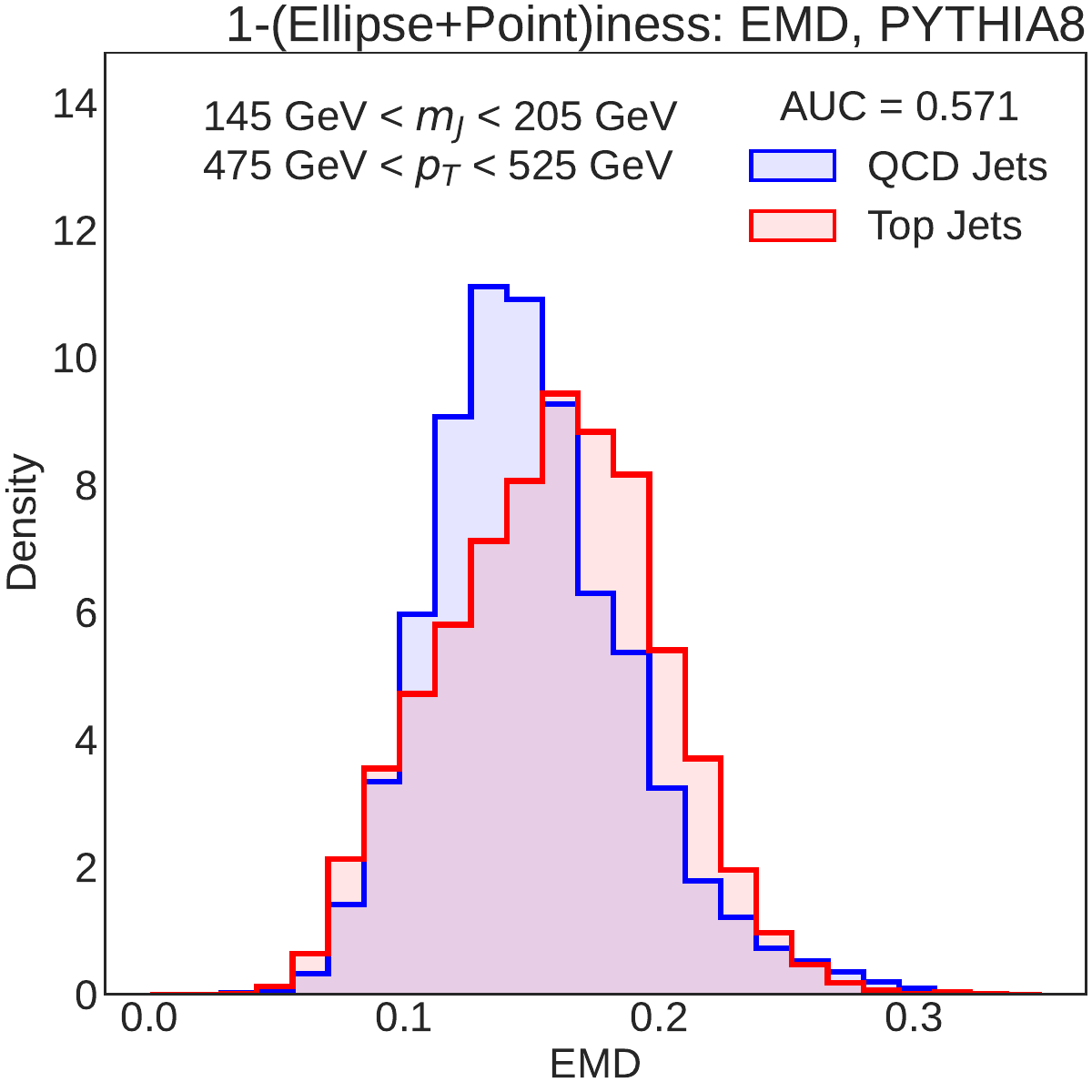}
        \label{fig:emd_point_ellipse1}
    }
    \subfloat[]{
         \includegraphics[width=0.32\textwidth]{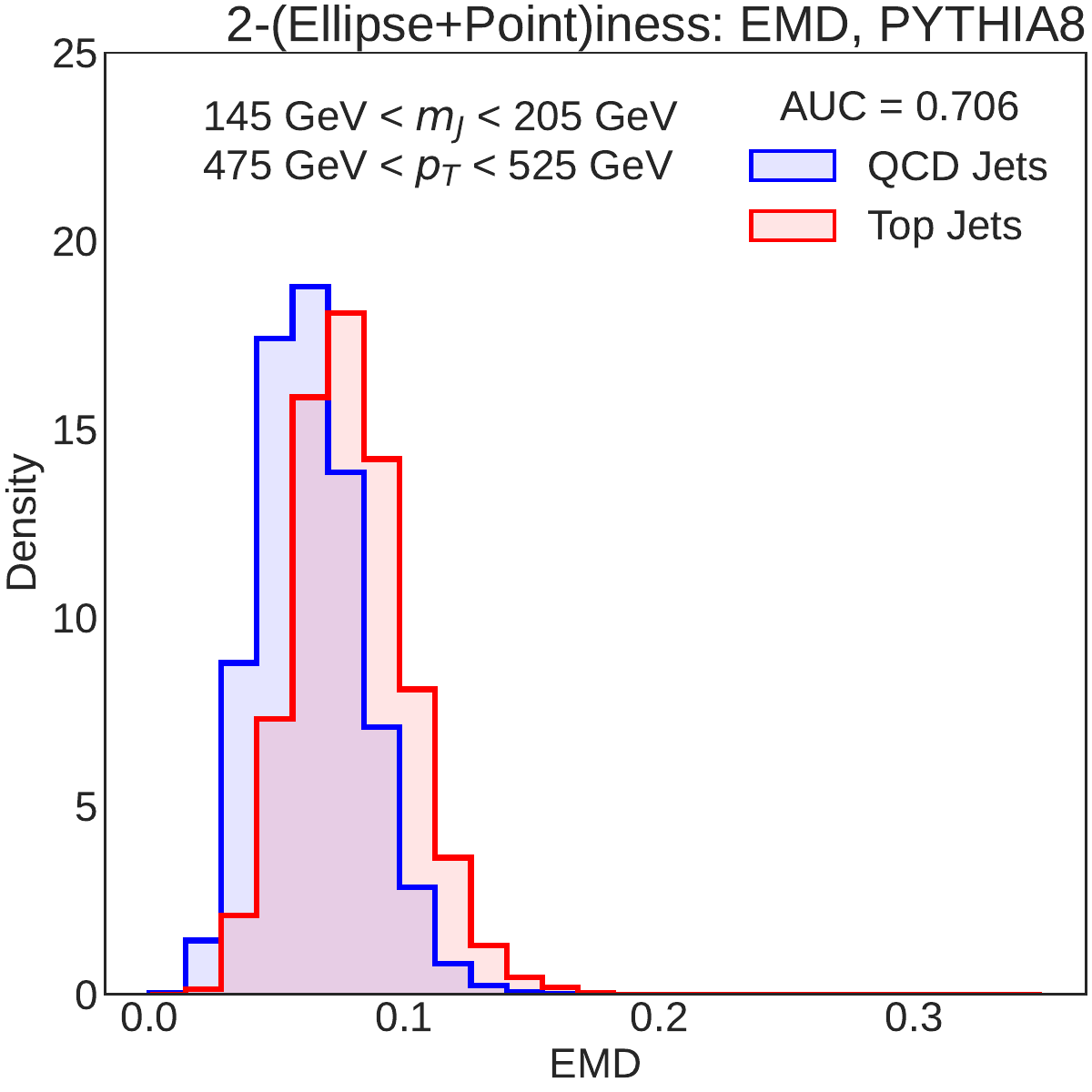}
        \label{fig:emd_point_ellipse2}
    }
    \subfloat[]{
         \includegraphics[width=0.32\textwidth]{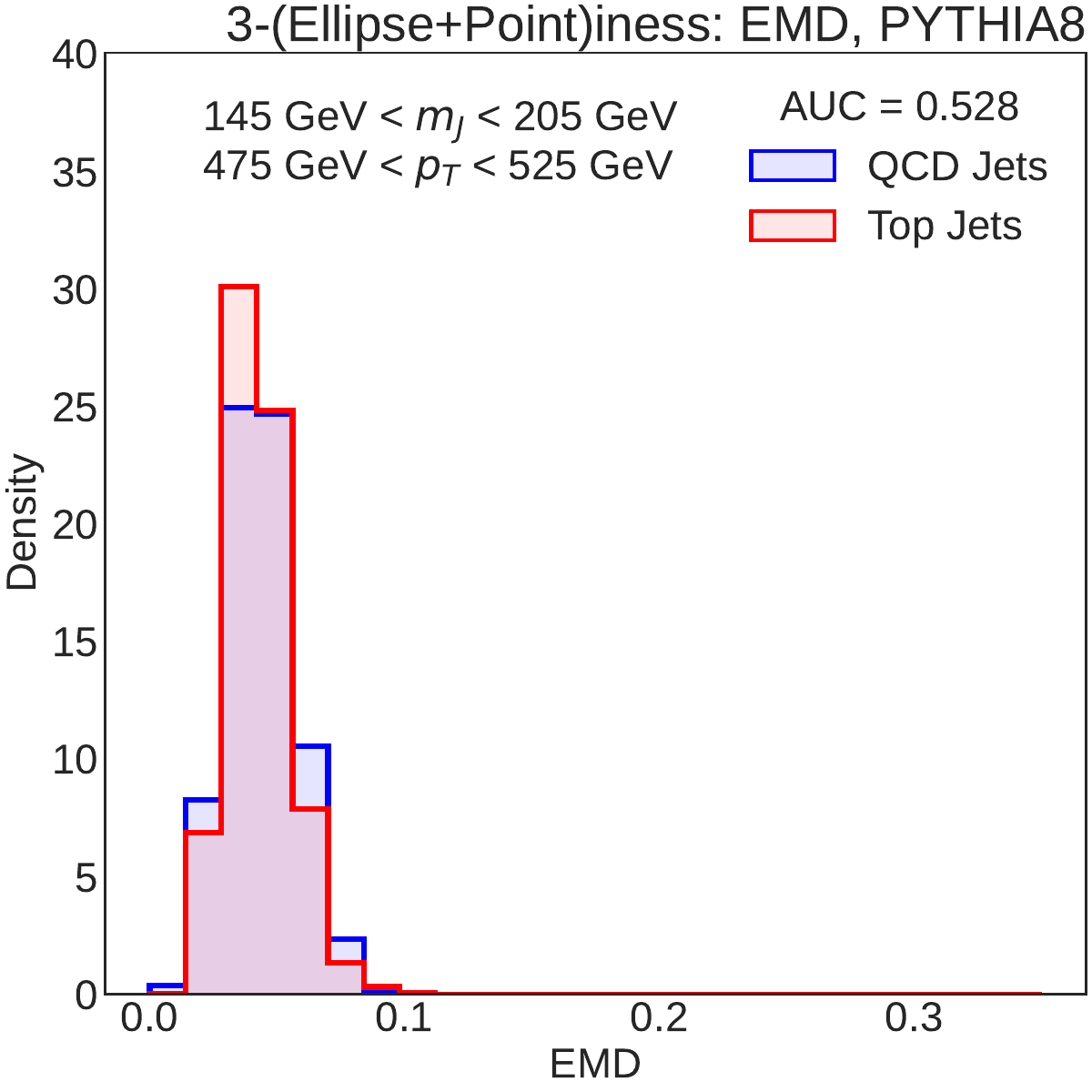}
        \label{fig:emd_point_ellipse3}
    }
    \caption{
        Distributions of the learned (a,d) 1-, (b,e) 2-, and (c,f) 3-ellipsiness (top row) and (ellipse$+$point)iness (bottom row) of the top (red) and QCD (blue) jet sample. 
        }
    \label{fig:emd_ellipse}
\end{figure*}

As a representative sample of the ``shapiness'' observable, we plot the $N$-ellipsiness and $N$-(ellipse$+$point)iness of our top and QCD jet samples in \Fig{emd_ellipse}.
The EMD distributions for the ring and disk variants are qualitatively similar to the ellipse, and thus our discussion of these distributions carry over to them. We notice that the $N$-(ellipse$+$point)iness is not much lower than the corresponding $N$-ellipsiness, indicating that (at least in the absence of pileup), subjets can indeed be approximated as roughly uniform. In the test event visualization in \Fig{n_ellipsiness}, we can see qualitatively that the found ellipses  have roughly the same center, comparing the $N$-ellipsiness and its corresponding point variant. We also note that the EMD decreases with $N$, as expected.

\begin{figure*}[tp]
    \centering
    \subfloat[]{
            \includegraphics[width=0.32\textwidth]{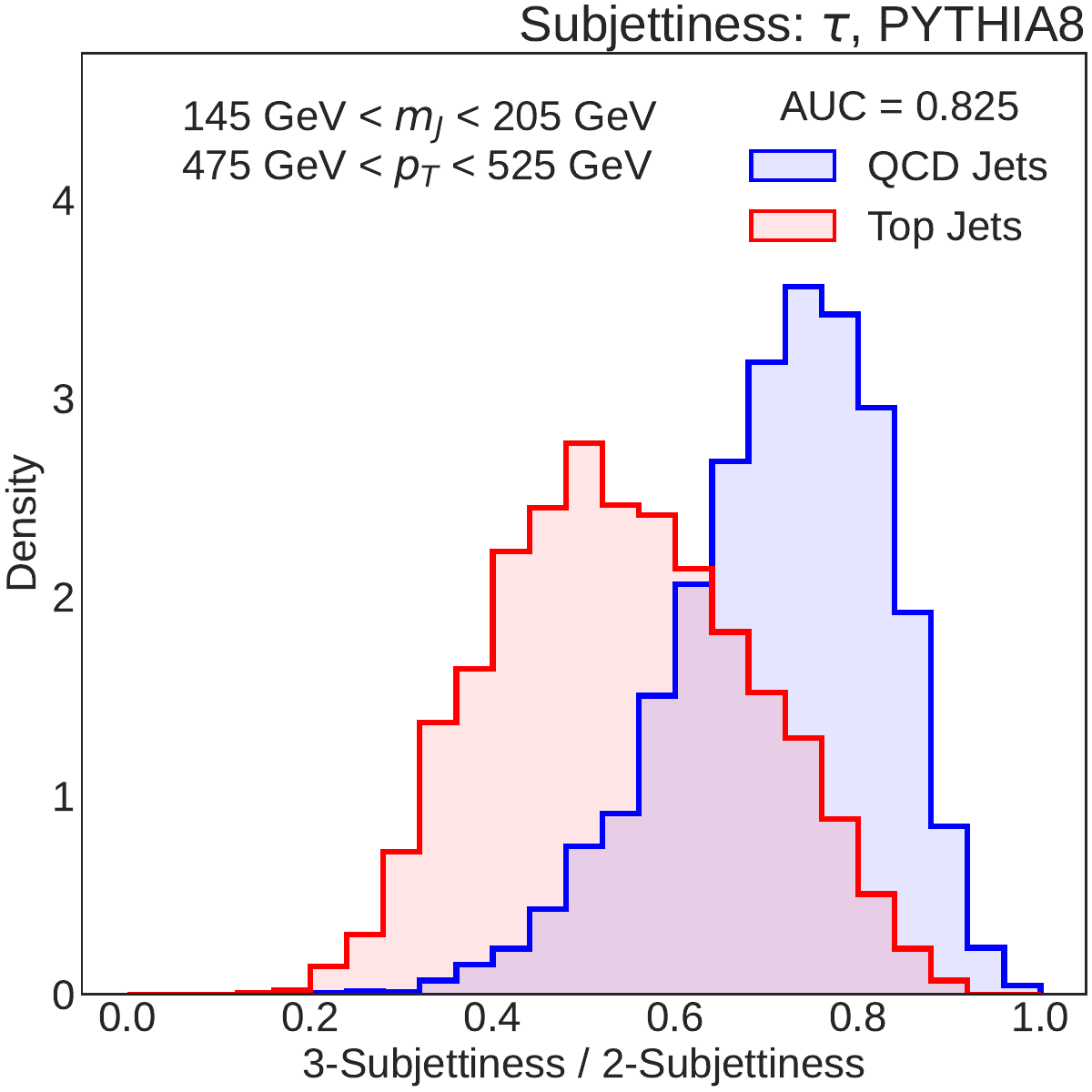}
        \label{fig:tau_subjet}
    }
    \vspace{1pt}
    \subfloat[]{
         \includegraphics[width=0.32\textwidth]{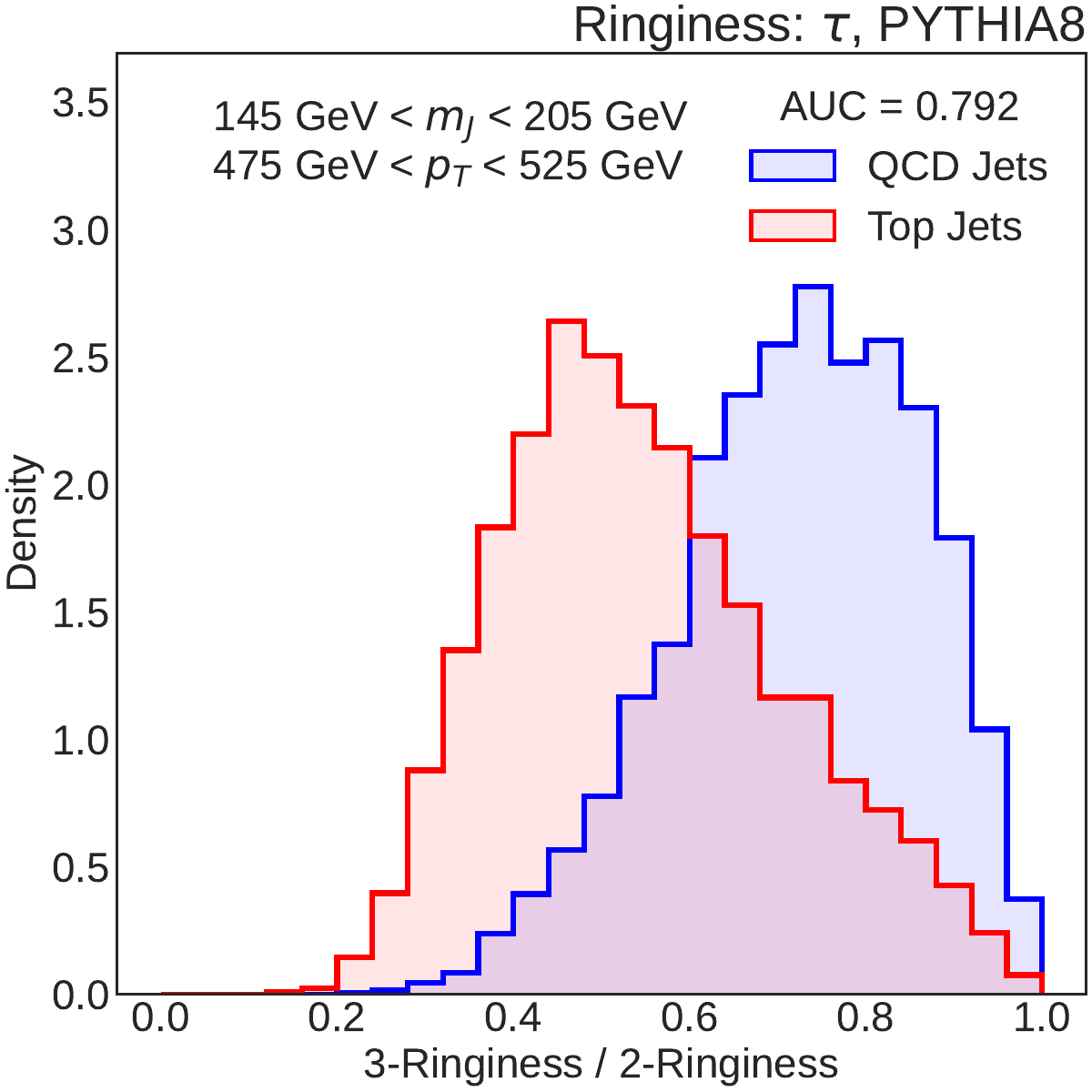}
        \label{fig:tau_ring}
    }
    \subfloat[]{
         \includegraphics[width=0.32\textwidth]{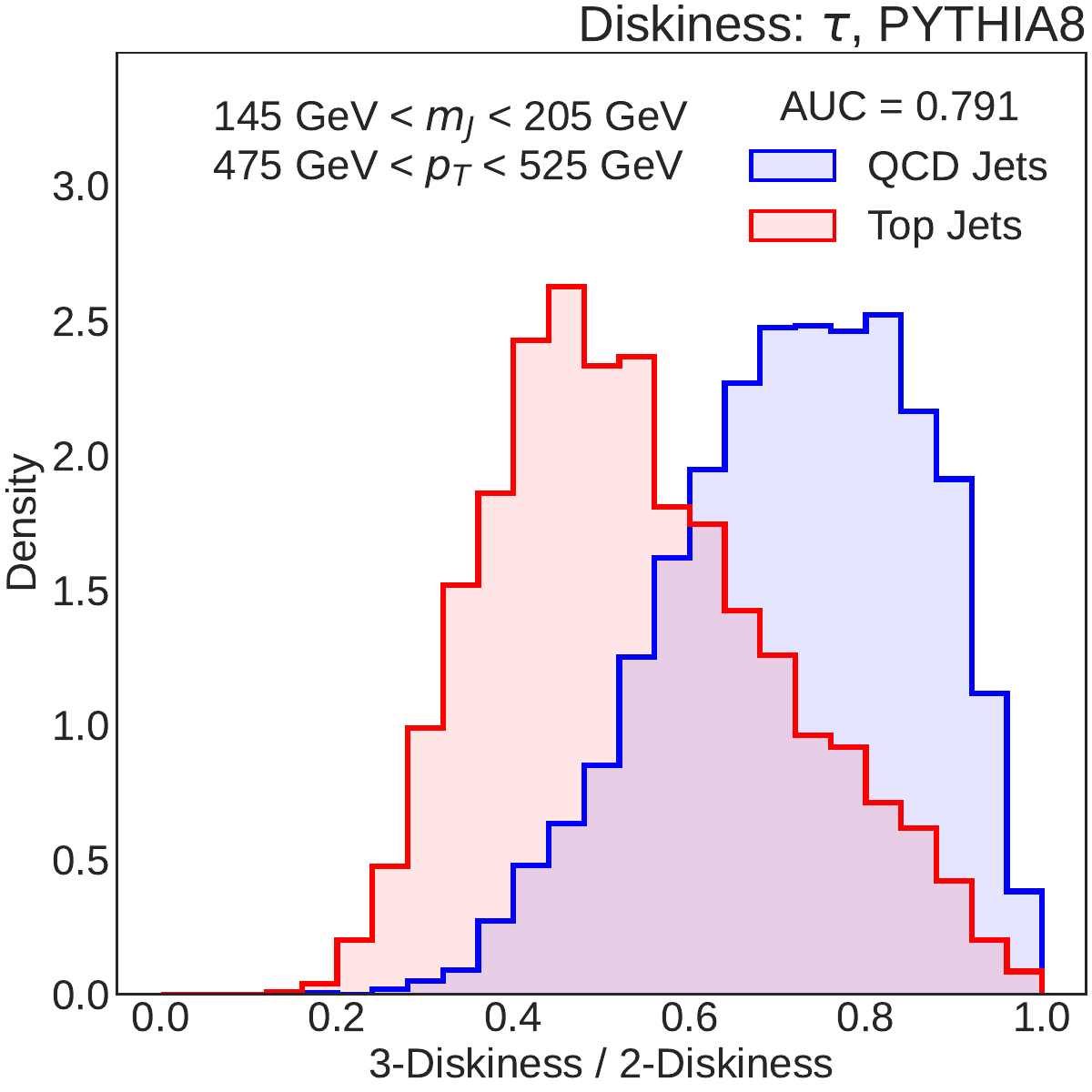}
        \label{fig:tau_disk}
    }
    \subfloat[]{
         \includegraphics[width=0.32\textwidth]{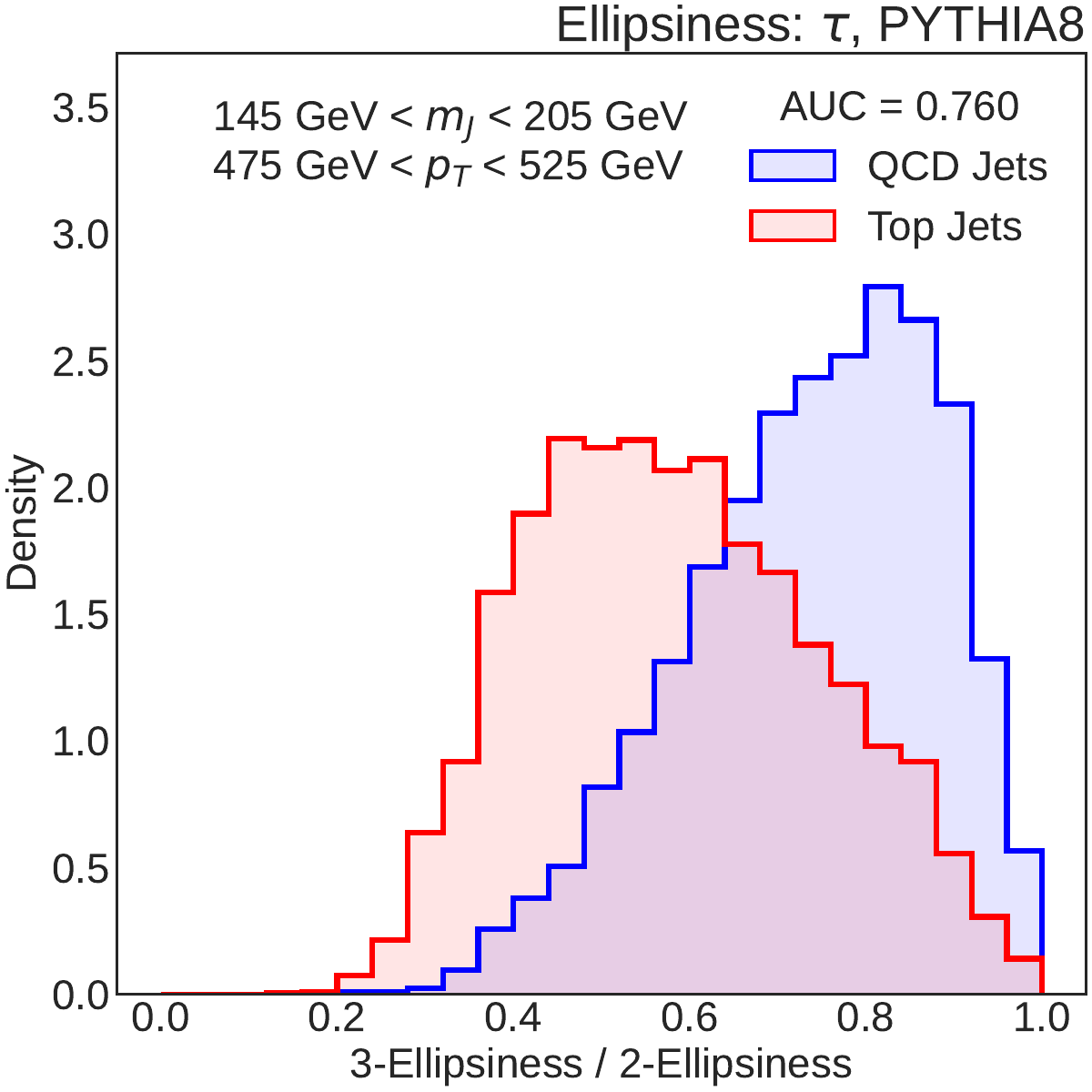}
        \label{fig:tau_ellipse}
    }
    \vspace{1pt}
    \subfloat[]{
         \includegraphics[width=0.32\textwidth]{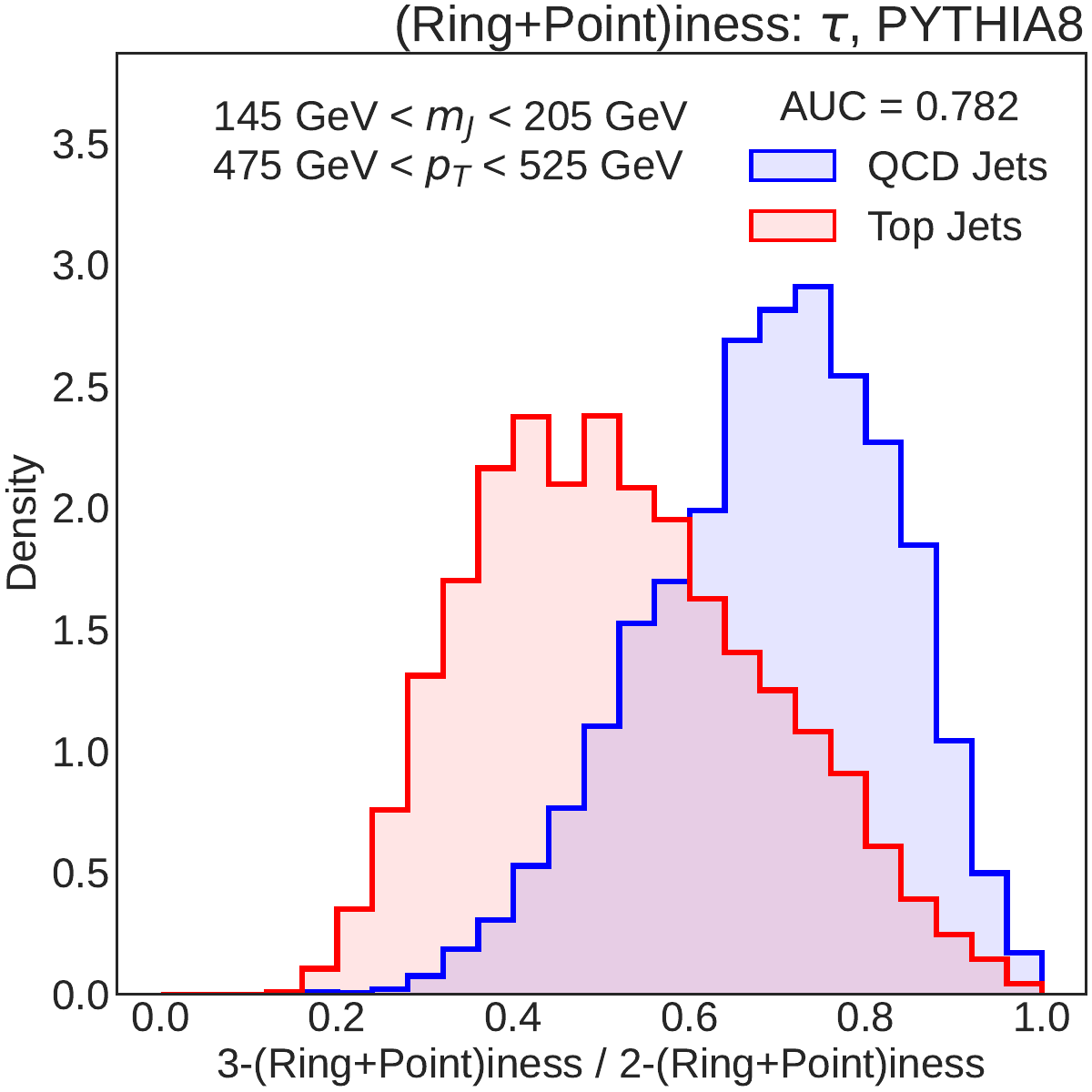}
        \label{fig:tau_point_ring}
    }
    \subfloat[]{
         \includegraphics[width=0.32\textwidth]{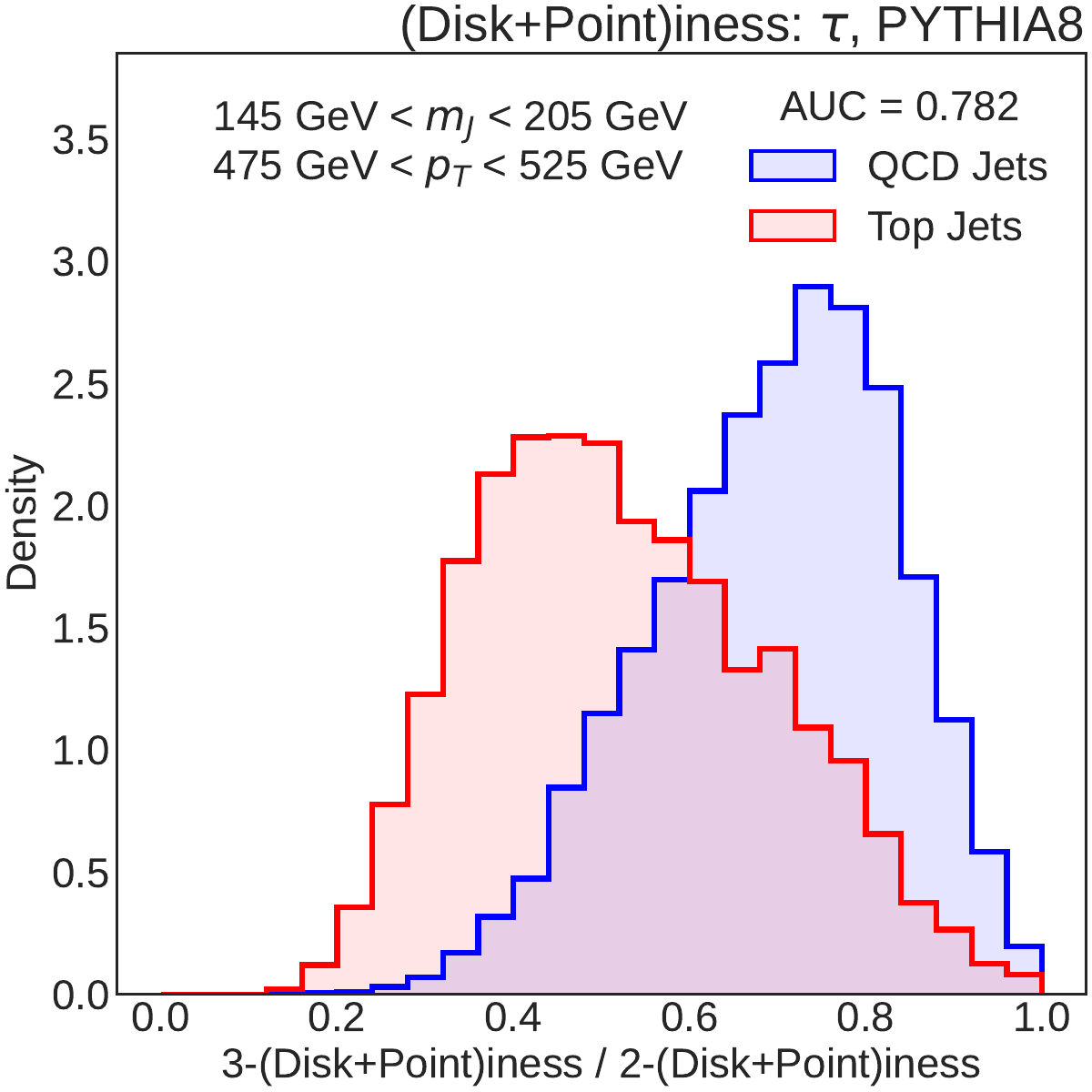}
        \label{fig:tau_point_disk}
    }
    \subfloat[]{
         \includegraphics[width=0.32\textwidth]{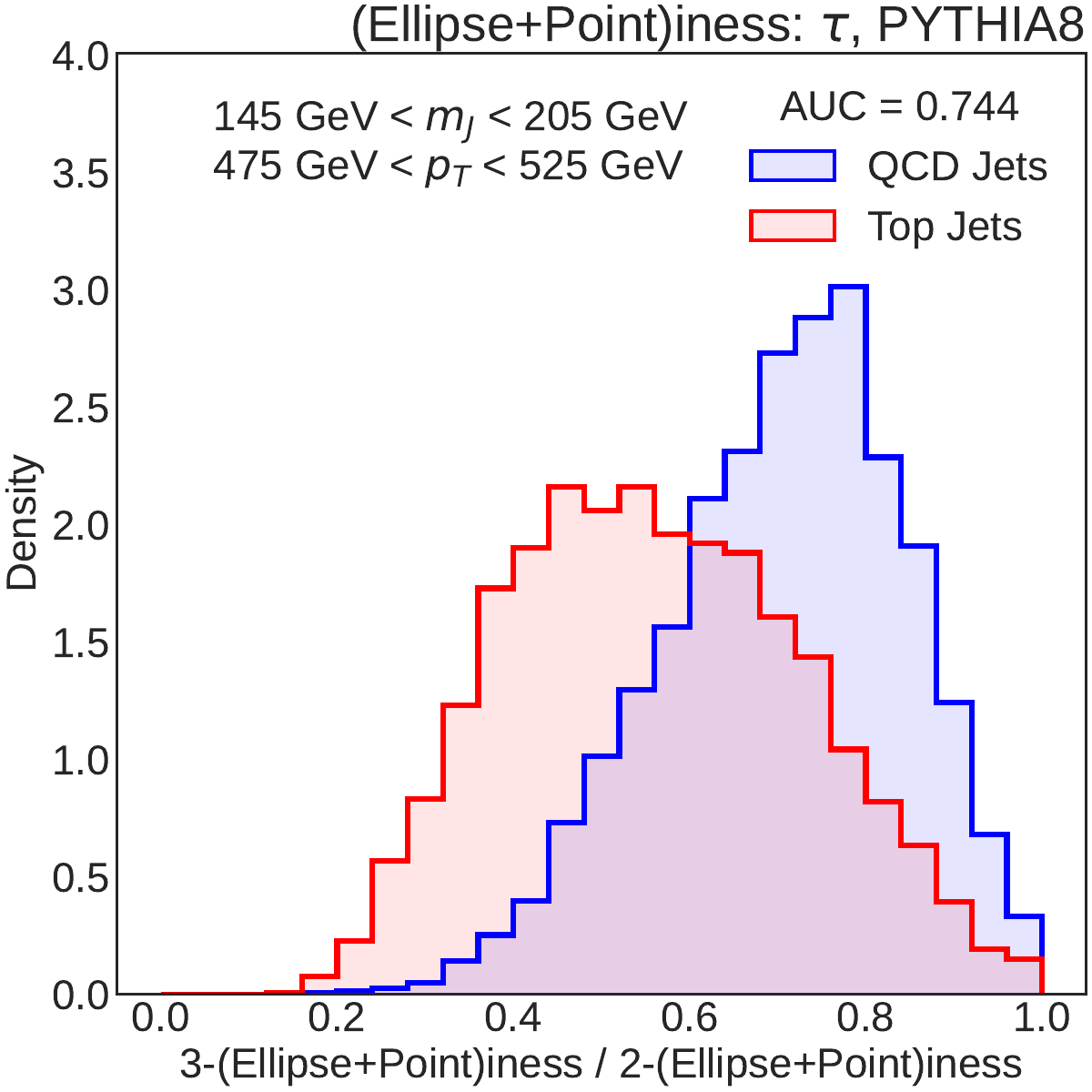}
        \label{fig:tau_point_ellipse}
    }
    \vspace{1pt}
    \caption{
        Top row: Distributions of the $N=3$ to $N=2$ ratio of the (a) $N$-subjettiness of the top (red) and QCD (blue) jet sample. Middle row: The same ratio, but for the $N$- (b) ringiness, (c) diskiness, and (d) ellipsiness. Bottom row: The same, but for the $N$- (e) (ring$+$point)iness, (f) (disk$+$point)iness, and (g) (ellipse$+$point)iness. 
        }
    \label{fig:tau}
\end{figure*}

In \Fig{tau}, we show the \emph{ratio} of the $N=3$ shapiness to the $N=2$ shapiness for each class of shape observable. For $N$-subjettiness, which we show for comparison, this is the classic observable $\tau_{32}$, which is known to be a good top vs.\ QCD jet discriminant~\cite{Thaler:2010tr}. First, we observe that the uniform ring, disk, and ellipse observable, along with the point variants, each have an AUC of approximately 0.75, which is still considerably less than $\tau_{32}$'s AUC of 0.825. One should expect that, using just the EMD alone, a more complexly parameterized shape should have a lower AUC than a simpler shape. In the extreme case, where the parameterization is flexible enough to reproduce any event in the dataset, the EMD will always be zero and have no discriminatory power. However, this is not the end of the story -- shape observables also include their learned parameters, and this information also contains multivariate discriminatory power. For the hypothetical infinitely flexible shape, the parameters contain the full event information even though the EMD is zero, and thus the combination of the shapiness and shape parameters together contain more information (and thus more discriminatory power) than just a simpler shape. We leave a full multivariate analysis of shape parameters for jet classification for potential future work.

\begin{figure*}[tp]
    \centering
    \subfloat[]{
         \includegraphics[width=0.32\textwidth]{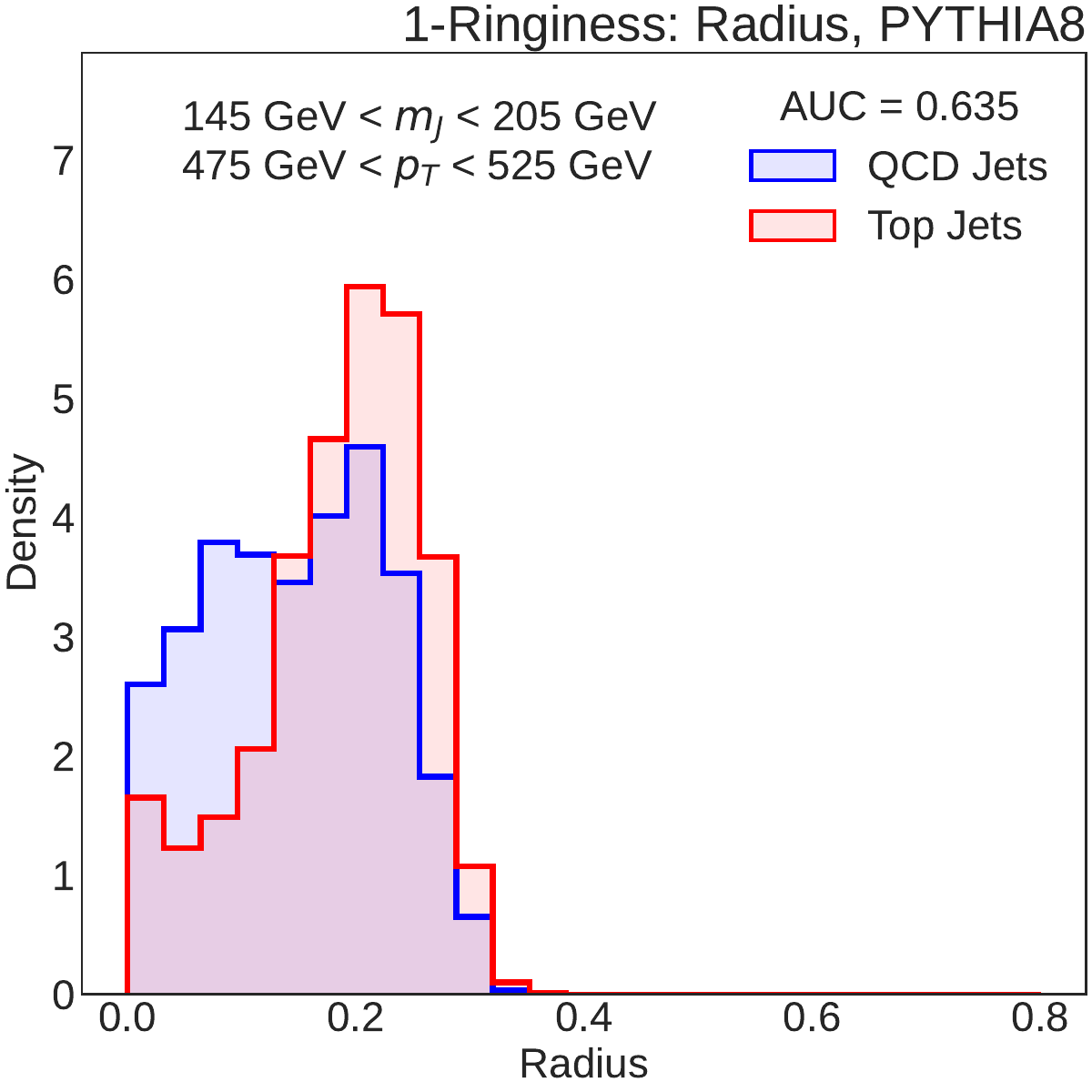}
        \label{fig:radius_a}
    }
    \subfloat[]{
         \includegraphics[width=0.32\textwidth]{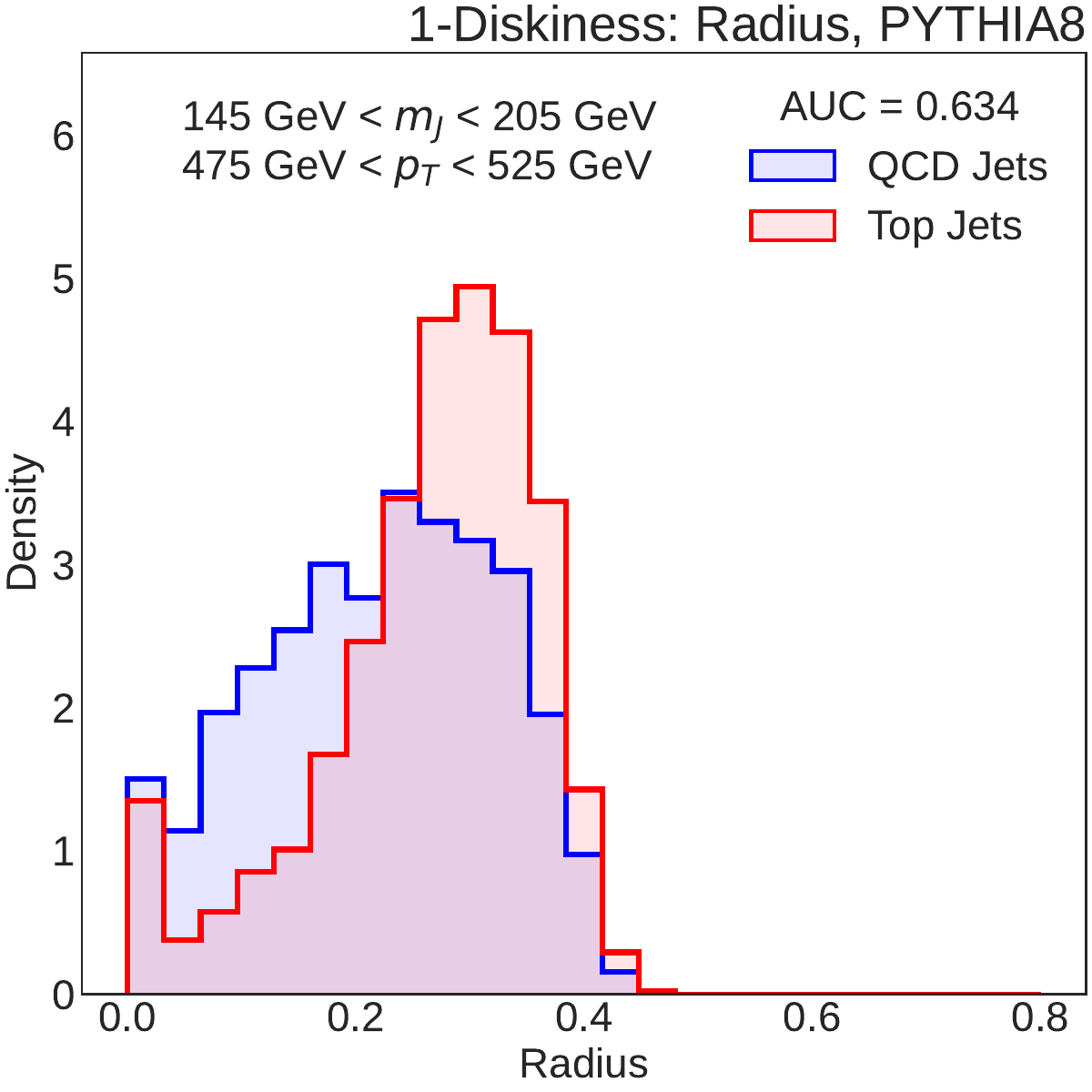}
        \label{fig:radius_b}
    }
    \subfloat[]{
         \includegraphics[width=0.32\textwidth]{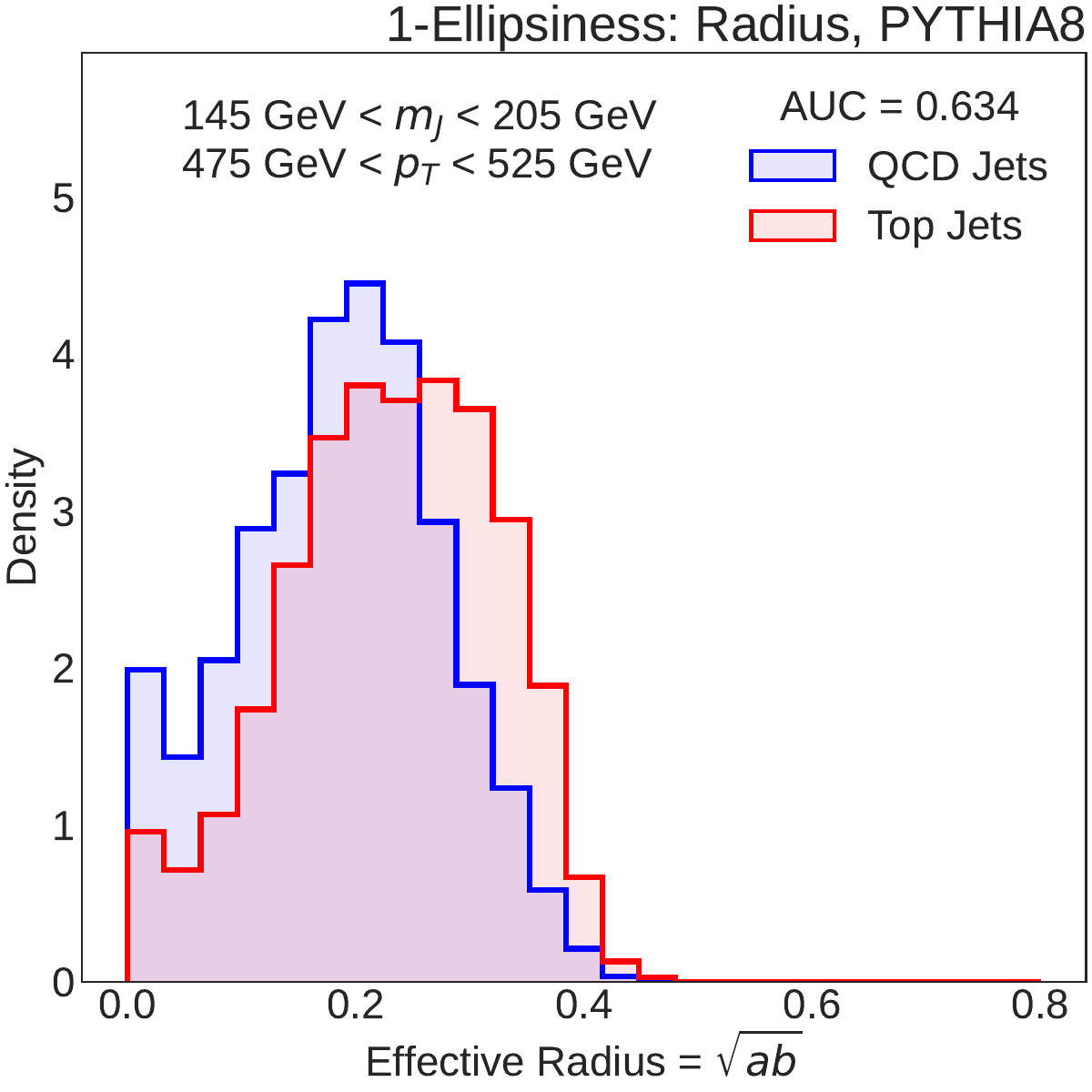}
        \label{fig:radius_c}
    }
    \vspace{1pt}
    \subfloat[]{
         \includegraphics[width=0.32\textwidth]{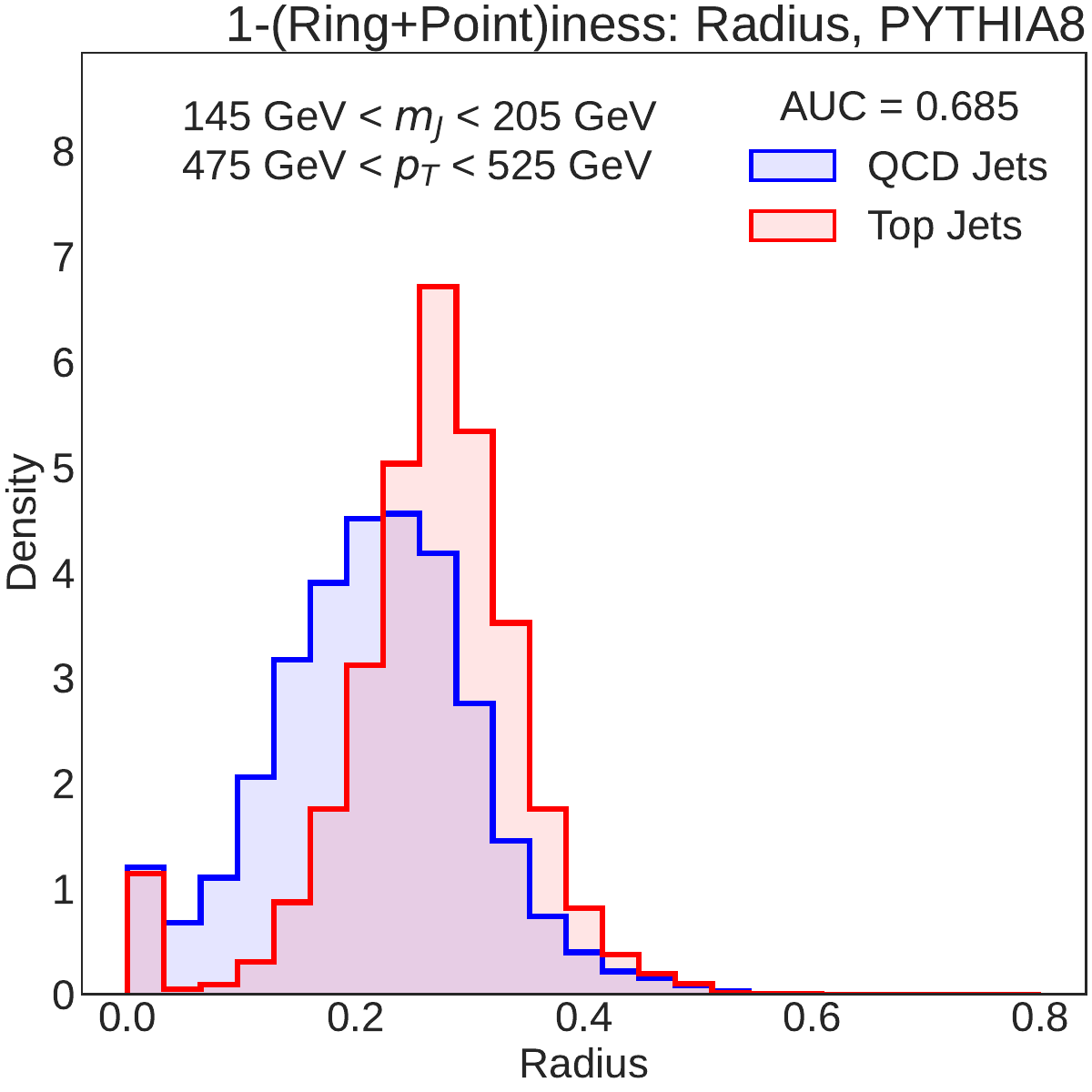}
        \label{fig:radius_d}
    }
    \subfloat[]{
         \includegraphics[width=0.32\textwidth]{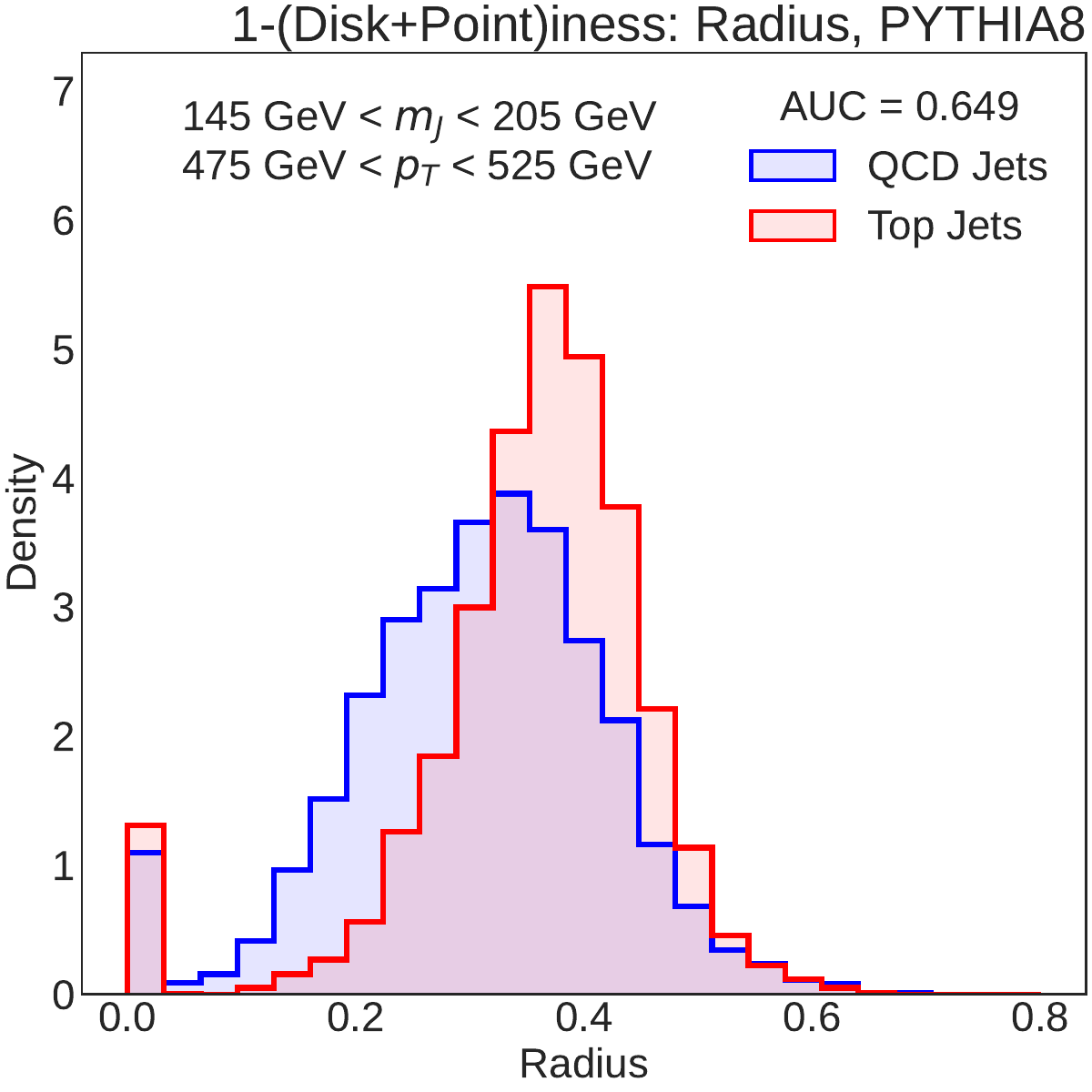}
        \label{fig:radius_e}
    }
    \subfloat[]{
         \includegraphics[width=0.32\textwidth]{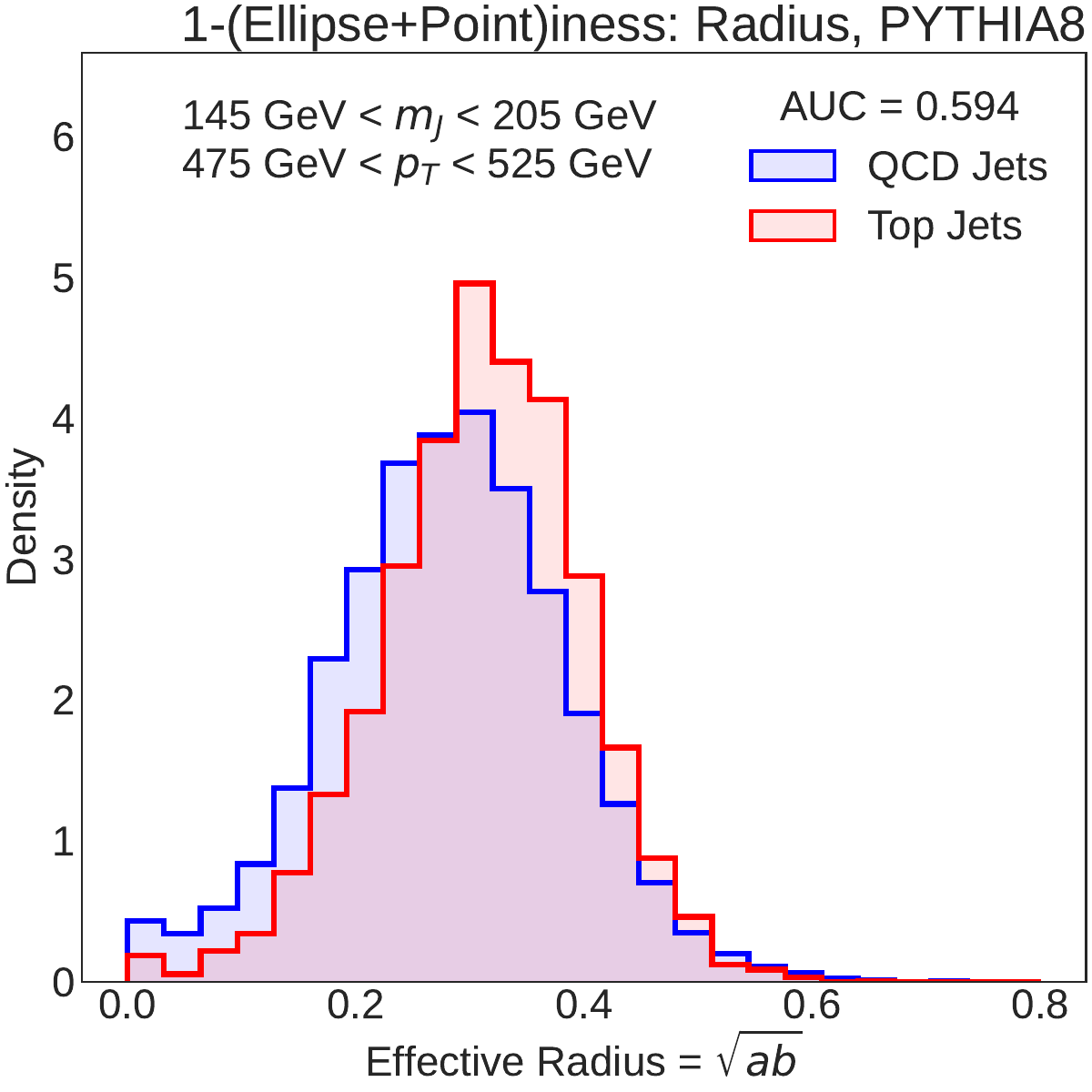}
        \label{fig:radius_f}
    }
    \caption{
        Top row: distributions of the learned radius parameter of the 1- (a) ringiness, (b) diskiness, and (c) ellipsiness of the top (red) and QCD (blue) jet sample. Bottom row: the radius parameter of the 1- (d) (ring$+$point)iness, (e) (disk$+$point)iness, and (f) (ellipse$+$point)iness of the same samples. For the ellipse, the effective radius is given by the geometric mean of the two axes. 
        }
    \label{fig:radius}
\end{figure*}

We now move on to analyze the learned shape parameters themselves. In \Fig{radius}, the learned radius parameters of the $N = 1$ shapes are shown.\footnote{For ellipses, we define an \emph{effective radius}, taken to be $\sqrt{ab}$. This is the radius a circle with the same area as the ellipse.} Considering the $1$-(disk$+$point)iness and $1$-(ellipse$+$point)iness, which as discussed earlier, we can think of as a ``jet algorithm'' looking for soft clusters of radius $R$ with collinear radiation inside, we see that both learn an average radius of approximately $R = 0.4$, which is approximately half the radius of the original AK8 jet. One can think of the uniformity condition as imposing an radius effective cutoff. If one assumes that the energy density falls to zero with distance from the jet center, then there is some critical radius for which any larger uniform shape is an increasingly worse approximation, and in principle this critical radius can be computed in perturbative QCD. 

For the point shape variants of the disk and ellipse, we note that there is only a small difference between the top and QCD jets -- that is to say, the soft component of a top jet is about as ``wide'' as the soft component of a QCD jet, as seen by disks and ellipses. However, as top jets tend to have their energy distributed across 3 prongs, rather than 1 prong as in QCD jets, we should expect clusters of radiation away from the central one for top jets. Since rings are thin, we should expect rings to be able to better capture these localized prongs than area-filling disks and ellipses, which can be visualized in \Fig{n_ringiness}. This is especially apparent in \Fig{radius_d} for the $1$-(ring$+$point)iness shape, where top jets are significantly larger than QCD jets. Interestingly, we note a spike at $R = 0$ in all of these distributions. This means that \Shaper has determined that the best shape for these events is in fact point-like, and reduces to the $N$-subjettiness. This spike is reduced for the ellipse shapes, which implies that there are events better modeled by either an eccentric ellipse or a point rather than a uniform disk.

\begin{figure*}[tp]
    \centering
    \subfloat[]{
         \includegraphics[width=0.32\textwidth]{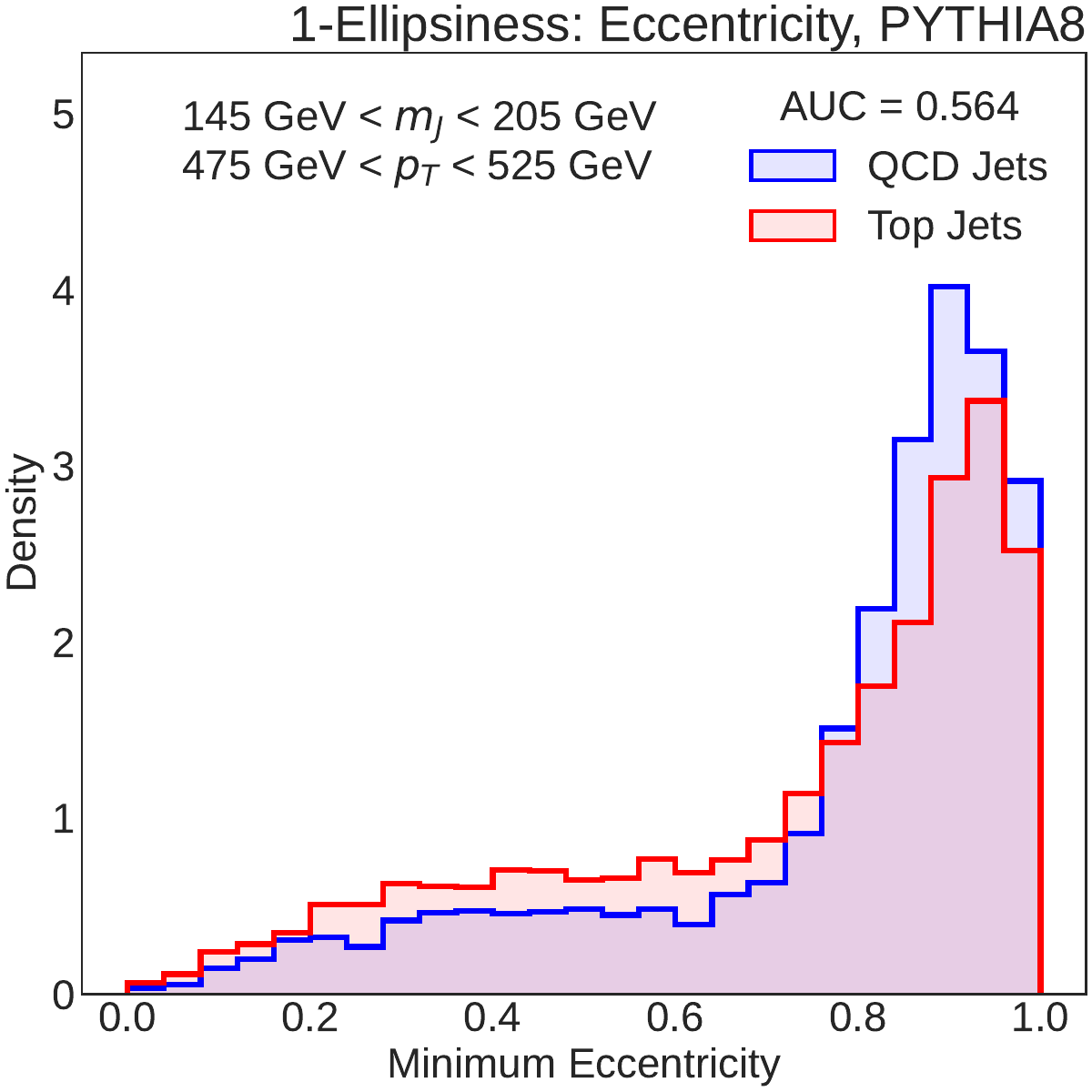}
        \label{fig:eccentricity_ellipse1}
    }
    \subfloat[]{
         \includegraphics[width=0.32\textwidth]{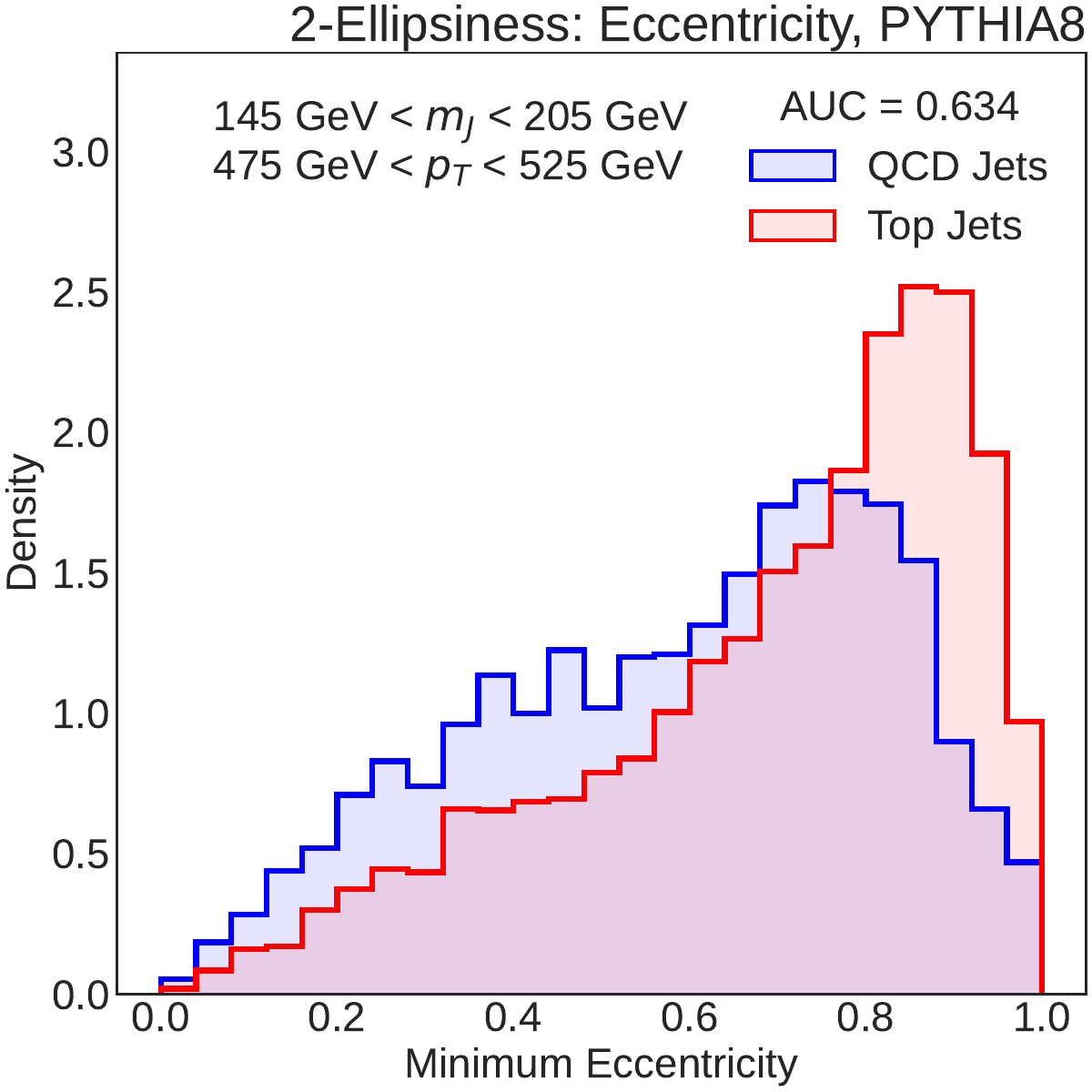}
        \label{fig:eccentricity_ellipse2}
    }
    \subfloat[]{
         \includegraphics[width=0.32\textwidth]{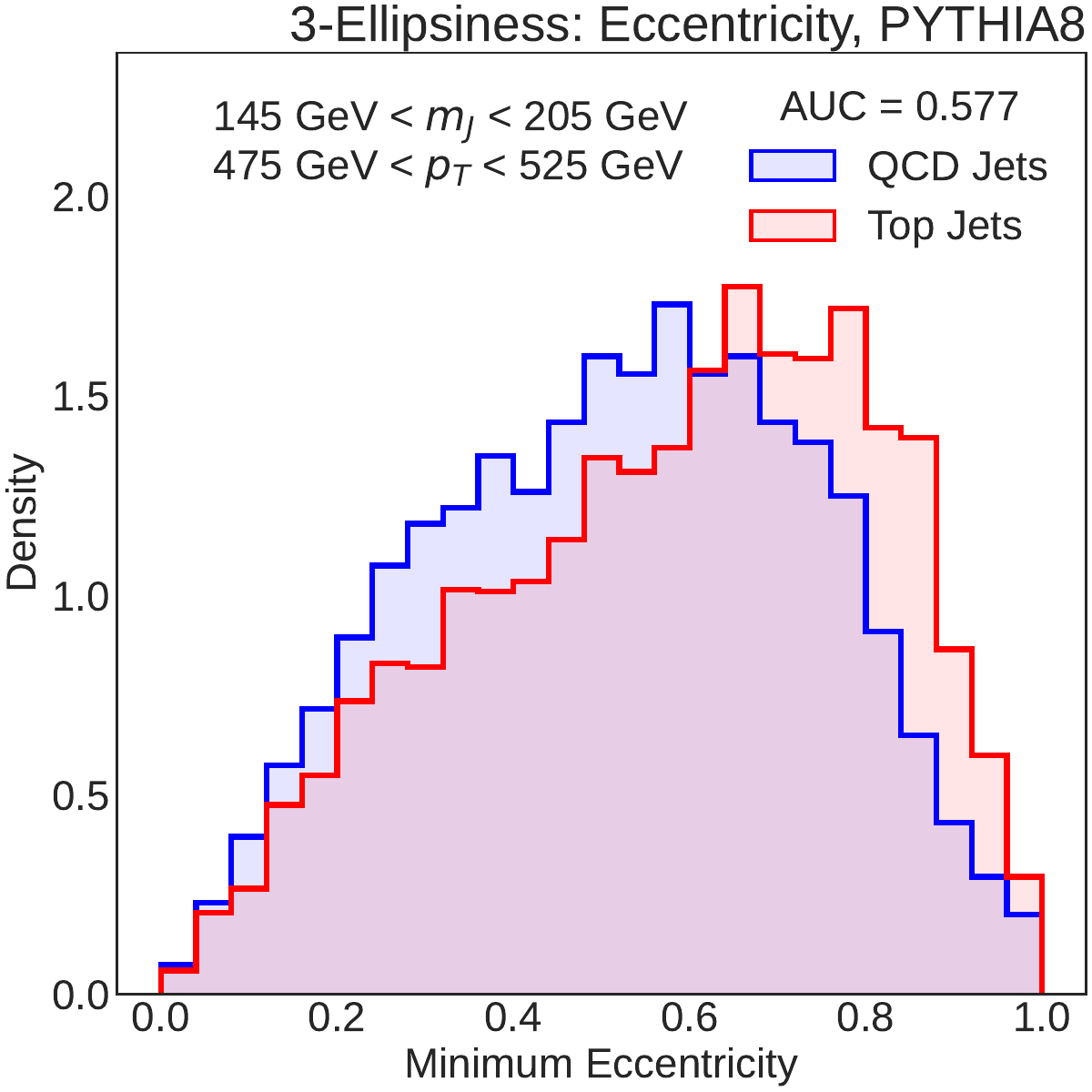}
        \label{fig:eccentricity_ellipse3}
    }
    \vspace{1pt}
    \subfloat[]{
         \includegraphics[width=0.32\textwidth]{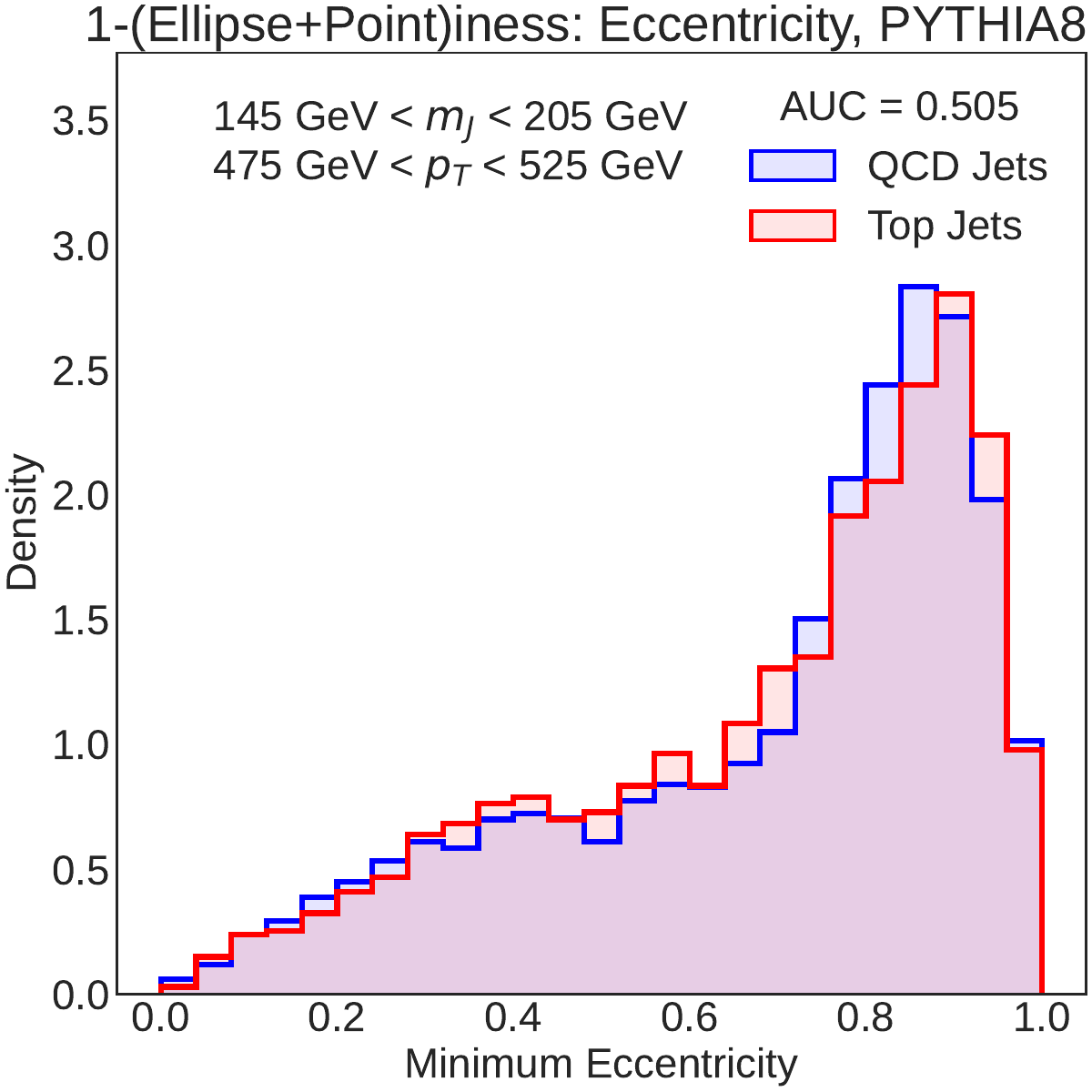}
        \label{fig:eccentricity_point_ellipse1}
    }
    \subfloat[]{
         \includegraphics[width=0.32\textwidth]{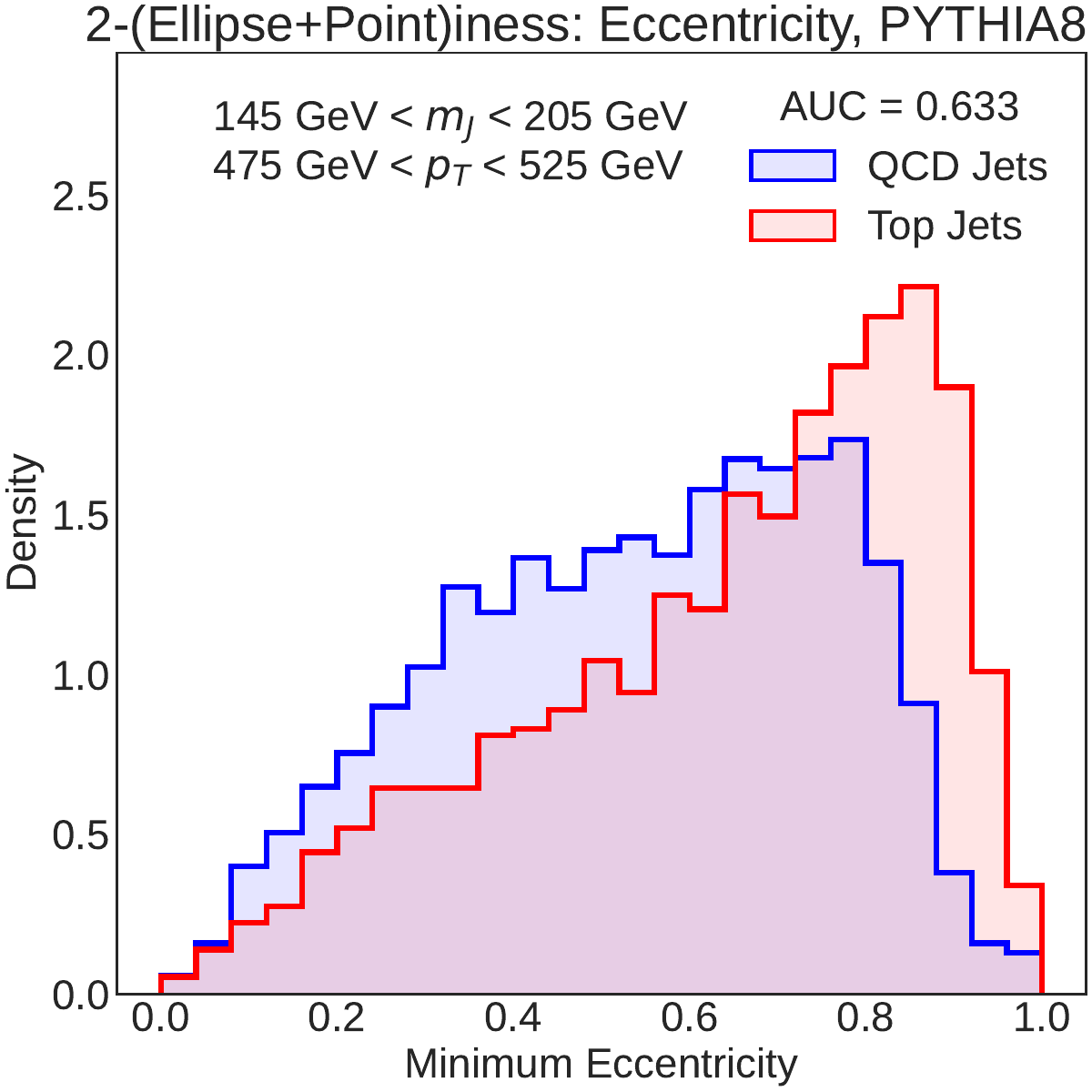}
        \label{fig:eccentricity_point_ellipse2}
    }
    \subfloat[]{
         \includegraphics[width=0.32\textwidth]{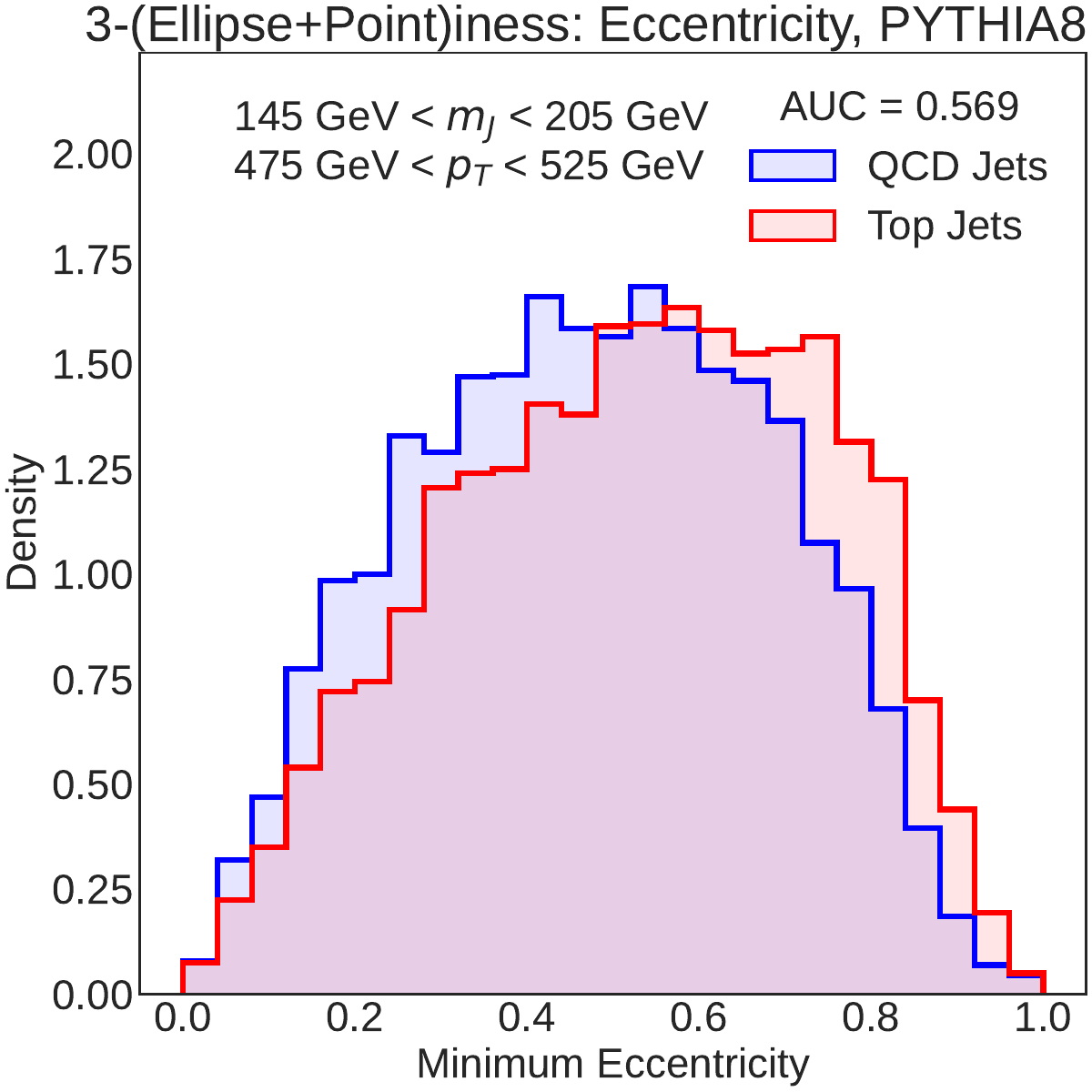}
        \label{fig:eccentricity_point_ellipse3}
    }
    \caption{
        Distributions of the learned minimum eccentricities for (a,d) 1-, (b,e) 2-, and (c,f) 3-ellipsiness (top row) and (ellipse$+$point)iness (bottom row) of the top (red) and QCD (blue) jet sample. 
        }
    \label{fig:eccentricity_ellipse}
\end{figure*}

 As a last example of the information that shape observables and shape parameters can extract, we look at the learned eccentricity of ellipses. In \Fig{eccentricity_ellipse}, we show the \emph{minimum eccentricity} across the $N$ ellipses for the $1$-, $2$-, and $3$- ellipsiness and (ellipse$+$point)iness for our top and QCD jet datasets, where the eccentricity is given by $\sqrt{1 - \min(a,b)/\max(a,b)}$. We can immediately see that a value of $\min(e) = 0$ is rare -- our jets are much better described by ellipses than they are described by disks. With increasing $N$, the distributions of $\min(e)$ tend to shift leftwards for both the top and QCD jet samples. This indicates that the $N$-'th ellipse, which probes more substructure than the $(N-1$)-'th ellipse, is often less eccentric.

 The eccentricity distributions for top jets are slightly different from QCD jets. This is a nontrivial result; top jets and QCD jets have very different prong structures (as demonstrated in \Fig{tau}), so the angular distribution of radiation in the rapidity-azimuth plane is less circular for top jets than QCD jets. Thus, jet eccentricity could be used as a discriminant.

The above plots show just a few examples of the information that can be extracted using generalized shape observables. It is important to emphasize there is much more information beyond what we have shown here available for analysis, such as the learned weights of the points versus uniform shapes, the radii of shapes beyond the leading jet, and so on. This is to say nothing of the multivariate information contained in the correlations between the EMDs and shape parameters, both within the same shape and between different shapes. We leave a full quantitative study of the information contained within shapes for potential future work.

\subsection{Pileup-Mitigating Shapes}\label{sec:pileup}

Finally, we show a use case for shape observable composition by applying our custom shape observables to the task of pileup mitigation. We consider the top jet mass spectrum as an example, which is sharply peaked near $m_{\rm top}^2 \approx (175\,\GeV)^2$.  In our modified dataset, jets are contaminated with uniform pileup radiation, which significantly biases and smears the top mass peak.  There exist many techniques to ``groom'' away extra contamination, such as area subtraction and soft drop~\cite{Cacciari_2008, Cacciari_2008_2, Soyez:2012hv, Larkoski:2014wba, Dasgupta_2013}, but these techniques often have external hyperparamters characterizing the contamination density (such as $z_{\rm cut}$ in soft drop).

In order to mitigate this bias even when the contamination density is unknown, we propose to build shapes that \emph{factorize} the event into a uniform component, $\mathcal{U}$, and a structure component, $\E$. As described in \Sec{plus_pileup}, we accomplish this by building observables of the form:
\begin{align}
  \O^\mathcal{I}_i = \O_i \oplus \mathcal{I},  
\end{align}
 where $\O_i$ is any shape observable and $\mathcal{I}$ is the jet isotropy.\footnote{Unlike~\Refers{Cesarotti:2020hwb,Cesarotti_2021}, we sample the isotropy using random points on the plane, with resampling each epoch, rather than use a uniform grid, for better statistical coverage of the plane.} We can then use the energy flow associated to $\O_i$ to calculate a ``pileup-corrected'' mass, discarding the energy flow associated to $\mathcal{I}$. The weight $z_2$ associated with $\mathcal{I}$ is then an estimate of the fraction of event energy due to pileup -- the contamination density is an extracted observable, \textit{not} an assumed hyperparameter!

\begin{figure*}[tp]
    \centering
    \subfloat[]{
        \includegraphics[width=0.475\textwidth]{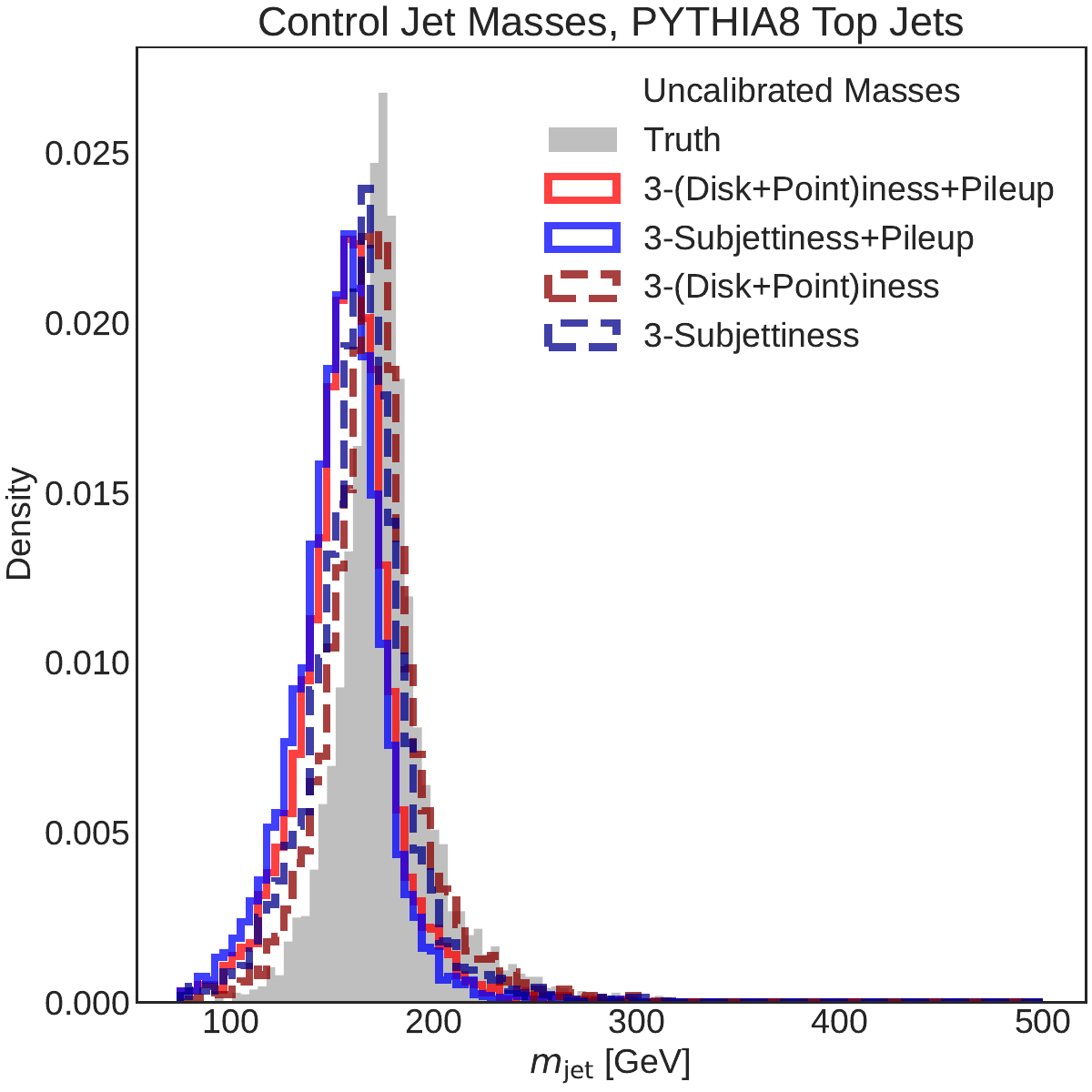}
        \label{fig:pileup_control}
    }
    \subfloat[]{
         \includegraphics[width=0.475\textwidth]{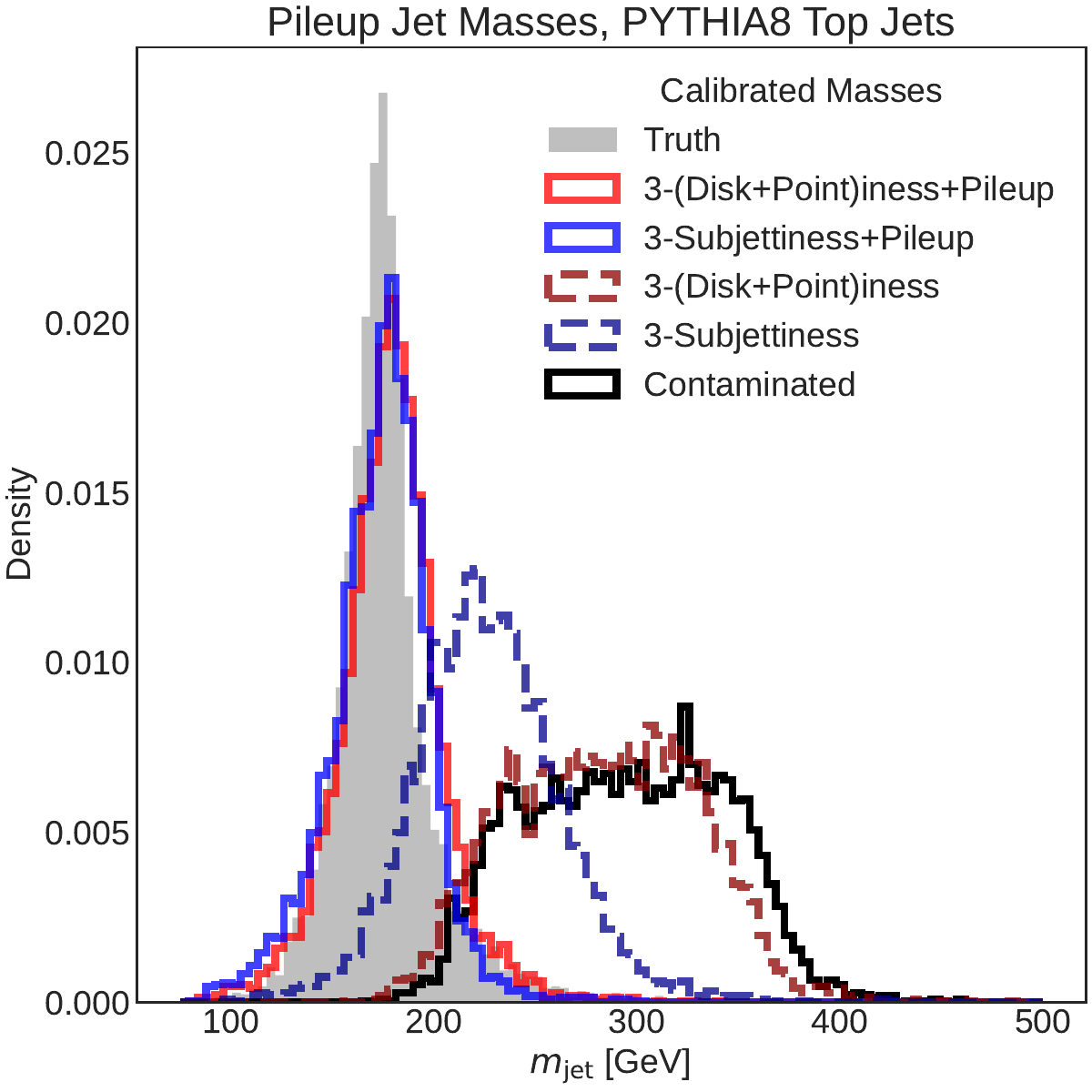}
        \label{fig:pileup}
    }
    \caption{
        The shape-jet mass for several different choices of shape observable, evaluated on (a) top jets without pileup, used to derive the calibration, and (b) top jets with pileup, calibrated using \Tab{biases}. Two different shape observables are used -- the 3-jettiness in blue and 3-(disk$+$point)iness in red. For each shape observable, the ordinary version is plotted with dark dashed lines while the pileup-mitigating variant is plotted with bright solid lines. 
        }
    \label{fig:pileup_observables}
\end{figure*}

 For this study, we consider $\mathcal{O}_i^{\mathcal{I}}$, where $\mathcal{O}_i$ is either the 3-subjettiness or the 3-(disk$+$point)\-iness, motivated by the 3-prong nature of top jets. We also consider the uncorrected observables, $\mathcal{O}_i$, without the uniform background for comparison. For each observable, the ``shape-jet mass'' $m_J$ is calculated as the sum as the (massless) four vectors of particles comprising the shape. This can be numerically approximated by sampling the disk with 200 particles. To calibrate the shape-jet masses, we calculate the shape-jet mass on top jets without the addition of pileup for each observable. In \Fig{pileup_control}, we plot the shape-jet mass corresponding to each observable as evaluated on top jets. Each observable slightly undershoots the top mass peak -- the exact discrepancies are given in \Tab{biases}. These values are used to shift the means of each shape-jet mass curve to the right for the remainder of this study, such that the means give the truth mean top mass when evaluated on uncontaminated  jets.

In \Fig{pileup}, we show the result of an empirical study using our pileup-mitigating shapes on our pileup-contaminated top jet sample. We also plot, for comparison, the mass corresponding to the \emph{uncorrected} observables $\mathcal{O}_i$, with no uniform background. These distributions are all calibrated using the values in \Tab{biases}. We see clearly that the two pileup-mitigating shapes indeed remove a significant amount of bias due to pileup, and bring $m_{\rm jet}$ much closer to $m_{\rm top}$, suggesting that this is a viable strategy for pileup removal. Note, interestingly, that the distribution for the uncorrected 3-(disk$+$point)iness jet mass is similar to the uncorrected jet mass -- this is because to best approximate the entire event, disks will grow large in an attempt to capture all of the pileup, since the disks themselves are also uniform radiation patterns.

\begin{table}[tp]
\centering
\begin{tabular}{|c|c|}
\hline\hline
\bf Observable & \bf Bias [GeV]  \\
\hline\hline
  3-(Disk$+$Point)iness$+$Pileup & $-18.7$ \\
  3-Subjettiness$+$Pileup & $-23.1$ \\
  3-(Disk$+$Point)iness  & $-5.5$ \\
  3-Subjettiness & $-12.5$ \\
\hline\hline
\end{tabular}
\caption{\label{tab:biases}
The difference between the mean shape-jet mass and the mean of the truth top jet distribution in \Fig{pileup_control}, for each of the four observables under consideration. These are used to calibrate the calculated mass distributions for each shape.
}
\end{table}

\begin{figure*}[tp]
    \centering
    \subfloat[]{
         \includegraphics[width=0.475\textwidth]{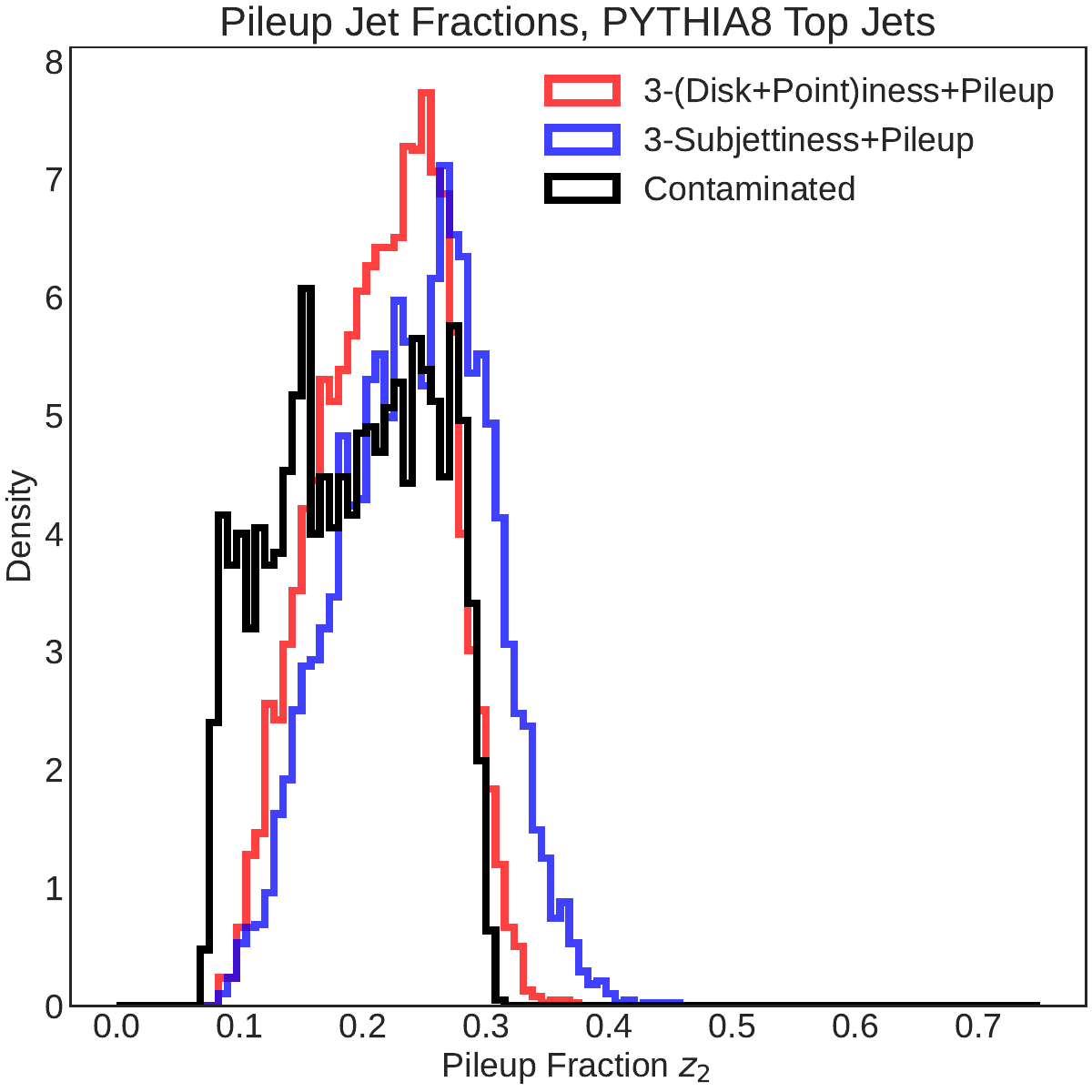}
        \label{fig:pileup_fraction}
    }
    \subfloat[]{
        \includegraphics[width=0.475\textwidth]{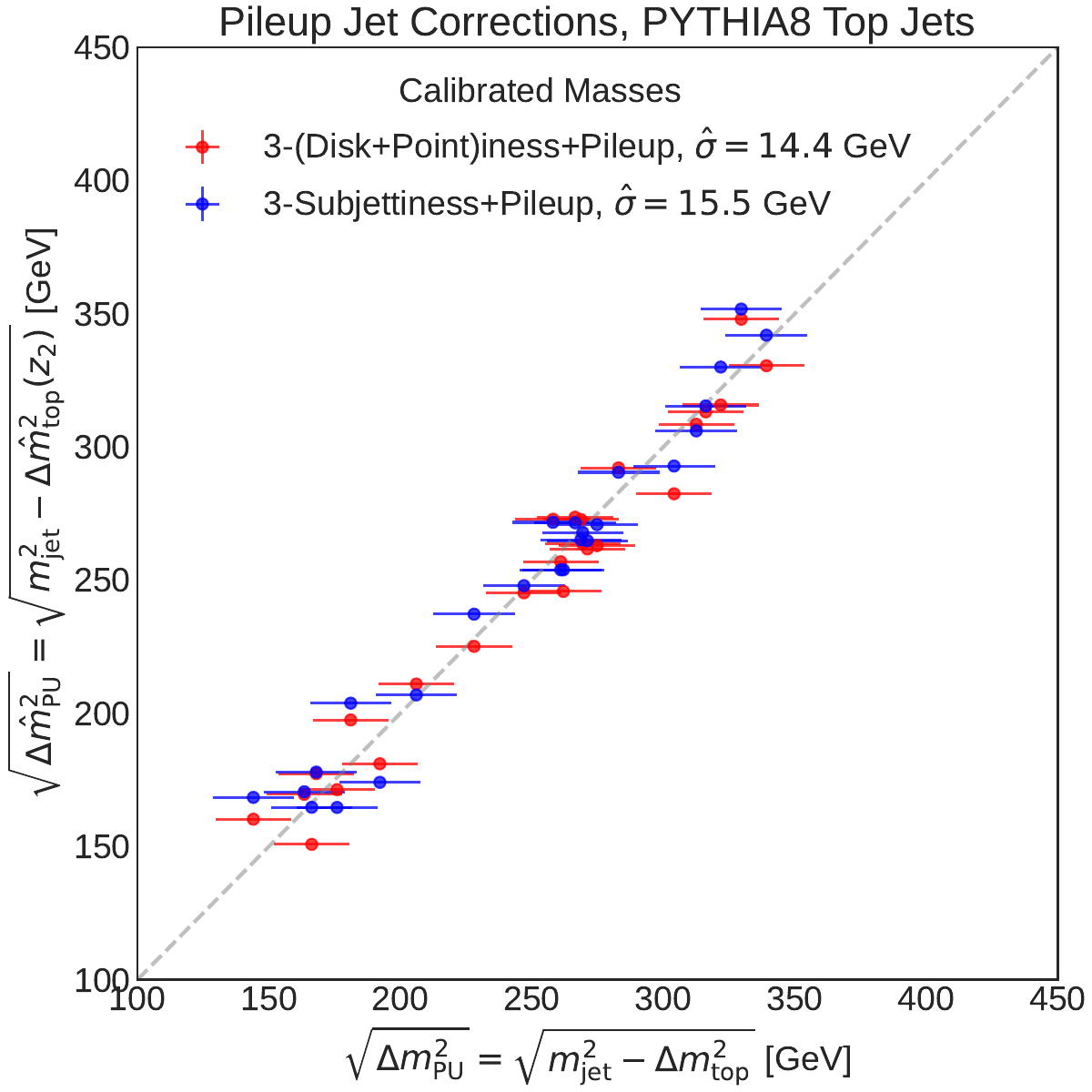}
        \label{fig:pileup_bias}
    }
    \caption{
        (a) The learned pileup fractions $z_2$ for the 3-jettiness in blue and 3-(disk$+$point)iness in red. The actual fraction of energy due to pileup is in black. (b) The mass corrections to the top mass due to pileup, for the 3-jettiness in blue and 3-(disk$+$point)\-iness in red. On the $x$-axis, the change in jet mass $(\Delta m^2_{\rm PU})^{1/2}$ before and after contamination is plotted. On the $y$-axis, the value of $(\Delta \hat{m}^2_{\rm PU})^{1/2}$ as extracted from the corrected top mass is plotted for a small test set. As a proxy for the resolution of the estimate, we take $\hat{\sigma}$ to be the standard deviation of the residuals.
        }
    \label{fig:pileup_corrections}
\end{figure*}

In \Fig{pileup_fraction}, we show the distribution of extracted pileup energy fraction values, corresponding to the $z_2$ shape parameter. We see that there is qualitative agreement between the extracted pileup energy fractions and the true one, though the shapes tend to overestimate the pileup density slightly. This is consistent with \Fig{pileup}, where the shapes slightly underestimate the top mass without the calibration. In \Fig{pileup_bias}, we use $z_2$ to compute the value of the extracted mass bias $\Delta \hat{m}^2_{\rm PU} = m^2_{\rm jet} - \hat{m}^2_{\rm top}(z_2)$, compared to the ``true'' mass bias obtained by comparing the jet mass before and after contamination. To approximate the resolution of our estimator, we take $\hat{\sigma}^2 = {\rm Var} \left[\sqrt{\hat{m}^2_{\rm PU}} - \sqrt{m^2_{\rm PU}}\right]$ as the average Gaussian uncertainty. We can see that both pileup-mitigating observables estimate the pileup mass bias correctly on average, and that the 3-(disk$+$point)iness$+$pileup has a slightly better resolution than the 3-subjetiness$+$pileup, which is to be expected as the former has more freedom to better capture features of jets. 

\section{Conclusions and Outlook}
\label{sec:conclusions}

In this work, we generalized the notion of event and jet shapes into \emph{shape observables}, which are a wide class of observables that can probe the geometric structure of collider events. We introduced a natural measure-theoretic language for describing events as energy flows, which encodes properties such as IRC safety as inherent topological information. Using this construction, we showed that the Wasserstein metric arises naturally from the requirement of geometric faithfulness. This post-hoc justifies its past use to define observables, IRC safety, and geometry on the space of events~\cite{Komiske_2019,2020,Cesarotti:2020hwb,Cai:2021hnn}.

We showcased how to define \emph{arbitrary} shape observables, which can be specified solely by parameterizing the manifold of shapes one wants to fit to. Importantly, we can extract both the ``shapiness'' value (how much the event looks like the shape) and the ``shape parameters'' (which shape is the best fit). This is a very intuitive picture -- if one wants to ask how ``ellipsy'' a jet is, for example, all one needs to do is parameterize the space of ellipses to define the ellipsiness observable. The \Shaper framework makes it easy to build new shape observables and evaluate them on events. It leverages the Sinkhorn approximation of the Wasserstein metric to enable fast, parallelizable, and differentiable $\epsilon$-approximations of shape observables. 

With the \Shaper framework in place, the natural question arises: \emph{Which shapes should we study?} We introduced a few examples of shapes motivated by jet substructure, including rings, disks, and ellipses. Using the \Shaper prescription, it is easy to modify these shapes to probe additional jet structures, by including $\delta$-functions to capture collinear radiation or uniform backgrounds to capture pileup. Our empirical studies showed how these custom observables might serve as an effective and simple pileup mitigation strategy in the study of top jets, where the amount of pileup does not need to be assumed beforehand. Many further analyses are possible: for example, in heavy quark decays, collinear bremsstrahlung is suppressed (the so-called ``dead-cone effect''~\cite{Dokshitzer:1991fc, Dokshitzer:1991fd, PhysRevLett.69.3025}), leading to ring-like or annulus-like, rather than point-like, jets. Moreover, the ability to take derivatives with respect to entire events could allow one to compute observable sensitivities, such as pileup sensitivity~\cite{Soyez:2012hv}.
The list of observables we have provided is certainly non-exhaustive, and we hope that the \Shaper framework can be used by the community to explore a wide range of brand new observables.

While \Shaper can be used to analyze \emph{any} shape observable, and thus has broad generality, it is not necessarily the fastest way to compute or approximate any \emph{specific} observable. For example, much faster, exact algorithms for computing $N$-jettiness and related observables are available in the \FastJet~\cite{Cacciari:2011ma} package, which make use of the fact that the double optimization in \Eq{shape_observable} can be simplified. Most of the new observables we have defined appear to be irreducible, though, making their analysis and theoretical treatment complicated. In principle, this class of observables is perturbatively calculable in QCD, at least numerically.
We anticipate that these generalized shapes might provide a useful groundwork for further theoretical studies of QCD distributions.

\section*{Code and Data}

The code for the general-use \Shaper framework can be found at \url{https://github.com/rikab/SHAPER}, along with installation instructions.
This directory also contains the analysis code for the top/QCD plots in \Sec{empirical}. 
In this directory is also a tutorialized example notebook, showing how to use the \Shaper framework.

Our dataset is a modified version of the top/QCD jet benchmark dataset described in~\Refers{Butter:2017cot,Kasieczka:2019dbj}. The original version of this dataset can be found at \url{https://desycloud.desy.de/index.php/s/llbX3zpLhazgPJ6}.

\section*{Acknowledgments}

Special thanks goes to Samuel Alipour-fard for helping to come up with the acronym \Shaper, and for useful discussions about pileup. We thank Cari Cesarotti and Matthew LeBlanc for useful discussions about event isotropy, and Ouail Kitouni, Niklas Nolte, and Mike Williams for useful discussions on EMD estimation with Kantorovich potentials. Finally, we would like to thank Eugene Wigner of \Refer{Wigner:1960kfi} for inspiring the title of \Sec{EMD}, and Mark Kac of \Refer{10.2307/2313748} for inspiring the title of this paper. 

DB, ASD, RG, and JT are supported by the National Science Foundation under Cooperative Agreement PHY-2019786 (\href{http://iaifi.org/}{The NSF AI Institute for Artificial Intelligence and Fundamental Interactions}). ASD's research was also funded by the President’s PhD Scholarship at Imperial College London and supported by the EPSRC Centre for Doctoral Training in Mathematics of Random Systems: Analysis, Modelling and Simulation (EP/S023925/1). RG and JT are additionally supported by the U.S. DOE Office of High Energy Physics under grant number DE-SC0012567. AT's research is supported by NSF DMS 2208392.

\appendix

\section{Energy Flows as Measures}\label{app:measure_theory}

In this appendix, we review aspects of measure theory and topology related to energy flows, as introduced in \Sec{energyflow}.
To begin, we define a \emph{measure}:
\begin{definition}
Given a set $\X$ and a $\sigma$-algebra on $\X$ denoted $\sigma(\X)$, a (positive) \textbf{measure} $\E$ over $\sigma(\X)$ is a function $\E:\sigma(\X)\to\mathbb{R}$ satisfying:
\begin{itemize}
    \item \textbf{Non-Negativity:} For all $X \in \sigma(\X)$, we have $\E(X) \geq 0$.
    \item \textbf{Null Set:} For the empty set, we have $\E(\emptyset) = 0$.
    \item \textbf{Additivity:} For a (countable) collection of disjoint sets $X_i \in \sigma(\X)$, we have:
    \begin{align}
        \E\left(\bigcup_i X_i\right) = \sum_i \E(X_i).
    \end{align}
\end{itemize}
If, additionally, we have $\E(\X) = 1$, we say $\E$ is a \textbf{probability measure}.
\end{definition}

Associating energy flows with measures is very natural: an energy flow answers the question ``How much energy did I detect in the calorimeters located within the subregion $X$ of my detector $\X$?'', and this answer satisfies all of the above properties.
We refer to $\E(X)$ as the energy flow, whereas we call the object
\begin{align}
\label{eq:energy_flow_density}
    \E(x) = \sum_i E_i \delta(x-x_i)
\end{align}
as the \emph{energy flow density} or \emph{measure density}.
This language (while different from \Refer{2020}) is natural, since integrating over \Eq{energy_flow_density} yields the
energy flow:
\begin{align}
    \E(X) = \int_X dx\, \E(x). 
\end{align}

Under this definition, the energy flow is independent of the choice of coordinates $x$ used on $\X$, unlike the energy flow density. In particular, the form of energy flow is completely invariant under exactly 0-energy emissions or exactly collinear splittings, and additionally is also invariant under particle relabellings. Throughout this paper, we restrict ourselves to measures that can be written as the integral of a well-behaved associated density, which is the case for all physically realized energy flows. We refer to measures whose density is a finite sum of weighted $\delta$-functions as \emph{atomic measures}.

Measures can be used to formalize what we mean by integration. 
We begin by defining $\expval{\E, \phi}$ as the integral of an integrable function $\phi:\mathcal{X}\to\mathbb{R}$ against an energy flow $\E$.\footnote{The integral of $\phi$ over $\E$ is often denoted $\int d\E\,\phi$. When $\E$ is the Lebesgue measure, uniform over $\X$, this becomes the ordinary integral.}
Assuming coordinates $x$ on $\mathcal{X}$, we define this quantity as the ordinary integral over the associated energy flow density:\footnote{We will be satisfied here defining integration in terms of the ordinary Lebesgue integration on real numbers, which is always possible if an associated density exists.}
\begin{align}
    \expval{\E, \phi}  \equiv \int_\mathcal{X}dx\, \E(x) \, \phi(x).
\end{align}
In the special case where $\E$ is a probability measure, this quantity is the expectation value $\mathbb{E}_{X\sim\E}\left[\phi(X)\right]$ of the random variable $\phi(X)$ sampled over $\E$. The notation $\expval{\E, \phi}$ highlights the inner product structure of a measure $\E$ ``acting'' on the space of functions on $\X$. Notably, this inner product is bilinear -- though when considering positive measures, one must be careful to not produce energy flows where the energy is anywhere negative, as these are nonphysical. 

Next, we discuss the important topological features of measures. First, we define \emph{weak* convergence} (also sometimes referred to as \emph{convergence in law}) of a sequence of measures:
\begin{definition}
    A sequence of measures $\E_n$ \textbf{converges with respect to the weak* topology} to a measure $\E$, which we denote $\E_n \to \E$, if for any continuous test function $\phi$ on $\X$, the sequence of real numbers $\expval{\E_n, \phi}$ converges to the real number $\expval{\E, \phi}$.
\end{definition}
That is, we say a sequence of measures converges if every single expectation value converges. Associated to this definition of convergence is the \emph{weak* topology}, which is the (weakest) topology for which the map $\expval{\cdot, \phi}$ is considered continuous for all continuous $\phi$. Note that this definition does \emph{not} make use of any metric on measures; it simply inherits the metric and topological structure of the real numbers. Building off of this, we can now define what it means for any function on energy flows to be continuous:
\begin{definition}
    A function $F$ on the space of positive measures is \textbf{continuous with respect to the weak* topology} if, for every convergent sequence of measures $\E_n \to \E$, the sequence $F(\E_n)$ converges to $F(\E)$. 
\end{definition}

Continuity is a very powerful tool for dealing with energy flows. First, it immediately implies that continuous measures can be arbitrarily well approximated by atomic measures with an increasing number of samples. Moreover, if a function $F$ on measures is continuous with respect to the weak* topology, this further implies it is only necessary to specify how $F$ acts on atomic measures, since the action on all other measures is fixed by weak* continuity.

Intuitively, weak continuity captures the idea that a continuous distribution is well approximated by a discrete one with ``enough samples'' -- as the number of sample increases, we approach the continuous one. Note, however, we have not yet defined what it means for two energy flows to be ``close'' to each other, so we cannot yet describe what happens to functions under small energy flow perturbations. This requires a metric on the space of energy flows that respects the weak* topology, which as argued in \Sec{EMD} and further justified in \App{construction}, can be taken to be the Wasserstein metric.

We finally introduce one last piece of notation. For any two energy flows $\E_1$ and $\E_2$, the \emph{joint energy flow} $\E = \E_1 \otimes \E_2$ is the energy flow satisfying for any sets $X, Y \subseteq \X$:
\begin{align}
    \E(X, Y) = \int_{X\times Y} dx \, dy\, \E_1(x) \, \E_2(y).
\end{align}
The energy flows $\E_1$ and $\E_2$ are referred to as the \emph{marginals} of $\E$. They can be written as $\E_1(X) = \E(X, \mathcal{X})$ and $\E_2(Y) = \E(\mathcal{X}, Y)$, with densities $\E_1(x) = \int_\mathcal{X}dy\, \E(x,y)$ and $\E_2(y) = \int_\mathcal{X}dx\, \E(x,y)$ respectively.

\section{Constructing the Wasserstein Metric}\label{app:construction}

In \Sec{shapes}, we proposed to write event and jet shapes $\O_\M$ using some universal loss function $\mathcal{L}$ on energy flows, and in \Sec{irc}, we showed that IRC safety implies that must $\mathcal{L}$ must be weakly continuous. Then, in \Sec{wasserstein}, we claimed that the condition of faithfully lifting the ground metric means that $\mathcal{L}$ must be the Wasserstein metric, and showed examples of how similar metrics do not satisfy this property. In this appendix, we show how the Wasserstein metric arrives constructively. First, in \App{loss}, we briefly review the properties we use to construct $\mathcal{L}$, before outlining the construction in \App{wasserstein_appendix}.

\subsection{Shaping Up the Loss}\label{app:loss}

To start, we list the properties we would like the universal loss function $\mathcal{L}$ to satisfy. First, we would like $\mathcal{L}$ to be a proper metric on the space of energy flows, which implies the following usual properties of metrics:
\begin{enumerate}
    \item \textbf{Finiteness}: We require that $\mathcal{L}$ is finite, even when evaluated on atomic measures. This immediately rules out the KL divergence~\cite{kullback1951information} and its variants, including log-likelihoods, used in \Refer{Mackey:2015hwa} to define ``fuzzy jet'' observables. These functions are only finite on continuous measures with support almost everywhere. 
    \item \textbf{Positivity and Closure}: We require that $\mathcal{L}(\E, \E') \geq 0$. Moreover, we require that  $\mathcal{L}(\E, \E') = 0$ if and only if $\E = \E'$. This captures the notion of the ``optimal shape'' as discussed in \Sec{shapes}.
    \item \textbf{Symmetry}: We require that $\mathcal{L}(\E, \E') = \mathcal{L}(\E', \E)$. This is to say, if the energy flow $\E$ looks like $\E'$, then the energy flow $\E'$ looks like $\E$ as well. 
    \newcounter{enumTemp}
    \setcounter{enumTemp}{\theenumi}
\end{enumerate}
We will not specifically require the triangle inequality, as we never use it throughout this work. Of course, the Wasserstein metric does indeed satisfy the triangle inequality when the appropriate powers of $1/\beta$ are included, making it a proper metric.

Not only is $\mathcal{L}$ a metric, but it must be continuous to the weak* topology to be IRC safe, as established in \Sec{irc}:
\begin{enumerate}
    \setcounter{enumi}{\theenumTemp}
    \item \textbf{Weak Contintuity/IRC Safety}: We require that $\mathcal{L}$ is weakly continuous in both of its arguments. This not only allows for continuous energy flows to be arbitrarily well approximated by atomic energy flows, but heavily constrains the energy- and position- dependence of $\mathcal{L}$.
    \setcounter{enumTemp}{\theenumi}
\end{enumerate}

Finally, as discussed in \Sec{wasserstein}, we demand that $\mathcal{L}$ \emph{faithfully lifts the ground metric}.
This allows the space of events to inherit the geometry of the ground metric space with no distortions or warping:
\begin{enumerate}
    \setcounter{enumi}{\theenumTemp}
    \item \textbf{Faithfully Lifts the Ground Metric}: We require that if $\E$ and $\E'$ are atomic measures with a single normalized particle at positions $x$ and $y$ respectively, then $\mathcal{L}(\E,\E')$ is proportional to $d(x,y)^\beta$, where $\beta$ is some fixed positive power.
    
    Moreover, we require that $\mathcal{L}$ \emph{faithfully} lifts the ground metric. For any measure $\E$, we can ``translate'' the measure by $t$ to $\E_t$, defined by the density $\E(x-t)$. We then require that $\mathcal{L}(\E, \E_t)$ is proportional to $d(0,t)^\beta$. 
\end{enumerate}

If the ground metric is translationally invariant, which is the case for Euclidean metrics, property 5 implies that $\mathcal{L}$ must also be translationally invariant -- if both $\E$ and $\E'$ are translated by a vector $t$, the metric distance between them is unchanged.

\subsection{Why Wassterstein?}\label{app:wasserstein_appendix}

Having defined the properties we would like our universal loss function to satisfy, we can now finally construct $\mathcal{L}(\E,\E')$. 

To begin, weak continuity allows us to drastically simplify the problem. Continuous energy flows may be approximated arbitrarily well by atomic energy flows, and thus it suffices to build our function only on atomic energy flows.
This allows us to argue that $\mathcal{L}$ can only depends on single powers of the distance $d(x,y)^\beta$.
In particular, lifting the ground metric (though not necessarily faithfully, yet) forbids terms of the form $d(x,y)^\beta d(x',y')^\beta$ or other higher-order distance correlations, since in the single-particle case this would result in losses of the form $d^{\beta_1} + d^{\beta_2} + ...$ with several differing exponents. 
This implies the following form for $\mathcal{L}$ when evaluated on uniform atomic flows $\E \sim \sum_i \delta_{x_i}$ and $\E' \sim \sum_j \delta_{y_j}$:
\begin{align}
    \mathcal{L}(\E,\E') &= \Pi\left(\sum_{i,j} \pi_{ij} \, d(x_i, y_j)^\beta \right) \nonumber \\
    & + K\left(\sum_{i_1,i_2}k_{i_1i_2} \, d(x_{i_1},x_{i_2})^\beta \right) + K\left(\sum_{j_1,j_2}k_{j_1j_2}\, d(y_{j_1},y_{j_2})^\beta \right) \nonumber\\
    & + F\left(\sum_i f_i \, d(x_i, g(x_i))^\beta\right) + F\left(\sum_j f_j \, d(\Tilde{g}(y_j), y_j)^\beta\right), \label{eq:appendix_generic_loss}
\end{align}
where $\Pi$, $K$, and $F$ are universal functions, $\pi$, $k$, and $f$ are coefficients that implicitly depend on the two measures, and $g$ is a function from the domain of $\E$ to the domain of $\E'$ (and vice-versa for $\Tilde{g}$). Note that this form is symmetric under swapping $\E$ and $\E'$. Without loss of generality, $k_{ii'}$ can be chosen to be a symmetric traceless matrix. 

Still considering the single particle case, lifting the ground metric also implies that the functions $\Pi$ and $F$ must be linear functions (though the single particle case provides no information about $K$), as to reproduce the desired $d(x,y)^\beta$ behavior. Moreover, to still maintain only terms proportional to $d^\beta$ even in the multiparticle case, $K$ must also be linear. This linearity means that we can combine the arguments of each of the $\Pi$, $K$, and $F$ functions.

Recall that energy flows are additive objects, and the energy flows defined by the densities $\E = (E_1 + E_2)\delta_x$ and $\E = E_1\delta_x + E_2\delta_x$ must be identical. This implies that each term of the terms in \Eq{appendix_generic_loss} must be individually linear in the energy weights of $\E$ and $\E'$, respectively.
This implies the following constraints, up to an overall proportionality factor:
\begin{align}
    \sum_i \pi_{ij} &= E_j', \label{eq:pii_constraint}  \\
    \sum_j \pi_{ij} &= E_i,  \label{eq:pij_constraint} \\
    \sum_{i_1} k_{i_1 i_2} &= E_{i_2}, & \sum_{i_2} k_{i_1 i_2} &= E_{i_1}, \label{eq:ki_constraint} \\
    \sum_{j_1} k_{j_1 j_2} &= E_{j_2}', & \sum_{j_2} k_{j_1 j_2} &= E_{j_1}', \label{eq:kj_constraint}\\
    f_i &= E_i, & f_j &= E_j'
\end{align}

Now, consider the two particle case, as in \Sec{wasserstein}, and
define the energy flows for vectors $a$ and $t$:
\begin{align}
    \E(x) &\sim \frac{1}{2}\delta_0 + \frac{1}{2}\delta_a, \\
    \E'(x) &\sim \frac{1}{2}\delta_t + \frac{1}{2}\delta_{a+t}.
\end{align}
When $t\to0$, the two energy flows are the same, and when $a\to0$, the energy flows reduce to the single particle case.
Evaluating \Eq{appendix_generic_loss}, we obtain:
\begin{align}
    \mathcal{L}(\E,\E') &= \Pi\left(\pi_{1_x1_y} d(0,t)^\beta + \pi_{1_x2_y} d(0,a+t)^\beta + \pi_{1_x1_y} d(a,t)^\beta + \pi_{2_x2_y} d(a,a+t)^\beta\right) \nonumber \\
    & + K\left(2k_{1_x2_x}d(0,a)^\beta + 2k_{1_y2_y}d(t,a+t)^\beta \right) \nonumber\\
    & + \frac{1}{2}F\left(d(0, g(0))^\beta + d(a, g(a))^\beta + d(\Tilde{g}(t), t)^\beta + d(\Tilde{g}(a+t), a+t)^\beta  \right). \label{eq:general_2particle}
\end{align}

In the limit $t\to0$, this expression must evaluate to zero by closure, for any choice of $a$. This is possible if the off-diagonal components of $\pi$ and $k$ either cancel or are individually zero. Note that the function $g$ must select a particle $y_j$ given $x_i$ (and vice-versa for $\Tilde{g}$). Importantly, however, it must do this in a label-independent way, since the labels $i$ and $j$ are completely arbitrary. The only way to do so is to select $y_j$ based on its distance from $x_i$ -- in principle, we can choose $g$ to select the furthest particle, or the closest particle, or even the particle closest to the average distance amongst all particles. However, the only choice consistent with closure, as can be seen in \Eq{general_2particle}, is to choose the \emph{closest} particle $y_j$ to the given $x_j$. This allows us to write the third line of \Eq{appendix_generic_loss} as:
\begin{align}
    \mathcal{L}(\E,\E') = [...] + F\left(\sum_i E_i \min_{y_j\in{\rm Supp}(\E')}d(x_i, y_j)^\beta + \sum_j E_j' \min_{x_i\in{\rm Supp}(\E)}d(x_i, y_j)^\beta\right). \label{eq:appendix_chamfer}
\end{align}
\Eq{appendix_chamfer} is exactly the form of a Chamfer distance from \Eq{chamfer}! 
The Chamfer distance is the average closest distance from points in $\E$ to $\E'$, plus the average closest distance from points in $\E'$ to $\E$.
Following the explicit counter-example given in \Eq{chamfer_computation}, the Chamfer distance does \emph{not} lift the ground metric \emph{faithfully}, and therefore, $F$ must be zero.

Returning to \Eq{general_2particle},  we now want to show that $K=0$.
Since $k$ is traceless by assumption, the constraints \Eqs{ki_constraint}{kj_constraint} imply that $k_{1_x2_x} = E_1 E_2$ and $k_{1_y2_y} = E_1' E_2'$. Assuming that $K \neq 0$, closure plus \Eqs{pii_constraint}{pij_constraint} allows us to solve for $\pi_{ij}$. From this, we deduce that $\pi_{ij}$ reduces to $E_i E_j'$ as $t \to 0$ (with proportionality constant $\Pi = -2K$). Furthermore, $\pi_{ij}$ has to be constant in $t$, since otherwise $\mathcal{L}$ would depend on higher powers of the metric and therefore fail to lift the ground metric. Given this, \Eq{appendix_generic_loss} reduces to:
\begin{align}
    \mathcal{L}(\E,\E') = K\left(\sum_{i_1,i_2}E_{i_1}E_{i_2}d(x_{i_1},x_{i_2})^\beta  + \sum_{j_1,j_2}E_{j_1}'E_{j_2}'d(y_{j_1},y_{j_2})^\beta - 2\sum_{ij}E_iE_j'd(x_i,y_j)^\beta\right)
\end{align}
However, this implies that $\mathcal{L}$ is exactly the maximum mean discrepancy (MMD) of \Eq{MMD}! 
The MMD can be thought of as the average potential energy of a system of springs with potential $V(r) \sim r^\beta$ connecting particles between $\E$ and $\E'$, minus the self-energy of springs connecting particles within $\E$ and within $\E'$.
The explicit counter-example given in \Eq{MMD_computation} shows that this cannot be faithful, \emph{except in the special case of $\beta = 2$}, which is not sufficient for our purposes.\footnote{Generically, physical systems with potentials $V(r) \sim r^\beta$ experience screening, most notably electrostatic screening when $\beta = -1$. However, screening does not occur for ideal springs, i.e. $\beta = 2$. Screening and (un)faithfulness are related -- as $\E$ and $\E'$ move relative to eachother, the potential energy between them will not necessarily scale with $r^\beta$ in extended systems due to screening.} It follows that $K$ must also be zero, and we have significantly constrained the form of our metric.

It now remains only to find the form of $\pi$. To do this, we now consider an energy flow $\E$, consisting of $N$ particles each with energy $\frac{1}{N}$. By closure, we again must have $\mathcal{L}(\E,\E)  = 0$, which occurs when $\pi_{ij}$ is $\frac{1}{N^2}$ times the $N\times N$ identity matrix. 

Next, consider $\E'$, an exact copy of $\E$, except the particles have been re-indexed, sending the original particle $x_i$ to $x_{j = \sigma(i)}$. As measures are invariant under reordering, $\mathcal{L}$ must still be zero, though $\pi$ will be (proportional to) some $N\times N$ permutation matrix.  We can write this as:
\begin{align}
    \mathcal{L}(\E, \E') &= \frac{1}{N}\sum_i d(x_i, y_{\sigma(i)})^\beta, \nonumber \\
    &= \sum_{ij} \pi_{ij} \, d(x_i, y_j)^\beta,
\end{align}
This will only be zero if $\pi_{ij}$ is the \emph{correct} permutation matrix that undoes the index shuffling. If one instead guesses that $\pi$ is a different permutation matrix, the point $y_{\sigma(i)}$ will not lie on top of the point $x_i$, leading to $\mathcal{L} > 0$.

Thus, even if we did not know how exactly the particle labels on $\E'$ were shuffled, all we would have to do to find the correct $\pi$ is to search through all possible permutation matrices, and take the one that gives the minimum answer:
\begin{align}
    \mathcal{L}(\E, \E') &= \min_{\sigma:[1,N]\to[1,N]}\frac{1}{N}\sum_{i}d(x_i, y_{\sigma(i)})^\beta \nonumber \\
    &= \min_{\pi_{ij} \in S_N} \frac{1}{N^2} \sum_{ij} \pi_{ij}d(x_i, y_j)^\beta, \label{eq:almost_emd}
\end{align}
where $S_N$ is the set of all $N\times N$ permutation matrices. Moreover, the index-shuffling trick allows us to see that \Eq{almost_emd} faithfully lifts the ground metric. If $\E'$ is translated by a vector $t$, then $\pi$ should still be the identity matrix before shuffling so that our result is proportional to $d(0,t)^\beta$. If we guess the wrong permutation matrix, while it may be possible for some distances to close (that is, $d(x_i, y_{\sigma(i)})$ to be less than $d(0,t)$), this will be made up for by other particles being further apart than $d(0,t)$, and so $\mathcal{L}$ will be larger.
For values of $\beta \geq 1$, this last fact follows from the triangle inequality. 
This can be seen explicitly in the two-particle case:   If $|t| \gg |a|$, it follows that:
 \begin{align}
     d(0,t)^\beta \leq \frac{1}{2}d(0,t-a)^\beta + \frac{1}{2}d(0,t+a)^\beta,
 \end{align}
for $\beta \geq 1$. Therefore \Eq{almost_emd} is a faithful metric for $\beta \geq 1$.

The problem in \Eq{almost_emd} is referred to as the \emph{combinatorial Monge problem}, which is the precursor to the Wasserstein metric. We can think of \Eq{almost_emd} as finding the optimal map $\pi$ to ``transport'' the points $x$ to $y$, minimizing the total distance$^\beta$ needed to travel. These arguments are enough to fully fix the form of $\mathcal{L}$ -- even if $\E$ and $\E'$ are different energy flows, as long as they both have $N$ particles with uniform weights, the minimization over permutation matrices is valid.

 To finish our construction, we  use the additivity of measures and the weak* topology to generalize \Eq{almost_emd} to any energy flow. Since measures are additive, any atomic energy flow with \emph{rational} energy weights can be thought of particles of a base denomination weight stacked on top of each other -- for example, the two-particle energy flow $\E \sim 0.381\delta_{x_1} + 0.619\delta_{x_2}$ is exactly equal to the energy flow given by 381 particles at site $x_1$ and 619 particles at site $x_2$, each with energy $0.001$. One can then solve the combinatorial Monge problem using this uniform energy flow, and ``collapse'' the corresponding redundant subspace in $\pi$ to produce the optimal transport map. In the case that energy weights are irrational, weak continuity  ensures that these may be constructed as limits of rationally-weighted energy flows, so this is not an issue. Similarly, any continuous energy flow can be written this way.

Thus, the loss \Eq{almost_emd} is our universal loss function! Accounting for arbitrary weights and particle numbers, as discussed above, this becomes the well-known Wasserstein metric. The $\beta$-Wasserstein metric, or ``Energy/Earth Mover's Distance'' (EMD), is the metric on the space of measures that is positive, closed, metrizes the weak convergence, and faithfully lifts the ground metric. Repeating \Eq{EMD} for convenience, the EMD between two measures $\E$ and $\E'$ is given by:
\begin{align} \label{eq:EMD_appendix}
    &\EMD^{(\beta,R)}(\E, \E') = \min_{\pi \in \M(\mathcal{X}\times\mathcal{X})}\left[\frac{1}{\beta R^\beta}\expval{\pi, d(x,y)^{\beta}}\right] + |\Delta E_{\rm tot}|, \nonumber \\
    &\pi(\mathcal{X}, Y) \leq \E'(Y) ,\quad \pi(X, \mathcal{X}) \leq \E(X), \quad \pi(\mathcal{X}, \mathcal{X}) = \min(E_{\rm tot}, E'_{\rm tot}).
\end{align}
 The parameter $R > 0$ sets a distance scale for the EMD. The additional energy difference term, $|\Delta E_{\rm tot}| = |E_{\rm tot} - E'_{\rm tot}|$, contributes whenever the two energy flows do not have the same total energy. Our argument in this section does not fix this term uniquely, and in fact there are many different approaches to unbalanced metrics~\cite{Piccoli_2013, Cai:2021hnn, Liero_2016}. 
In the common case that both $\E$ and $\E'$ are atomic measures with $M$ and $N$ particles with energies $E_i$ and $E'_j$ respectively, the EMD takes the form:
\begin{align} \label{eq:EMD_atomic_appendix}
    &\EMD^{(\beta,R)}(\E, \E') = \min_{\pi_{ij} \geq 0}\left[\frac{1}{\beta R^\beta}\sum_{i=1}^M\sum_{j=1}^N \pi_{ij} \, d_{ij}^\beta\right] + |\Delta E_{\rm tot}|, \nonumber \\
    &\sum_{i=1}^M \pi_{ij} \leq E'_j , \quad \sum_{j=1}^N \pi_{ij} \leq E_i, \quad \sum_{j=1}^N \pi_{ij} = \min(E_{\rm tot}, E'_{\rm tot}).
\end{align}

\bibliographystyle{JHEP}
\bibliography{refs}

\providecommand{\href}[2]{#2}\begingroup\raggedright\begin{thebibliography}{100}

\bibitem{PhysRevLett.39.1587}
E.~Farhi, \emph{Quantum chromodynamics test for jets},
  \href{https://doi.org/10.1103/PhysRevLett.39.1587}{\emph{Phys. Rev. Lett.}
  {\bfseries 39} (Dec, 1977) 1587--1588}.

\bibitem{Barber:1979bj}
D.~P. Barber et~al., \emph{{Tests of Quantum Chromodynamics and a Direct
  Measurement of the Strong Coupling Constant $\alpha_S$ at
  $\sqrt{s}=30$-{GeV}}},
  \href{https://doi.org/10.1016/0370-2693(79)90092-3}{\emph{Phys. Lett.}
  {\bfseries 89B} (1979) 139--144}.

\bibitem{Dasgupta:2003iq}
M.~Dasgupta and G.~P. Salam, \emph{{Event shapes in e+ e- annihilation and deep
  inelastic scattering}},
  \href{https://doi.org/10.1088/0954-3899/30/5/R01}{\emph{J. Phys. G}
  {\bfseries 30} (2004) R143},
  [\href{https://arxiv.org/abs/hep-ph/0312283}{{\ttfamily hep-ph/0312283}}].

\bibitem{Dissertori:2008cn}
G.~Dissertori, A.~Gehrmann-De~Ridder, T.~Gehrmann, E.~W.~N. Glover, G.~Heinrich
  and H.~Stenzel, \emph{{e+ e- ---> 3 jets and event shapes at NNLO}},
  \href{https://doi.org/10.1016/j.nuclphysbps.2008.09.072}{\emph{Nucl. Phys.
  Proc. Suppl.} {\bfseries 183} (2008) 2--7},
  [\href{https://arxiv.org/abs/0806.4601}{{\ttfamily 0806.4601}}].

\bibitem{Almeida:2008yp}
L.~G. Almeida, S.~J. Lee, G.~Perez, G.~F. Sterman, I.~Sung and J.~Virzi,
  \emph{{Substructure of high-$p_T$ Jets at the LHC}},
  \href{https://doi.org/10.1103/PhysRevD.79.074017}{\emph{Phys. Rev. D}
  {\bfseries 79} (2009) 074017},
  [\href{https://arxiv.org/abs/0807.0234}{{\ttfamily 0807.0234}}].

\bibitem{Gur-Ari:2011cjr}
G.~Gur-Ari, M.~Papucci and G.~Perez, \emph{{Classification of Energy Flow
  Observables in Narrow Jets}},
  \href{https://arxiv.org/abs/1101.2905}{{\ttfamily 1101.2905}}.

\bibitem{BRANDT196457}
S.~Brandt, C.~Peyrou, R.~Sosnowski and A.~Wroblewski, \emph{The principal axis
  of jets — an attempt to analyse high-energy collisions as two-body
  processes},
  \href{https://doi.org/https://doi.org/10.1016/0031-9163(64)91176-X}{\emph{Physics
  Letters} {\bfseries 12} (1964) 57--61}.

\bibitem{DERUJULA1978387}
A.~{De Rujula}, J.~Ellis, E.~Floratos and M.~Gaillard, \emph{Qcd predictions
  for hadronic final states in e+e- annihilation},
  \href{https://doi.org/https://doi.org/10.1016/0550-3213(78)90388-7}{\emph{Nuclear
  Physics B} {\bfseries 138} (1978) 387--429}.

\bibitem{Berger:2003iw}
C.~F. Berger, T.~Kucs and G.~F. Sterman, \emph{{Event shape / energy flow
  correlations}}, \href{https://doi.org/10.1103/PhysRevD.68.014012}{\emph{Phys.
  Rev. D} {\bfseries 68} (2003) 014012},
  [\href{https://arxiv.org/abs/hep-ph/0303051}{{\ttfamily hep-ph/0303051}}].

\bibitem{Berger:2004xf}
C.~F. Berger and L.~Magnea, \emph{{Scaling of power corrections for
  angularities from dressed gluon exponentiation}},
  \href{https://doi.org/10.1103/PhysRevD.70.094010}{\emph{Phys. Rev. D}
  {\bfseries 70} (2004) 094010},
  [\href{https://arxiv.org/abs/hep-ph/0407024}{{\ttfamily hep-ph/0407024}}].

\bibitem{Becher:2008cf}
T.~Becher and M.~D. Schwartz, \emph{{A precise determination of $\alpha_s$ from
  LEP thrust data using effective field theory}},
  \href{https://doi.org/10.1088/1126-6708/2008/07/034}{\emph{JHEP} {\bfseries
  07} (2008) 034}, [\href{https://arxiv.org/abs/0803.0342}{{\ttfamily
  0803.0342}}].

\bibitem{Banfi:2014sua}
A.~Banfi, H.~McAslan, P.~F. Monni and G.~Zanderighi, \emph{{A general method
  for the resummation of event-shape distributions in e+e- annihilation}},
  \href{https://doi.org/10.1007/JHEP05(2015)102}{\emph{JHEP} {\bfseries 05}
  (2015) 102}, [\href{https://arxiv.org/abs/1412.2126}{{\ttfamily 1412.2126}}].

\bibitem{Althoff:1983ew}
{\scshape TASSO} collaboration, M.~Althoff et~al., \emph{{Jet Production and
  Fragmentation in e+ e- Annihilation at 12-GeV to 43-GeV}},
  \href{https://doi.org/10.1007/BF01547419}{\emph{Z. Phys.} {\bfseries C22}
  (1984) 307--340}.

\bibitem{Abrams:1989ez}
{\scshape MARK-II} collaboration, G.~S. Abrams et~al., \emph{{First
  Measurements of Hadronic Decays of the $Z$ Boson}},
  \href{https://doi.org/10.1103/PhysRevLett.63.1558}{\emph{Phys. Rev. Lett.}
  {\bfseries 63} (1989) 1558}.

\bibitem{Li:1989sn}
{\scshape AMY} collaboration, Y.~K. Li et~al., \emph{{Multi - hadron event
  properties in $e^+e^-$ annihilation at $\sqrt{s} = 52$ GeV to 57-GeV}},
  \href{https://doi.org/10.1103/PhysRevD.41.2675}{\emph{Phys. Rev.} {\bfseries
  D41} (1990) 2675}.

\bibitem{Buskulic:1995aw}
{\scshape ALEPH} collaboration, D.~Buskulic et~al., \emph{{Measurement of
  alpha-s from scaling violations in fragmentation functions in e+ e-
  annihilation}}, \href{https://doi.org/10.1016/0370-2693(95)01380-6,
  10.1016/0370-2693(95)00917-A}{\emph{Phys. Lett.} {\bfseries B357} (1995)
  487--499}.

\bibitem{Adriani:1992gs}
{\scshape L3} collaboration, O.~Adrian et~al., \emph{{Determination of alpha-s
  from hadronic event shapes measured on the Z0 resonance}},
  \href{https://doi.org/10.1016/0370-2693(92)90463-E}{\emph{Phys. Lett.}
  {\bfseries B284} (1992) 471--481}.

\bibitem{Braunschweig:1990yd}
{\scshape TASSO} collaboration, W.~Braunschweig et~al., \emph{{Global Jet
  Properties at 14-{GeV} to 44-{GeV} Center-of-mass Energy in $e^+ e^-$
  Annihilation}}, \href{https://doi.org/10.1007/BF01552339}{\emph{Z. Phys.}
  {\bfseries C47} (1990) 187--198}.

\bibitem{Abe:1994mf}
{\scshape SLD} collaboration, K.~Abe et~al., \emph{{Measurement of alpha-s
  (M(Z)**2) from hadronic event observables at the Z0 resonance}},
  \href{https://doi.org/10.1103/PhysRevD.51.962}{\emph{Phys. Rev.} {\bfseries
  D51} (1995) 962--984},
  [\href{https://arxiv.org/abs/hep-ex/9501003}{{\ttfamily hep-ex/9501003}}].

\bibitem{Heister:2003aj}
{\scshape ALEPH} collaboration, A.~Heister et~al., \emph{{Studies of QCD at e+
  e- centre-of-mass energies between 91-GeV and 209-GeV}},
  \href{https://doi.org/10.1140/epjc/s2004-01891-4}{\emph{Eur. Phys. J.}
  {\bfseries C35} (2004) 457--486}.

\bibitem{Abdallah:2003xz}
{\scshape DELPHI} collaboration, J.~Abdallah et~al., \emph{{A Study of the
  energy evolution of event shape distributions and their means with the DELPHI
  detector at LEP}},
  \href{https://doi.org/10.1140/epjc/s2003-01198-0}{\emph{Eur. Phys. J.}
  {\bfseries C29} (2003) 285--312},
  [\href{https://arxiv.org/abs/hep-ex/0307048}{{\ttfamily hep-ex/0307048}}].

\bibitem{Achard:2004sv}
{\scshape L3} collaboration, P.~Achard et~al., \emph{{Studies of hadronic event
  structure in $e^{+} e^{-}$ annihilation from 30-GeV to 209-GeV with the L3
  detector}}, \href{https://doi.org/10.1016/j.physrep.2004.07.002}{\emph{Phys.
  Rept.} {\bfseries 399} (2004) 71--174},
  [\href{https://arxiv.org/abs/hep-ex/0406049}{{\ttfamily hep-ex/0406049}}].

\bibitem{Abbiendi:2004qz}
{\scshape OPAL} collaboration, G.~Abbiendi et~al., \emph{{Measurement of event
  shape distributions and moments in e+ e- ---> hadrons at 91-GeV - 209-GeV and
  a determination of alpha(s)}},
  \href{https://doi.org/10.1140/epjc/s2005-02120-6}{\emph{Eur. Phys. J.}
  {\bfseries C40} (2005) 287--316},
  [\href{https://arxiv.org/abs/hep-ex/0503051}{{\ttfamily hep-ex/0503051}}].

\bibitem{Abdesselam:2010pt}
A.~Abdesselam et~al., \emph{{Boosted Objects: A Probe of Beyond the Standard
  Model Physics}},
  \href{https://doi.org/10.1140/epjc/s10052-011-1661-y}{\emph{Eur. Phys. J. C}
  {\bfseries 71} (2011) 1661},
  [\href{https://arxiv.org/abs/1012.5412}{{\ttfamily 1012.5412}}].

\bibitem{2020}
P.~T. Komiske, E.~M. Metodiev and J.~Thaler, \emph{The hidden geometry of
  particle collisions},
  \href{https://doi.org/10.1007/jhep07(2020)006}{\emph{Journal of High Energy
  Physics} {\bfseries 2020} (Jul, 2020) }.

\bibitem{Komiske_2019}
P.~T. Komiske, E.~M. Metodiev and J.~Thaler, \emph{Metric space of collider
  events}, \href{https://doi.org/10.1103/physrevlett.123.041801}{\emph{Physical
  Review Letters} {\bfseries 123} (jul, 2019) }.

\bibitem{192468}
S.~Peleg, M.~Werman and H.~Rom, \emph{A unified approach to the change of
  resolution: space and gray-level},
  \href{https://doi.org/10.1109/34.192468}{\emph{IEEE Transactions on Pattern
  Analysis and Machine Intelligence} {\bfseries 11} (1989) 739--742}.

\bibitem{10.5555/938978.939133}
Y.~Rubner, C.~Tomasi and L.~J. Guibas, \emph{A metric for distributions with
  applications to image databases},  in \emph{Proceedings of the Sixth
  International Conference on Computer Vision}, ICCV '98, (USA), p.~59, IEEE
  Computer Society, 1998.

\bibitem{Rubner2004TheEM}
Y.~Rubner, C.~Tomasi and L.~J. Guibas, \emph{The earth mover's distance as a
  metric for image retrieval},
  \href{https://doi.org/0.1023/A:1026543900054}{\emph{International Journal of
  Computer Vision} {\bfseries 40} (2004) 99--121}.

\bibitem{Pele2008ALT}
O.~Pele and M.~Werman, \emph{A linear time histogram metric for improved sift
  matching},  in \emph{ECCV}, 2008,
  \href{https://doi.org/10.1007/978-3-540-88690-7\_37}{DOI}.

\bibitem{tangentEMD}
O.~Pele and B.~Taskar, \emph{The tangent earth mover’s distance}, .

\bibitem{wasserstein1969markov}
L.~N. Wasserstein, \emph{Markov processes over denumerable products of spaces
  describing large systems of automata}, {\emph{Problems of Information
  Transmission} {\bfseries 5} (1969) 47--52}.

\bibitem{dobrushin1970prescribing}
R.~L. Dobrushin, \emph{Prescribing a system of random variables by conditional
  distributions}, {\emph{Theory of Probability \& Its Applications} {\bfseries
  15} (1970) 458--486}.

\bibitem{Komiske:2019jim}
P.~T. Komiske, R.~Mastandrea, E.~M. Metodiev, P.~Naik and J.~Thaler,
  \emph{{Exploring the Space of Jets with CMS Open Data}},
  \href{https://doi.org/10.1103/PhysRevD.101.034009}{\emph{Phys. Rev. D}
  {\bfseries 101} (2020) 034009},
  [\href{https://arxiv.org/abs/1908.08542}{{\ttfamily 1908.08542}}].

\bibitem{Collins:2021pld}
J.~H. Collins, \emph{{An Exploration of Learnt Representations of W Jets}},  9,
  2021, \href{https://arxiv.org/abs/2109.10919}{{\ttfamily 2109.10919}}.

\bibitem{Park:2022zov}
S.~E. Park, P.~Harris and B.~Ostdiek, \emph{{Neural Embedding: Learning the
  Embedding of the Manifold of Physics Data}},
  \href{https://arxiv.org/abs/2208.05484}{{\ttfamily 2208.05484}}.

\bibitem{CrispimRomao:2020ejk}
M.~Crispim Rom\~ao, N.~F. Castro, J.~G. Milhano, R.~Pedro and T.~Vale,
  \emph{{Use of a generalized energy Mover's distance in the search for rare
  phenomena at colliders}},
  \href{https://doi.org/10.1140/epjc/s10052-021-08891-6}{\emph{Eur. Phys. J. C}
  {\bfseries 81} (2021) 192},
  [\href{https://arxiv.org/abs/2004.09360}{{\ttfamily 2004.09360}}].

\bibitem{Cai:2020vzx}
T.~Cai, J.~Cheng, N.~Craig and K.~Craig, \emph{{Linearized optimal transport
  for collider events}},
  \href{https://doi.org/10.1103/PhysRevD.102.116019}{\emph{Phys. Rev. D}
  {\bfseries 102} (2020) 116019},
  [\href{https://arxiv.org/abs/2008.08604}{{\ttfamily 2008.08604}}].

\bibitem{Cai:2021hnn}
T.~Cai, J.~Cheng, K.~Craig and N.~Craig, \emph{{Which metric on the space of
  collider events?}},
  \href{https://doi.org/10.1103/PhysRevD.105.076003}{\emph{Phys. Rev. D}
  {\bfseries 105} (2022) 076003},
  [\href{https://arxiv.org/abs/2111.03670}{{\ttfamily 2111.03670}}].

\bibitem{Cesarotti:2020hwb}
C.~Cesarotti and J.~Thaler, \emph{{A Robust Measure of Event Isotropy at
  Colliders}}, \href{https://doi.org/10.1007/JHEP08(2020)084}{\emph{JHEP}
  {\bfseries 08} (2020) 084},
  [\href{https://arxiv.org/abs/2004.06125}{{\ttfamily 2004.06125}}].

\bibitem{Cesarotti_2021}
C.~Cesarotti, M.~Reece and M.~J. Strassler, \emph{The efficacy of event
  isotropy as an event shape observable},
  \href{https://doi.org/10.1007/jhep07(2021)215}{\emph{Journal of High Energy
  Physics} {\bfseries 2021} (jul, 2021) }.

\bibitem{ATLAS:2022jwu}
{\scshape ATLAS} collaboration, G.~Aad et~al., \emph{{Measurements of multijet
  event isotropies using optimal transport with the ATLAS detector}},
  \href{https://arxiv.org/abs/2305.16930}{{\ttfamily 2305.16930}}.

\bibitem{sinkhorn_1966}
R.~Sinkhorn, \emph{A relationship between arbitrary positive matrices and
  stochastic matrices},
  \href{https://doi.org/10.4153/CJM-1966-033-9}{\emph{Canadian Journal of
  Mathematics} {\bfseries 18} (1966) 303–306}.

\bibitem{cuturi2013sinkhorn}
M.~Cuturi, \emph{Sinkhorn distances: Lightspeed computation of optimal
  transport}, {\emph{Advances in neural information processing systems}
  {\bfseries 26} (2013) }.

\bibitem{CLASON2021124432}
C.~Clason, D.~A. Lorenz, H.~Mahler and B.~Wirth, \emph{Entropic regularization
  of continuous optimal transport problems},
  \href{https://doi.org/https://doi.org/10.1016/j.jmaa.2020.124432}{\emph{Journal
  of Mathematical Analysis and Applications} {\bfseries 494} (2021) 124432}.

\bibitem{feydy2019interpolating}
J.~Feydy, T.~S{\'e}journ{\'e}, F.-X. Vialard, S.-i. Amari, A.~Trouve and
  G.~Peyr{\'e}, \emph{Interpolating between optimal transport and mmd using
  sinkhorn divergences},  in \emph{The 22nd International Conference on
  Artificial Intelligence and Statistics}, pp.~2681--2690, 2019.

\bibitem{Kitouni:2022qyr}
O.~Kitouni, N.~Nolte and M.~Williams, \emph{{Finding NEEMo: Geometric Fitting
  using Neural Estimation of the Energy Mover's Distance}},
  \href{https://arxiv.org/abs/2209.15624}{{\ttfamily 2209.15624}}.

\bibitem{Butter:2017cot}
A.~Butter, G.~Kasieczka, T.~Plehn and M.~Russell, \emph{{Deep-learned Top
  Tagging with a Lorentz Layer}},
  \href{https://doi.org/10.21468/SciPostPhys.5.3.028}{\emph{SciPost Phys.}
  {\bfseries 5} (2018) 028},
  [\href{https://arxiv.org/abs/1707.08966}{{\ttfamily 1707.08966}}].

\bibitem{Kasieczka:2019dbj}
A.~Butter et~al., \emph{{The Machine Learning landscape of top taggers}},
  \href{https://doi.org/10.21468/SciPostPhys.7.1.014}{\emph{SciPost Phys.}
  {\bfseries 7} (2019) 014},
  [\href{https://arxiv.org/abs/1902.09914}{{\ttfamily 1902.09914}}].

\bibitem{Krohn:2009zg}
D.~Krohn, J.~Thaler and L.-T. Wang, \emph{{Jets with Variable R}},
  \href{https://doi.org/10.1088/1126-6708/2009/06/059}{\emph{JHEP} {\bfseries
  06} (2009) 059}, [\href{https://arxiv.org/abs/0903.0392}{{\ttfamily
  0903.0392}}].

\bibitem{Mackey:2015hwa}
L.~Mackey, B.~Nachman, A.~Schwartzman and C.~Stansbury, \emph{{Fuzzy Jets}},
  \href{https://doi.org/10.1007/JHEP06(2016)010}{\emph{JHEP} {\bfseries 06}
  (2016) 010}, [\href{https://arxiv.org/abs/1509.02216}{{\ttfamily
  1509.02216}}].

\bibitem{Mukhopadhyaya:2023rsb}
B.~Mukhopadhyaya, T.~Samui and R.~K. Singh, \emph{{Dynamic Radius Jet
  Clustering Algorithm}},  \href{https://arxiv.org/abs/2301.13074}{{\ttfamily
  2301.13074}}.

\bibitem{Larkoski:2023nye}
A.~J. Larkoski, D.~Rathjens, J.~Veatch and J.~W. Walker, \emph{{Jet SIFT-ing: a
  new scale-invariant jet clustering algorithm for the substructure era}},
  \href{https://arxiv.org/abs/2302.08609}{{\ttfamily 2302.08609}}.

\bibitem{Stewart:2015waa}
I.~W. Stewart, F.~J. Tackmann, J.~Thaler, C.~K. Vermilion and T.~F. Wilkason,
  \emph{{XCone: N-jettiness as an Exclusive Cone Jet Algorithm}},
  \href{https://doi.org/10.1007/JHEP11(2015)072}{\emph{JHEP} {\bfseries 11}
  (2015) 072}, [\href{https://arxiv.org/abs/1508.01516}{{\ttfamily
  1508.01516}}].

\bibitem{Georgi:1977sf}
H.~Georgi and M.~Machacek, \emph{{A Simple QCD Prediction of Jet Structure in
  e+ e- Annihilation}},
  \href{https://doi.org/10.1103/PhysRevLett.39.1237}{\emph{Phys. Rev. Lett.}
  {\bfseries 39} (1977) 1237}.

\bibitem{Larkoski:2014uqa}
A.~J. Larkoski, D.~Neill and J.~Thaler, \emph{{Jet Shapes with the Broadening
  Axis}}, \href{https://doi.org/10.1007/JHEP04(2014)017}{\emph{JHEP} {\bfseries
  04} (2014) 017}, [\href{https://arxiv.org/abs/1401.2158}{{\ttfamily
  1401.2158}}].

\bibitem{Stewart:2010tn}
I.~W. Stewart, F.~J. Tackmann and W.~J. Waalewijn, \emph{{N-Jettiness: An
  Inclusive Event Shape to Veto Jets}},
  \href{https://doi.org/10.1103/PhysRevLett.105.092002}{\emph{Phys. Rev. Lett.}
  {\bfseries 105} (2010) 092002},
  [\href{https://arxiv.org/abs/1004.2489}{{\ttfamily 1004.2489}}].

\bibitem{Banfi_2004}
A.~Banfi, G.~P. Salam and G.~Zanderighi, \emph{Resummed event shapes at
  hadron-hadron colliders},
  \href{https://doi.org/10.1088/1126-6708/2004/08/062}{\emph{Journal of High
  Energy Physics} {\bfseries 2004} (sep, 2004) 062--062}.

\bibitem{Banfi:2010xy}
A.~Banfi, G.~P. Salam and G.~Zanderighi, \emph{{Phenomenology of event shapes
  at hadron colliders}},
  \href{https://doi.org/10.1007/JHEP06(2010)038}{\emph{JHEP} {\bfseries 06}
  (2010) 038}, [\href{https://arxiv.org/abs/1001.4082}{{\ttfamily 1001.4082}}].

\bibitem{Catani:1993hr}
S.~Catani, Y.~L. Dokshitzer, M.~H. Seymour and B.~R. Webber,
  \emph{{Longitudinally invariant $K_t$ clustering algorithms for hadron hadron
  collisions}}, \href{https://doi.org/10.1016/0550-3213(93)90166-M}{\emph{Nucl.
  Phys. B} {\bfseries 406} (1993) 187--224}.

\bibitem{Ellis:1993tq}
S.~D. Ellis and D.~E. Soper, \emph{{Successive combination jet algorithm for
  hadron collisions}},
  \href{https://doi.org/10.1103/PhysRevD.48.3160}{\emph{Phys. Rev. D}
  {\bfseries 48} (1993) 3160--3166},
  [\href{https://arxiv.org/abs/hep-ph/9305266}{{\ttfamily hep-ph/9305266}}].

\bibitem{Dokshitzer:1997in}
Y.~L. Dokshitzer, G.~D. Leder, S.~Moretti and B.~R. Webber, \emph{{Better jet
  clustering algorithms}},
  \href{https://doi.org/10.1088/1126-6708/1997/08/001}{\emph{JHEP} {\bfseries
  08} (1997) 001}, [\href{https://arxiv.org/abs/hep-ph/9707323}{{\ttfamily
  hep-ph/9707323}}].

\bibitem{Wobisch:1998wt}
M.~Wobisch and T.~Wengler, \emph{{Hadronization corrections to jet
  cross-sections in deep inelastic scattering}},  in \emph{{Workshop on Monte
  Carlo Generators for HERA Physics (Plenary Starting Meeting)}}, pp.~270--279,
  4, 1998, \href{https://arxiv.org/abs/hep-ph/9907280}{{\ttfamily
  hep-ph/9907280}}.

\bibitem{Cacciari:2008gp}
M.~Cacciari, G.~P. Salam and G.~Soyez, \emph{{The anti-$k_t$ jet clustering
  algorithm}}, \href{https://doi.org/10.1088/1126-6708/2008/04/063}{\emph{JHEP}
  {\bfseries 04} (2008) 063},
  [\href{https://arxiv.org/abs/0802.1189}{{\ttfamily 0802.1189}}].

\bibitem{Ellis:1992qq}
S.~D. Ellis, Z.~Kunszt and D.~E. Soper, \emph{{Jets at hadron colliders at
  order $\alpha-s^{3:}$ A Look inside}},
  \href{https://doi.org/10.1103/PhysRevLett.69.3615}{\emph{Phys. Rev. Lett.}
  {\bfseries 69} (1992) 3615--3618},
  [\href{https://arxiv.org/abs/hep-ph/9208249}{{\ttfamily hep-ph/9208249}}].

\bibitem{PhysRevLett.70.713}
D.~Abe, F.~Amidei, C.~Anway-Weiss, G.~Apollinari, M.~Atac, P.~Auchincloss,
  A.~R. Baden et~al., \emph{Measurement of jet shapes in pp collisions at
  sqrt{s} =1.8 tev},
  \href{https://doi.org/10.1103/PhysRevLett.70.713}{\emph{Phys. Rev. Lett.}
  {\bfseries 70} (Feb, 1993) 713--717}.

\bibitem{Ellis_2010}
S.~D. Ellis, C.~K. Vermilion, J.~R. Walsh, A.~Hornig and C.~Lee, \emph{Jet
  shapes and jet algorithms in {SCET}},
  \href{https://doi.org/10.1007/jhep11(2010)101}{\emph{Journal of High Energy
  Physics} {\bfseries 2010} (nov, 2010) }.

\bibitem{Thaler:2010tr}
J.~Thaler and K.~Van~Tilburg, \emph{{Identifying Boosted Objects with
  N-subjettiness}}, \href{https://doi.org/10.1007/JHEP03(2011)015}{\emph{JHEP}
  {\bfseries 03} (2011) 015},
  [\href{https://arxiv.org/abs/1011.2268}{{\ttfamily 1011.2268}}].

\bibitem{Dasgupta:2001sh}
M.~Dasgupta and G.~P. Salam, \emph{{Resummation of nonglobal QCD observables}},
  \href{https://doi.org/10.1016/S0370-2693(01)00725-0}{\emph{Phys. Lett. B}
  {\bfseries 512} (2001) 323--330},
  [\href{https://arxiv.org/abs/hep-ph/0104277}{{\ttfamily hep-ph/0104277}}].

\bibitem{Banfi:2010pa}
A.~Banfi, M.~Dasgupta, K.~Khelifa-Kerfa and S.~Marzani, \emph{{Non-global
  logarithms and jet algorithms in high-pT jet shapes}},
  \href{https://doi.org/10.1007/JHEP08(2010)064}{\emph{JHEP} {\bfseries 08}
  (2010) 064}, [\href{https://arxiv.org/abs/1004.3483}{{\ttfamily 1004.3483}}].

\bibitem{Tkachov:1995kk}
F.~V. Tkachov, \emph{{Measuring multi - jet structure of hadronic energy flow
  or What is a jet?}},
  \href{https://doi.org/10.1142/S0217751X97002899}{\emph{Int. J. Mod. Phys. A}
  {\bfseries 12} (1997) 5411--5529},
  [\href{https://arxiv.org/abs/hep-ph/9601308}{{\ttfamily hep-ph/9601308}}].

\bibitem{Sveshnikov:1995vi}
N.~A. Sveshnikov and F.~V. Tkachov, \emph{{Jets and quantum field theory}},
  \href{https://doi.org/10.1016/0370-2693(96)00558-8}{\emph{Phys. Lett. B}
  {\bfseries 382} (1996) 403--408},
  [\href{https://arxiv.org/abs/hep-ph/9512370}{{\ttfamily hep-ph/9512370}}].

\bibitem{Korchemsky:1997sy}
G.~P. Korchemsky, G.~Oderda and G.~F. Sterman, \emph{{Power corrections and
  nonlocal operators}}, \href{https://doi.org/10.1063/1.53732}{\emph{AIP Conf.
  Proc.} {\bfseries 407} (1997) 988},
  [\href{https://arxiv.org/abs/hep-ph/9708346}{{\ttfamily hep-ph/9708346}}].

\bibitem{Basham:1978zq}
C.~L. Basham, L.~S. Brown, S.~D. Ellis and S.~T. Love, \emph{Energy
  correlations in electron-positron annihilation in quantum chromodynamics:
  Asymptotically free perturbation theory},
  \href{https://doi.org/10.1103/PhysRevD.19.2018}{\emph{Phys. Rev. D}
  {\bfseries 19} (Apr, 1979) 2018--2045}.

\bibitem{Cherzor:1997ak}
P.~S. Cherzor and N.~A. Sveshnikov, \emph{{Jet observables and energy momentum
  tensor}},  in \emph{{12th International Workshop on High-Energy Physics and
  Quantum Field Theory (QFTHEP 97)}}, pp.~402--407, 9, 1997,
  \href{https://arxiv.org/abs/hep-ph/9710349}{{\ttfamily hep-ph/9710349}}.

\bibitem{Tkachov:1999py}
F.~V. Tkachov, \emph{{A Theory of jet definition}},
  \href{https://doi.org/10.1142/S0217751X02009977}{\emph{Int. J. Mod. Phys. A}
  {\bfseries 17} (2002) 2783--2884},
  [\href{https://arxiv.org/abs/hep-ph/9901444}{{\ttfamily hep-ph/9901444}}].

\bibitem{Korchemsky:1999kt}
G.~P. Korchemsky and G.~F. Sterman, \emph{{Power corrections to event shapes
  and factorization}},
  \href{https://doi.org/10.1016/S0550-3213(99)00308-9}{\emph{Nucl. Phys. B}
  {\bfseries 555} (1999) 335--351},
  [\href{https://arxiv.org/abs/hep-ph/9902341}{{\ttfamily hep-ph/9902341}}].

\bibitem{Belitsky:2001ij}
A.~V. Belitsky, G.~P. Korchemsky and G.~F. Sterman, \emph{{Energy flow in QCD
  and event shape functions}},
  \href{https://doi.org/10.1016/S0370-2693(01)00899-1}{\emph{Phys. Lett. B}
  {\bfseries 515} (2001) 297--307},
  [\href{https://arxiv.org/abs/hep-ph/0106308}{{\ttfamily hep-ph/0106308}}].

\bibitem{Berger:2002jt}
C.~F. Berger et~al., \emph{{Snowmass 2001: Jet energy flow project}},
  {\emph{eConf} {\bfseries C010630} (2001) P512},
  [\href{https://arxiv.org/abs/hep-ph/0202207}{{\ttfamily hep-ph/0202207}}].

\bibitem{Bauer:2008dt}
C.~W. Bauer, S.~P. Fleming, C.~Lee and G.~F. Sterman, \emph{{Factorization of
  e+e- Event Shape Distributions with Hadronic Final States in Soft Collinear
  Effective Theory}},
  \href{https://doi.org/10.1103/PhysRevD.78.034027}{\emph{Phys. Rev. D}
  {\bfseries 78} (2008) 034027},
  [\href{https://arxiv.org/abs/0801.4569}{{\ttfamily 0801.4569}}].

\bibitem{Hofman:2008ar}
D.~M. Hofman and J.~Maldacena, \emph{{Conformal collider physics: Energy and
  charge correlations}},
  \href{https://doi.org/10.1088/1126-6708/2008/05/012}{\emph{JHEP} {\bfseries
  05} (2008) 012}, [\href{https://arxiv.org/abs/0803.1467}{{\ttfamily
  0803.1467}}].

\bibitem{Mateu:2012nk}
V.~Mateu, I.~W. Stewart and J.~Thaler, \emph{{Power Corrections to Event Shapes
  with Mass-Dependent Operators}},
  \href{https://doi.org/10.1103/PhysRevD.87.014025}{\emph{Phys. Rev. D}
  {\bfseries 87} (2013) 014025},
  [\href{https://arxiv.org/abs/1209.3781}{{\ttfamily 1209.3781}}].

\bibitem{Belitsky:2013xxa}
A.~V. Belitsky, S.~Hohenegger, G.~P. Korchemsky, E.~Sokatchev and A.~Zhiboedov,
  \emph{{From correlation functions to event shapes}},
  \href{https://doi.org/10.1016/j.nuclphysb.2014.04.020}{\emph{Nucl. Phys. B}
  {\bfseries 884} (2014) 305--343},
  [\href{https://arxiv.org/abs/1309.0769}{{\ttfamily 1309.0769}}].

\bibitem{ramdas2017wasserstein}
A.~Ramdas, N.~Garc{\'\i}a~Trillos and M.~Cuturi, \emph{On wasserstein
  two-sample testing and related families of nonparametric tests},
  {\emph{Entropy} {\bfseries 19} (2017) 47}.

\bibitem{10.5555/1622943.1622971}
H.~G. Barrow, J.~M. Tenenbaum, R.~C. Bolles and H.~C. Wolf, \emph{Parametric
  correspondence and chamfer matching: Two new techniques for image matching},
  in \emph{Proceedings of the 5th International Joint Conference on Artificial
  Intelligence - Volume 2}, IJCAI'77, (San Francisco, CA, USA), p.~659–663,
  Morgan Kaufmann Publishers Inc., 1977.

\bibitem{Soyez:2018opl}
G.~Soyez, \emph{{Pileup mitigation at the LHC: A theorist\textquoteright{}s
  view}}, \href{https://doi.org/10.1016/j.physrep.2019.01.007}{\emph{Phys.
  Rept.} {\bfseries 803} (2019) 1--158},
  [\href{https://arxiv.org/abs/1801.09721}{{\ttfamily 1801.09721}}].

\bibitem{PhysRevD.65.092002}
{\scshape CDF Collaboration} collaboration, T.~Affolder, H.~Akimoto,
  A.~Akopian, M.~G. Albrow, P.~Amaral, D.~Amidei et~al., \emph{Charged jet
  evolution and the underlying event in proton-antiproton collisions at 1.8
  tev}, \href{https://doi.org/10.1103/PhysRevD.65.092002}{\emph{Phys. Rev. D}
  {\bfseries 65} (Apr, 2002) 092002}.

\bibitem{Agocs:2010ft}
A.~G. Agocs, G.~G. Barnafoldi and P.~Levai, \emph{{Jets and Underlying Events
  at LHC Energies}},
  \href{https://doi.org/10.1088/1742-6596/270/1/012017}{\emph{J. Phys. Conf.
  Ser.} {\bfseries 270} (2011) 012017},
  [\href{https://arxiv.org/abs/1011.5363}{{\ttfamily 1011.5363}}].

\bibitem{https://doi.org/10.48550/arxiv.1506.02557}
D.~P. Kingma, T.~Salimans and M.~Welling, \emph{Variational dropout and the
  local reparameterization trick},  2015.
\newblock 10.48550/ARXIV.1506.02557.

\bibitem{Dokshitzer:1991fc}
Y.~L. Dokshitzer, V.~A. Khoze and S.~I. Troian, \emph{{Particle spectra in
  light and heavy quark jets}},
  \href{https://doi.org/10.1088/0954-3899/17/10/003}{\emph{J. Phys. G}
  {\bfseries 17} (1991) 1481--1492}.

\bibitem{Dokshitzer:1991fd}
Y.~L. Dokshitzer, V.~A. Khoze and S.~I. Troian, \emph{{On specific QCD
  properties of heavy quark fragmentation ('dead cone')}},
  \href{https://doi.org/10.1088/0954-3899/17/10/023}{\emph{J. Phys. G}
  {\bfseries 17} (1991) 1602--1604}.

\bibitem{PhysRevLett.69.3025}
B.~A. Schumm, Y.~L. Dokshitzer, V.~A. Khoze and D.~S. Koetke, \emph{Average
  charged multiplicity of events containing heavy quarks in e+e- annihilation},
  \href{https://doi.org/10.1103/PhysRevLett.69.3025}{\emph{Phys. Rev. Lett.}
  {\bfseries 69} (Nov, 1992) 3025--3028}.

\bibitem{Cacciari_2008}
M.~Cacciari and G.~P. Salam, \emph{Pileup subtraction using jet areas},
  \href{https://doi.org/10.1016/j.physletb.2007.09.077}{\emph{Physics Letters
  B} {\bfseries 659} (jan, 2008) 119--126}.

\bibitem{Cacciari_2008_2}
M.~Cacciari, G.~P. Salam and G.~Soyez, \emph{The catchment area of jets},
  \href{https://doi.org/10.1088/1126-6708/2008/04/005}{\emph{Journal of High
  Energy Physics} {\bfseries 2008} (apr, 2008) 005--005}.

\bibitem{Soyez:2012hv}
G.~Soyez, G.~P. Salam, J.~Kim, S.~Dutta and M.~Cacciari, \emph{{Pileup
  subtraction for jet shapes}},
  \href{https://doi.org/10.1103/PhysRevLett.110.162001}{\emph{Phys. Rev. Lett.}
  {\bfseries 110} (2013) 162001},
  [\href{https://arxiv.org/abs/1211.2811}{{\ttfamily 1211.2811}}].

\bibitem{Larkoski:2014wba}
A.~J. Larkoski, S.~Marzani, G.~Soyez and J.~Thaler, \emph{{Soft Drop}},
  \href{https://doi.org/10.1007/JHEP05(2014)146}{\emph{JHEP} {\bfseries 05}
  (2014) 146}, [\href{https://arxiv.org/abs/1402.2657}{{\ttfamily 1402.2657}}].

\bibitem{Dasgupta_2013}
M.~Dasgupta, A.~Fregoso, S.~Marzani and G.~P. Salam, \emph{Towards an
  understanding of jet substructure},
  \href{https://doi.org/10.1007/jhep09(2013)029}{\emph{Journal of High Energy
  Physics} {\bfseries 2013} (sep, 2013) }.

\bibitem{alt_17}
J.~Altschuler, J.~Niles-Weed and P.~Rigollet, \emph{Near-linear time
  approximation algorithms for optimal transport via sinkhorn iteration},  in
  \emph{Advances in Neural Information Processing Systems} (I.~Guyon, U.~V.
  Luxburg, S.~Bengio, H.~Wallach, R.~Fergus, S.~Vishwanathan et~al., eds.),
  vol.~30, Curran Associates, Inc., 2017,
  \href{https://proceedings.neurips.cc/paper/2017/file/491442df5f88c6aa018e86dac21d3606-Paper.pdf}{https://proceedings.neurips.cc/paper/2017/file/491442df5f88c6aa018e86dac21d3606-Paper.pdf}.

\bibitem{sink_n2ln(n)}
P.~Dvurechensky, A.~Gasnikov and A.~Kroshnin, \emph{Computational optimal
  transport: Complexity by accelerated gradient descent is better than by
  sinkhorn’s algorithm},  in \emph{Proceedings of the 35th International
  Conference on Machine Learning} (J.~Dy and A.~Krause, eds.), vol.~80 of
  \emph{Proceedings of Machine Learning Research}, pp.~1367--1376, PMLR, 10--15
  Jul, 2018,
  \href{https://proceedings.mlr.press/v80/dvurechensky18a.html}{https://proceedings.mlr.press/v80/dvurechensky18a.html}.

\bibitem{Sink_ICML_19}
T.~Lin, N.~Ho and M.~I. Jordan, \emph{On the efficiency of the sinkhorn and
  greenkhorn algorithms and their acceleration for optimal transport},
  {\emph{International Conference on Machine Learning} (2019) }.

\bibitem{sink_nlogn}
J.~Altschuler, F.~Bach, A.~Rudi and J.~Niles-Weed, \emph{Massively scalable
  sinkhorn distances via the nystr\"{o}m method},  in \emph{Advances in Neural
  Information Processing Systems} (H.~Wallach, H.~Larochelle, A.~Beygelzimer,
  F.~d\textquotesingle Alch\'{e}-Buc, E.~Fox and R.~Garnett, eds.), vol.~32,
  Curran Associates, Inc., 2019,
  \href{https://proceedings.neurips.cc/paper/2019/file/f55cadb97eaff2ba1980e001b0bd9842-Paper.pdf}{https://proceedings.neurips.cc/paper/2019/file/f55cadb97eaff2ba1980e001b0bd9842-Paper.pdf}.

\bibitem{sink_n_comp}
M.~Scetbon and M.~Cuturi, \emph{Linear time sinkhorn divergences using positive
  features},  in \emph{Advances in Neural Information Processing Systems}
  (H.~Larochelle, M.~Ranzato, R.~Hadsell, M.~Balcan and H.~Lin, eds.), vol.~33,
  pp.~13468--13480, Curran Associates, Inc., 2020,
  \href{https://proceedings.neurips.cc/paper/2020/file/9bde76f262285bb1eaeb7b40c758b53e-Paper.pdf}{https://proceedings.neurips.cc/paper/2020/file/9bde76f262285bb1eaeb7b40c758b53e-Paper.pdf}.

\bibitem{wass_n3_input_dist}
K.~Atasu and T.~Mittelholzer, \emph{Linear-complexity data-parallel earth
  mover’s distance approximations},  in \emph{Proceedings of the 36th
  International Conference on Machine Learning} (K.~Chaudhuri and
  R.~Salakhutdinov, eds.), vol.~97 of \emph{Proceedings of Machine Learning
  Research}, pp.~364--373, PMLR, 09--15 Jun, 2019,
  \href{https://proceedings.mlr.press/v97/atasu19a.html}{https://proceedings.mlr.press/v97/atasu19a.html}.

\bibitem{NEURIPS2019_9015}
A.~Paszke, S.~Gross, F.~Massa, A.~Lerer, J.~Bradbury, G.~Chanan et~al.,
  \emph{Pytorch: An imperative style, high-performance deep learning library},
  in \emph{Advances in Neural Information Processing Systems 32} (H.~Wallach,
  H.~Larochelle, A.~Beygelzimer, F.~d\textquotesingle Alch\'{e}-Buc, E.~Fox and
  R.~Garnett, eds.), pp.~8024--8035.
\newblock Curran Associates, Inc., 2019.

\bibitem{gartner_06}
J.~Matousek and B.~Gartner, \emph{Understanding and Using Linear Programming}.
\newblock Springer Berlin, Heidelberg, 2006.

\bibitem{Vilani_03}
C.~Vilani, \emph{Topics in Optimal Transportation}.
\newblock American Mathematical Society, 2003.

\bibitem{feydy_20}
J.~Feydy, \emph{Geometric data analysis, beyond convolutions}.
\newblock ENS Paris-Saclay, 2020.

\bibitem{sinkhorn1967diagonal}
R.~Sinkhorn, \emph{Diagonal equivalence to matrices with prescribed row and
  column sums}, {\emph{The American Mathematical Monthly} {\bfseries 74} (1967)
  402--405}.

\bibitem{sinkhorn1967concerning}
R.~Sinkhorn and P.~Knopp, \emph{Concerning nonnegative matrices and doubly
  stochastic matrices}, {\emph{Pacific Journal of Mathematics} {\bfseries 21}
  (1967) 343--348}.

\bibitem{KOSOWSKY1994477}
J.~Kosowsky and A.~Yuille, \emph{The invisible hand algorithm: Solving the
  assignment problem with statistical physics},
  \href{https://doi.org/https://doi.org/10.1016/0893-6080(94)90081-7}{\emph{Neural
  Networks} {\bfseries 7} (1994) 477--490}.

\bibitem{Bertsekas}
D.~Bertsekas, \emph{The auction algorithm: A distributed relaxation method for
  the assignment problem},
  \href{https://doi.org/10.1007/BF02186476}{\emph{Annals of Operations
  Research} {\bfseries 14} (12, 1988) 105--123}.

\bibitem{https://doi.org/10.48550/arxiv.1412.6980}
D.~P. Kingma and J.~Ba, \emph{Adam: A method for stochastic optimization},
  2014.
\newblock 10.48550/ARXIV.1412.6980.

\bibitem{tankala2020k}
P.~Tankala, A.~Tasissa, J.~M. Murphy and D.~Ba, \emph{K-deep simplex: Deep
  manifold learning via local dictionaries}, {\emph{arXiv preprint
  arXiv:2012.02134} (2020) }.

\bibitem{mueller2022geometric}
M.~Mueller, S.~Aeron, J.~M. Murphy and A.~Tasissa, \emph{Geometric sparse
  coding in wasserstein space}, {\emph{arXiv preprint arXiv:2210.12135} (2022)
  }.

\bibitem{https://doi.org/10.48550/arxiv.1309.1541}
W.~Wang and M.~A. Carreira-Perpinan, \emph{Projection onto the probability
  simplex: An efficient algorithm with a simple proof, and an application},
  2013.
\newblock 10.48550/ARXIV.1309.1541.

\bibitem{Sjostrand:2014zea}
T.~Sj\"ostrand, S.~Ask, J.~R. Christiansen, R.~Corke, N.~Desai, P.~Ilten
  et~al., \emph{{An introduction to PYTHIA 8.2}},
  \href{https://doi.org/10.1016/j.cpc.2015.01.024}{\emph{Comput. Phys. Commun.}
  {\bfseries 191} (2015) 159--177},
  [\href{https://arxiv.org/abs/1410.3012}{{\ttfamily 1410.3012}}].

\bibitem{deFavereau:2013fsa}
{\scshape DELPHES 3} collaboration, J.~de~Favereau, C.~Delaere, P.~Demin,
  A.~Giammanco, V.~Lema\^\i{}tre, A.~Mertens et~al., \emph{{DELPHES 3, A
  modular framework for fast simulation of a generic collider experiment}},
  \href{https://doi.org/10.1007/JHEP02(2014)057}{\emph{JHEP} {\bfseries 02}
  (2014) 057}, [\href{https://arxiv.org/abs/1307.6346}{{\ttfamily 1307.6346}}].

\bibitem{Cacciari:2011ma}
M.~Cacciari, G.~P. Salam and G.~Soyez, \emph{{FastJet User Manual}},
  \href{https://doi.org/10.1140/epjc/s10052-012-1896-2}{\emph{Eur. Phys. J. C}
  {\bfseries 72} (2012) 1896},
  [\href{https://arxiv.org/abs/1111.6097}{{\ttfamily 1111.6097}}].

\bibitem{flamary2021pot}
R.~Flamary, N.~Courty, A.~Gramfort, M.~Z. Alaya, A.~Boisbunon, S.~Chambon
  et~al., \emph{Pot: Python optimal transport}, {\emph{Journal of Machine
  Learning Research} {\bfseries 22} (2021) 1--8}.

\bibitem{CMS:2016lmd}
{\scshape CMS} collaboration, V.~Khachatryan et~al., \emph{{Jet energy scale
  and resolution in the CMS experiment in pp collisions at 8 TeV}},
  \href{https://doi.org/10.1088/1748-0221/12/02/P02014}{\emph{JINST} {\bfseries
  12} (2017) P02014}, [\href{https://arxiv.org/abs/1607.03663}{{\ttfamily
  1607.03663}}].

\bibitem{ATLAS:2020cli}
{\scshape ATLAS} collaboration, G.~Aad et~al., \emph{{Jet energy scale and
  resolution measured in proton\textendash{}proton collisions at
  $\sqrt{s}=13$~TeV with the ATLAS detector}},
  \href{https://doi.org/10.1140/epjc/s10052-021-09402-3}{\emph{Eur. Phys. J. C}
  {\bfseries 81} (2021) 689},
  [\href{https://arxiv.org/abs/2007.02645}{{\ttfamily 2007.02645}}].

\bibitem{Wigner:1960kfi}
E.~Wigner, \emph{{The unreasonable effectiveness of mathematics in the natural
  sciences}}, \href{https://doi.org/10.1002/cpa.3160130102}{\emph{Commun. Pure
  Appl. Math.} {\bfseries 13} (2, 1960) 1--14}.

\bibitem{10.2307/2313748}
M.~Kac, \emph{Can one hear the shape of a drum?}, {\emph{The American
  Mathematical Monthly} {\bfseries 73} (1966) 1--23}.

\bibitem{kullback1951information}
S.~Kullback and R.~A. Leibler, \emph{On information and sufficiency},
  {\emph{The annals of mathematical statistics} {\bfseries 22} (1951) 79--86}.

\bibitem{Piccoli_2013}
B.~Piccoli and F.~Rossi, \emph{Generalized wasserstein distance and its
  application to transport equations with source},
  \href{https://doi.org/10.1007/s00205-013-0669-x}{\emph{Archive for Rational
  Mechanics and Analysis} {\bfseries 211} (sep, 2013) 335--358}.

\bibitem{Liero_2016}
M.~Liero, A.~Mielke and G.~Savar{\'{e} }, \emph{Optimal transport in
  competition with reaction: The hellinger--kantorovich distance and geodesic
  curves}, \href{https://doi.org/10.1137/15m1041420}{\emph{{SIAM} Journal on
  Mathematical Analysis} {\bfseries 48} (jan, 2016) 2869--2911}.

\end{thebibliography}\endgroup

\end{document}